\documentclass[sn-mathphys,Numbered,eqsecnum]{sn-jnl}

\usepackage{graphicx,epstopdf}
\usepackage[square,numbers]{natbib}
\usepackage{multirow}
\usepackage{longtable}
\usepackage{amsmath,amssymb,amsfonts}
\usepackage{mathtools}
\usepackage{mathrsfs}
\usepackage[title]{appendix}
\usepackage{xcolor}
\usepackage{bbm,bm}
\usepackage{manyfoot}
\usepackage{float}
\usepackage{subfigure}
\usepackage{bigints}
\def\rvec{{\bf r}}

\def\hvec{{\bf h}}
\def\kvec{{\bf k}}
\def\pvec{{\bf p}}

\def\qvec{{\bf q}}

\def\he#1{$^{#1}$He}

\def\Re{{\cal R}e}
\def\Im{{\cal I}m}
\def\I{{\rm i}}

\def\Tr{{\cal T}r}
\def\KF{k_{\rm F}}
\def\SF{S_{\rm F}}
\def\tF{t_{\rm F}}
\def\1{\mathbbm{1}}

\def\bra#1{\bigl\langle{ #1} \bigr|}
\def\ket#1{\bigl|{ #1} \bigr\rangle}
\def\Bra#1{\Bigl\langle{ #1} \Bigr|}

\def\ovlp#1#2{\bigl\langle{ #1}\big|{#2} \bigr\rangle}
\def\Ovlp#1#2{\Bigl\langle{ #1}\Big|{#2} \Bigr\rangle}
\def\pd{{\phantom{\dagger}}} \def\rvec {{\bf r}} \def\pvec {{\bf p}}
\def\etal{{\em et al.\/}\ }
\def\ie{{\em i.e.\/}\ }

\def\sij#1#2{{\bm\sigma}_{#1}\cdot{\bm\sigma}_{#2}}
\def\sqij#1#2{\left({\bm\sigma}_{#1}\cdot\hat\qvec\right)
  \left({\bm\sigma}_{#2}\cdot\hat\qvec\right)}

\def\LS{\widetildeto{{\bf L}\!\cdot\!{\bf S}}{{\bf L}\!\cdot\!{\bf S}}}
\def\LSp{\widetildeto{{\rm LS}}{{\rm LS}}'}
\def\LSsup{{\rm(LS)}}

\def\bsigma{{\bm\sigma}}
\def\balpha{{\bm\alpha}}

\def\EF{e_{\rm F}}
\def\KF{k_{\rm F}}

\def\SF{S_{\rm F}}
\def\S{{\bf S}}

\def\creat#1{a_{#1}^{\dagger}}  
\def\annil#1{a_{#1}^{\pd}} 
\def\bcreat#1{b_{#1}^{\dagger}}  
\def\bannil#1{b_{#1}^{\pd}} 

\def\VLSq{{\tilde V}_{\rm p-h}^{\LSsup}}

\def\boxit#1{
        \centerline{\vbox{\hsize=6.0truein\hrule\hbox{\vrule\kern5pt
        \vbox{\kern5pt\noindent #1\smallskip
        \kern5pt}\kern5pt\vrule}\hrule}
}}
\makeatletter
\def\mathcenterto#1#2{\mathclap{\phantom{#1}\mathclap{#2}}\phantom{#1}}
\let\old@widetilde\widetilde
\def\widetildeto#1#2{\mathcenterto{#2}{\old@widetilde{\mathcenterto{#1}{#2\,}}}}
\let\old@widehat\widehat
\def\widehatto#1#2{\mathcenterto{#2}{\old@widehat{\mathcenterto{#1}{#2\,}}}}
\makeatother

\def\he#1{$^{#1}$He}

\numberwithin{equation}{section}

\begin{document}

\title[Variational Theory and Parquet Diagrams for Nuclear Systems]
      {Variational Theory and Parquet Diagrams for\\
        \phantom{AAAA}Nuclear Systems:\\
A Comprehensive Study of Neutron Matter}

      \author[1,2]{\fnm{Eckhard}\sur{Krotscheck}}
\author[1]{\fnm {Jiawei}\sur{Wang}}
\affil[1]{Department of Physics, University at Buffalo, SUNY,
Buffalo NY 14260}
\affil[2]{Institut f\"ur Theoretische Physik, Johannes
Kepler Universit\"at, A 4040 Linz, Austria}

\abstract{
The main task of microscopic many-body theory is to provide an
understanding and an explanation of properties of macroscopic systems
from no other information than the properties of the underlying
Hamiltonian, the particle statistics, and the macroscopic geometry of
the system.

Variational wave functions have been a very successful technique for
examining the ground state and dynamic properties of dense quantum
fluids. Specifically the ``optimized (Fermi)-hypernetted chain
((F)HNC-EL) hierarchy of approximations reproduces, for simple
interactions like electrons and quantum fluids, the most basic
features of a self-bound many-body system, namely binding, saturation,
and spinodal decomposition.

The relationships between Green's functions based perturbation theory
and variational wave functions have been clarified in much detail in
the past. In the language of Feynman diagrams, a self-consistent
summation of ring- and ladder-diagrams is the minimum requirement for
a satisfactory microscopic description of these basic features.

Realistic nucleon-nucleon interactions pose further problems: The most
difficult aspect of the nuclear many-body problem is caused by the
form of the microscopic nucleon-nucleon interaction which depends at
least on the spin, isospin, the relative orientation and angular
momentum of the interacting particles. Simple variational methods as
used for electrons and quantum fluids are inadequate.  Correlation
functions that have the same structure as the interactions lead to
so-called ``commutator diagrams'' that have, so far, been mostly
neglected.  We have shown in the past that these corrections can be
very important if the two-body interactions are very different in
different spin- or isospin channels.

To deal with the problem of realistic nuclear interactions we have
combined techniques of the Jastrow-Feenberg variational method and the
local parquet-diagram theory.  In the language of diagrammatic
perturbation theory, ``commutator diagrams'' can be identified with
non-parquet diagrams. We examine the physical processes described by
these terms and include the relevant diagrams in a way that is
suggested by the Jastrow-Feenberg approach. We show that the
corrections from non-parquet contributions are, at short distances,
larger than all other many-body effects.

We examine here neutron matter as a prototype of systems with
state-dependent interactions. Calculations are carried out for
neutrons interacting via the so-called $v_8$ version of four popular
interactions. We determine the structure and effective interactions
and apply the method to the calculation of the energetics, structure
and dynamic properties such as the single-particle self-energy and the
dynamic response functions as well as BCS pairing in both singlet and
triplet states. We find that many-body correlations lead to a strong
reduction of the spin-orbit interaction, and, therefore, to a
suppression of the $^3P_2$ and $^3P_2$-$^3F_2$ gaps. We also
find pairing in $^3P_0$ states; the strength of the pairing gap
depends sensitively on the potential model employed.

} 

\keywords{Nuclear Matter, Neutron Matter, Microscopic Many-Body Theory}

\maketitle

\newpage
\section{Introduction: The long path from microscopic interactions
  to macroscopic observables}
\label{sec:genericeos}

Methods of theoretical many-body physics tend to be rather technical
and, as a consequence, occasionally hide the physical content. The
difficulty of the problem sometimes suggests the use of approaches
that are guided by technical feasibility, but are of uncontrolled
validity. Before digging into the technical details of the many-body
problem, we will therefore in this section have a look at the physical
effects one is trying to describe, and what the consequences for a
microscopic theory are. To that end, we will focus on self-bound
systems, specifically nuclear matter and the helium liquids, because
these are the ones that best illuminate \cite{JWE93} the situation.

\subsection{Self-bound extended Fermi systems} 

\begin{figure}[H]
  \centerline{\includegraphics[width=0.5\textwidth,angle=-90]{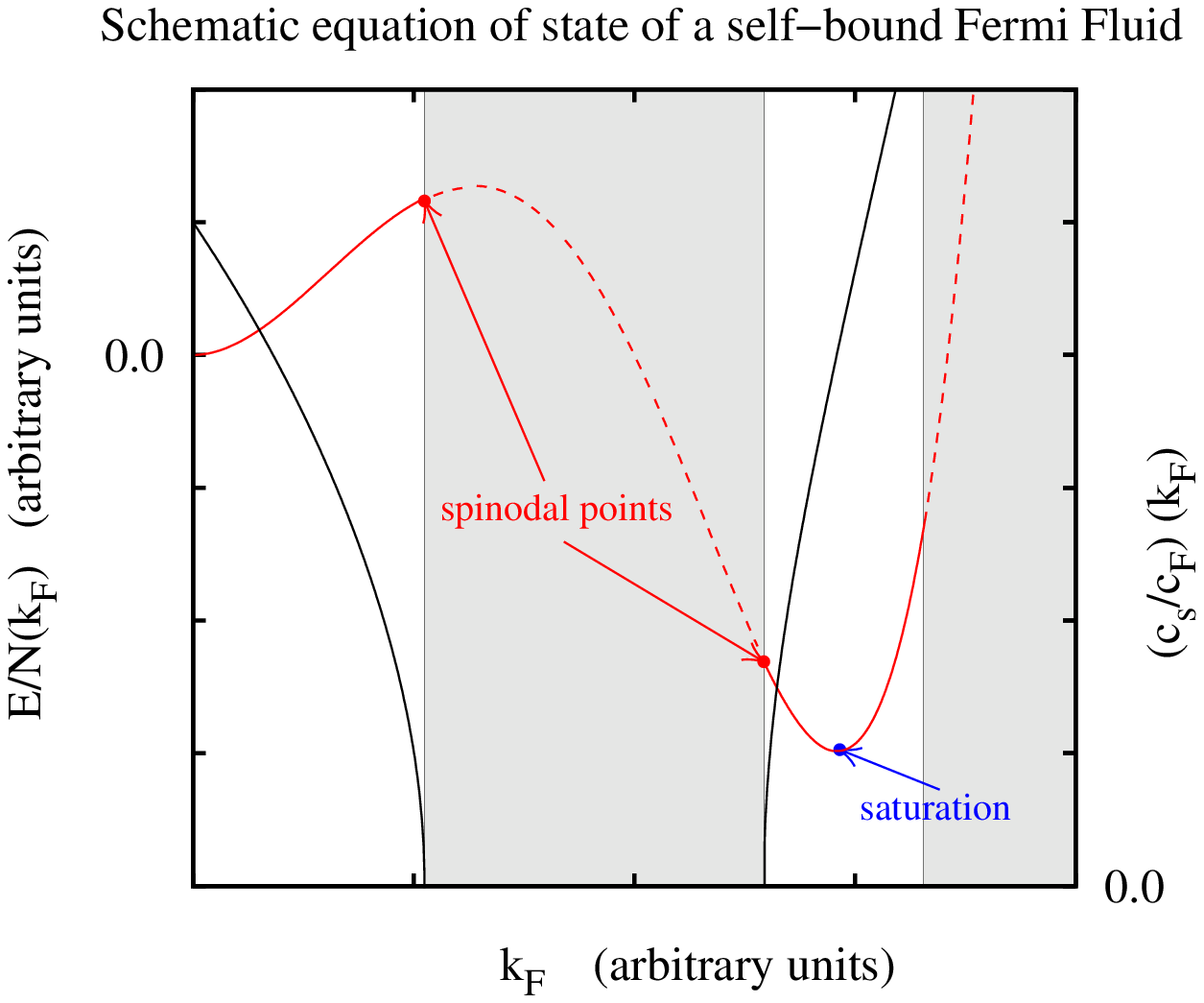}}
  \caption{The figure shows the schematic equation of state of a
    self-bound Fermi liquid, such as nuclear matter or \he3, at zero
    temperature as described in the text. The left gray-shaded area is
    the density regime where no homogeneous system exists, and the
    right gray-shaded area is the density regime where nucleons might
    dissolve or where \he3 becomes solid.  The left scale and the red
    curve depict the energy per particle, the black curves and the
    right scale depict the isothermal speed of sound. From
    Ref. \citenum{fullbcs}.  \label{fig:schematic}}
\end{figure}

Fig. \ref{fig:schematic} shows a schematic equation of state of an
extended self-bound many-fermion system like nuclear matter or liquid
\he3. There are several obvious features: At very low density, the
energy per particle is basically that of a free Fermi gas, $E/N =
\frac{3\hbar^2\KF^2}{10m}$.  Since the system is self-bound, there
must be an energy minimum at finite density. Evidently, between the
low density limit and the saturation density there must be at least
two points where the hydrodynamic speed of sound,
\begin{equation}
mc_s^2 = \frac{d}{d\rho}\rho^2 \frac{d}{d\rho}\frac{E}{N}\,.
\label{eq:mcfromeos}
\end{equation}
vanishes; these points are called ``spinodal points''. There is no
homogeneous phase between these two points; the dashed line in
Fig. \ref{fig:schematic} connecting these two points is only for
reference.  Finally, at high densities, there should be yet another
phase transition; in \he3 the system would solidify, in nuclear
systems the internal structure of the nucleons becomes relevant.

The physical mechanisms behind binding and, at high density,
saturation, are clear, as well as the set of Feynman diagrams that
describes these effects: Binding and saturation are caused by the
medium-ranged interparticle attraction and short-ranged repulsion. The
simplest way of dealing with these effects is by summing the ladder
diagrams, leading to the time-honored Brueckner or
Brueckner-Bethe-Goldstone theory
\cite{Bru54}-\nocite{Bru55,Bru55a,BruecknerLesHouches,BetheGoldstone57}\cite{Goldstone57}
which has been the workhorse of nuclear theory for decades.

A second important effect is encountered at low densities: As the
density is lowered to about 2/3 of the saturation density, the system
becomes unstable against infinitesimal density fluctuations. This is
one of the spinodal points mentioned above where the speed of sound
goes to zero.  A second spinodal point exists at very low density
where the interparticle attraction begins to overcome the Pauli
pressure.

The take-away of the analysis of Fig. \ref{fig:schematic} is that
perturbation theory, which is an expansion of the energy in the
interaction strength, will necessarily diverge in the gray-shaded
area. Moreover, since the equation of state of the homogeneous phase
ends at the spinodal points it cannot be an analytic function of the
density.

At the spinodal points, the system becomes unstable against small
fluctuations.  These are dealt with in linear response theory
\cite{PinesNoz,ThoulessBook} which implies, in its simplest version,
the calculation of the ring diagrams to all orders. A divergence of
the sum of all ring diagrams then indicates a physical instability,
the spinodal instabilities emphasized above are just one example. In
nuclear physics, the diagnosis of physical or spurious instabilities
is a known application of the random-phase
approximation~\cite{KoL2010,LBP2024}. If we want to formulate a theory
that has no solution if the system is unstable, it must therefore
include the summation of all ring diagrams.

Thus, one is led to the conclusion \cite{JWE93} that the summation of
all ring-- and ladder--diagrams of perturbation theory is the least
one needs to do for a consistent description of the equation of state
of a self-bound many-body system over the whole density regime. This
set of diagrams is called the set of ``parquet'' diagrams.

While the insight into what is needed is quite obvious, the execution
is far from trivial. A comprehensive treatment of diagrammatic
perturbation theory has been formulated by Baym and Kadanoff
\cite{BaymKad}. From that seminal work it is clear that each two-body
vertex depends on sixteen variables (two incoming and two outgoing
energy and momentum sets). Taking energy and momentum conservation as
well as isotropy into account reduces this number to ten, which still
present a formidable task. One must seek approximations, but such
steps are normally ambiguous without further guidance.

The purpose of this paper is a comprehensive review of some of our
recent work \cite{v3eos}-\nocite{v3twist,v3bcs,v4}\cite{v43p2} on the
nuclear many-body problem. We shall also extend our calculation to
more modern interactions and to dynamic effects. As in our previous
work we will focus our applications on neutron matter because we feel
that the technical manipulations are best analyzed in this
case. Applications of the $v_8$ model for the nuclear interactions
tend to over-bind symmetric nuclear matter significantly
\cite{Baldo2012}, they would therefore be of limited value.

\subsection{The nucleon-nucleon interaction} 

Accurate representations of the nucleon-nucleon potentials
\cite{Reid68}\nocite{Bethe74,AV18,Reid93,RevModPhys.81.1773}-\cite{Machleidt}
are constructed from both first principles and to fit the interactions
in each partial wave to scattering data and deuteron binding
energies. For the purpose of identifying specific physical effects and
for high-level many-body calculations, an interaction given in the
form of a sum of local functions times operators acting on the spin,
isospin, and possibly the relative angular momentum variables of the
individual particles is preferred
\cite{Day81,AV18,Wiri84,RevModPhys.81.1773,Machleidt}, \ie one writes
\begin{equation}
\hat v (i,j) = \sum_{\alpha=1}^n v_\alpha(r_{ij})\,
        \hat O_\alpha(i,j),
\label{eq:vop}
\end{equation}
where $r_{ij}=\left|\rvec_i-\rvec_j\right|$ is the distance between
particles $i$ and $j$. According to the number of operators $n$, the
potential model is referred to as a $v_n$ interaction. Semi-realistic
models for nuclear matter keep at least the six to eight base
operators
\begin{eqnarray}
\hat O_1(i,j;\hat\rvec_{ij})
        &\equiv& \hat O_{\rm c} = \1\,,
\nonumber\\
\hat O_3(i,j;\hat\rvec_{ij})
        &\equiv& {\bsigma}_i \cdot {\bsigma}_j\,,
\nonumber\\
\hat O_5(i,j;\hat\rvec_{ij})
&\equiv& S_{ij}(\hat\rvec_{ij})\nonumber\\
      &\equiv& 3({\bm\sigma}_i\cdot \hat\rvec_{ij})
      ({\bsigma}_j\cdot \hat\rvec_{ij})-{\bsigma}_i \cdot {\bsigma}_j\,,
      \nonumber\\
  \hat O_7(i,j;\rvec_{ij},\pvec_{ij})
  &\equiv&{\bf L}\cdot{\bf S} = \rvec_{ij}\times\pvec_{ij}\cdot\S\,,
  \nonumber\\
      \hat O_{2\alpha}(i,j;\hat\rvec_{ij}) &=& \hat O_{2\alpha-1}(i,j;\hat\rvec_{ij})
      {\bm\tau}_i\cdot{\bm\tau}_j\,,
  \label{eq:Vop1}
\end{eqnarray}
where $\S\equiv\frac{1}{2}(\bsigma_i+\bsigma_j)$ is the total spin,
and $\pvec_{ij}=\frac{1}{2}(\pvec_i-\pvec_j)$ is the relative momentum
operator of the pair of particles. In the following, we will also use
the notation $\alpha \in \{({\rm cc}), (\rm{c}\tau), (\sigma{\rm c}),
(\sigma\tau), (Sc), (S\tau) , (LSc) , (LS\tau) \}$ for
$\alpha=1$--$8$.  In neutron matter, the operators are projected to
the isospin $T=1$ channel, \ie, we have
\begin{equation}
  O_\alpha(i,j,\hat\rvec_{ij}) \rightarrow  O_\alpha(i,j,\hat\rvec_{ij})+O_{\alpha+1}(i,j,\hat\rvec_{ij})
\label{eq:Vop1nm}
\end{equation}
for odd $\alpha$ and $O_\alpha(i,j,\hat\rvec_{ij})=0$ for even
$\alpha$.  The new set of interaction channels will be $\alpha \in
\{({\rm c}), (\sigma), (S), (LS)\}$.

We will in this paper explore four interactions: A ``classic'' one,
which we will denote as Reid 68, is the Reid potential \cite{Reid68}
in the operator representation of Ben Day \cite{Day81} with the
spin-orbit force of Eqs. (20) and (30) of Ref. \citenum{Reid68};
another potential which has found many applications is the Argonne
potential \cite{AV18}, of which we will use the $v_6'$ and $v_8'$
versions. As a representative of the most modern, ``chiral effective
field theory'' (CEFT) interactions, we adopt a local version in
next-to-next-to-leading order (N2LO) with a cutoff of $1.0$~fm
\cite{PhysRevC.90.054323}, which is also available in the above
operator basis.  Finally, we employ a somewhat more recent version of
the Reid interaction \cite{Reid93}, which we will denote Reid 93.
 
The Reid 93 potential is given in a partial wave decomposition
only. It is evidently impossible to represent all individual partial
waves with only eight (or, in neutron matter, four) interaction
components as required by the operator representation \eqref{eq:vop}.
Therefore, we have constructed in Ref. \citenum{v43p2} an operator
form by reproducing the tensor force and the four lowest angular
momentum components. The procedure is described in Appendix
\ref{app:v8}.

Figs. \ref{fig:vbare} show the bare interactions for the four
potential models to be studied here for isospin=1. Attention is directed
towards the very different shape of the interaction in spin-singlet and
spin-triplet configurations highlighted in the left figure.
\begin{figure}[H]
  \centerline{
  \includegraphics[width=0.35\textwidth,angle=-90]{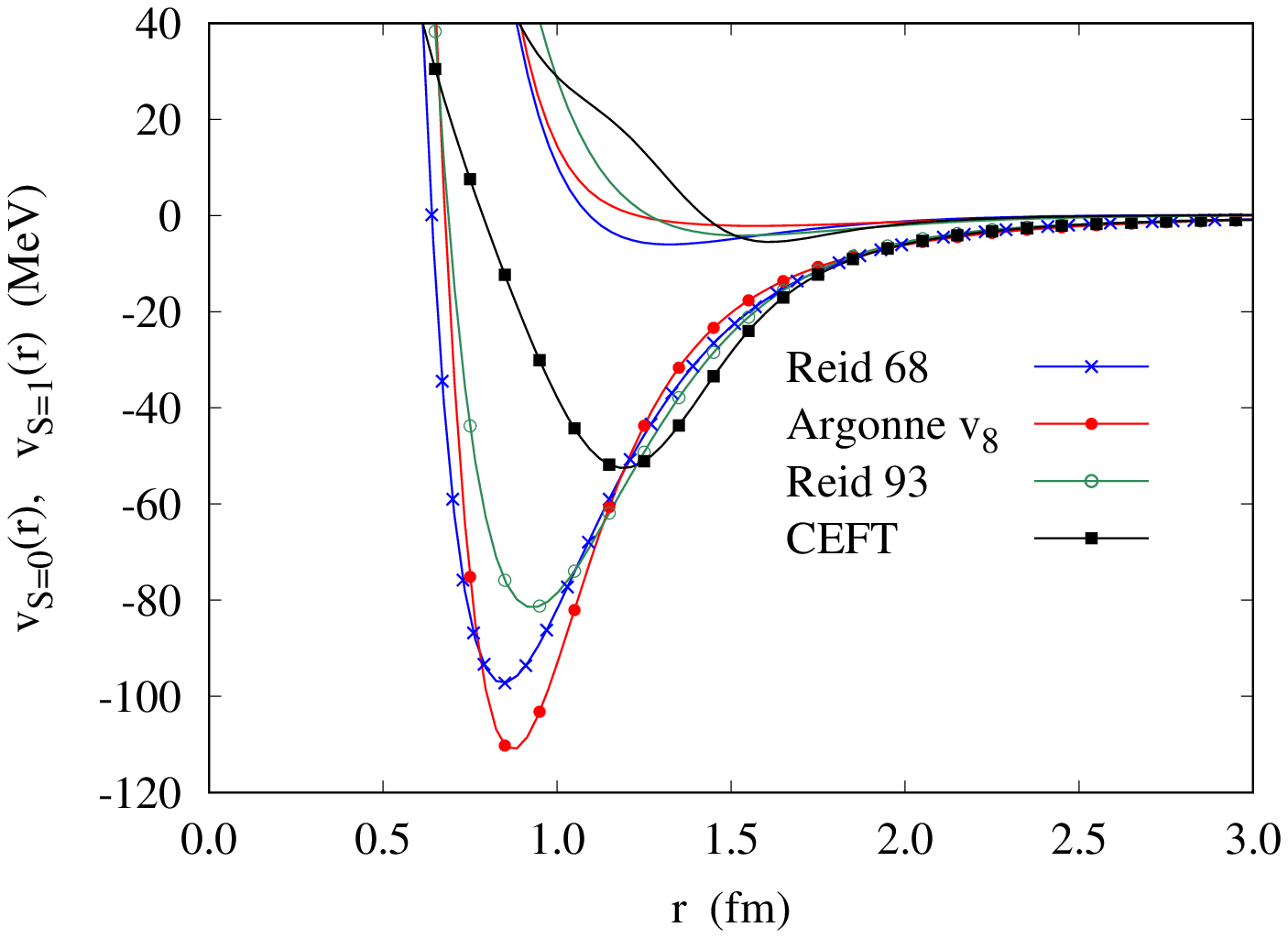}
  \includegraphics[width=0.35\textwidth,angle=-90]{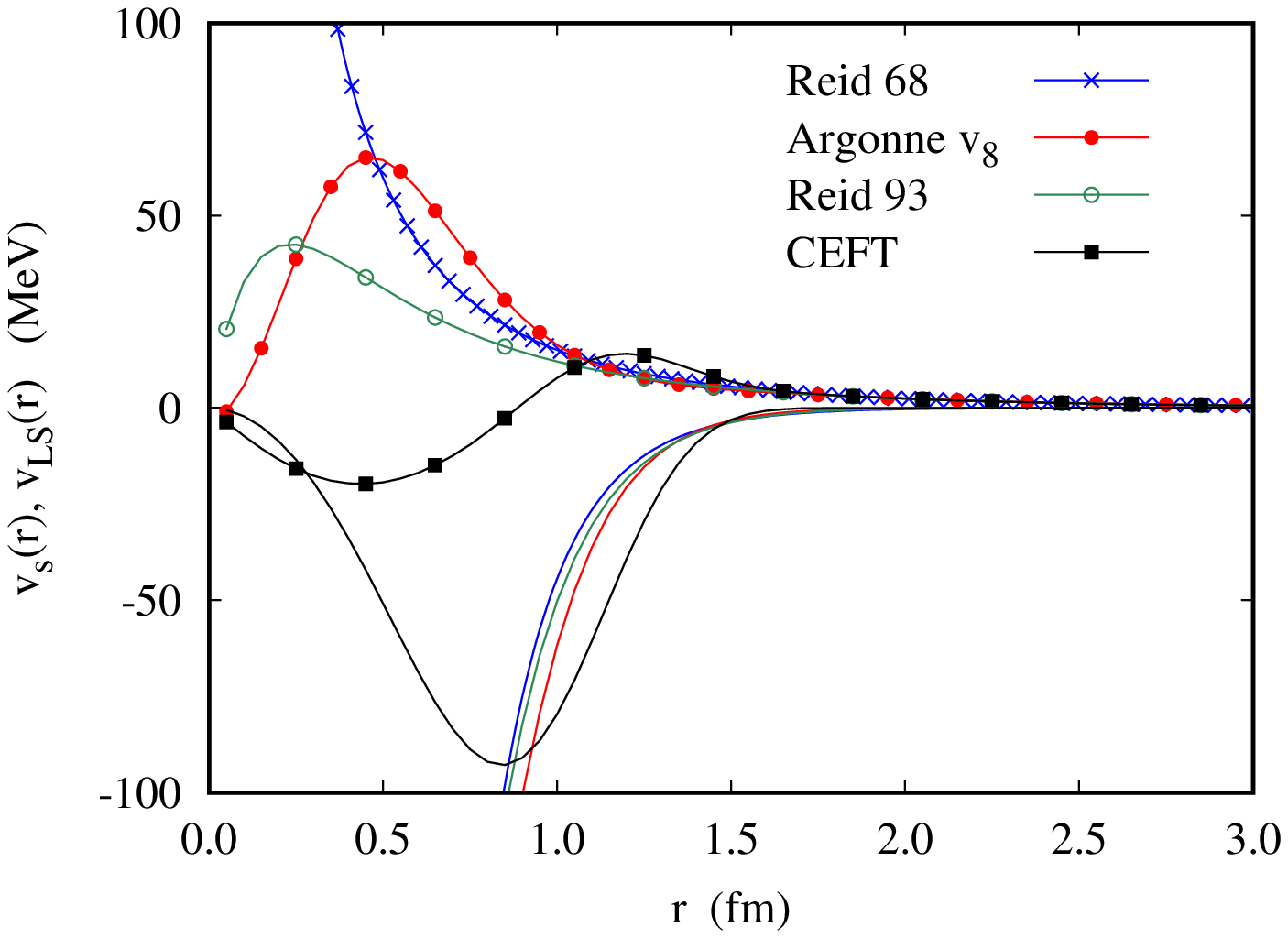}
  }
  \caption{(color online) The left figure shows the central
    interactions in the spin-singlet (curves with markers) and the
    spin-triplet case for the above four potentials studied here. The
    right figure shows the tensor (curves with markers) and spin-orbit
    interactions for the same four interactions. \label{fig:vbare}}
  \end{figure}

All of the various parameterizations exhibit repulsion at short
distances. A simple way to deal with this is to soften the bare
interaction, for example by means of the similarity renormalization
group.  For a review of this line of work, particularly relevant for
finite nuclei, see, {\em e.g.}, Ref. \citenum{Hegert2020}. The strong
repulsion is, {\em per-se\/} not a problem; recall that the helium
liquids \he3 and \he4 exhibit a much stronger short-ranged repulsion
but can still be described with high accuracy by modern many-body
techniques \cite{JordiEncyclopedia,Encyclopedia}.  We also need to
stress here again that for self-bound systems with a saturation regime
such as shown in Fig. \ref{fig:schematic}, perturbation theory does
not converge in the whole (density -- interaction strength) regime
regardless of the softness of the interaction used. That problem too
has been solved in quantum fluids; the really difficult aspect is the
nuclear many-body problem in the complicated interaction. This paper
addresses exactly this problem.

Our paper is organized as follows. In the next section
\ref{sec:central} we give a brief review of the variational theory and
the local parquet-diagram theory for simple central interactions.
Variational theory begins with an {\em ansatz\/} for the many-body
wave function of the form
\begin{equation} 
  \ket{\Psi_0} = \frac{\hat{F}_N\ket{\Phi_0}}{\bra{\Phi_0}\hat
    F^\dagger_N \hat F^{\phantom{\dagger}}_N \ket{\Phi_0}^{1/2}}\,
\label{eq:PsiFPhi}
\end{equation}
where $\hat{F}_N$ is the $N$-body correlation operator, and the model
state $\ket{\Phi_0}$ describes the statistics and, if appropriate, the
geometry of the system; for fermions, $\ket{\Phi_0}$ is normally taken
to be a Slater determinant.  For simple, state-independent
interactions as appropriate for electrons or many atomic quantum
fluids, the Jastrow-Feenberg ansatz \cite{FeenbergBook} for the
correlation operator of an $N$-particle system
\begin{eqnarray}
  \hat F_N(\rvec_1,\ldots\rvec_N) &=&
        \exp\frac{1}{2}\left\{
        \sum_{i<j} u_2({\bf r}_i,{\bf r}_j)
        + \sum_{i<j<k} u_3({\bf r}_i,{\bf r}_j,{\bf r}_k)
        + \ldots\right\}\nonumber\\
        &=&\prod_{i<j}f_2({\bf r}_i,{\bf r}_j)\prod_{i<j<k}f_3({\bf r}_i,{\bf r}_j,{\bf r}_k)\times ...
\label{eq:Jastrow}
\end{eqnarray}

Parquet-diagram summations begin with standard Green's functions
perturbation theory \cite{FetterWalecka} and sum the ring and ladder
diagrams self-consistently \cite{parquet1,parquet2}. Variational
theory begins with the {\em ansatz\/} \eqref{eq:Jastrow}, and carries
out cluster expansions and re-summations. The ``hypernetted chain
(HNC)'' hierarchy of approximations has turned out to permit the
optimization of the correlations $f_2({\bf r}_i,{\bf r}_j)$ by
unconstrained minimization of the energy expectation value (``HNC-EL''
method).  It was proven to lead, in the simplest case, to equations
identical to those of parquet diagrams \cite{parquet1,parquet2},
leading the authors of that work to the assessment that ``the
formulation of the many-body problem is a matter of language but not a
matter of substance''. We shall use the terms ``parquet-diagram
summations'' and ``HNC-EL method'' interchangeably.
Section \ref{sec:central} gives only a rudimentary overview of the
methods; for very detailed derivation and analysis, the reader is
referred to Refs. \citep{cbcs} and \citep{fullbcs}.

Section \ref{sec:nucleons} turns to the implementation of the ideas of
parquet-diagram summation and the variational method to nuclear
interactions of the $v_8$ type. Superficially, the approaches become
very different: The Jastrow-Feenberg ansatz \eqref{eq:Jastrow} is
insufficient for dealing with realistic nucleon-nucleon interactions
of the form (\ref{eq:vop}). A plausible generalization of the wave
function (\ref{eq:Jastrow}) is the symmetrized operator product
\cite{FantoniSpins,IndianSpins},
\begin{equation}
        \Psi_0^{{\rm SOP}}
        = {\cal S} \Bigl[ \prod^N_{i,j=1 \atop i<j} \hat f (i,j)\Bigr] \Phi_0
        \equiv F_N \Phi_0\,,
\label{eq:f_prodwave}
\end{equation}
where
\begin{equation}
  \hat f(i,j) = \sum_{\alpha=1}^n f_\alpha(r_{ij})
  \hat O_\alpha(i,j)\,,
  \label{eq:fop}
\end{equation}
and ${\cal S}$ stands for symmetrization, which takes all possible
orderings of $\hat f(i,j)$ into account. It is necessary because the
operators $\hat O_\alpha(i,j)$ and $\hat O_\beta(i,k)$ may not
commute.  The symmetrization generates so-called ``commutator
diagrams'', which do not appear in parquet diagram summations
\cite{SmithSpin}. Rather, the theory becomes in the Bose limit
identical to an approximation of the variational method
\cite{FantoniSpins} where all commutators are neglected.  The
conclusion to be drawn from this is not that ``commutator
corrections'' are just a technical nuisance of the variational
approach, but rather that these corrections are non-parquet
diagrams. The effect of these contributions, which have so far been
mostly overlooked or ignored, can be quite dramatic \cite{SpinTwist}.

Another aspect of section \ref{sec:nucleons} is the summation of all
ring diagrams for a spin-orbit interaction in a closed form. We will
show that this can be accomplished by introducing just 3 more
operators in addition to those defined in Eqs. \eqref{eq:Vop1}; the
relevance of these operators will be discussed.

The remaining sections \ref{sec:static}-\ref{sec:BCS} discuss
implementations of our methods for various quantities of interest:
Sec. \ref{sec:static} focuses on ground state properties. By
comparing different ways to calculate the hydrodynamic speed of sound
\eqref{eq:mcfromeos} we can make statements on the rate of convergence
of our calculations for different interactions. We will discuss
specifically the short-ranged structure of the system and demonstrate
that different interactions come to rather different predictions.
We will also show that the spin-orbit force is suppressed by the
strong repulsion in the spin-triplet channels.

Section \ref{sec:dynamics} then turn to the calculation of dynamic
properties such as the dynamic (spin-) response and single-particle
excitations. We implement here, for the first time in nuclear systems,
the ``dynamic many-body theory (DMBT)'' which has led to predictions
of the dynamic response of quantum fluids of unprecedented accuracy
without the need of phenomenological input
\cite{Encyclopedia,eomIII,2p2h,Nature_2p2h}. The theory may be
understood as a formulation of the ``second RPA'' (SRPA)
\cite{PhysRev.126.2231}-\nocite{SRPA83,SRPA87,Wambach88}\cite{PhysRevC.81.024317}
which is built on the ground state and not an uncorrelated Slater
determinant. The section also includes the calculation of the
effective mass and single-particle lifetimes.

The final section \ref{sec:BCS} addresses the problem of BCS pairing
in nuclear matter. We will show that the predictions for the size of
the $S$-wave gap are quite robust under the change of the
interactions. $P$-wave pairing is more subtle; the strongly repulsive
core of the interaction in spin-triplet states causes a relatively
large ``correlation hole'' and, as a consequence, a suppression of the
spin-orbit interaction. Therefore, pairing in $^3P_2$-$^3F_2$ is
suppressed and pairing in $^3P_0$ states enhanced.

A summary reviews briefly our findings; some technical aspects are
addressed in the appendices.

\section{Microscopic Theory: State-independent interactions}
\label{sec:central}

Before we proceed with the exposition of a comprehensive many-body
theory for realistic nuclear forces, and in the interest of making
this manuscript self-contained, let us review some formal points and
basic applications of diagrammatic methods. Among others, we wish to
highlight the complications presented by fermionic systems with
exchange effects as compared to bosonic systems.

\subsection{The Jastrow-Feenberg method}

One of the reasons for the success of the Jastrow-Feenberg method is
that it provides a reasonable upper bound for the ground state energy
expectation value
\begin{equation}
E_0 = \bra{\Psi_0}H\ket{\Psi_0}\,.
  \label{eq:energyexp}
\end{equation}
It has therefore been applied in both semi-analytic calculations
\cite{FeenbergBook} and in early Monte Carlo calculations
\cite{KalosLevVer,CeperleyVMC} and is still being used as an
importance sampling function for diffusion and Green's Functions Monte
Carlo computations \cite{CeperleyRMP,JordiQFSBook}.

In semi-analytic calculations it is in practice impossible to handle
expectation values for the $N-$body operator ${F}_N$ without
approximations.  Originally, expectation values with respect to the
Jastrow-Feenberg wave function \eqref{eq:Jastrow} are expanded in
terms of the correlation function
\begin{equation}
h(r_{ij}) \equiv
f^2(r_{ij})-1\label{eq:hdef}\,.\end{equation}
The individual cluster contributions are then summed to
infinite order to give rise to the ``hypernetted chain (HNC)''
\cite{Morita58,LGB} equations or their fermion version (FHNC)
\cite{Mistig,Fantoni}.  These are used to eliminate the correlation
function altogether and to formulate physical quantities like the
ground state energy entirely in terms of the observable pair
distribution $g(|\rvec-\rvec'|)$,
\begin{equation}
  g(|\rvec-\rvec'|) = \frac{\int d^3r_1\ldots d^3r_N
    \psi^*(\rvec_1,\ldots\rvec_N)\sum_{i \ne j}\delta(\rvec_i-\rvec)\delta(\rvec_j-\rvec')\psi(\rvec_1,\ldots\rvec_N)}
  {\int d^3r_1\ldots d^3r_N\left|\psi(\rvec_1,\ldots\rvec_N)\right|^2}
  \label{eq:gofr}
\end{equation}
and the static structure function 
\begin{equation}
  S(q) = 1 + \rho\int d^3r (g(r)-1)e^{\I \qvec\cdot\rvec}\label{eq:Sfromg}. 
\end{equation}
For systems with simple interactions such as the
helium liquids and electrons, the Jastrow-Feenberg {\em ansatz\/} for
the correlated wave function is sufficient.  In many cases,
\eqref{eq:Jastrow} can be truncated at the level of $u_2({\bf
  r}_i,{\bf r}_j)$. For the very dense systems three-body correlations
can contribute up to 10 percent of the binding energy
\cite{ChaC,EKthree,polish}.

The (F)HNC equations have the specific feature that they permit, in
every level of approximation, the optimization of the correlation
functions by functionally minimizing the energy expectation
\begin{equation}
 \frac{\delta E_0}{\delta f_n}({\bf r}_1,\ldots,{\bf r}_n) = 0,
\label{eq:euler}
\end{equation}
in which case the method is referred to as the
(Fermi-)Hypernetted-Chain-Euler-Lagrange (F)HNC-EL procedure.

One is then led to a set of non-linear equations relating the pair
distribution function and the static structure function $S(q)$ to the
interparticle interaction $v(r)$
\cite{FeenbergBook,Chuckreview,LanttoSiemens,EKVar}.

It was quickly realized that the Jastrow-Feenberg-Euler-Lagrange
procedure had exactly the desirable features of reproducing the
high-density saturation and a low-density spinodal instability
outlined above. For bosons, Sim, Buchler, and Woo \cite{Woo70} came
therefore to the conclusion that ``...it appears that the optimized
Jastrow function is capable of summing all rings and ladders, and
partially all other diagrams, to infinite order.''  This being more a
matter of observation than of rigorous derivation, Jackson, Lande and
Smith went back to diagrammatic perturbation theory and showed in a
series of papers \cite{parquet1,parquet2,parquet3} that indeed the
HNC-EL equations represented an approximate summation of all parquet
diagrams, and determined exactly what these approximations were. With
that, the HNC-EL procedure has been identified as a very practical way
to perform parquet-diagram summations. The types of approximations
were singled out by the upper bound property of $E_0$ as the best
possible for the computational price one was willing to pay.
Moreover, while the derivation of the relevant equations is somewhat
complicated, the resulting equations to be solved numerically were
just the Bethe-Goldstone (BG) equation and the random phase
approximation (RPA) equation that had been solved individually for
decades.

The above holds rigorously only for bosons. A similar systematic
diagrammatic equivalence between the fermion version and
parquet-diagram summations has not been carried out. Rather, the
equivalence has been established for specific sets of diagrams: rings,
ladders, and self-energy insertions \cite{fullbcs}, as we will
demonstrate below.

In the following, we will make use of a dimensionless Fourier 
transform defined by including a density factor $\rho$, \ie
\begin{equation}
  \tilde f(q) = \rho\int d^3r e^{\I \qvec\cdot\rvec}f(r)\label{eq:ft}\,.
\end{equation}
We shall also use the notation $\left[\ldots\right]^{\cal F}$ for the
above operation.  For further reference, let us also write the
relation of $S(q)$ with the density-density response function
$\chi(q,\omega)$,
\begin{equation} 
 S(q) = -\int_0^\infty \frac{d\hbar\omega}{\pi} \Im \chi(q,\omega) 
\label{eq:Sfromchi}.
 \end{equation} 

\subsection{The simplest example: Bosons}
\label{ssec:bosons}
\subsubsection{Variational method}

For the case of Bosons, the optimization condition \eqref{eq:euler}
can be expressed in different forms as a relationship between the pair
distribution function $g(r)$, the static structure function $S(q)$,
and the microscopic interaction $v(r)$.  The static structure function
$S(q)$ is expressed in terms of a Bogoliubov equation
\cite{FeenbergBook,Chuckreview}
\begin{equation}
  S(q) = \frac{1}{\sqrt{1+\displaystyle\frac{2\tilde V_{\rm p-h}(q)}{t(q)}}}
  \label{eq:BoseRPA}
\end{equation}
in terms of a self-consistently determined ``particle-hole''
interaction $V_{\rm p-h}$.
\begin{eqnarray}
  V_{\rm p-h}(r) &=& g(r)\left[v(r) +  V_e(r)\right]
  + \frac{\hbar^2}{m}\left|\nabla\sqrt{g(r)}\right|^2
  + \left[g(r)-1\right]w_{\rm I}(r)\,,\label{eq:BoseVph}\\
  \tilde w_{\rm I}(q) &=& -t(q)
\left[1-\frac{1}{ S(q)}\right]^2
\left[S(q)+\frac{1}{2}\right]\,.
  \label{eq:Bosewind}
\end{eqnarray}
where $t(q) = \hbar^2 q^2/2m$ is the kinetic energy of a free
particle.  The only unknown quantity is the term $V_e(r)$. In
the language of Jastrow-Feenberg theory, it accounts for the contribution
from ``elementary diagrams'' and multiparticle correlations
\cite{EKthree}.

A few algebraic manipulations show that the pair distribution function
satisfies the coordinate-space equation \cite{LanttoSiemens}
\begin{equation}
  \frac{\hbar^2}{m}\nabla^2\sqrt{g(r)} = \left[v(r) +  V_e(r) +
    w_{\rm I}(r)\right]\sqrt{g(r)}
  \label{eq:BoseBG}\,.
  \end{equation}
Eq.~(\ref{eq:BoseBG}) is recognized as the boson Bethe-Goldstone
equation in terms of the interaction $v(r) +  V_e(r) + w_{\rm I}(r)$. 

Eq. \eqref{eq:BoseRPA} also leads us back to the discussion of our
schematic equation of state, Fig. \ref{fig:schematic}. Evidently,
$\tilde V_{\rm p-h}(0+) > 0$ is a condition for existence of solutions
of the set of HNC-EL equations, but it is not {\em a-priori\/} clear
that $\tilde V_{\rm p-h}(0+)$ is related to $mc_s^2$ as derived from
the equation of state via Eq. \eqref{eq:mcfromeos}. In fact, if
the energy $E_0$ is calculated exactly for a Jastrow correlation
operator \eqref{eq:Jastrow} and the correlation function $f(r)$ is
optimized at all densities, we have \cite{EKVar}
\begin{equation}
  mc_s^2 = \tilde V_{\rm p-h}(0+) - \frac{\rho^2}{N}\int d^3r d^3r'
  \frac{\delta^2 E_0}{\delta f(r)\delta f(r')}\frac{d f(r)}{d\rho}\frac{d f(r')}{d\rho}\,.
  \end{equation}
The term containing the second variation of the energy is zero only if
also three- and four- body correlations functions are included
and optimized. In other words, $m c_s^2$ as derived from the equation
of state and $\tilde V_{\rm p-h}(0+)$ agree only in an exact theory.

\subsubsection{Parquet diagram summations}

Since we will heavily rely on the derivations and localization
procedures of parquet-diagram theory, let us briefly review the
relevant steps for a bose system:

To make the connection between the HNC-EL method and diagrammatic
perturbation theory, we note that the Bogoliubov equation
(\ref{eq:BoseRPA}) is derived from a random-phase approximation (RPA)
equation for the density-density response function
 \begin{eqnarray}
    \chi(q,\omega) &=&
    \frac{\chi_0(q,\omega)} {1-\tilde V_{\rm
        p-h}(q)\chi_0(q,\omega)}\label{eq:chiRPA}\,\\
    S(q) &=& -\int_0^\infty \frac{d\hbar\omega}{\pi}
    \frac{\chi_0(q,\omega)} {1-\tilde V_{\rm
        p-h}(q)\chi_0(q,\omega)}\label{eq:SRPA}
 \end{eqnarray}
in terms of a {\em local\/} and {\em energy-independent\/} particle-hole
interaction $\tilde V_{\rm p-h}(q)$. Here
\begin{equation}
        \chi_0(q,\omega) =
        \displaystyle \frac{2 t(q)}
        { (\hbar\omega+\I\eta)^2-
        t^2(q)} \,,
\label{eq:Chi0Bose}
\end{equation}
is the density-density response function of a non-interacting Bose system.

The interaction enters the above RPA equation in the form of a {\em
  local\/} and {\em energy-independent\/} particle-hole interaction
$\tilde V_{\rm p-h}(q)$, a key quantity in making contact between the
variational and diagrammatic approaches. The procedure
of localized parquet-diagram summations \cite{parquet1,parquet2} is then

\begin{itemize}
\item{} Eqs.~(\ref{eq:chiRPA}), (\ref{eq:Chi0Bose}) define an
  {\em energy-dependent\/} effective interaction
 \begin{equation}
   \widetilde W(q,\omega) = \frac{\tilde V_{\rm p-h}(q)} {1-\tilde V_{\rm
       p-h}(q)\chi_0(q,\omega)}\,.
   \label{eq:Wnonlocal}
 \end{equation}
\item{}
An {\em energy-independent\/} effective interaction $\widetilde W(q)$ is then
defined such that it leads to the same $S(q)$, {\ie \/}
\begin{eqnarray}
S(q) &=&  -\int_0^\infty \frac{d\hbar\omega}{\pi} \Im
 \left[\chi_0(q,\omega)+ \chi^2_0(q,\omega) \widetilde W(q,\omega)\right]
 \nonumber\\
  &\overset{!}{=}& 
-\int_0^\infty \frac{d\hbar\omega}{\pi} \Im
\left[\chi_0(q,\omega)+ \chi^2_0(q,\omega) \widetilde W(q)\right]\,,
\label{eq:Wlocal}
\end{eqnarray}
where the last line defines $\widetilde W(q)$. Carrying out the
frequency integration leads to
\begin{equation}
  \widetilde W(q) = -t(q)(S(q)-1)\label{eq:BoseWind}\,,
\end{equation}
which defines the {\em induced interaction}
\begin{equation}
  \tilde w_I(q) = \widetilde W(q) - \tilde V_{\rm p-h}(q)
\label{eq:wind}
\end{equation}
\item{} Supplement, in the Bethe-Goldstone equation
  the bare interaction $v(r)$ by this static induced interaction
  as indicated in Eq. \eqref{eq:BoseBG}.
\item{} Demand that the pair distribution function $g(r)$
  obtained from the Bethe-Goldstone equation and the static
  structure function \eqref{eq:SRPA} are connected by
  Eq. \eqref{eq:Sfromg}.
\item{} The energy can be calculated by coupling-constant integration
     of the potential energy alone \cite{hellmann1933,Feynman1939}
\begin{equation}
\frac{E}{N} = \frac{\rho}{ 2}
\int d^3r\, v(r)\int_0^1 d\lambda g_\lambda(r),
\label{eq:FH}\end{equation}
where $g_\lambda(r)$ is the pair distribution function calculated for
a potential strength $\lambda v(r)$. It turns out \cite{parquet3} that
the energy obtained by the coupling-constant integration is identical
to the variational energy functional $E_0$.
\end{itemize}

The correction $V_e(r)$ represents, in the language of Feynman
diagrams, corrections to the localization procedure of the parquet
equation and the set of fully irreducible diagrams
\cite{TripletParquet}. The equivalence between the $V_e(r)$ as
obtained from parquet-diagram summations and from elementary diagrams
and triplet correlations has been proven in Ref. \citep{MixMonster}.
The identification of $\tilde V_{\rm p-h}(0+)$ with the hydrodynamic
speed of sound is less straightforward, equality holds again only in
an exact theory \cite{parquet5}.
 
\subsection{Fermions with state-independent interactions}
\label{ssec:fermions}

The procedure of deriving cluster expansions for Fermions
\cite{IwamotoYamada57,HartoghTolhoek57a,HartoghTolhoek57b,Aviles58}
and summations of infinite series \cite{Fantoni74,Mistig} follows
largely that for bosons. Besides the correlation functions
$h(r_{ij})$, the cluster expansions contain
``exchange functions''
\begin{equation}
  \ell(r\KF) = \frac{\nu}{N}\sum_\kvec n(\kvec) e^{\I\kvec\cdot\rvec}\,.
  \label{eq:eldef}
\end{equation}
where
\begin{equation}
  n(\kvec) = \begin{cases}1 & |\kvec|\le\KF\\
    0 &  |\kvec|>\KF
  \end{cases}
  \label{eq:n0def}
\end{equation}
is the momentum distribution of a non-interacting Fermi system in its
ground state.  These exchange lines appear in closed loops
$\ell(r_{12}\KF)\ell(r_{23}\KF)\dots\ell(r_{n1}\KF)$.

The relationship between the Jastrow-Feenberg variational theory and
parquet-diagram summations is somewhat more complicated for Fermions
due to the multitude of exchange diagrams.
We discuss here only the simplest case (dubbed the ``FHNC//0-EL''
approximation) which highlights this relationship most clearly.
As we shall see, this versions sums, in an approximation to be determined
below, all ring and ladder diagrams of perturbation theory. Higher order
versions, referred to as ``FHNC//n-EL'' include, with increasing complexity,
the RPA-exchange diagrams as well as self-energy corrections to the
single-particle propagator. In particular, self-energy corrections
can be identified with the ``cyclic chain'' diagrams of the FHNC summations
\cite{fullbcs}.

\subsubsection{Ring Diagrams\label{ssec:rings}}
In the FHNC//0 approximation, the generalization of
Eq.~(\ref{eq:BoseRPA}) is
\begin{equation}
  S(q) = \frac{\SF(q)}{\sqrt{1 + 2 \displaystyle\frac{\SF^2(q)}{t(q)}
      \tilde  V_{\rm p-h}(q)}}\,.
\label{eq:FermiPPA}
\end{equation}
with 
\begin{equation}
  \SF(q) = \begin{cases}
     \displaystyle \frac{3q}{4\KF}-\frac{q^3}{16\KF^3},
                & q < 2\KF ;\\
      1,      & q \ge 2\KF\,
      \end{cases}
\label{eq:SqFermi} 
\end{equation}
is the static structure function of a non-interacting Fermi gas.  In
the same approximation, the particle-hole interaction is
\begin{equation}
  V_{\rm p-h}(r) = \left[1+\Gamma_{\!\rm dd}(r)\right]
  \left[v(r) +  V_e(r)\right]
  + \frac{\hbar^2}{m}\left|\nabla\sqrt{1+\Gamma_{\!\rm dd}(r)}\right|^2
  + \Gamma_{\!\rm dd}(r)w_{\rm I}(r)\,.\label{eq:FermiVph}
\end{equation}
where $1+\Gamma_{\!\rm dd}(r)$ is the ``direct correlation function''
\begin{equation}
  \tilde \Gamma_{\!\rm dd}(q)= \frac{S(q)-\SF(q)}{\SF^2(q)}\,.
\label{eq:Gammadd}
\end{equation}
Higher order versions of the FHNC//n-EL hierarchy lead to a set of
equations that is structurally the same as Eqs. \eqref{eq:FermiPPA}
and \eqref{eq:FermiVph}, only the ingredients are more complicated
\cite{polish}.

To make the connection to Eq. \eqref{eq:SRPA} for fermions, we
identify $\chi_0(q,\omega)$ with the Lindhard function
\begin{equation}
  \chi_0(q;\omega) = \frac{1}{N}\Tr_\bsigma\sum_\hvec\frac{2(t(p)-t(h))}
  {(\hbar\omega+\I\eta)^2 - (t(p)-t(h))^2}\,,\qquad \pvec=\hvec+\qvec
  \,.\label{eq:chi0}
\end{equation} 
In general, we will denote ``hole'' states that lie inside the Fermi
sea by $\hvec$, $\hvec'$, $\hvec_i$ and ``particle'' state that lie
outside the Fermi sea by $\pvec$, $\pvec'$, $\pvec_i$.  Vectors
$\kvec$, $\kvec'$, $\qvec$, $\qvec'$ are unrestricted.  Consistent
with the convention (\ref{eq:ft}) we have defined the Lindhard
function (and the response functions to be discussed in section
\ref{ssec:response}) slightly different than usual
\cite{FetterWalecka}, namely such that it has the dimension of an
inverse energy.

Define now the Fermi-sea average of any function $f(\pvec,\hvec)$
depending on a ``hole momentum'' $|\hvec|<\KF$ and a ``particle
momentum'' $\pvec=\hvec+\qvec$ with $|\pvec|>\KF$,
\begin{eqnarray}
\left\langle f(\pvec,\hvec) \right\rangle(q)
&=& \frac{\sum_{\hvec} \bar n(|\hvec+\qvec|) n(h) f(\hvec+\qvec,\hvec)}
      {\sum_{\hvec} \bar n(|\hvec+\qvec|) n(h)}\label{eq:favq}\\
&=& \frac{1}{S_{\rm F}(q)}\int \displaystyle\frac{d^3h}{V_{\rm F}}
\bar n(|\hvec+\qvec|) n(h)  f(\hvec+\qvec,\hvec)\,,\nonumber
\label{eq:favg}
\end{eqnarray}
where $\bar n(q)
= 1-n(q)$, and $V_{\rm F}$ is the volume of the Fermi sphere.  In
particular, we find
\begin{equation}
\left\langle t(|\hvec+\qvec|)-t(h)\right\rangle(q)
= \frac{t(q)}{\SF(q)} \equiv \tF(q)\label{eq:tqcoll}
\end{equation}
Applying this hole-state averaging procedure to the Lindhard
function \eqref{eq:chi0} defines a ``collective'' Linhard function
\begin{equation}
        \chi_0^{\rm coll}(q,\omega) =
        \displaystyle \frac{2 t(q)}
        { (\hbar\omega+\I\eta)^2-
        \tF^2(q)}
\label{eq:Chi0Coll}
\end{equation}
and, consequently, to the collective approximation for the
density-density response function
\begin{equation}
        \chi^{\rm coll}(q,\omega) =
        \displaystyle \frac{2 t(q)}
        { (\hbar\omega+\I\eta)^2-
          t_{\rm F}^2(q)-
            2t(q)\tilde V_{\rm p-h}(q)} \,.
\label{eq:ChiColl}
\end{equation}

This approximation is also referred to as a ``one-pole
approximation'' or ``mean spherical approximation'.  Alternative
rationalizations of the collective approximation for the Lindhard
function may be found in Ref.~\citenum{fullbcs}.  In the case
\eqref{eq:ChiColl}, the frequency integration \eqref{eq:SRPA} can be
carried out analytically, which leads to the static structure function
Eq. (\ref{eq:FermiPPA}).

Low-lying excitations in a Fermi fluid are adequately described by Landau's
Fermi-Liquid theory \cite{LandauFLP1,LandauFLP2,BaymPethick}. The speed
of sound is
\begin{equation}
  mc_s^2 = mc_{\rm F}^{*2} + \tilde V_{\rm p-h}(0+) \equiv  mc_{\rm F}^{*2}(1+F_0^S)
\label{eq:FermimcfromVph}
\end{equation}
where $c_{\rm F}^{*2} = \frac{\hbar^2\KF^2}{3mm^*}$ is the speed of
sound of the non-interacting Fermi gas with the effective mass $m^*$,
and $F_0^s$ is Landau's Fermi liquid parameter.  Requiring a positive
compressibility leads to Landau's stability condition $F_0^s > -1$.
Solutions of the FHNC-EL equation exist if the expression under the
square root of Eq. (\ref{eq:FermiPPA}) is positive, which leads to the
stability condition $F_0^s > -4/3$. This result is a manifestation of
the fact that the wave function (\ref{eq:Jastrow}) is not exact. Of
course, the same {\em caveat\/} about the quantitative connection
between $\tilde V_{\rm p-h}(0+)$ and the hydrodynamic speed of sound
applies as in the Bose case: The relationship
\eqref{eq:FermimcfromVph} between $ mc_s^2$ as derived from the
equation of state by means of Eq. \eqref{eq:mcfromeos} and from
Eq. \eqref{eq:FermimcfromVph} is an identity only in an exact
evaluation of all quantities involved.

\subsubsection{Ladders}
\label{ssec:ladders}

The next objective is to clarify the relationship between the
coordinate-space equation determining the short--ranged structure of
the Jastrow-Feenberg correlations with the Bethe-Goldstone equation.
The purpose of the Bethe-Goldstone equation is to calculate the {\em
  minimal\/} set of diagrams in perturbation theory that must be
summed to generate a finite correction to the energy for hard-core
interactions.

We begin with the Bethe-Goldstone equation as formulated in
Eqs.~(2.1), (2.2) of Ref.~\citep{BetheGoldstone57}. For our purposes
the equation is rewritten in terms of the pair wave function $\psi$
using Eq. (2.9) of Ref.~\citep{BetheGoldstone57}. This leads to an
integral equation which we write in the form
\begin{equation}
  \left[t(k) + t(k') -t(h)-t(h')\right]\bra{\kvec,\kvec'}\psi-1\ket{\hvec,\hvec'}
  =- \bar n(k)\bar n(k')\bra{\kvec,\kvec'}v\psi
    \ket{\hvec,\hvec'}\,.\label{eq:fullpsi2}
\end{equation}

The pair wave function $\psi$ depends on both the
center of mass and the relative momentum of the interacting
particles. Originally, the problem was simplified by setting the
center of mass momentum to zero such that the pair wave function
depends only on the relative momentum.

\begin{equation}
\bra{\kvec,\kvec'}\psi\ket{\hvec,\hvec'} = \frac{1}{N}\tilde\psi(q)
\delta_{\kvec+\kvec',\hvec+\hvec'}\,.\label{eq:psilocal}
\end{equation}

As a consequence one gets, for a local interaction
\[
\bra{\kvec,\kvec'}v\psi\ket{\hvec,\hvec'} = \frac{1}{N}
\left[v(r)\psi(r)\right]^{\cal F}(q)\,\qquad\qvec=\kvec-\hvec,.
\]
To obtain an equation that leads to such a local pair wave function,
the kinetic energy coefficient $\left[t(k) + t(k') -t(h)-t(h')\right]$
must also be approximated by a function of momentum transfer $q$.
This can be accomplished by setting the center of mass momentum to zero.
In the same spirit, we may also apply the averaging procedure
(\ref{eq:favg}) to the above energy coefficient:
\begin{equation}
  t(|\hvec+\qvec|) -t(h)
  \approx \left\langle t(|\hvec+\qvec|) -t(h)
  \right\rangle(q) = \tF(q)\,.
\end{equation}
That gives, in coordinate space
\begin{equation}
  \left[-\frac{\hbar^2}{m}\nabla^2 + v(r)\right]\psi(r)
  = \left[2\tF(q)(\SF(q)-1)(\tilde\psi(q)-\delta(q))\right]^{\cal F}(r)\,.
  \label{eq:BGSchr}
\end{equation}

The pair wave function $\psi(\rvec)$ should be related to the direct
correlation functions by
\begin{equation}
  \left|\psi(\rvec)\right|^2 = 1+\Gamma_{\!\rm dd}(r)\label{eq:psiofr}\,.
\end{equation}

Returning to FHNC-EL, the manipulation of Eqs. \eqref{eq:FermiPPA},
\eqref{eq:FermiVph} and \eqref{eq:Fermiwind} leads to a slightly
different equation determining the short--ranged structure of the pair
wave function \cite{fullbcs}
\begin{equation}
  \psi^*(\rvec)\left[-\frac{\hbar^2}{2m}\nabla^2 + v(\rvec) + V_e(\rvec)
    + w_{\rm I}(\rvec)\right]
  \psi(\rvec)
  =\left[\tF(k)(\SF(k)-1)\hat\Gamma_{\!\rm dd}(k)\right]^{\cal F}(r)
  \label{eq:ELSchr}
\end{equation}
Eq.~(\ref{eq:BGSchr}) is similar to, but not identical with,
(\ref{eq:ELSchr}), which is obtained by the further assumption
$\psi^2(r)-1 \ll 1$. The identification between the two expressions
(\ref{eq:BGSchr}) and the coordinate-space form of the Euler equation
of the FHNC-EL method is not as precise as in the case of the ring
diagrams; but note that FHNC-EL//0 contains more than just
particle-particle ladders, also including particle-hole and hole-hole
ladders \cite{RIP79}.

It is often argued at this point that effects of the surrounding
medium can be incorporated by changing the spectrum $t(k)$ of
non-interacting fermions to a non-trivial single-particle spectrum
$e(k)$. Such corrections can be relatively easily implemented. They do
not solve the problem that the pair wave function is long-ranged;
there are different ways to enforce a finite ``healing distance''
\cite{BetheBrandowPetschek}. More thorough investigations (see Ref.
\citep{JWE93} and also the discussion in the introductory section
\ref{sec:genericeos}) came to the insight that other effects are much
more important, the induced interaction $w_I(r)$ guarantees, among others,
a finite healing of the correlation hole. We therefore take here a
pragmatic approach and simply take a non-interacting spectrum.

\subsubsection{Ladder Rungs}
\label{ssec:rungs}

The induced interaction  $w_{\rm I}(\rvec)$ entering
Eq. \eqref{eq:ELSchr} remains to be determined.
Following again Refs.~\citep{parquet1,parquet2}, we define a {\em
  local\/} effective interaction through the condition
(\ref{eq:Wlocal}),
\begin{eqnarray}
S(q) &=&\SF(q) - \widetilde W(q)\int_0^\infty \frac{d\hbar\omega}{\pi}\,
\Im  \chi_0^2(q,\omega) \,.
\label{eq:Scond}
 \end{eqnarray}
For further reference, let
\begin{equation} 
  \int_0^\infty \frac{d\hbar\omega}{\pi}\Im  \chi_0^2(q,\omega)
  \equiv \frac{\SF^3(q)}{t(q)\lambda(q)}\,;
   \label{eq:lambdadef}
\end{equation}
where $\lambda(q)=1$ if we replace $\chi(q,\omega) \approx \chi^{\rm coll}(q,\omega)$.
Then we have in the local approximation
\begin{equation}
  \widetilde W(q) = -\tF(q)\frac{S(q)-\SF(q)}{\SF^2(q)}\equiv
  -\tF(q)\tilde\Gamma_{\rm dd}(q)\label{eq:Wfermi}
\end{equation}
and from \eqref{eq:FermiPPA}
\begin{equation}
  \tilde w_{\rm I}(q)= -t(q)
\left[\frac{1}{\SF(q)}-\frac{1}{ S(q)}\right]^2
\left[\frac{S(q)}{\SF(q)}+\frac{1}{2}\right]
  \label{eq:Fermiwind}
\end{equation}
which is immediately seen to reduce to Eq. \eqref{eq:Bosewind}
for $\SF(q) = 1$. 

\subsubsection{Exchange corrections}
\label{ssec:exchanges}

\begin{figure}[H]
    \centerline{\includegraphics[width=0.8\columnwidth]{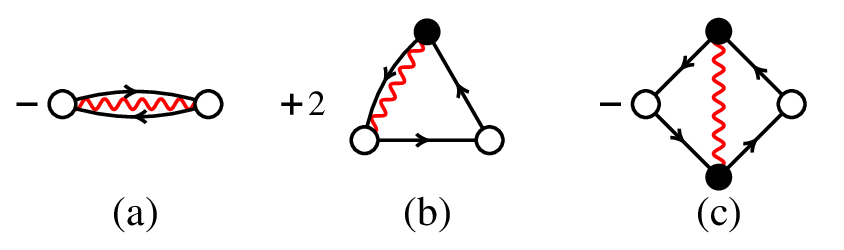}}
    \caption{The figure shows the diagrammatic representation of the
      lowest order exchange corrections $V_{\rm ee}(r)$ containing
      exactly one correlation line. We use the conventions of
      correlated wave functions \cite{Johnreview}: Dots represent
      particle coordinates, black dots imply a density factor and the
      integration over the coordinate. Lines connecting represent
      correlations or interactions: For the interaction correction
      $V_{\rm ee}(r)$, the red wavy line is to be interpreted as the
      effective interaction $W(r_{ij})$. In the correlation correction
      $X_{\rm ee}(r)$, the wavy red line represents the function
      $\Gamma_{\rm dd}(r)$. The oriented solid lines represent
      exchange function $\ell(r_{ij}\KF)$, Eq. \eqref{eq:eldef}.
       \label{fig:eelink}}
\end{figure}

Exchange diagrams have important consequences for the effective
interactions, not only in nucleonic systems but also in \he3 where they are
responsible for the paramagnon mode \cite{Encyclopedia}. They must be
included even at low densities to achieve consistency between the
energetics and the quasiparticle interaction \cite{fullbcs}.  The
simplest version of the FHNC hierarchy that corrects for this
deficiency is FHNC//1, which includes the sum of the three exchange
diagrams shown in Fig. \ref{fig:eelink}.

The corrections shown in Fig. \ref{fig:eelink} can be understood, in
the language of Goldstone perturbation theory, as a combination of
self-energy insertions and vertex corrections \cite{v4}.

There is a strong cancellation between the three diagrams indicating
that these terms should either all be kept or all be neglected.  This
has been demonstrated in the very early developments of the FHNC
technique \cite{Kro79}, we will demonstrate in section
\ref{ssec:exchange} that the same statement holds in the presently
much more elaborate computation.

The relevant modification from the full FHNC-EL equations as
formulated in Ref.~\citep{polish} involves keeping only the exchange
term $V_{\rm ee}(k)$.  The Euler equation becomes
\begin{equation}
     S(q) = \frac{\SF(q) + \tilde X_{\rm ee}(q)}
     {\sqrt{1+\displaystyle\frac{2\SF^2(q)}{t(q)}\tilde V_{\rm p-h}(q)}}\,.
       \label{eq:Stemp}
   \end{equation}
where the particle-hole interaction is modified by
\begin{equation}
  \tilde V_{\rm p-h}(q) \rightarrow \tilde V_{\rm p-h}(q) +
  \tilde V_{\rm ex}(q)\,,\quad \tilde V_{\rm ex}(q)\equiv
  \frac{\tilde V_{\rm ee}(q)}{\SF^2(q)}
  \label{eq:Vphexc}
\end{equation}
and where $X_{\rm ee}(r_{12})$ and $V_{\rm ee}(r_{12})$ are represented
topologically
by the sum of the three diagrams shown in Fig. \ref{fig:eelink},

We have shown in Ref.~\citep{fullbcs} that na\"\i ve addition of
exchange diagrams is problematic because it leads to an
incorrect low-density limit of the pair correlations.
We have rectified this situation by a slight modification of the
Euler equation, namely
\begin{equation}
 S(q) =
 \SF(q)\sqrt{\frac{1+\displaystyle\frac{2\SF^2(q)}{t(q)}\tilde V_{\rm
       ex}(q)} {1+\displaystyle\frac{2\SF^2(q)}{t(q)}\tilde
       V_{\rm p-h}(q)}}\,.
   \end{equation}
The square-root term in the numerator may be identified with a
``collective RPA'' expression for the exchange contribution to the
static structure function (for state-independent interactions this is
equal to the spin-structure function),
   \begin{equation}
     S_{\rm ex}(q) = \frac{\SF(q)}{\sqrt{1+\displaystyle\frac{2\SF^2(q)}{t(q)}
         \tilde V_{\rm ex}(q)}}\,,\label{eq:SsigmaColl}
   \end{equation}
The expression (\ref{eq:Stemp}) is then obtained by
expanding $S_{\rm ex}(q)$ to first order in the interactions and
identifying
   \[\tilde X_{\rm ee}(q) \approx -\frac{\SF^3(q)}{t(q)} \tilde V_{\rm ee}(q)\,.\]

We have commented above on the fact that, with the qualification that 
the Jastrow-Feenberg wave function is not exact, the positivity of 
the term under the square root in the denominator is related to the 
stability against density fluctuations. Likewise, the positivity of 
the numerator is connected with the stability against spin-density 
fluctuations.

In time-dependent Hartree-Fock theory \cite{ThoulessBook}, the
diagrams shown in Fig.~\ref{fig:eelink} correspond to the
particle-hole ladder diagrams, driven by the {\em exchange\/} term of
the particle-hole interaction
\begin{equation}
  W_{\rm ex}(\hvec,\hvec';\qvec)
  = \Omega\bra{\hvec+\qvec,\hvec'-\qvec} W \ket{\hvec',\hvec}\,.
  \label{eq:Wex}
\end{equation}
This non-local term supplements the RPA sum by the RPA-exchange (or
particle-hole ladder) summation.  The connection to the (local) FHNC
expression (\ref{eq:Vphexc}) is made by realizing that this expression
is obtained from the exact expression (\ref{eq:Wex}) by exactly the
same hole-state averaging process as was introduced in
Eq.~(\ref{eq:favg}):
\begin{equation}
  V_{\rm ex}(q) = \frac{\tilde V_{\rm ee}(q)}{S_{\rm F}^2(q)}
  = \left\langle  W_{\rm ex}(\hvec,\hvec';\qvec)\right\rangle(q)\,.
\end{equation}

\subsubsection{Energy\label{ssec:fenergy}}

In calculating the energy, we can again simply follow the analysis of
Smith and Jackson, inserting exchange corrections where appropriate.
We must keep in mind that there is no finite truncation scheme of the
FHNC equations such that acceptable expressions for the pair
distribution function and the static structure function are the
Fourier transforms of each other.  That is, having obtained an
optimized static structure function $S(q)$, one must construct the
pair distribution function $g(r)$ by appropriate combination of
correlation diagrams and exchanges. In the case of state-independent
correlations, the simplest expression for $g(r)$ is
\begin{eqnarray}
  g(r) &=& \left[1+\Gamma_{\!\rm dd}(r)\right]\left[g_{\rm F}(r) + C(r)\right]\\
  \tilde C(q) &=& \left[\SF^2(q)-1\right]\tilde \Gamma_{\!\rm dd}(q)
  + (\Delta \tilde X_{\rm ee})(q)\,.
  \label{eq:gofrFHNC}
\end{eqnarray}
where $g_{\rm F}(r) = 1-\frac{1}{2}\ell^2(r\KF)$ is the pair
distribution function of non-interacting fermions. In the FHNC//1
approximation, $\SF(q)$ is replaced by $\SF(q) + \tilde X_{\rm
  ee}(q)$, and $(\Delta \tilde X_{\rm ee})(q)$, which is represented
by the sum of diagram (b) and (c) shown in Fig.~\ref{fig:eelink} is
added to $\tilde C(q)$. Summarizing, we obtain
\begin{eqnarray}
  \frac{E}{N}
  &=&\frac{T_{\rm F}}{N} + \frac{E_{\rm R}}{N} +
  \frac{E_{\rm Q}}{N}
\,,\nonumber \\
\frac{E_{\rm R}}{N} &=& \frac{\rho }{2}\int\! d^3r\>
\bigl[g_{\rm F}(r) + C(r)\bigr]\biggl[(1+\Gamma_{\!\rm dd}(r))v(r)\nonumber\\
 && \qquad + \frac{\hbar^2}{m}\left|\nabla\sqrt{1+\Gamma_{\!\rm dd}(r)}\right|^2\biggr]\,,
\label{eq:ER}\\
\frac{E_{\rm Q}}{N} &=& \frac{1}{4}\int\!\frac{d^3q}{(2\pi)^3\rho}\>
t(q)\tilde\Gamma_{\!\rm dd}^2(q)\left[S^2_{\rm F}(q)/S(q)-1\,\right]\,,
\nonumber\\
\label{eq:EQ}
\end{eqnarray}
where $T_{\rm F}$ is the kinetic energy of the non-interacting Fermi gas.

\section{Variational and parquet theory for realistic nuclear interactions}
\label{sec:nucleons}

\subsection{The problem of variational wave functions}

Let us now turn to realistic nuclear systems that are described by
interactions of the form \eqref{eq:vop}. A plausible variational wave
function is then written in the form \eqref{eq:f_prodwave} with
\eqref{eq:fop}. We have already commented on the problem that the
symmetrization of the correlation operator generates ``commutator
diagrams''.  The most relevant example to see how commutator
contributions affect the evaluation of system properties is the potential
energy, which can be written in the form
\begin{equation}
  \frac{\left\langle V \right\rangle}{N} =
    \frac{\rho}{2}\int d^3r \sum_\alpha g_\alpha(r) v_\alpha(r)
    \frac{1}{\nu^2}\Tr_{12}\, O_\alpha^2(1,2)\,, 
\label{eq:epot}
\end{equation}
where $\nu$ is the degree of degeneracy of the
single-particle states, $\Tr_{12}$ indicates the trace over spin
(and, when applicable, isospin) variables of particles 1 and 2.
The pair distribution function now also depends on the operators
\eqref{eq:Vop1}:

\begin{equation} \rho^2 g_\alpha(|\rvec-\rvec'|)=
  \frac{\Bra{\Psi_0^{{\rm
  SOP}}}\sum_{i<j}\delta(\rvec-\rvec_i)\delta(\rvec'-\rvec_j) \hat
  O_\alpha(i,j)\ket{\Psi_0^{{\rm
  SOP}}}}{\displaystyle\frac{1}{\nu^2}\Tr_{12}\, \hat O_\alpha^2(1,2)
  \Ovlp{\Psi_0^{{\rm SOP}}}{\Psi_0^{{\rm SOP}}}}\,.
\label{eq:gop}
\end{equation}
When the symmetrization is carried out, the pair
distribution functions have the form
\begin{equation} g_\alpha(r)
  =\sum_{\beta\gamma} f_\beta(r)
  f_\gamma(r)F^{(\alpha)}_{\beta\gamma}(r)\,.
\label{eq:gsymmetrized}
\end{equation}
where the $F^{(\alpha)}_{\beta\gamma}(r)$ are coupling coefficients
that are functionals of the correlation functions $f_\alpha(r)$.  The
analytic structure of the $F^{(\alpha)}_{\beta\gamma}(r)$ is
complicated and no summation exists that comes anywhere close to the
diagrammatic richness of the (F)HNC summations for state-independent
correlations. Fortunately, to make our point, one does not need to go
into the gory details of calculating these coefficients. The only
important feature for our discussion is that they are non-diagonal in
the operator coefficients $\alpha$, $\beta$, and
$\gamma$. Specifically, the spin-triplet pair distribution function
has a component proportional to the spin-singlet correlation function
and vice versa.  To highlight the relevance of this, we have shown in
Fig. \ref{fig:vbare} the singlet and triplet components of our four
interaction models.  Common to all of them is the strongly repulsive
core in spin-triplet states and the much smaller core as well as the
attractive well in spin-singlet states.  Hence, the effect of
commutator diagrams can be very dramatic if the correlation functions
are determined, for example, from a low-order Schr\"odinger-like
equation because a correlation function derived from a singlet
interaction would multiply the triplet potential and vice versa.

The relationship \eqref{eq:gsymmetrized} displays the clear message
that the short-ranged structure of the pair distribution function in
each spin channel is not just determined by the short-ranged structure
of the bare interaction in the same spin-channel, but rather by a mixture
of all spin channels. That message raises several questions:
\begin{itemize}
\item{} It must, of course, be established that the effect is not just
  an artifact of the symmetrized operator product wave function
  \eqref{eq:f_prodwave}.
\item{} Once this is established, the {\em physical mechanism\/} causing
  the coupling of different spin states must be clarified.
\item{} Finally, a practical way to include these effects in a
    many-body calculation must be established.
\end{itemize}

Light on the matter can be shed by going back to the work of Smith and
Jackson \cite{SmithSpin} on their localized parquet-diagram summations
for a fictive system of bosonic nucleons interacting via a $v_6$
interaction. It turned out that the equations derived in that work are
identical to those one would obtain in a bosonic version of the
summation method of Fantoni and Rosati \cite{FantoniSpins}, which
simply ignored the fact that the individual operators
$\hat O(i,j)$ in the correlation operator \eqref{eq:f_prodwave} do not
commute.

The conclusion of this comparison is that the problem of commutator
corrections does not go away in the parquet-diagram summations, but it
suggests that {\em commutator contributions are to be identified with
  ``beyond parquet'' diagrams.\/}

Looking at non-parquet contributions in terms of Goldstone diagrams
offers a very plausible interpretation of ``commutator diagrams'' and
the physical mechanisms they describe. For the purpose of discussion,
let us consider a simplified example of a weak interaction of the
momentum-space form
\begin{equation}
  \hat v(q) = \tilde v_{(c)}(q) + \tilde v_{(\sigma)}(q){\bsigma}_i \cdot {\bsigma}_j
  = \tilde v_s(q)P_s + \tilde v_t(q)P_t
  \label{eq:v2def}
  \end{equation}
where $P_s=(1-{\bsigma}_i \cdot {\bsigma}_j)/4$ and
$P_t=(3+{\bsigma}_i \cdot {\bsigma}_j)/4$ are the projection operators
on spin-singlet and spin-triplet states, respectively.

\begin{figure}[H]
  \centering
    \includegraphics[width=0.95\columnwidth]{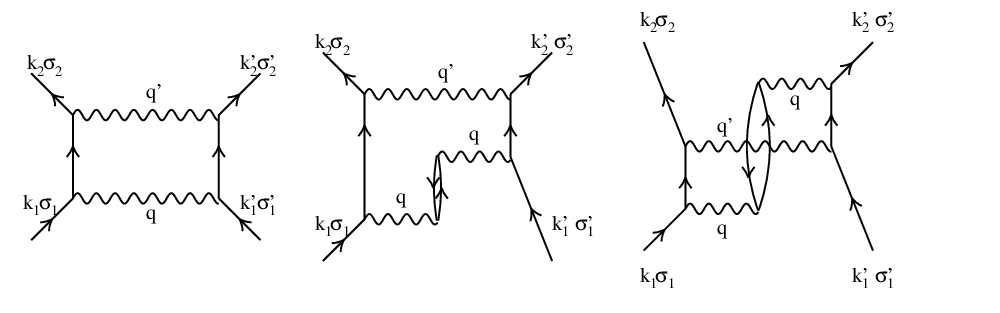}
    \caption{The figure shows the simplest second-order
      ladder diagrams including a ``twisted chain'' correction. The
      left diagram is the ordinary two-body ladder that is summed by
      the Bethe-Goldstone equation. The middle diagram is where one of
      the bare interactions is replaced by $\tilde w_I(q)$, and the
      right one is the simplest contribution to the totally
      irreducible interaction.}\label{fig:ladders}
\end{figure}

We show in Fig. \ref{fig:ladders} the simplest possibility.  The
leftmost diagram is the ordinary second-order ladder diagram as summed
by the Bethe-Goldstone equation. The middle diagram shows the simplest
case where the bare interaction is replaced by the first term of the
induced interaction or, in practice, its local approximation $\hat
w_I(q)$.
\begin{equation}
  \hat w_2(q,\omega)=\tilde v_{(c)}^2(q)\chi_0(q,\omega) +
  \tilde v_{(\sigma)}^2(q)\chi_0(q,\omega){\bsigma}_i \cdot {\bsigma}_j
  \rightarrow  \hat w_2(q)\label{eq:wind2}\,.
\end{equation}
In both cases, if a pair of particles enters with quantum numbers
$\ket{\kvec_1,\kvec_1',S}$ and in a specific spin (singlet or triplet)
configuration {\em then it remains in that spin configuration
  throughout the process.}

The third diagram has the same components, the energy denominators
are, of course, different. It is by definition not a parquet diagram
and represents the simplest contribution to the totally irreducible
interaction $\hat V_e(r)$. 

Working out all possible spin configurations one obtains for the third
diagram the two terms
\begin{subequations} \label{eq:vdiag3}
\begin{eqnarray}
  & & \int \frac{d^3q}{(2\pi)^3\rho}
  \frac{1}{E(\kvec,\qvec)}\left[V_s(q)W_s(\kvec-\qvec)P_s+V_t(q)W_t(\kvec-\qvec)P_t\right]
  \label{eq:vI2}\\
  &+& 4\int \frac{d^3q}{(2\pi)^3\rho}\frac{1}{E(\kvec,\qvec)}
  V_{(\sigma)}(q)W_{(\sigma)}(\kvec-\qvec){\bsigma}_i \cdot {\bsigma}_j
\label{eq:vline2}
\end{eqnarray}
\end{subequations} 
where $E(\kvec,\qvec)$ is the energy denominator appropriate for the
process.

We have written Eq. \eqref{eq:vdiag3} in a suggestive form: 
The first line in the expression, \eqref{eq:vI2} is the second diagram of
Fig. \ref{fig:ladders}. The third diagram represents the sum
of both terms, \eqref{eq:vI2} {\em plus\/} \eqref{eq:vline2}.
The second line, \eqref{eq:vline2}, should therefore be identified
with the commutator of the two diagrams. It carries our message: If the
interaction has a spin-component, then this term will act in both the
spin-singlet and the spin-triplet case. If the interaction in the
spin-singlet and the spin-triplet channel are very different, (see the
left pane of Fig. \ref{fig:vbare}) the effect can be large.

We proceed now with the implementation of the ideas of local parquet
diagrams for a $v_8$ interaction.

\subsection{Parquet diagram summations for state-dependent interactions}
\label{ssec:parquet}
We generalize here the analysis of the previous section to
state-dependent interactions of the form \eqref{eq:vop}. Implementing
a regular parquet summation at the level of a $v_6$ interaction within
the restricted set of exchange diagrams discussed in
Sec. \ref{ssec:exchanges} is relatively straightforward \cite{v3eos}.
All quantities introduced in section \ref{sec:central} are replaced by
linear combinations of radial functions times the first six operators
$O_\alpha(i,j,\hat\rvec_{ij})$ spelled out in Eqs. \eqref{eq:Vop1},
for example
\begin{equation}
  V_{\rm p-h}(r_{ij}) \rightarrow \hat V_{\rm p-h}(i,j)= \sum_{\alpha=1}^n
  V_{\rm p-h}^{(\alpha)}(r_{ij})O_\alpha(i,j,\hat \rvec_{ij})\label{eq:Vphop}
\end{equation}
or
\begin{equation}
  \Gamma_{\rm dd}(r_{ij}) \rightarrow \hat \Gamma_{\rm dd}(i,j)= \sum_{\alpha=1}^n
  \Gamma_{\rm dd}^{(\alpha)}(r_{ij})O_\alpha(i,j,\hat \rvec_{ij})\,.\label{eq:Gddop}
\end{equation}

Interactions containing operators with relative momentum dependence
such as the spin-orbit interaction require additional work. We follow
a popular assumption that the particle-hole interaction $\hat V_{\rm
  p-h}$ has the same operator structure as the $v_8$ interaction
\cite{JPG41_2014}-\nocite{PhysRep536,AnnPhys214,NPA627,NPA658,%
  NPA658,PhysRevC.80.024314,PhysRevC.84.059904,PhysRevC.89.044302}\cite{PhysRevC.100.064301}. We
stress that this assumption is consistent with the localization
procedure \eqref{eq:favg} and necessary to make the summation of the
parquet diagrams practical. It should be valid when {\em average\/}
quantities are of interest, but there is no guarantee that this
assumption is generally justified: The most general particle-hole
interaction includes terms that explicitly depend on the
center-of-mass momentum and that can couple different total spin
states. We will also show below that the summation of chain diagrams
with a spin-orbit interaction leads to terms that couple spin-singlet
and spin-triplet states, see Eqs. \eqref{eq:VLSp}.

For the summation of chain diagrams, it is convenient to introduce the
``longitudinal'' and ``transverse'' operators \cite{FNN82}. We need a
few additional operators when the spin-orbit interaction is included:
\begin{subequations}
\label{subeq:Qalpha} 
\begin{eqnarray}
 \hat Q_1 &\equiv&  \1\,,\label{eq:Qdefa}\\
 \hat Q_3 &\equiv& \hat L \equiv \sqij12\,,\label{eq:Qdefb}\\
 \hat Q_5 &\equiv&  \hat T \equiv \sij12-\sqij12\,.\label{eq:Qdef}\label{eq:Qdefc}\\
 \hat Q_7 &\equiv& \frac{1}{\KF^2}\left[(\hat\qvec\times\hvec)\cdot\bsigma\right]
  \left[(\hat\qvec\times\hvec')\cdot\bsigma'\right]\,,\label{eq:Qdef7}\\
  \hat Q_9 &\equiv& \frac{2}{\KF^2}
  (\hat\qvec\times\hvec)\cdot(\hat\qvec\times\hvec')\,\label{eq:Qdef9}\\
  \LS &\equiv&
  \frac{\I}{\KF}\hat
  \qvec\times(\hvec-\hvec')\cdot\S\label{eq:LSdef}\,,\\
  \LSp &\equiv& \frac{\I}{2\KF}\hat\qvec\times(\hvec+\hvec')
\cdot(\bsigma-\bsigma')\label{eq:VLSp} \,.
\end{eqnarray}
\end{subequations}
We have here defined the spin-orbit operator $\LS$ and its
antisymmetric-spin counterpart $\LSp$ dimensionless to keep the
consistency with other operators. These act only on the spin variables
and depend parametrically on the direction of momentum transfer $\hat
\qvec$ as well as the relative hole wave numbers
$\Delta\hvec\equiv\hvec-\hvec' $. The antisymmetric operator $\LSp$ is
the only term in the effective interaction that couples spin-singlet
and spin-triplet states and could be interesting in special physical
systems.

The ladder diagrams, on the other hand, are best formulated in terms
of the projection operators
\begin{align}
  \hat P_{s\phantom{+}} &\equiv \frac{1}{4}\left(\1-{\bm\sigma}_1\cdot{\bm\sigma}_2\right)\,,
  \nonumber\\
  \hat P_{t+} &\equiv \frac{1}{6}\left(3\1+\sij12+S_{12}(\hat\rvec)
  \right)\,,\label{eq:projectors}\\
  \hat P_{t-} &\equiv \frac{1}{12}\left(3\1+\sij12-2S_{12}(\hat\rvec)
  \right)\,.
  \nonumber
  \end{align}

These satisfy the relations $\hat P_i \hat P_j = \hat P_i\delta_{ij}$
and $\hat P_s + \hat P_{t+} + \hat P_{t-} = \1$. For a $v_8$
interaction we need additionally the spin-orbit operator in
\eqref{eq:Vop1}.

\subsubsection{Ring diagrams\label{ssec:phG}}

Particle-hole matrix elements are best calculated in the operator basis
$\{\1,{\bm\sigma}_1\cdot\bm{\sigma}_2,S_{12}(\hat\rvec),  \LS\}$
\begin{equation}
    \bra{\hvec+\qvec,\hvec'-\qvec}v_\alpha(1,2)O_\alpha(1,2)
    \ket{\hvec,\hvec'}= \frac{1}{N}\tilde v_{\alpha}(q) O_\alpha(q)\,.
\end{equation}
For a $v_8$ interaction of the form \eqref{eq:vop}, the matrix elements
of particle-hole interaction in the different channels are 
\begin{equation}
  \tilde v_{\alpha}(q) =\begin{cases}
  \phantom{-}\rho \int d^3r v_{\alpha}(r)
  j_0(qr)&\ \mathrm{for}\ \alpha = 1\ldots 4,\\
   -\rho \int d^3r v_{\alpha}
   (r)j_2(qr)&\ \mathrm{for}\ \alpha = 5,6\,.\\
   \phantom{-}\frac{\rho}{2}\int d^3r v_{\alpha}(r) r\KF j_1(qr)&\ \mathrm{for}\ \alpha = 7,8\,.
   \end{cases}
\end{equation}
$\tilde v_{\rm (LS)}(q)$ is defined such that has the dimension of an
energy. Recall also that we have defined the operator $\LS$
dimensionless; as a consequence $\tilde V_{\rm p-h}^{\rm (LS)}(q)$ has
the dimension of an energy.  The momentum space representation of the
tensor operator is obtained by replacing $\hat r_{ij}\rightarrow\hat
\qvec$.  From here on, we shall generally mean the momentum space
representations \eqref{eq:Qdef}, \eqref{eq:LSdef} when we refer to the
operators $\hat O_\alpha$, $\hat Q_\alpha$ or $\LS$.

The basic operation of the summation of the ring diagrams is
to evaluate the convolution product of two operators
$\hat A \equiv \hat A(\qvec,\hvec,\hvec',\bsigma,\bsigma')$ and
$\hat B\equiv \hat B(\qvec,\hvec,\hvec',\bsigma,\bsigma')$ 
which is defined as
\begin{eqnarray}
  &&\left[\hat A*\chi_0*\hat B\right](\qvec,\hvec,\hvec',\bsigma,\bsigma')
  \nonumber\\
  &=&\frac{1}{N}\Tr_{\bsigma''} \sum_{\hvec''}
  \hat A(\qvec,\hvec,\hvec'',\bsigma,\bsigma'')\chi_0(\qvec,\hvec'';\omega)
  \hat B(\qvec,\hvec'',\hvec',\bsigma'',\bsigma')\,,
  \label{eq:convol}
\end{eqnarray}
where
\begin{equation}
  \chi_0(\qvec,\hvec;\omega) =\frac{2(t(p)-t(h))}
      {(\hbar\omega-\I\eta)^2-(t(p)-t(h))^2}\,\qquad\pvec=\hvec+\qvec
      \label{eq:Linhard0}\,.
      \end{equation}

The operators $\{\hat Q_\alpha\}$ satisfy the convolution properties
\begin{equation}
  \left[\hat Q_\alpha*\chi_0*\hat Q_\beta\right] =
  \begin{cases}\phantom{\frac{1}{2}}\chi_0(q;\omega)
    \hat Q_\alpha\delta_{\alpha\beta} &\ \text{for}\ \alpha =
    1,3,5,\\ \phantom{\frac{1}{2}}\chi_0^{(\perp)}(q;\omega)\hat
    Q_\alpha\delta_{\alpha\beta}&\ \text{for}\ \alpha =
    7,9\,,
  \end{cases}\label{eq:Qortho}
\end{equation}
where we introduced the ``transverse Lindhard function''
\begin{equation}
  \chi_0^{(\perp)}(q;\omega)=\frac{1}{N}\Tr_\bsigma\sum_\hvec
  \frac{|\hat \qvec\times\hvec|^2}{\KF^2}
  \chi_0(\qvec,\hvec;\omega)\,.
  \label{eq:chi0trans}
\end{equation}
The explicit form of $\chi_0^{(\perp)}(q;\omega)$ is given in Appendix
of Ref, \citep{v4}.

Due to the convolution properties \eqref{eq:Qortho}, one can get in
the basis defined by the operators $\{Q_\alpha\}$ the response
functions and the static structure functions of the form
\eqref{eq:SRPA} in each operator channel for the first six operators.
This is no longer the case if the spin-orbit interaction is included
since the $\LS$ term mixes different channels.

The manipulations to derive an effective interaction including a
spin-orbit potential are rather involved; they have been carried out
in Ref. \citep{v4}.  In terms of the operators \eqref{eq:Qdef},
\eqref{eq:LSdef} and \eqref{eq:VLSp}, the effective interaction
consists of two contributions: One that contains only even powers of
the spin-orbit interaction and one that contains odd powers.  For a
compact representation, we define {\em energy-dependent\/}
particle-hole interactions in the central and the transverse channel
\begin{subequations}
\begin{eqnarray}
  \tilde V_{\rm p-h}^{\rm (c)}(q;\omega)&\equiv& \tilde V_{\rm p-h}^{\rm (c)}(q)+
  \frac{1}{4}\chi_0^{(\perp)}(q;\omega)\left[\VLSq(q)\right]^2
  \,,\label{eq:Vcredef}\\
  \tilde V_{\rm p-h}^{\rm (T)}(q;\omega)&\equiv&
  \tilde V_{\rm p-h}^{\rm (T)}(q)+
  \frac{1}{8}\chi_0^{(\perp)}(q;\omega)\left[\VLSq(q)\right]^2
  \,.\label{eq:VTredef}
\end{eqnarray}
\end{subequations}

The longitudinal component $\tilde V_{\rm p-h}^{\rm (L)}(q)$ does not
couple to the spin-orbit operator, it has therefore no
energy-dependent correction. We nevertheless define, for the purpose
of a symmetric notation, $\tilde V_{\rm p-h}^{\rm (L)}(q;\omega)\equiv
\tilde V_{\rm p-h}^{\rm (L)}(q)$.

The effective interactions for $\alpha=1\,\dots 6$ can contain only
even numbers of spin-orbit operators, we can write these as
\begin{equation}
  \hat W^{(\rm even)}(\qvec,\hvec,\hvec',\bsigma,\bsigma';\omega) =
  \sum_{\substack{\alpha\,{\rm odd}}}^9 \tilde W^{(\alpha)}(q;\omega)\hat Q_\alpha
\end{equation}
with
\begin{equation}
  \tilde W^{(\alpha)}(q;\omega)
  = \frac{\tilde V_{\rm p-h}^{(\alpha)}(q;\omega)}
  {1-\chi_0(q;\omega)\tilde V_{\rm p-h}^{(\alpha)}(q;\omega)}
  \label{eq:W135}
\end{equation}
for $\alpha=1,\ldots, 5$ and
  \begin{eqnarray}
    \tilde W^{(7)}(q;\omega) &=&\frac{1}{4}\frac{\left[\VLSq(q)\right]^2\chi_0(q;\omega)}{1-\chi_0(q;\omega)\tilde V_{\rm p-h}^{\rm (c)}(q;\omega)}\,,\\
    \tilde W^{(9)}(q;\omega) &=&\frac{1}{8}\frac{\left[\VLSq(q)\right]^2\chi_0(q;\omega)}
   {1-\chi_0(q;\omega)\tilde V_{\rm p-h}^{\rm (T)}(q;\omega)}\,.\label{eq:W79}
   \end{eqnarray}
  For odd numbers of spin-orbit operators we get
\begin{subequations}
\begin{align}
\hat W_{\rm LS}^{(\rm odd)}(\qvec;\omega)&=W^{\rm (LS)}(\qvec,\omega) \LS +
W^{(\rm LS')}(\qvec,\omega) \LSp\\
W^{(\rm LS)}(\qvec,\omega)  &= \frac{1}{2}
  \frac{\VLSq(q)}{1-\chi_0(q;\omega)\tilde V_{\rm p-h}^{\rm (c)}(q;\omega)}
  + \frac{1}{2}\frac{\VLSq(q)}{1-\chi_0(q;\omega)\tilde V_{\rm p-h}^{\rm (T)}(q;\omega)}\label{eq:VLSlocal}\\
  W^{(\rm LS')}(\qvec,\omega) &=
  \frac{1}{2}
  \frac{\VLSq(q)}{1-\chi_0(q;\omega)\tilde V_{\rm p-h}^{\rm (T)}(q;\omega)}
  - \frac{1}{2}\frac{\VLSq(q)}{1-\chi_0(q;\omega)\tilde V_{\rm p-h}^{\rm (c)}(q;\omega)}
 \label{eq:VLSodd}\,,
\end{align}
\end{subequations}

\subsubsection{The particle-hole interaction and ladder diagrams\label{ssec:ladderVph}}

For a $v_6$ interaction, the particle-hole interaction and the
parquet/Euler equation is the same as Eq. \eqref{eq:ELSchr}, only all
functions must be re-interpreted as operators, {\em
  cf. Eqs. \eqref{eq:Vphop} and \eqref{eq:Gddop}}. The tensor
component of the direct correlation function $\hat\Gamma_{\!\rm
  dd}(\rvec)$ is angular dependent, we must keep this angular
dependence in the evaluation of the kinetic energy term.  Defining in
the projector basis $\psi_\alpha(r) = \sqrt{1+\Gamma_{\!\rm
    dd}^{(\alpha)}(\rvec)}$ (see Eq. \eqref{eq:psiofr}) we get for the
kinetic energy term
\begin{equation}
  \left|\nabla \sum_\alpha \psi_\alpha(r)P_\alpha\right|^2=
  \sum_{\alpha} \left|\frac{d \psi_\alpha(r)}{dr}\right|^2\hat P_\alpha
  + \frac{36}{r^2}\psi_S^2(r)\left[\hat P_{t+} + 2
  \hat P_{t-}\right]
\label{eq:TensorKin}
\end{equation}
where $\psi_S(r) = (\psi_{t+}(r)-\psi_{t-}(r))$ is the component of
$\hat\psi(\rvec)$ in the tensor channel.
The particle--hole interaction \eqref{eq:FermiVph} then simply becomes
an operator
\begin{equation}
  \hat V_{\rm p-h}(r) = \hat\psi^*(r)
  \left[\hat v(r) +  \hat V_e(r)\right]\hat\psi(r)
  + \frac{\hbar^2}{m}\left|\nabla\hat\psi  (r)\right|^2
  + \hat \Gamma_{\!\rm dd}(r)\hat w_{\rm I}(r)\,.\label{eq:FermiVphhat}
\end{equation}
which separates into three independent components in the projector basis.

In the case of a $v_8$ interaction, the spin-orbit operator also
contributes to the particle-hole interaction though the bare interaction
\[\psi^*(\rvec) v_{\rm LS}(r){\bf L}\!\cdot\!{\bf S}\psi(\rvec)\]
  and from the spin-orbit component of the induced interaction
  \[\psi^*(\rvec)w_I^{\LSsup}(r){\bf L}\!\cdot\!{\bf S}
  \psi(\rvec)-w_I^{\LSsup}(r){\bf L}\!\cdot\!{\bf S}\,\]
The spin-orbit operator commutes with a spherically symmetric wave
function, hence we can write this term as
\begin{equation}
  V_{\rm p-h}^{\LSsup}(r) =
  \left[\psi^*(r)v_{\rm LS}(r)\psi(r)+w_{\rm I}^{\LSsup}(r)\Gamma_{\rm dd}(r)\right]
  \,.\label{eq:VphLS}
\end{equation}

A second contribution to the particle-hole interaction in the spin-orbit
channel comes from the modification of the wave function due to a
spin-orbit term in Eq. \eqref{eq:ELSchr}. That correction goes to zero
for both $r\rightarrow 0$ and $r\rightarrow \infty$
\cite{OBI76,Tafrihi2015}.  Considering the fact that, as we shall see,
the spin-orbit corrections are small anyway, we follow here the
strategy of Ref. \citep{OBI76} and disregard this term.

\subsubsection{Twisted chains corrections to the Bethe Goldstone equation
  \label{ssec:twist}}

The $v_6$ version of the particle-hole interaction maintains its
operator structure; some of its components have an energy-dependent
correction \eqref{eq:Vcredef} and \eqref{eq:VTredef} originating from
the spin-orbit potential which we will show to be small. The physical
mechanism of the twisted-chain corrections is to mix the spin-singlet
and the spin-triplet channels. On the other hand, the spin-orbit
operator only applies to the spin-triplet channel; thus we can focus
on the $v_6$ operator structure of the interactions.

We are now ready to develop a method that includes these processes
systematically.  The analysis in connection with
Fig. \ref{fig:ladders} assumed that the interaction are weak such that
perturbation theory is applicable.  To deal with strong, short-ranged
interactions we need to carry out at least summations of ladder
diagrams. We treat these terms by taking their topologies from
perturbation theory, and using the paradigms of Jastrow-Feenberg
theory for their evaluation.

\begin{itemize}
\item Local parquet theory replaces the energy-dependent induced
  interaction $\hat w_I(q,\hbar\omega)$ by a local function $\hat
  w_I(q)$. Then, chains carrying momentum $\qvec$ are the same except
  the energy denominators of the processes shown in Fig. \ref{fig:ladders}(middle) and Fig.
  \ref{fig:ladders}(right).
\item The bare interaction $\hat v(q)$ always comes along with the
  induced interaction $\hat w_I(q)$; we represent this combination 
  by a magenta wavy line.
\item The ``cross-going'' process must contain at least the chain of two
  interactions, {\em i.e.\/} it contains only the induced interaction
  $\hat w_I(q)$, we represent this by a blue wavy line.
\item Both the effective interaction lines $\hat v(q)+\hat w_I(q)$
  as well as the induced interaction lines $\hat w_I(q)$ can be parallel
  connected, the sums are solutions of a Bethe-Goldstone equation.
\end{itemize}

\begin{figure}[H]
  \centering
    \includegraphics[width=0.95\columnwidth]{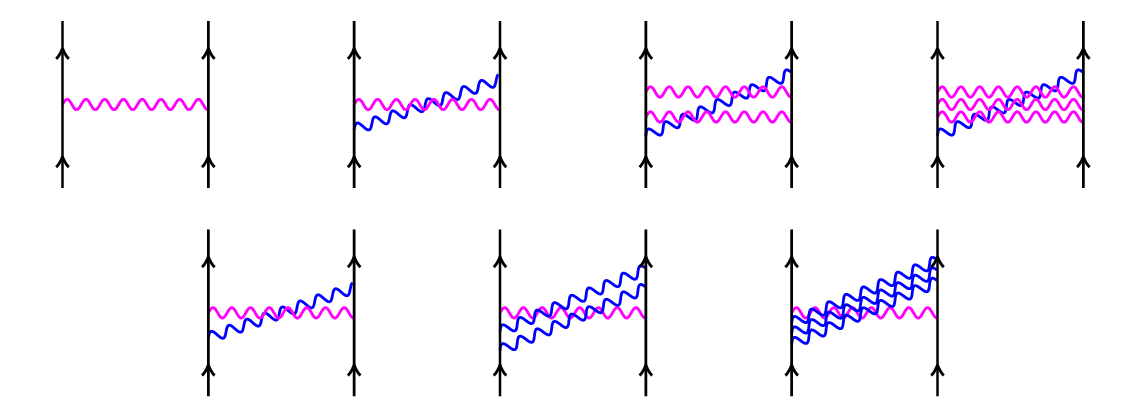}
    \caption{(color online) Examples of the diagrams summed by the
      integral equation
      \eqref{eq:vi}. The magenta wavy line represents the combination
      $\hat v(r)+\hat w_{\rm I}(r)$ and the blue line represents the
      induced interaction $\hat w_{\rm I}(r)$.  The magenta rungs can all
      be summed to the $\hat G(r)$-matrix whereas the blue rungs sum to
      $\hat G_w(r)$}  \label{fig:vn_twist}
 \end{figure}

Skipping the technical details which have been worked out in
Ref. \citep{v3twist}, we show in Figs. \ref{fig:vn_twist} the types
of diagrams that are summed that way.
The irreducible interaction is then obtained
by solving the integral equation
\begin{eqnarray}
    \hat V_e(\qvec)&=&
    -\frac{1}{2}\sum_{\alpha,\beta}
  \int \frac{d^3q'}{(2\pi)^3\rho}\left[\tilde G^{(\alpha)}(|\qvec-\qvec'|)
  -\tilde V_e^{(\alpha)}(|\qvec-\qvec'|)\right]\times\nonumber
  \\&&\phantom{-\frac{1}{2}}\times\frac{\tilde G_w^{(\beta)}(q')}{2\tF(q')}
  \times\Tr_1
  \left[\hat O_\beta(a,1)\left[\hat O_\alpha(a,b),
        \hat O_\beta(1,b)\right]\right]\,,\label{eq:vi}
\end{eqnarray}
where $\tilde G(\qvec)$ is the Fourier transform of $G(\rvec)$.  These
corrections are important when the interactions in the spin-singlet
and the spin-triplet case are very different which is the case for all
the interactions studied here, see Fig. \ref{fig:vbare}.

\subsubsection{Exchange corrections and the energy\label{ssec:echx}}

For state-dependent correlations and interactions, we can simply go
back to the definition (\ref{eq:Wex}) and interpret the interaction
$W(r)$ as an operator of the form (\ref{eq:Vop1}). The calculation for
the central and spin components go exactly as before.  The tensor
component needs special treatment which has been outlined in the
section \ref{ssec:exchanges}.

In the state-dependent case, $g(r)$ becomes an operator
in spin-space,
\begin{eqnarray}
  \hat g(r) &=& \left[1+\Gamma_{\!\rm dd}^{(s)}(r)\right]
    \left[1+C_s(r)-\ell^2(r\KF)\right]\hat P_s\nonumber\\
    &+& \left[1+\Gamma_{\!\rm dd}^{(t+)}(r)\right]
    \left[1+C_{t+}(r)+\ell^2(r\KF)\right]\hat P_{t+}\nonumber\\
    &+&\left[1+\Gamma_{\!\rm dd}^{(t-)}(r)\right]
    \left[1+C_{t-}(r)+\ell^2(r\KF)\right]\hat P_{t-}\nonumber\\
    &\equiv& g_s(r) \hat P_s + g_{t+}(r)\hat P_{t+}(r) + g_{t-}(r)\hat P_{t-}(r)
\,
\label{eq:gasmatrix}
\end{eqnarray}
with which we obtain the potential energy
\begin{eqnarray}
  &&\frac{\left\langle \hat V \right\rangle}{N}
  = \frac{\rho}{2}\Tr\int d^3r \hat v(r)\hat g(r)\nonumber\\
  &=&\frac{\rho}{4}\int d^3r \left[v_s(r) g_s(r) + 2v_{t+}(r)g_{t+}(r)
   + v_{t-}(r)g_{t-}(r)\right]\nonumber
\end{eqnarray}
The kinetic energy term in Eq.~(\ref{eq:ER}) is generalized to 
state-dependent correlations in the same way, without the 
$[1+\Gamma_{\!\rm dd}^{(\alpha)}(r)]$ factors.  Note, of course, 
that we need to keep the kinetic-energy correction spelled out in
Eq.~\eqref{eq:TensorKin}. Finally, the term $E_{\rm Q}$ is generalized to
\begin{eqnarray}
  \frac{E_{\rm Q}}{N} &=& \frac{1}{4}\int \frac{d^3q}{(2\pi)^3\rho} t(q)
  \sum_\alpha (\tilde\Gamma_{\!\rm dd}^{(\alpha)}(q))^2\left[S^2_{\rm F}(q)/S_\alpha(q)-1\,\right]
  \Tr O_{\alpha}^2(1,2)\,.
\end{eqnarray}

The energy correction from the spin-orbit potential are first of all
due to the implicit dependence of all quantities on the spin-orbit
potential, see Eqs. \eqref{eq:Vcredef}-\eqref{eq:W135}. These terms
all survive if the spin-orbit interaction is neglected. There is one
additional term that is of second order in the spin-orbit interaction,
a good estimate can be calculated by summing the corresponding
ring-diagrams. We have obtained for that in Ref. \citep{v4} by
performing the usual coupling constant integration \cite{rings}.
\begin{eqnarray}
&&\frac{\Delta E_{\rm ring}}{N}
 =-\frac{1}{4}\sum_{\alpha\in\{\rm c,T\}}
\int \frac{d^3q}{(2\pi)^3\rho} \int_0^\infty\frac{d\omega}{\pi}
\int_0^1 d\lambda
\Im\frac{\left[\tilde V_{\rm p-h}^{\LSsup}(q)\right]^2\chi_0(q;\omega)\chi_0^{(\perp)}(q,\omega)}
        {1-\lambda \tilde V^{(\alpha)}_{\rm p-h}(q,\omega)\chi_0(q;\omega)}\nonumber\\
        &=&\frac{1}{4}\sum_{\alpha\in\{\rm c,T\}}
        \int \frac{d^3q}{(2\pi)^3\rho} \int_0^\infty\frac{d\omega}{\pi}
        \frac{\left[\tilde V_{\rm p-h}^{\LSsup}(q)\right]^2\chi_0^{(\perp)}(q,\omega)}
             {\tilde V^{(\alpha)}_{\rm p-h}(q,\omega)}
             \ln\left[
               1-\tilde V^{(\alpha)}_{\rm p-h}(q,\omega)\chi_0(q;\omega)\right]\,.\nonumber\\
\label{eq:ELSring}
\end{eqnarray}

\section{Results: Energetics, structure, and effective interactions
  \label{sec:static}}
\label{ssec:EnergyResults}

\subsection{Equation of state and the compressibility\label{ssec:eos}}

\begin{figure}[H]
  \centerline{\includegraphics[width=0.6\textwidth,angle=270]{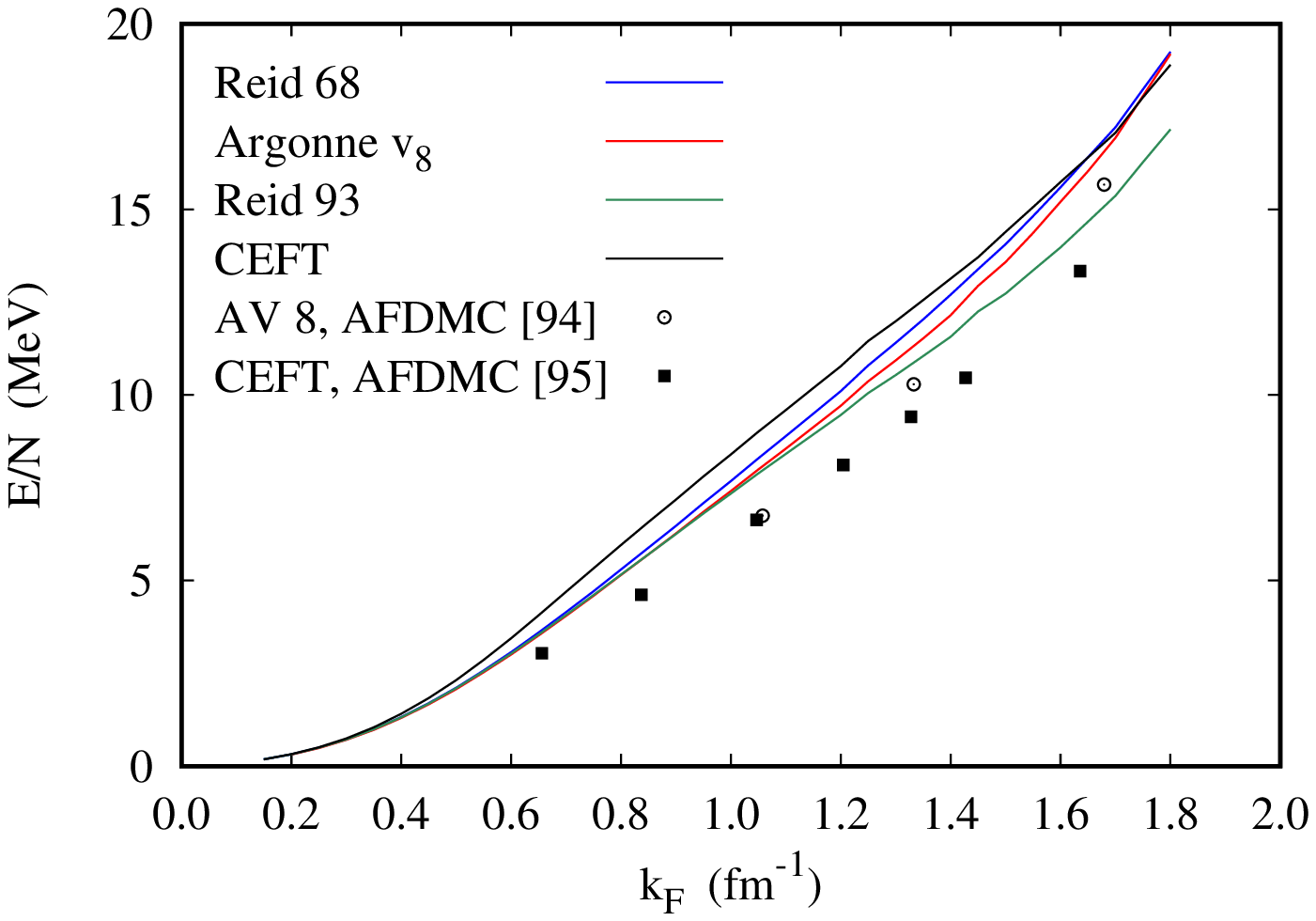}}
  \caption{The figure shows the equation of state for our four
    potential models, including twisted-chain and spin-orbit
    corrections.  Also shown are the AFMC calculations for the Argonne
    interaction \cite{Gandolfi2009a} and the CEFT potential
    \cite{PhysRevLett.111.032501} as indicated in the legend of the
    figure.\label{fig:eosplot}}
\end{figure}

Let us now turn to the results for the ground state energetics and
correlations and, in particular, the different predictions for these
quantities coming from different interactions. The first result of our
calculations is, of course, the equation of
state. Fig. \ref{fig:eosplot} shows our results and a comparison with
Auxiliary-Field Monte Carlo (AFMC) calculations for the Argonne and
the CEFT interactions.

For the CEFT interaction we have used the range parameter $R_0=1$;
note also that the AFMC calculations for the Argonne interaction used
the AV18 + Urbana IX interaction.  Our results for the equation of
state for the different interactions are almost identical and slightly
above the AFMC calculations. The reason for that might be that we have
used the simplest possible version of the FHNC-EL equations.

The most important input for the calculation of the dynamic structure
and the dynamic response function to be discussed in section
\ref{ssec:response} is the particle-hole interaction which is also one
of the central quantities of the parquet summation technique.  The
hydrodynamic speed of sound \eqref{eq:mcfromeos} is related to the
long-wavelength limit of the particle-hole interaction. In a Fermi
fluid, an additional contribution to the speed of sound comes from the
Pauli repulsion, see Eq. \eqref{eq:FermimcfromVph}. In neutron matter
we can mostly set $m^*=m$, see section \ref{ssec:selfen}. We note in
passing that the local approximation \eqref{eq:favg} gives the correct
contribution to $mc_s^2$ as defined in Eq. \eqref{eq:FermimcfromVph}
from the leading-order exchange diagrams \cite{EKVar}.

\begin{figure}[H]
  \centerline{\includegraphics[width=0.8\textwidth,angle=270]{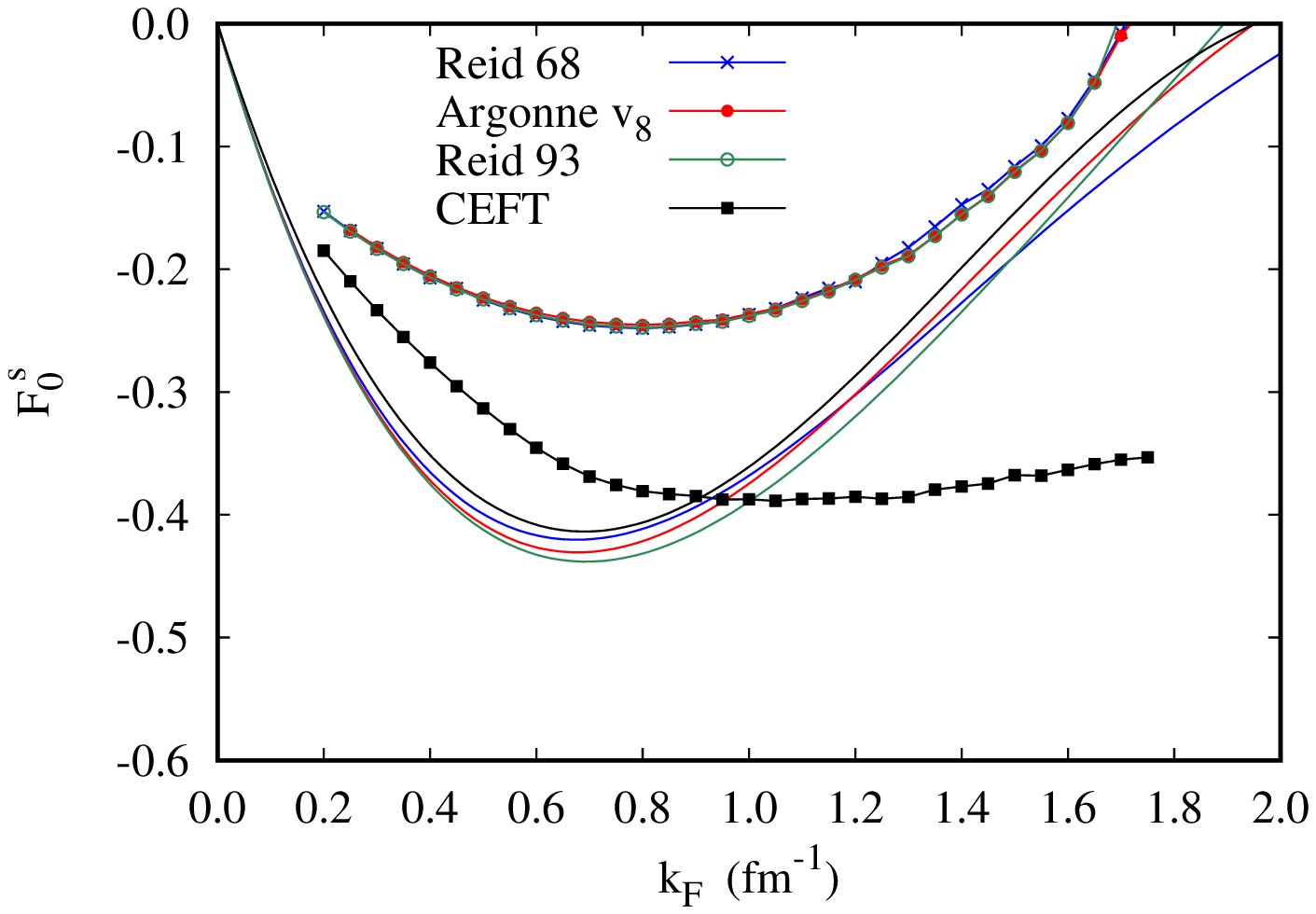}}
  \caption{The figure shows the Fermi-liquid parameter $F_0^s$ as
    calculated from the equation of state \eqref{eq:mcfromeos} (lines)
    and from the long wavelength limit of the excitations
    \eqref{eq:FermimcfromVph} (lines with markers) for our four
    potential models as indicated in the legend.
    \label{fig:F0Splot}}
  \end{figure}

As mentioned above, the relationships \eqref{eq:mcfromeos} and
\eqref{eq:FermimcfromVph} give identical predictions only in an exact
theory \cite{EKVar,parquet5,PhysRevA.110.052222}; good agreement is
typically reached only at very low densities. Even in the much simpler
system $^4$He, where four- and five-body elementary diagrams and
three-body correlations are routinely included, the two expressions
\eqref{eq:mcfromeos} and \eqref{eq:FermimcfromVph} can differ by up to
a factor of two \cite{lowdens}. It is possible to force the agreement
between the two calculations by slightly adjusting the exchange
correction \cite{v3eos} or, in \he4, the three-body correction to the
particle-hole interaction.  Such a procedure is, for example,
necessary to obtain the correct non-analytic behavior of the equation
of state in the vicinity of the spinodal point \cite{lowdens}; it
makes practically no difference on the equation of state.  We have
here refrained from such phenomenological modifications.
Fig. \ref{fig:F0Splot} shows the Fermi Liquid Parameter $F_0^s$ as
obtained in the two different ways; the difference corresponds to what
we know from \he3 and \he4.

So what can be learned from the comparisons in Fig. \ref{fig:F0Splot} ?
\begin{itemize}
\item{} Generally we can write the energy -- in the areas where a bulk
  system exists -- in a power series expansion
  \begin{equation}
    \frac{E}{N}=\sum_n \KF^n e_n\label{eq:EoverNexpand}
  \end{equation}
  from which we obtain for Eq. \eqref{eq:mcfromeos}
  \begin{equation}
    mc_s^2 = \frac{n}{3}\left(\frac{n}{3}+1\right)\KF^n e_n\,.
    \label{eq:mc2expand}
  \end{equation}
  Hence, $n$-th order corrections to the energy contributes a
  correction to $ mc_s^2$ that is roughly a factor of $n^2$ larger.
  One would therefore expect that any diagrammatic expansion of
  $mc_s^2$ converges more slowly than that for the energy.
\item{} For the variational wave function \eqref{eq:Jastrow} it is
  rather straightforward \cite{EKVar} to derive a cluster expansion
  for $\tilde V_{\rm p-h}(0+)$ and compare that expansion with what
  one obtains from \eqref{eq:mcfromeos}. The result is that the terms
  contributing to $\tilde V_{\rm p-h}(0+)$ are, in any level of the
  (F)HNC expansion, a proper subset of those contributing to $m c_s^2$
  as obtained from \eqref{eq:mcfromeos}. The analysis is significantly
  more complicated in Green's functions based theories
  \cite{parquet5,PhysRevA.110.052222} and has, so far, not led to a
  comparably clear result. One would, of course, expect a
  corresponding statement.
\end{itemize}

The conclusion of this analysis is that the calculation of $m c_s^2$
from \eqref{eq:mcfromeos} is generally more trustworthy than what is
obtained from the long-wavelength limit $\tilde V_{\rm
  p-h}(0+)$. Keeping in mind the argument made around
Eqs. \eqref{eq:EoverNexpand} and \eqref{eq:mc2expand} we can use the
difference between the two calculations of $F_0^s$ shown in
Fig. \ref{fig:F0Splot} as an estimate for the convergence of the
methods. From that we conclude that higher-order corrections for the
equation of state for the three interactions Reid 68, Argonne, and
Reid 93 are practically identical whereas those for CEFT are rather
different in sign albeit comparable in magnitude.

We conclude this subsection by remarking that the correction
\eqref{eq:ELSring} has turned out to be less than 0.1 MeV which is of
the order of a percent of the total energy.

\subsection{The short-ranged structure\label{ssec:gof0}}

The equation of state is, of course, important for understanding the
maximum masses of neutron stars. But energetics is just one result of
microscopic calculations and there are many other features of
interest.

The structure of strongly interacting systems like the ones considered
here is mostly determined by the short-ranged interaction. The
Brueckner-Bethe-Goldstone equation
\cite{Bru54,Bru55,Bru55a,BruecknerLesHouches,BetheGoldstone57,Goldstone57}
deals with these short-ranged correlations. Of course, the only
many-body effect included in the Bethe-Goldstone equation is Fermi
statistics, the equation reduces, for bosons, to a zero-energy
Schr\"odinger equation.

Modern nucleon-nucleon interactions \cite{PhysRevC.90.054323} have
much weaker cores such that finite-order perturbation theory should
give trustworthy results. Since short-ranged correlations are, in
principle, observable, it is therefore of interest how these depend on
the interaction and the importance of many-body methods and effects
beyond Fermi statistics.

\begin{figure}[H]
  \centerline{
    \includegraphics[width=0.35\textwidth,angle=270]{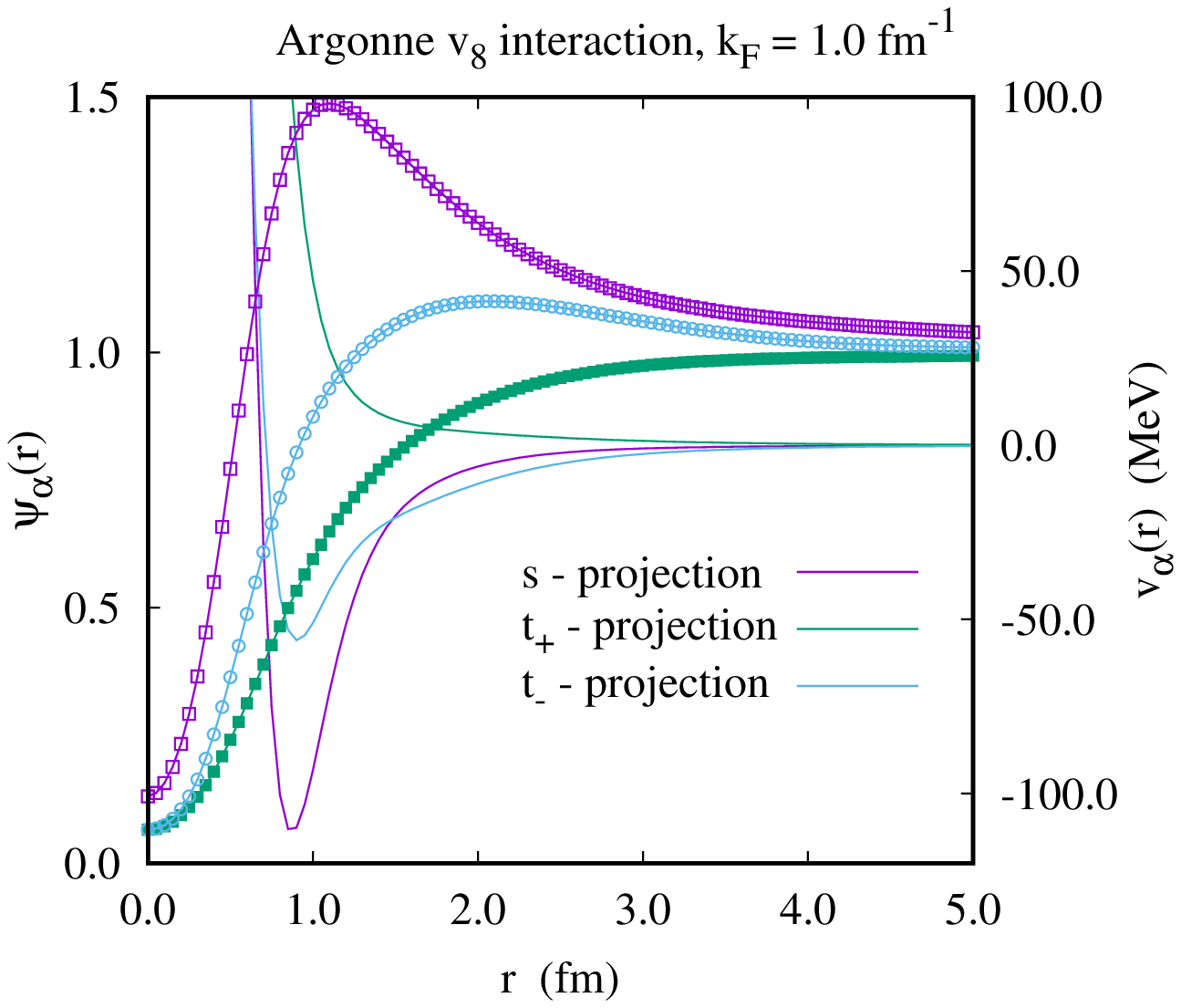}
    \includegraphics[width=0.35\textwidth,angle=270]{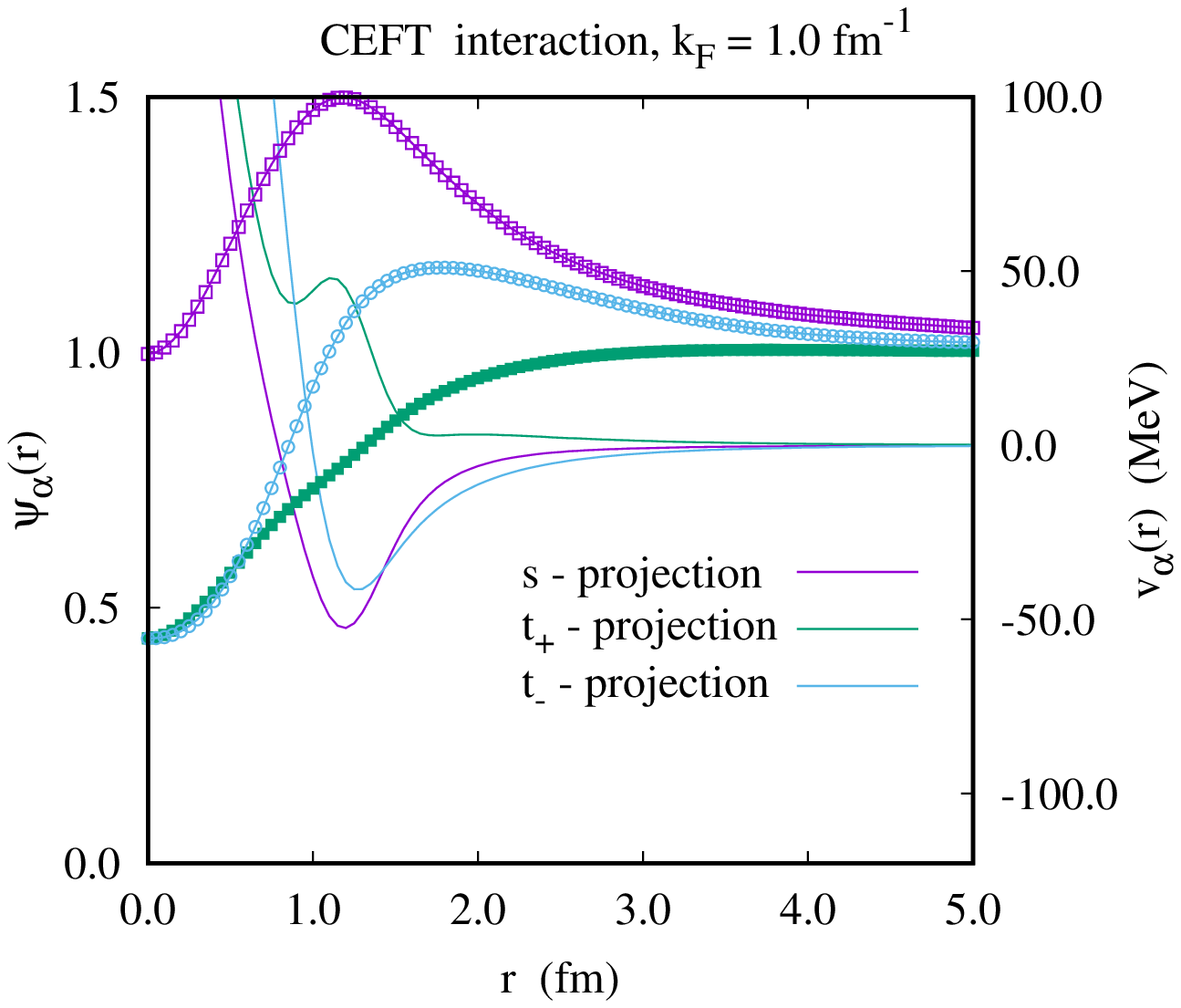}
  }
  \caption{The figure shows the components of the Argonne $v_8$ and
    the CEFT interaction in the projector channels $s$, $t+$ and $t-$
    (unmarked curves, right scale) and the pair wave functions
    $\psi_\alpha(r)$ (lines with markers, left scale) for the
    state-dependent correlations including twisted diagrams at
    $\KF = 1.0\,$fm$^{-1}$.
    \label{fig:Gddkf100}}
  \end{figure}

A first question is therefore how the state-dependent correlations
described in this chapter compare to the state-independent
calculations described in Ref. \cite{ectpaper}.
Fig. \ref{fig:Gddkf100} shows the pair wave function $\psi_\alpha(r) =
\sqrt{\Gamma_{\rm dd}^{(\alpha)}(r)+1}$ in the three projector
channels $s$, $t+$ and $t-$ along with the corresponding bare
interactions.  In that calculation we went beyond Jastrow-Feenberg
theory by including the twisted-chain diagrams derived in section
\ref{ssec:twist}; correlation functions $f_\alpha(r)$ (See
Eq. \eqref{eq:fop}) are therefore not defined.

The short-ranged behavior of the pair wave functions $\psi_\alpha(r)$
reflects indeed the short-ranged behavior of the interactions which
the $f_\alpha(r)$ would not.  We see that the state-dependent
correlations show exactly the expected behavior: A strong nearest
neighbor peak for the attractive $S$-wave channel and an increased
correlation hole in the repulsive triplet channels.  The
state-independent theory tries to compensate for the lack of
flexibility by generating what looks like an ``average'' correlation
function. Such an ``average'' correlation function does reasonably well
for the energetics, see Fig. 2 of Ref. \citep{v3eos}, in fact the
ground state energy expectation values for different calculations are
almost identical up to a moderate Fermi wave number of
$\KF=1.0\,$fm$^{-1}$. The conclusion of that is, of course, not that
the simple state-independent correlation function \eqref{eq:Jastrow}
provides an accurate description of the ground state correlations but
rather that the energy is insensitive to the quality of the wave
function.

Figs. \ref{fig:Gddkf100} also shows the comparison with results
obtained from the CEFT interaction. The most remarkable feature is
that the pair correlation function in the spin-singlet channel is,
unlike for the three other interactions, far from zero at the
origin. What remains is the strong nearest-neighbor peak of the
spin-singlet component and the large correlation hole for the
spin-triplet states. These are in fact remarkably similar for all
interactions.  We hasten to note that the value $\psi_s(r=0)\approx 1$
is accidental and does not persist at other densities.

\begin{figure}[H]
  \centerline{\includegraphics[width=0.35\textwidth,angle=270]{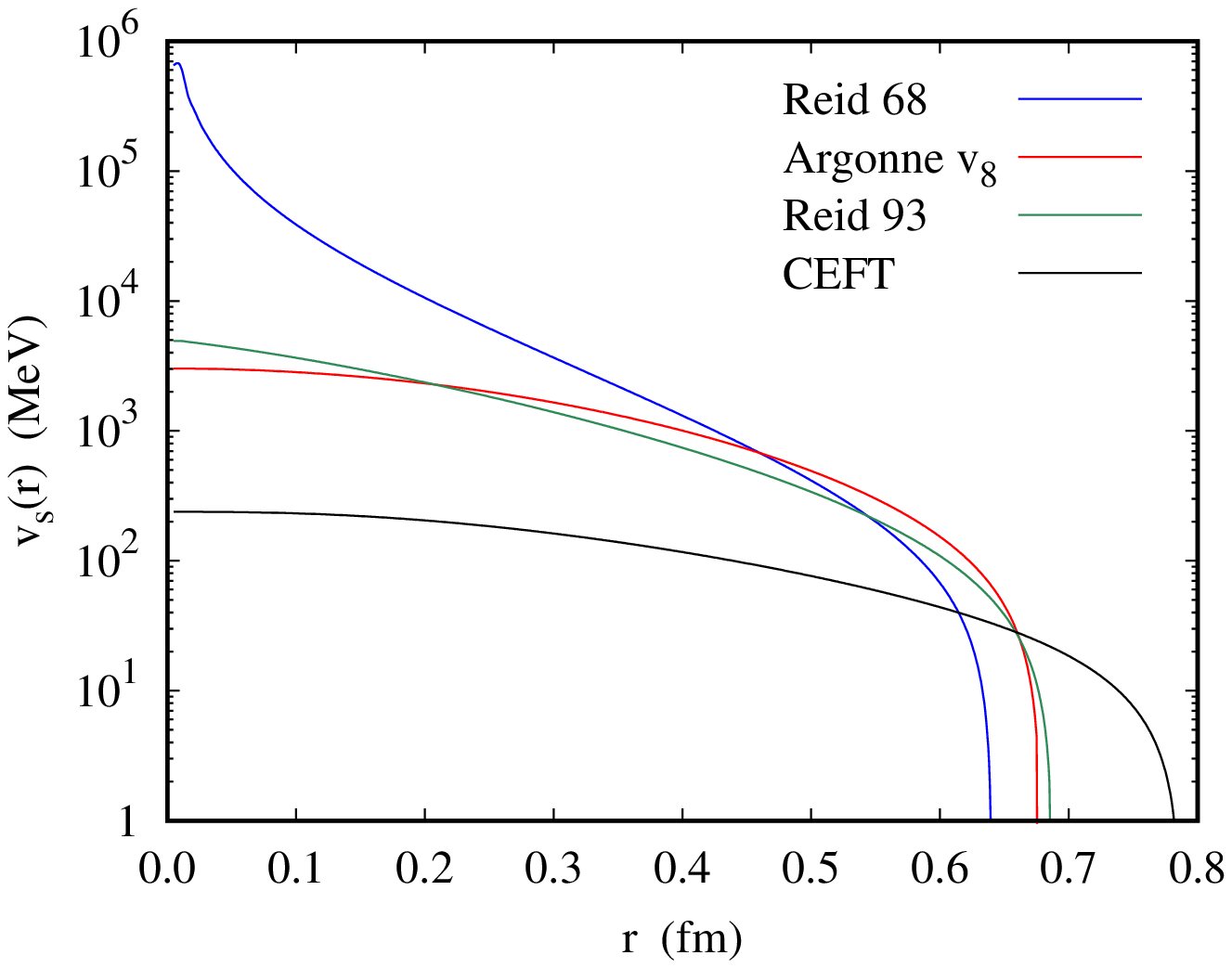}}
  \caption{The figure shows the short-ranged part of the bare
    singlet interaction on a logarithmic scale for our four
    potential models as indicated in the legend.
    \label{fig:vcshort}}
\end{figure}

The reason for this behavior is the much weaker repulsion of the
singlet CEFT interaction shown in Fig. \ref{fig:vcshort}. We wish to
point out again that the equation of state of the CEFT interaction is
almost indistinguishable from those of the other interactions although
both the interactions and, hence, the correlations, are very
different, see Figs. \ref{fig:eosplot} and \ref{fig:F0Splot}.

\begin{figure}[H]
  \centerline{
    \includegraphics[width=0.27\textwidth,angle=270]{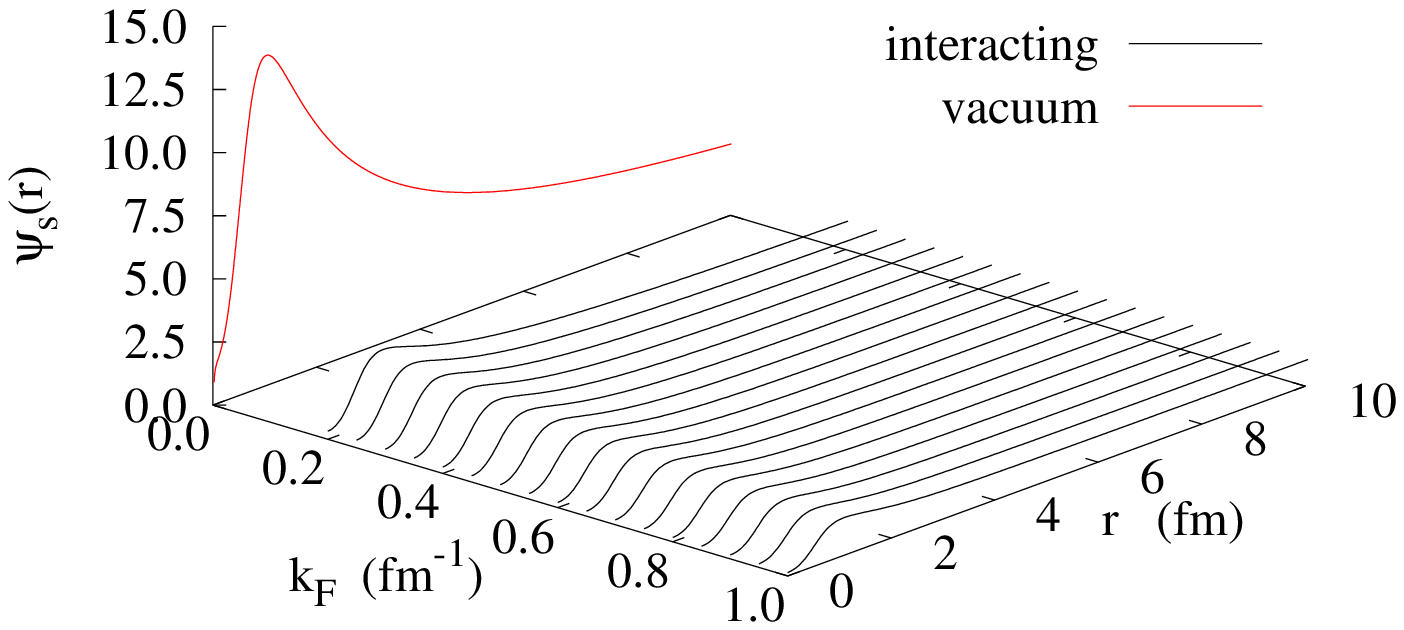}
    \includegraphics[width=0.27\textwidth,angle=270]{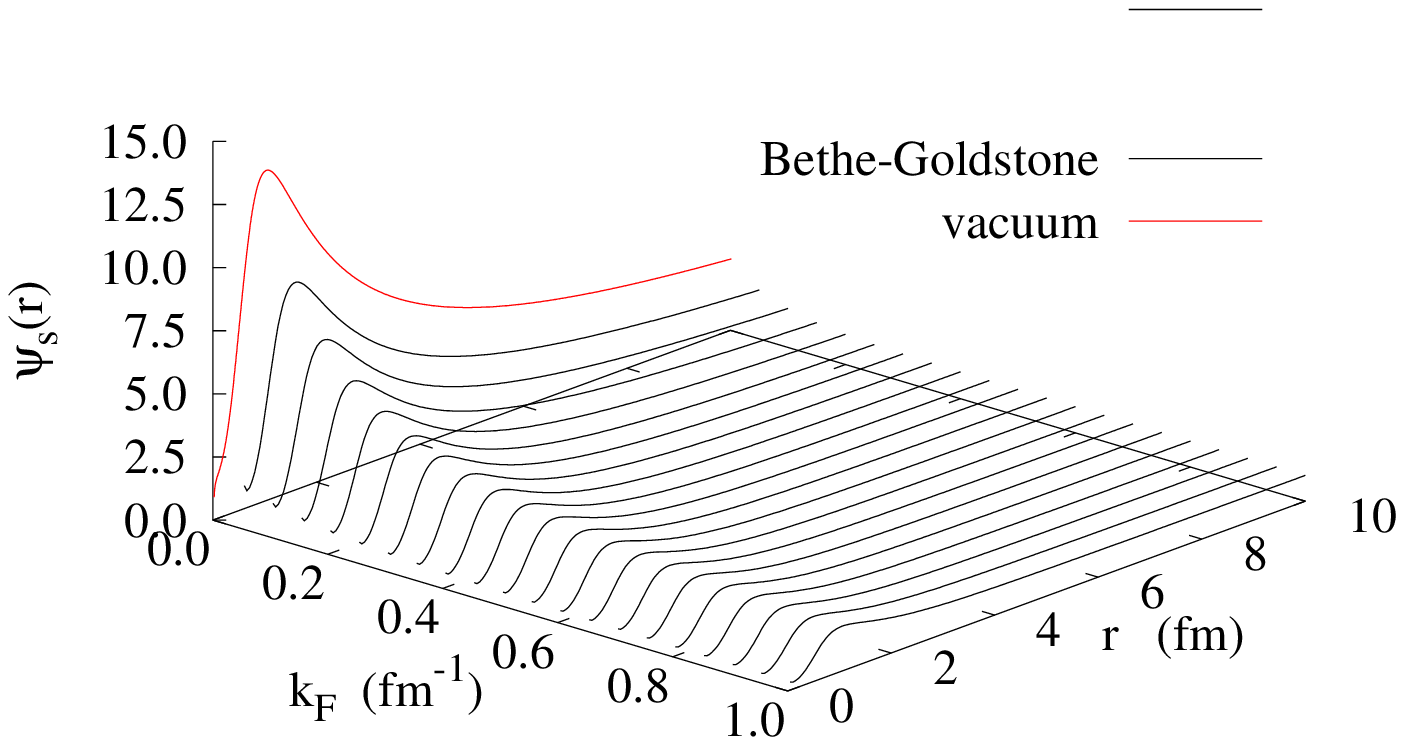}
  }
  \caption{The figures show the density dependence of the pair
    wave function in the $s$-projection for the Argonne $v_8$
    from the parquet calculation (left figure, black lines) 
    and from the localized Bethe-Goldstone equation (right figure,
    black lines). Also shown is the vacuum
    solution for the same interaction (red line).
    \label{fig:psi_s_3d}}
  \end{figure}

It is interesting to see the development of the pair wave function
as we go to low densities. In the zero-density limit, the pair wave
function should become the zero-energy solution of the Schr\"odinger
equation, the asymptotic behavior is
\begin{equation}
  \psi_s(r) = 1-\frac{a_0}{r}\qquad\mathrm{as}\qquad r\rightarrow\infty
  \label{eq:psilongvaccum}
\end{equation}
where $a_0$ is the vacuum scattering length. Many-body effects change
that behavior to \cite{cbcs}
\begin{equation}
  \psi_s(r) = 1-\frac{9\tilde V^{(s)}_{\rm p-h}(0+)}{16\EF}\frac{1}{r^2\KF^2}
  \qquad\mathrm{as}\qquad r\rightarrow\infty
  \label{eq:psilongmedium}
\end{equation}
where $\EF$ is the Fermi energy of the non-interacting system.
Figs. \ref{fig:psi_s_3d} show the density dependence of the singlet
pair wave function in the low density regime. For finite densities it
is clear that the nearest neighbor peak keeps increasing with
decreasing density; however it is still far from the zero-density
limit. This explains why much larger discretization volumes are needed
for calculations at very low densities; in Ref. \citep{cbcs} we had to
use cutoff radii that were about 100 times larger than the ones used
here to make reliable calculations.

Figs. \ref{fig:psi_s_3d} also show the solution of the local
Bethe-Goldstone Eq. \eqref{eq:BGSchr} for the bare interaction in the
same density regime. When the induced interaction is omitted, the
resulting pair wave function also has the asymptotic behavior
\eqref{eq:psilongvaccum} and is, hence, not suitable for many-body
calculations. In Brueckner-Bethe-Goldstone theory the problem is
handled by {\em ad-hoc\/} prescriptions to enforce a ``healing'' of
the pair wave function at finite distances, see for example
Ref. \citep{BetheBrandowPetschek}. We have not applied such
modifications. One can see that the nearest neighbor peak in the
pair wave function is typically twice as high as for the fully
correlated case. The results for the other interactions are rather
similar and not shown here.

One would normally think that many-body correlations become weaker
with lower density. Fig. \ref{fig:psi_s_3d} shows that this is not the
case, in fact correlations become stronger. The basic reason for this
is, of course, the rather large vacuum $S$-wave scattering length of the
nucleon-nucleon interaction $a_0 \approx -18.7\ $fm
\cite{PhysRevLett.83.3788}, \ie a neutron pair is close to developing
a bound state.  The induced interaction suppresses correlations by
damping the long-ranged structure at finite densities, see
Eq. \eqref{eq:psilongmedium}. The issue has led to many speculations
in pairing in low-density neutron matter and/or if a ``unitary limit''
can be approached.
\cite{MatsuoPRC73,MargueronPRC76,PhysRevC.79.054003,%
  PhysRevC.94.034004,NeutronBEC,PethickSchaeferSchwenk}.  We have
studied this problem in a simpler model \cite{cbcs} of purely central
interactions. In that work, we have concluded that at low densities a
dimerization can occur due to the divergence of the in-medium
scattering length. This scenario could be investigated in the future
within the much more sophisticated scheme presented here.

\subsection{Exchange diagrams\label{ssec:exchange}}

\begin{figure}[H]
  \centerline{
    \includegraphics[width=0.35\textwidth,angle=270]{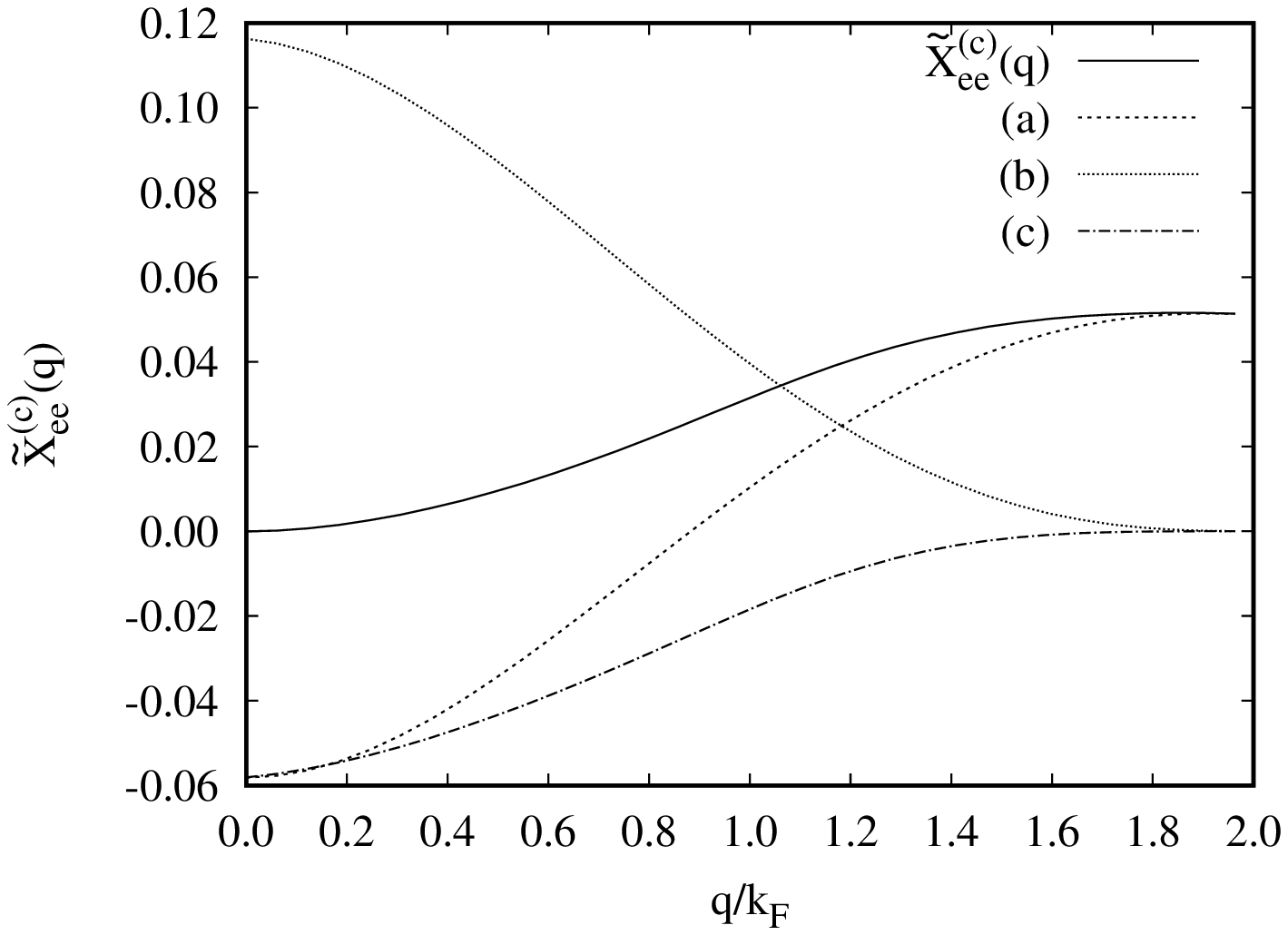}
    \includegraphics[width=0.35\textwidth,angle=270]{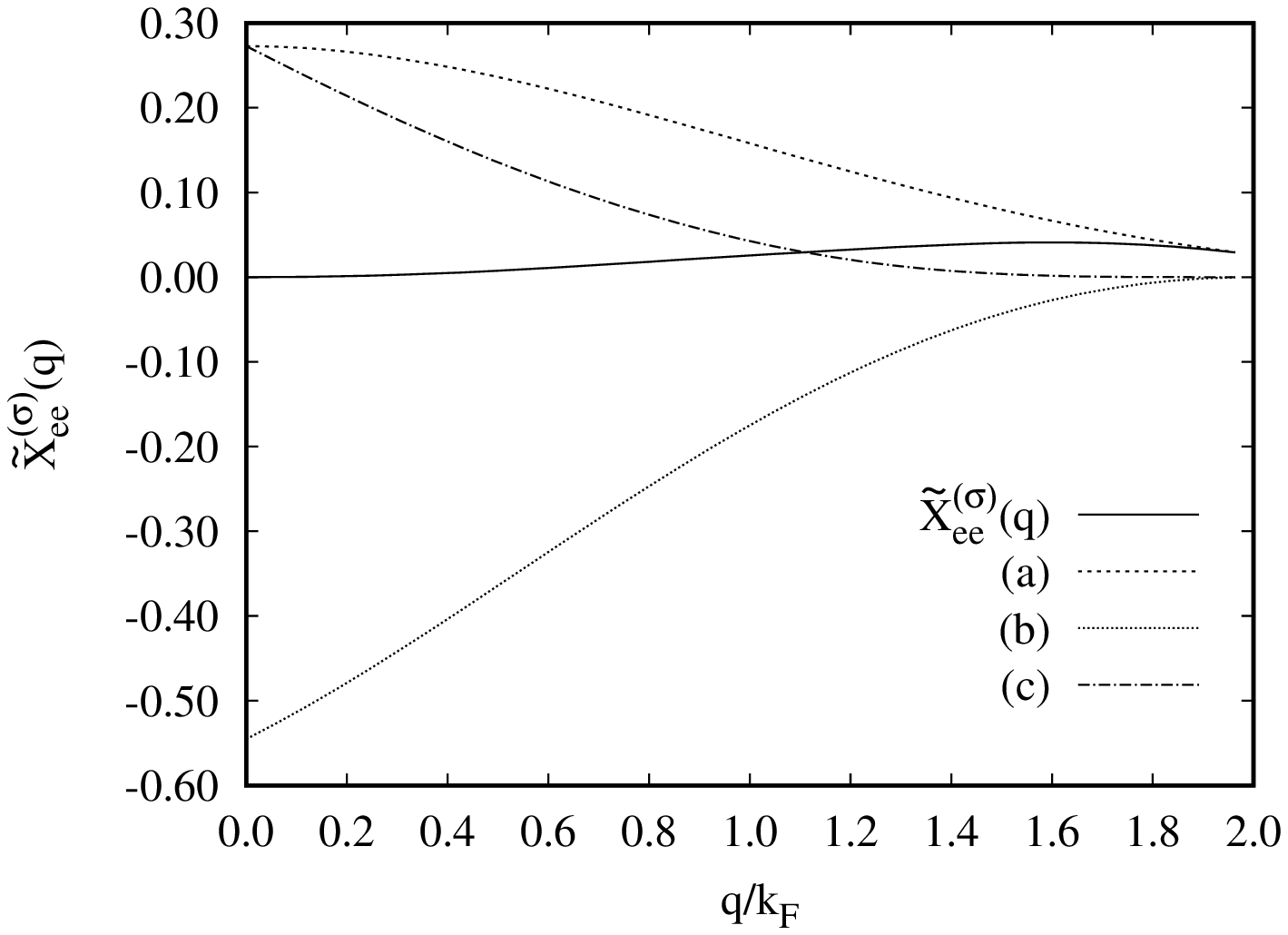}
  }
\caption{The figures show the Argonne $v_8$ interaction at $\KF =
  1.0\,\text{fm}^{-1}$, the individual contributions (a), (b) and
  (c) to $\tilde X_{ee}(q)$ originating from central (left figure)
  and spin (right figure) correlations.\label{fig:Xee}}
\end{figure}

We have in connection with Fig. \ref{fig:eelink} pointed out that
there is strong cancellation between the three diagrams: Diagram
\ref{fig:eelink}a goes, for $k\rightarrow 0+$, quadratically towards a
finite value; diagrams \ref{fig:eelink}b and \ref{fig:eelink}c go
linearly towards finite values such that the sum of all three terms is
quadratic as $k\rightarrow 0+$. Moreover, the terms \ref{fig:eelink}b
and \ref{fig:eelink}c vanish for $k\ge 2\KF$.  Figs. \ref{fig:Xee}
shows the individual contributions to the function $\tilde X_{ee}(q)$
coming from direct and spin correlations.  All three terms involving
the tensor correlations \cite{v3eos} are quadratic in the momentum and
not shown.  The take-away from this figure is that the cancellations
between the individual terms persist and it is certainly not
legitimate to split them apart.
  
\subsection{Effective interactions}
\subsubsection{The importance of twisted-chain corrections\label{ssec:vtwist}}

It remains the question of the importance of non-parquet diagrams.  We
have, so far, seen two different scenarios: In a model calculation for
a fictitious system of bosons with spins \cite{SpinTwist} the effect
was seen to be quite dramatic whereas we have here seen practically no
change in the energy.

\begin{figure}[H]
  \centerline{
    \includegraphics[width=0.35\textwidth,angle=270]{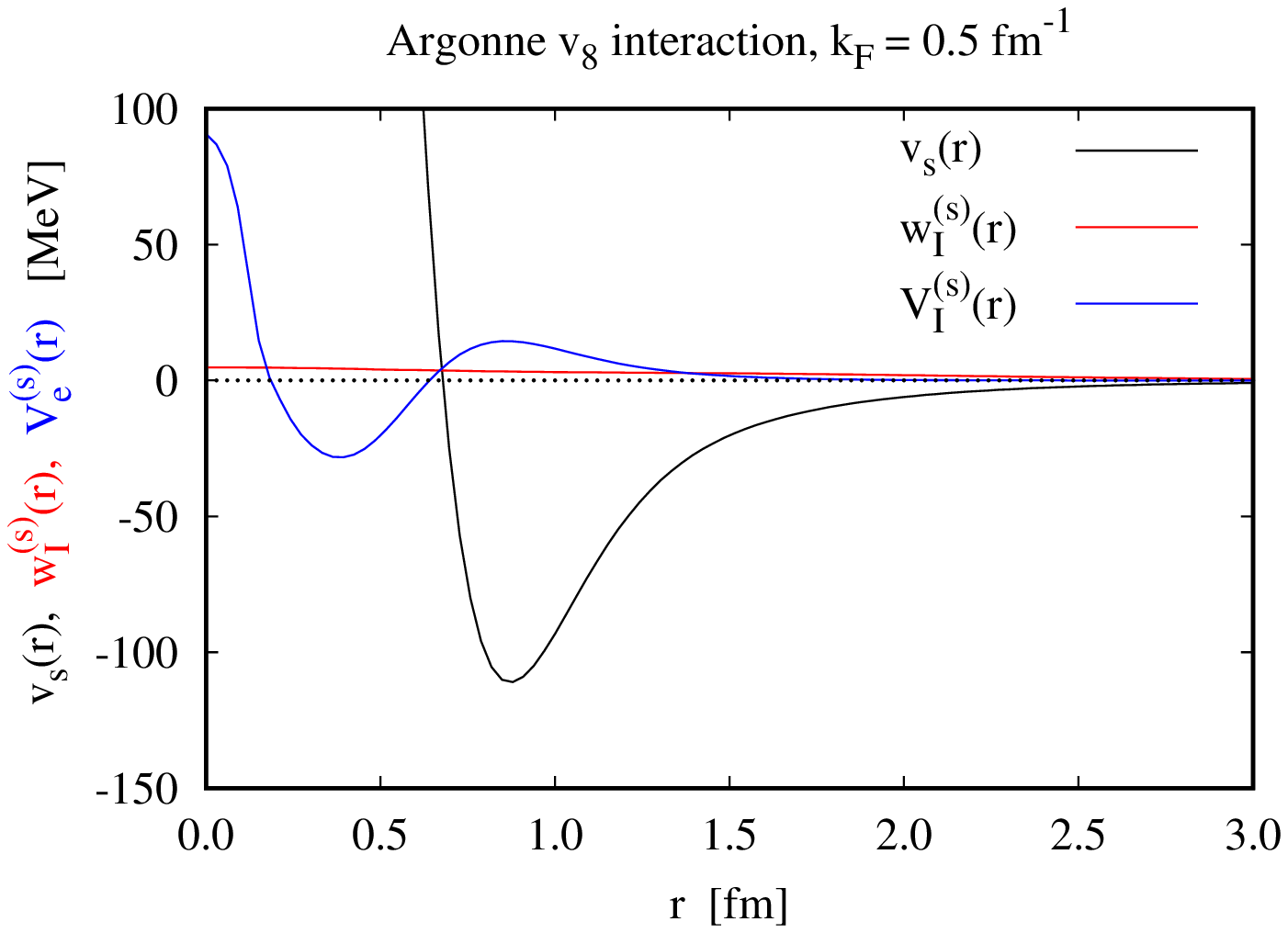}
    \includegraphics[width=0.35\textwidth,angle=270]{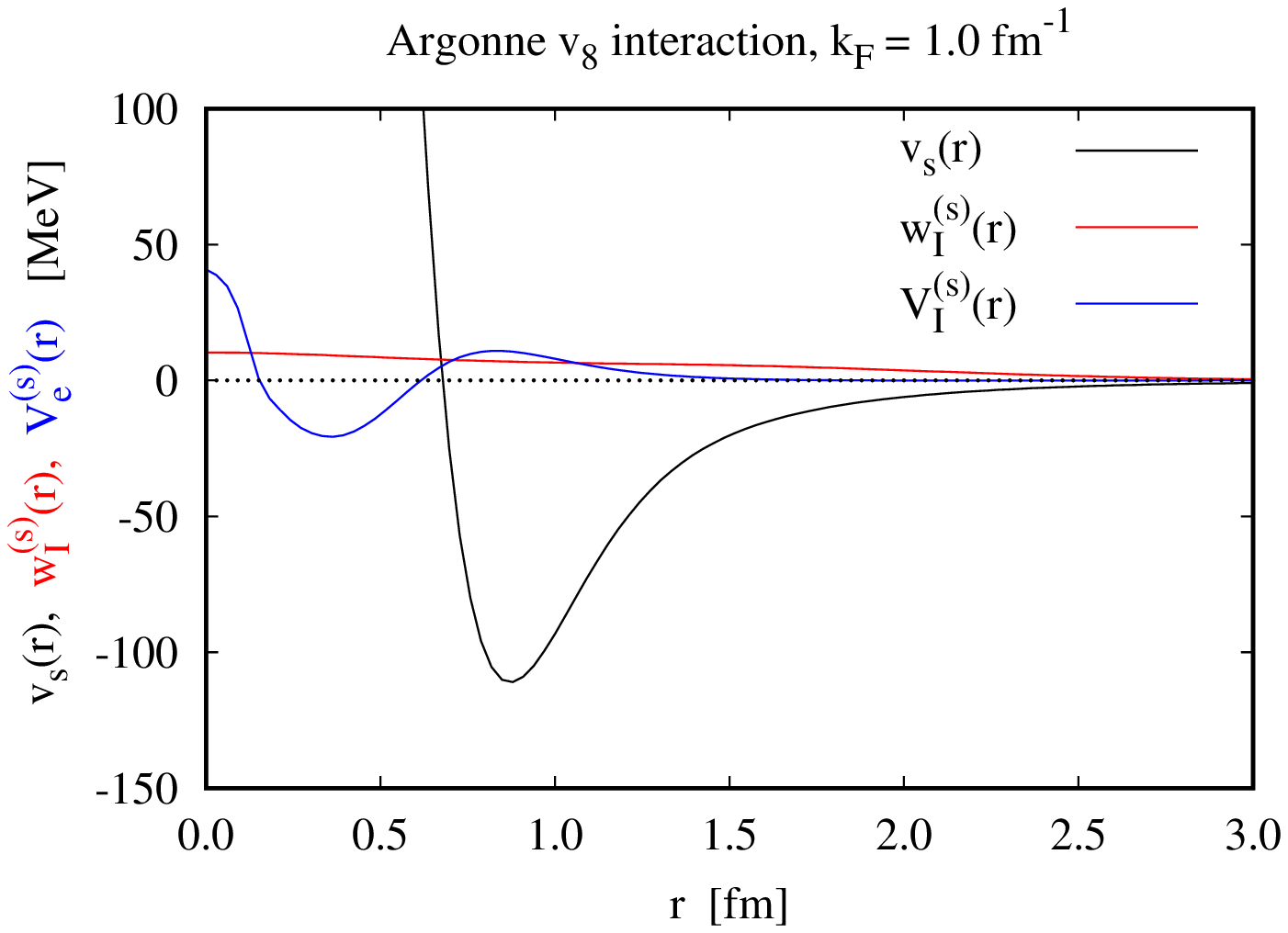}
  }
  \centerline{
    \includegraphics[width=0.35\textwidth,angle=270]{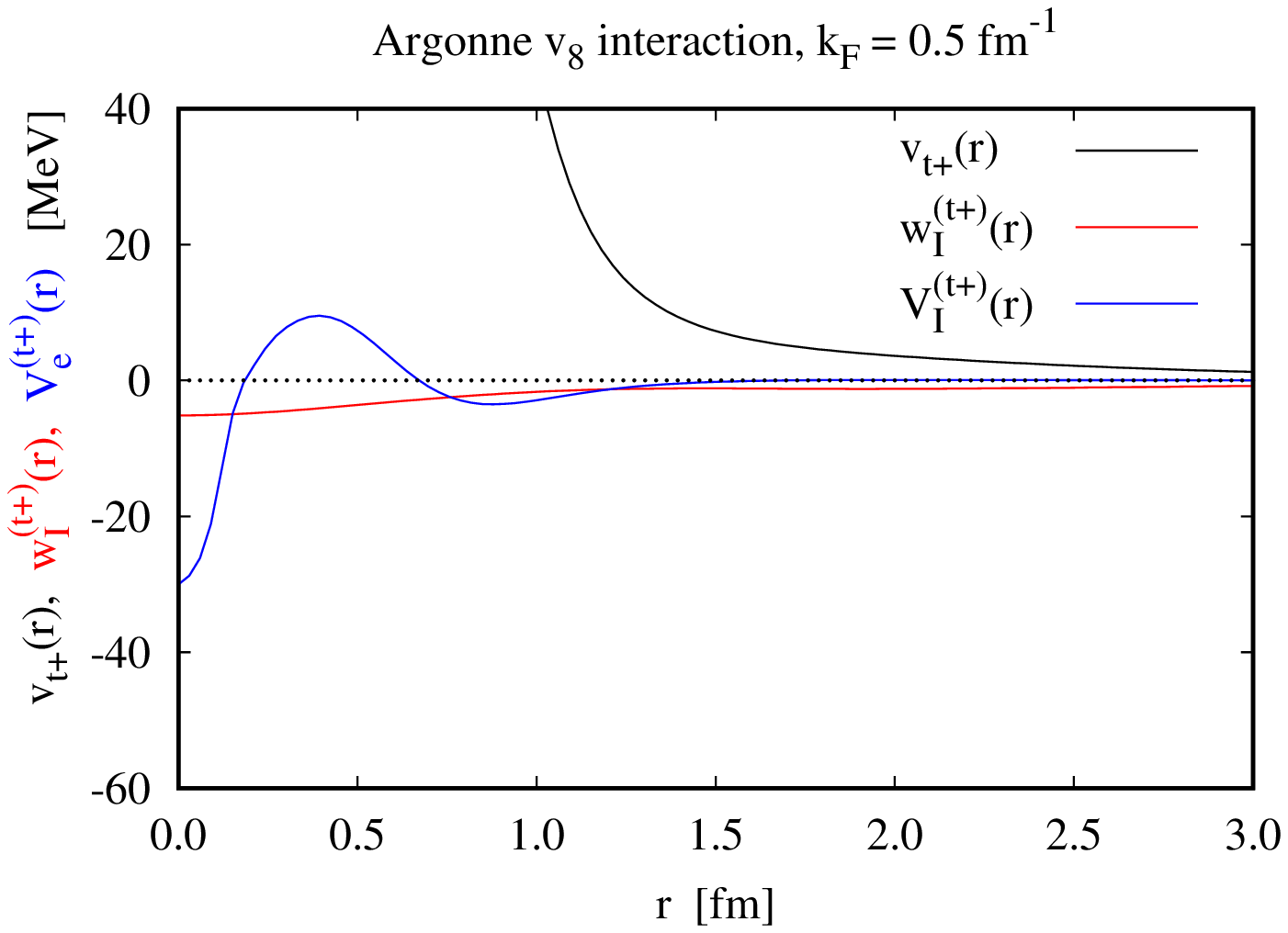}
    \includegraphics[width=0.35\textwidth,angle=270]{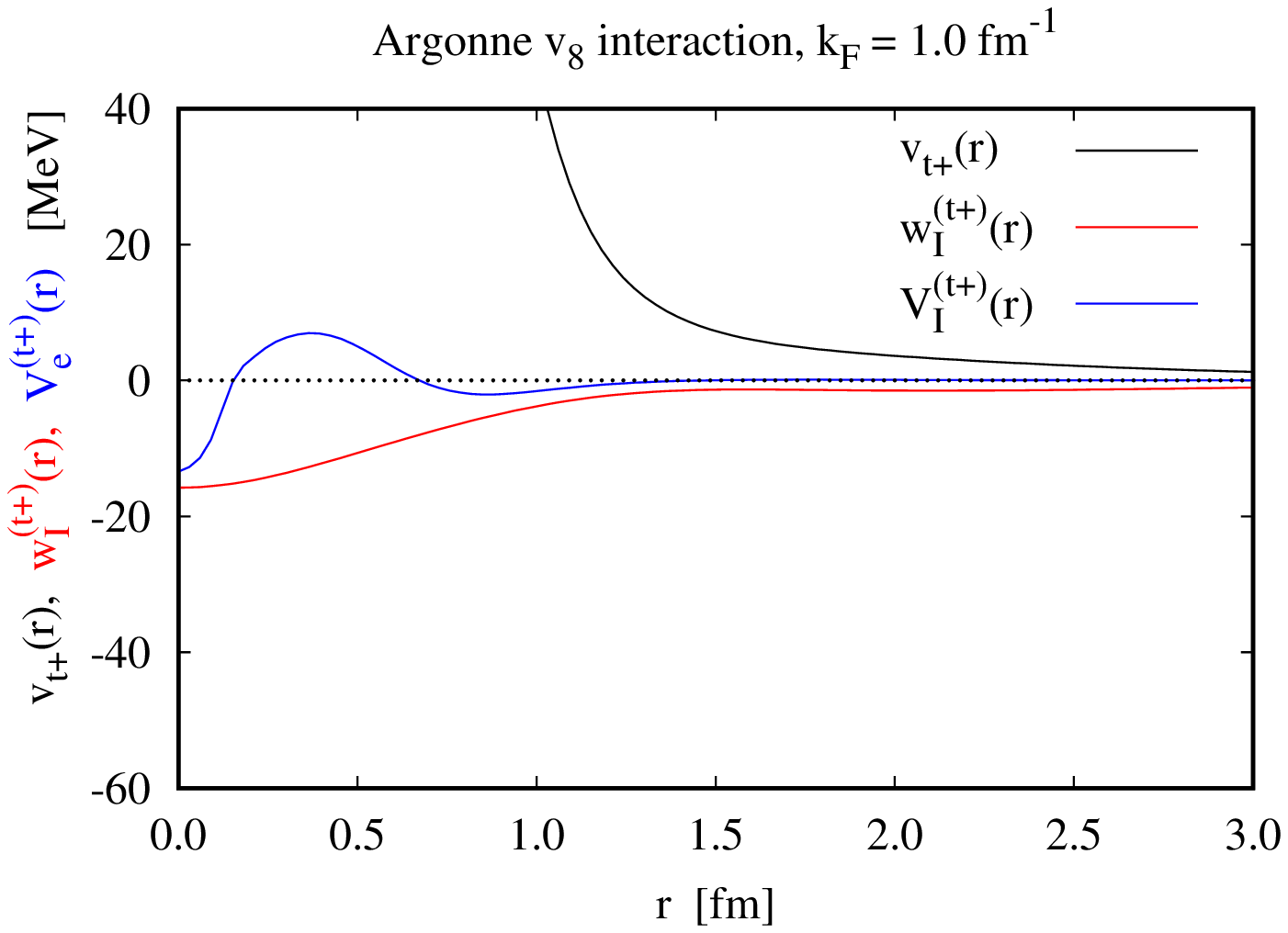}
  }
  \centerline{
    \includegraphics[width=0.35\textwidth,angle=270]{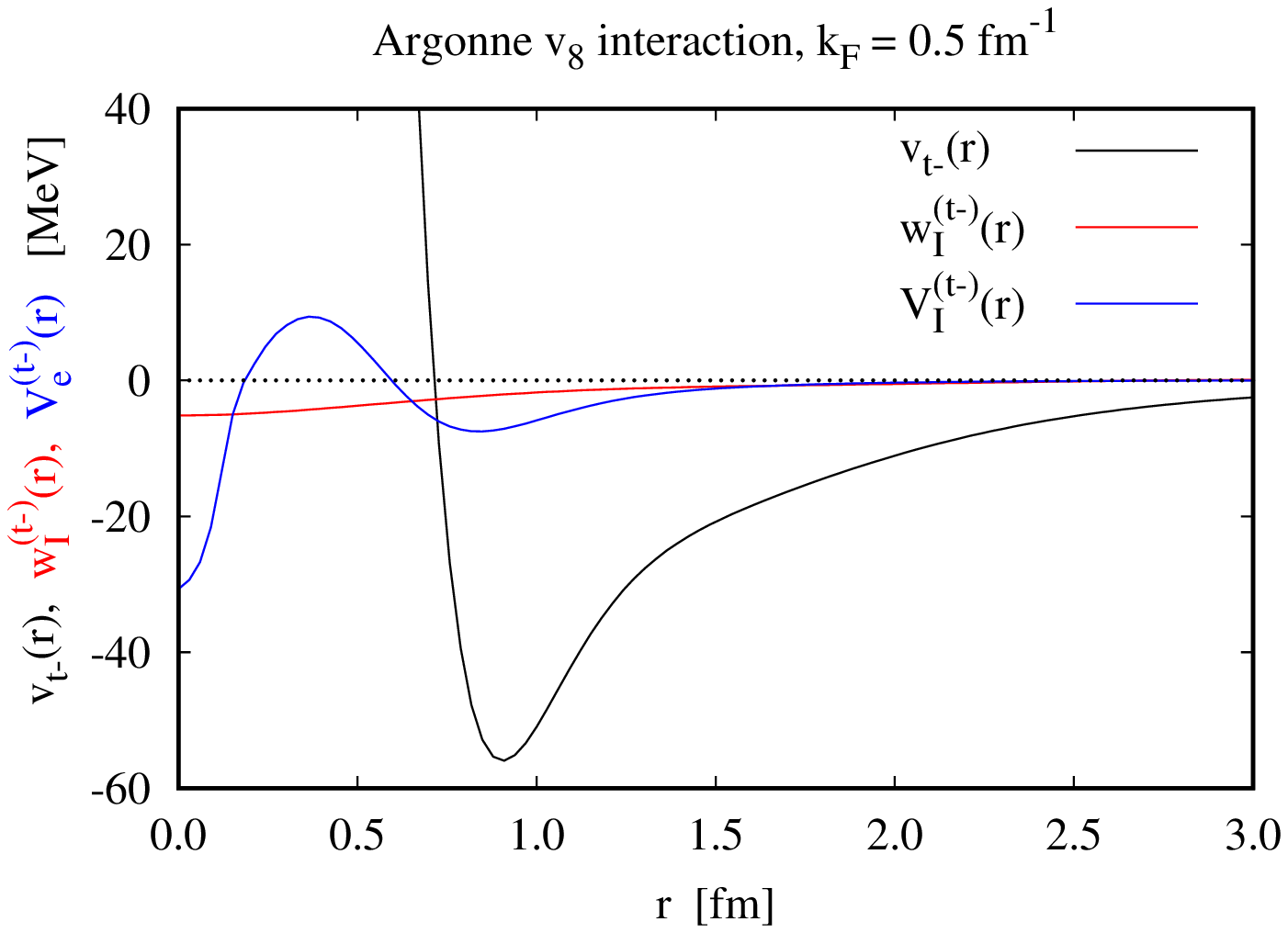}
    \includegraphics[width=0.35\textwidth,angle=270]{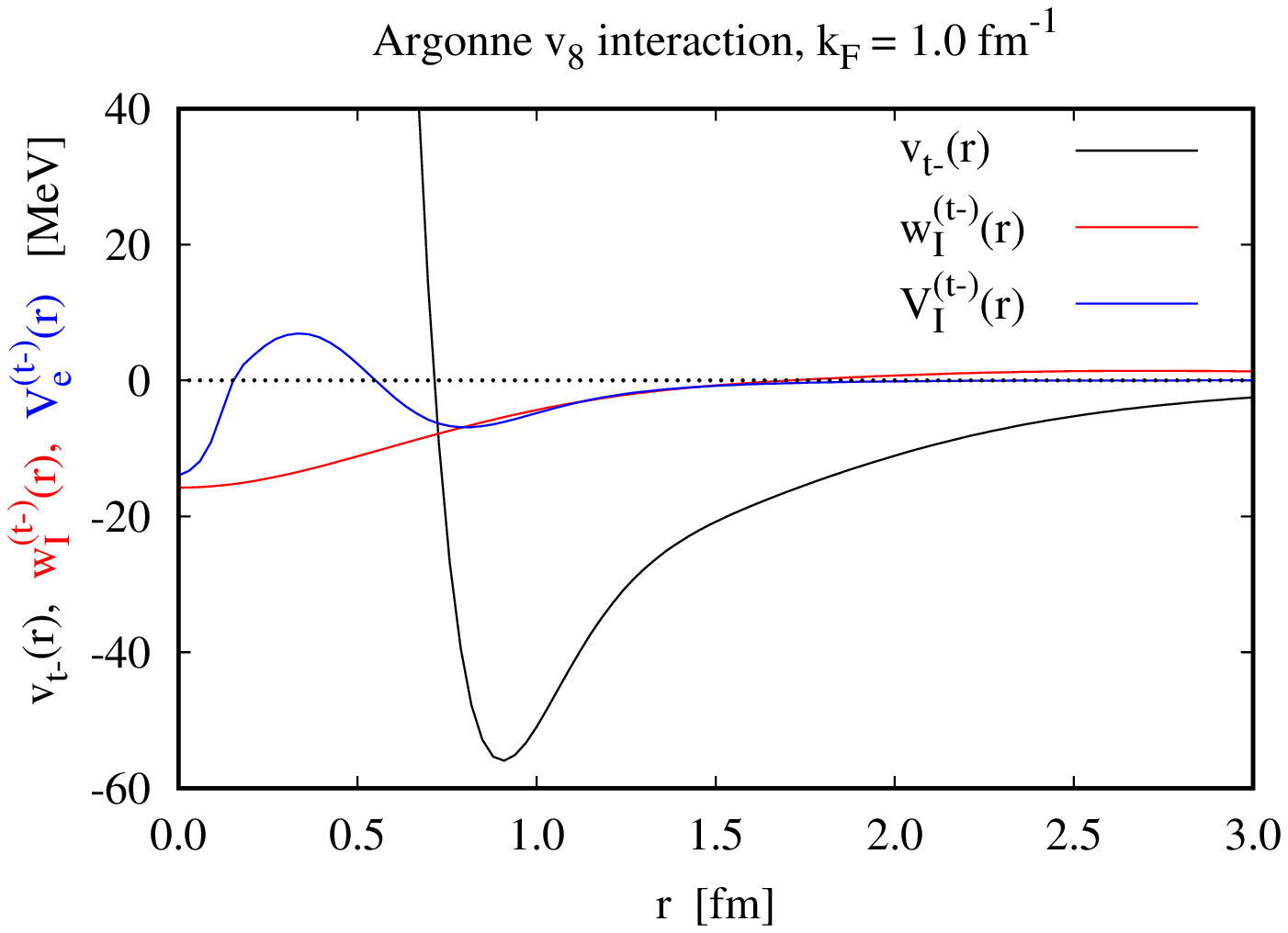}
  }
\caption{The figures show the bare Argonne $v_8$ interaction, the
  induced interaction $w_I^{(\alpha)}(r)$, and the non-parquet
  correction $V_I^{(\alpha)}(r)$ in the three projector channels at
  two different densities. Note that the scale is different for the
  singlet interactions.\label{fig:Vtwist}}
\end{figure}

Figs. \ref{fig:Vtwist} show a comparison of different many-body effects
contributing to the effective interactions in the three projector channels.
In general, we see that the induced interaction is a rather smooth
function whereas the $V_I^{(\alpha)}(r)$ are localized at short
distances. The most relevant regime is, of course, the area outside
the repulsive core.  In the singlet channel, we see indeed that the
non-parquet corrections practically double the repulsive induced
interaction around the potential minimum whereas the attraction is
enhanced in the triplet channels. This is a direct consequence of
mixing the large hard core of the triplet potentials with the much
more attractive singlet channel. The same is true in the $t-$ channel
where $V_{\rm I}(r)$ in the relevant regime is just as large as the
induced interaction. Only in the $t-$ channel both terms are
noticeable only far inside the core region. We will return to this
point in section \ref{sec:BCS} where we will look at these effects in
momentum space.

\subsubsection{The spin-orbit channel\label{ssec:spinorbit}}

We have derived in section \ref{ssec:phG} a closed-form expression for
the sum of all ring-diagrams including a spin-orbit effective
interaction. The derivation led to only three additional operators,
Eqs. \eqref{eq:Qdef7}, \eqref{eq:Qdef9} and \eqref{eq:VLSp}.  We have
identified a number of effects: One is the modification of the induced
interactions by energy dependent corrections \eqref{eq:Vcredef} and
\eqref{eq:VTredef}. The two contributions proportional to the
additional operators $Q_7$ and $Q_9$ are quadratic in the effective
spin-orbit interaction $\VLSq(q)$ and will be shown to be
negligible. The term proportional to $\LSp$ could be very interesting
because it is the only term in the effective interaction that couples
spin-singlet and spin-triplet states.

Attention is now directed to the definition \eqref{eq:FermiVphhat},
specifically Eq. \eqref{eq:VphLS}.  The only place where the bare
interaction occurs is the first term where the interaction operator is
multiplied by the pair wave function $\hat\psi(r)$. The spin-orbit
interaction acts only in spin-triplet states in which the central
interaction part is strongly repulsive and generates rather large
correlation hole. Hence, the short-ranged behavior of the bare
interaction in the spin-orbit channel is screened by the correlations
caused by the surrounding particles. This screening is manifested in
the factor $1+\Gamma^{(t\pm)}(r)$ in Eq. \eqref{eq:VphLS}. The first
term is also the driving term for the induced interaction, hence if
the first term is small the spin-orbit term of the induced interaction
will also be small.

The situation is very similar for the $v_8$ representation of all
four interactions studied here. In particular, the CEFT
interaction has a much weaker spin-orbit force, see
Fig. \ref{fig:vbare} whereas the spin-orbit component of the Reid
interaction is singular as $r\rightarrow 0$ and appears to be
much stronger than the bare spin-orbit interaction from the Argonne
potential. However, the short-ranged screening is also stronger,
resulting in an effective interaction that is rather similar for all
four potential models.

\begin{figure}[H]
  \centerline{\includegraphics[width=0.35\columnwidth,angle=-90]{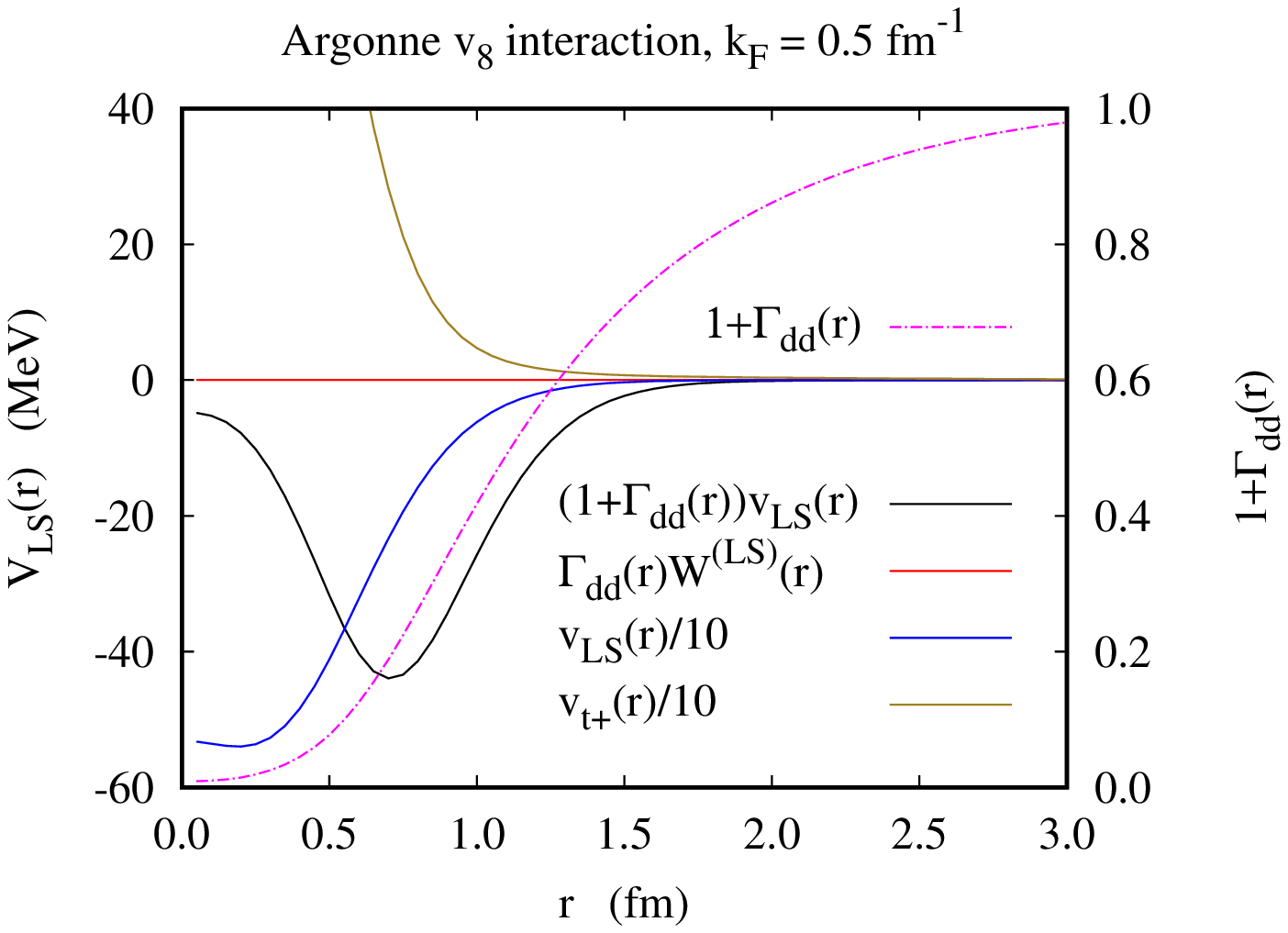}
    \includegraphics[width=0.35\columnwidth,angle=-90]{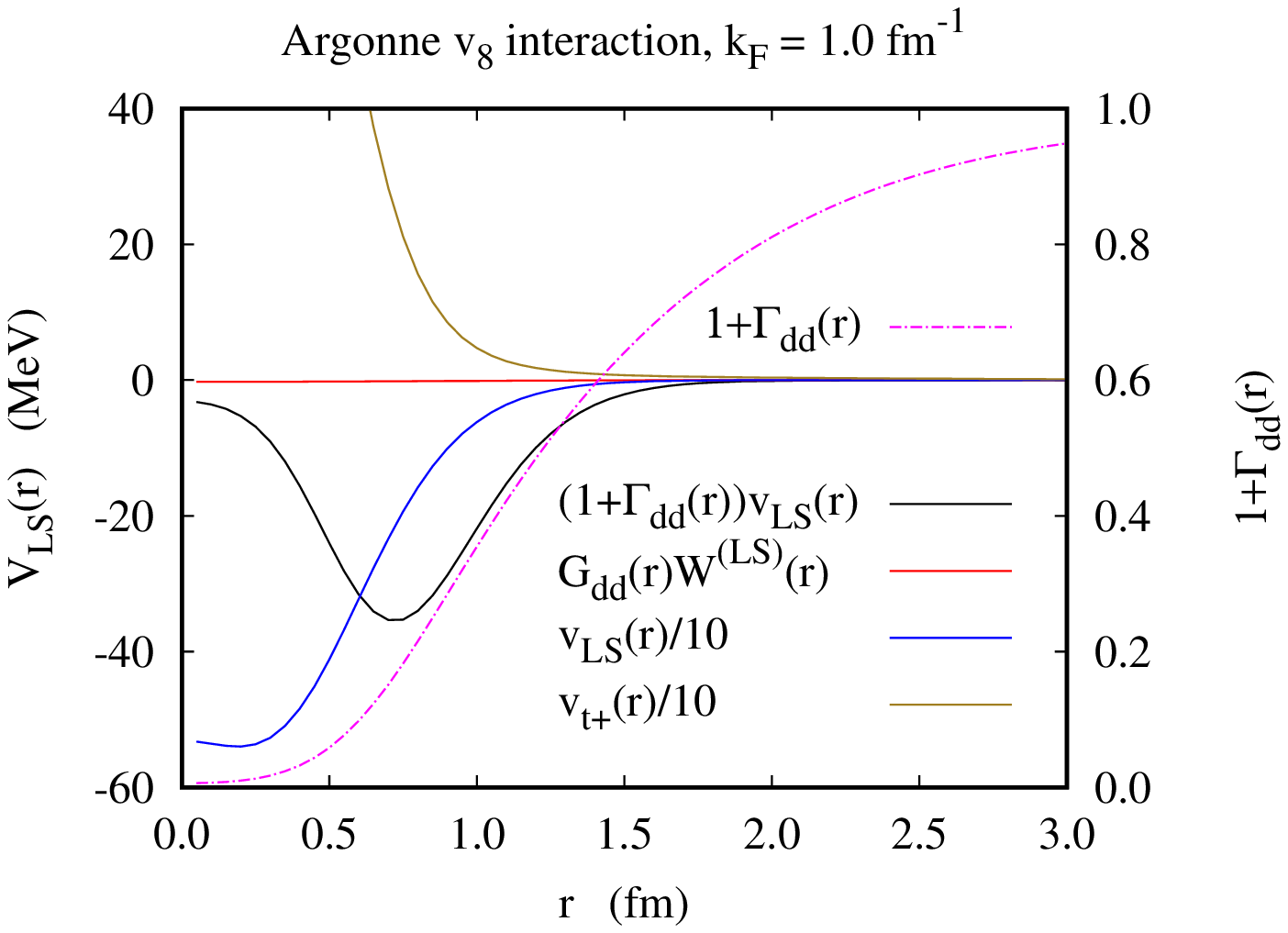}
  }
  \caption{The figures show, for $\KF = 0.5\,\text{fm}^{-1}$ and
    $\KF = 1.0\,\text{fm}^{-1}$
    the individual contributions to the effective spin-orbit
    interaction (left scale). We also show the direct correlation
    function $1+\Gamma_{\rm dd}^{(t+)}(r)$ in the $t+$ projector
    channel (magenta line, right scale) and the bare interactions
    $v_{\rm LS}(r)$ and $v_{t+}(r)$ of the Argonne potential. These
    were scaled by a factor of 0.1 to fit into the plot.
    \label{fig:vkf100argo}}
\end{figure}

\begin{figure}[H]
  \centerline{\includegraphics[width=0.35\columnwidth,angle=-90]{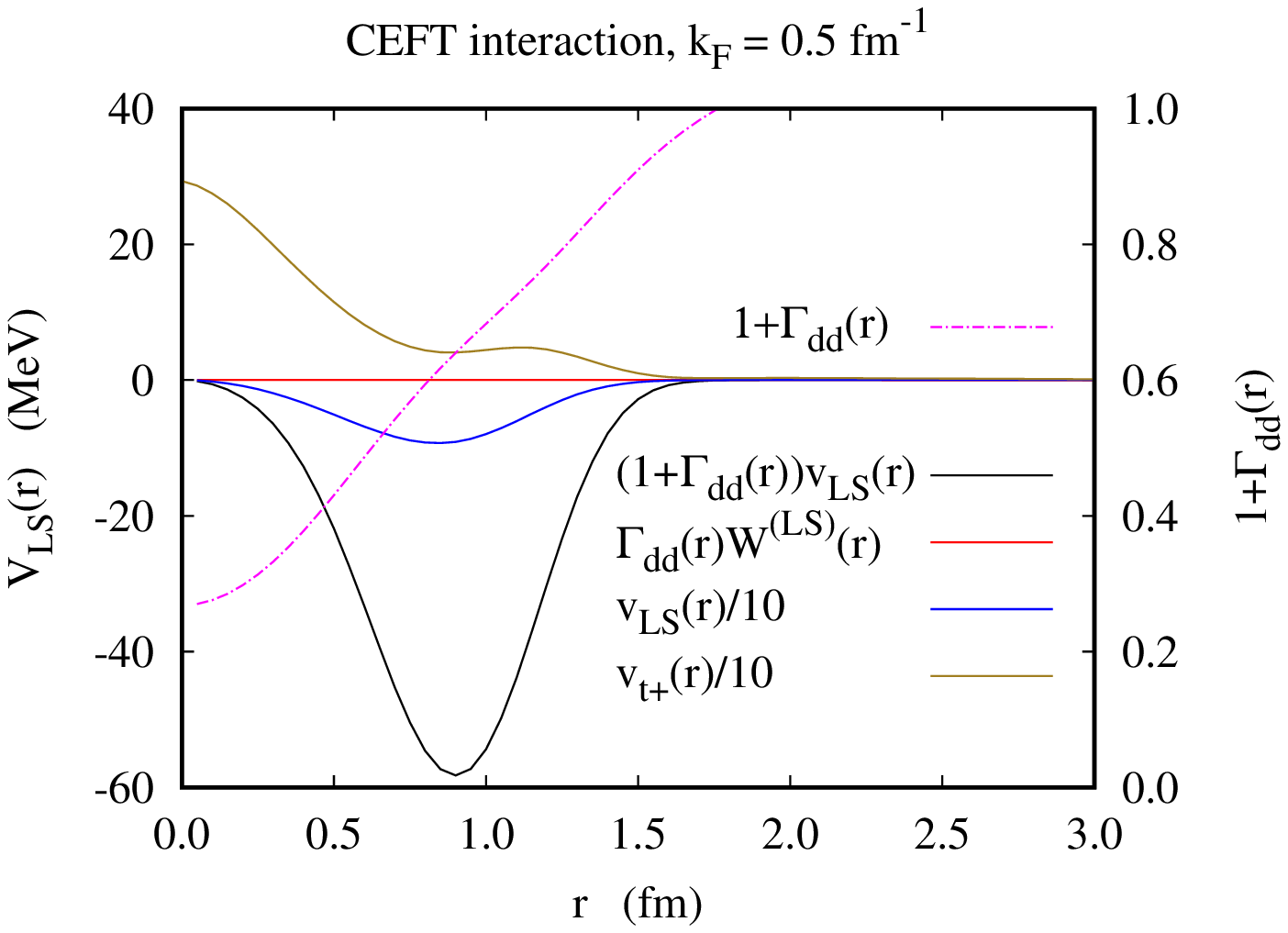}
    \includegraphics[width=0.35\columnwidth,angle=-90]{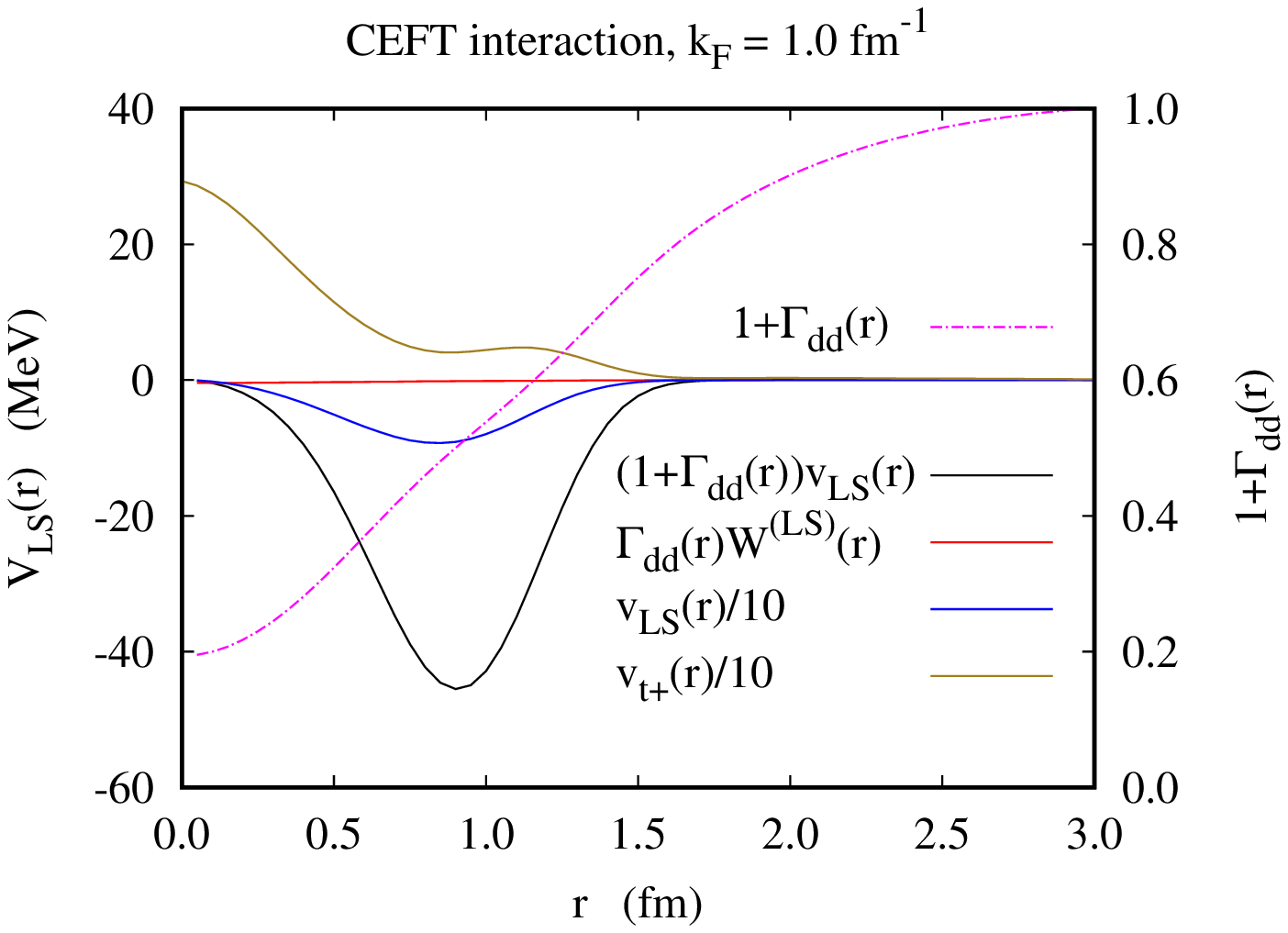}
  }
  \caption{Same as Fig. \ref{fig:vkf100argo} for the CEFT interaction.
    \label{fig:vkf100ceft}}
\end{figure}

Details on the different contributions to the effective
interactions are shown in Figs. \ref{fig:vkf100argo} and
\ref{fig:vkf100ceft}.  The figures show, at two representative densities
$\KF = 0.5\,\text{fm}^{-1}$ and $\KF = 1.0\,\text{fm}^{-1}$,
the decomposition \eqref{eq:VphLS} of the
particle-hole interaction into the short-ranged screening effect
$v_{\rm LS}(r)(1+\Gamma^{(t+)}_{\rm dd}(r))$ and the induced
interaction $\Gamma^{(t+)}_{\rm dd}(r)W_I^{\LSsup}(r)$. We also show,
for comparison, the bare interactions $v_{\rm LS}(r)$ and $v_{t+}(r)$.
The comparison offers an explanation for why the spin-orbit potential
is strongly suppressed by many-body correlations. The pair correlation
$1+\Gamma^{(t+)}_{\rm dd}(r)$ is predominantly determined by
$v_{t+}(r)$ which is strongly repulsive. Hence, the correlation
function tends to suppress the interaction.

The results for the Reid 68 and Reid 93 interactions are quite similar
to the ones shown above; in fact the direct correlation functions
$\Gamma_{\rm dd}(r)^{(t\pm)}$ are almost indistinguishable. The CEFT
interaction makes quantitatively somewhat different, but qualitatively
similar predictions.  These are shown in Figs. \ref{fig:vkf100ceft}.

Thus, our result is that the effective interaction has very little
resemblance to the bare interaction, but that a relatively simple
treatment of correlations is adequate to deal with many-body effects.

\subsection{Effective Interactions}

Effective interactions have so far played only the role of auxiliary
quantities in the calculations of the ground-state structure. They are,
however, essential input to the calculation of dynamic properties
like the dynamic response, single-particle excitations, and pairing
phenomena as discussed in the next two sections.

\begin{figure}[H]
  \subfigure[\label{fig:Vphargo_s}]%
            {\includegraphics[width=0.25\columnwidth,angle=-90]%
              {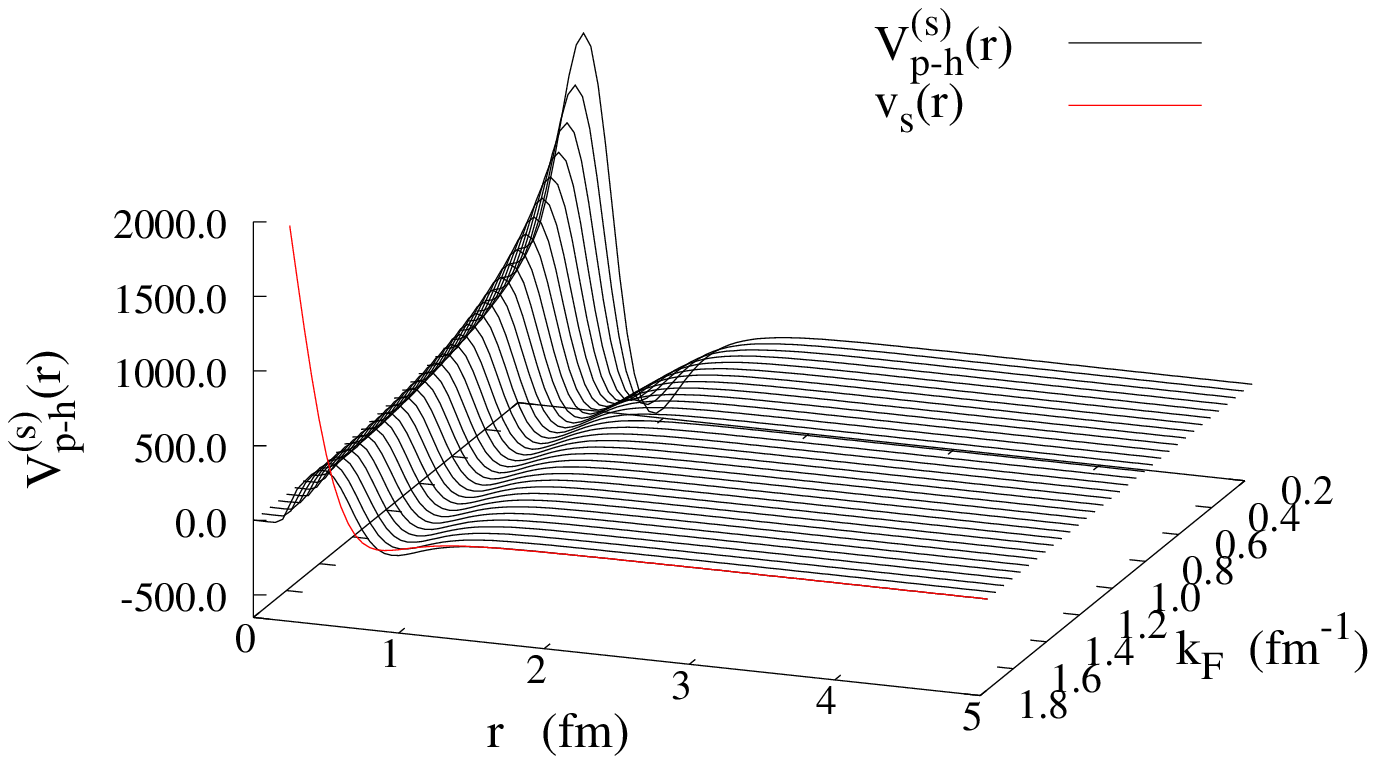}}\hfill
   \subfigure[\label{fig:Vphargo_tp}]%
            {\includegraphics[width=0.25\columnwidth,angle=-90]%
              {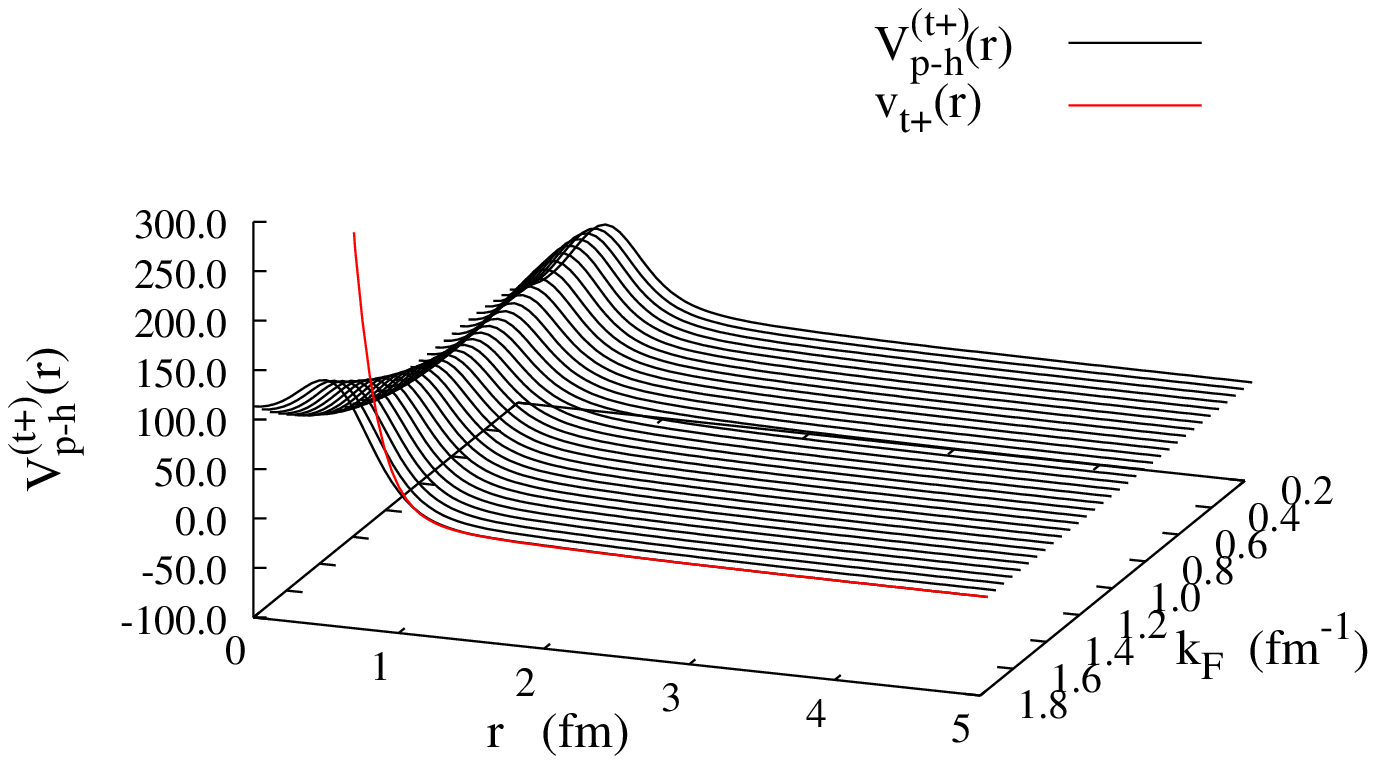}}\\
  \subfigure[\label{fig:Vphargo_tm}]%
            {\includegraphics[width=0.25\columnwidth,angle=-90]%
              {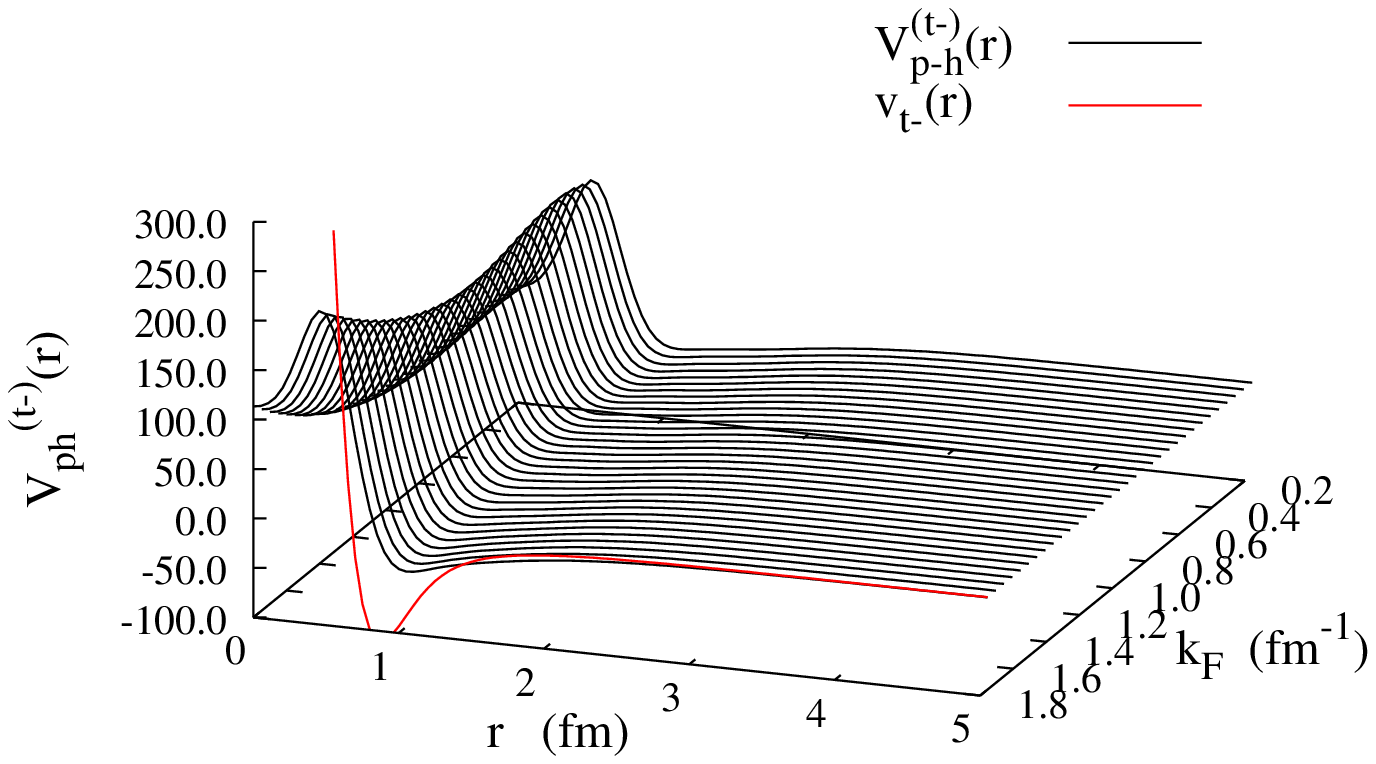}}\hfill
  \subfigure[\label{fig:Vphargo_ls}]%
            {\includegraphics[width=0.25\columnwidth,angle=-90]%
              {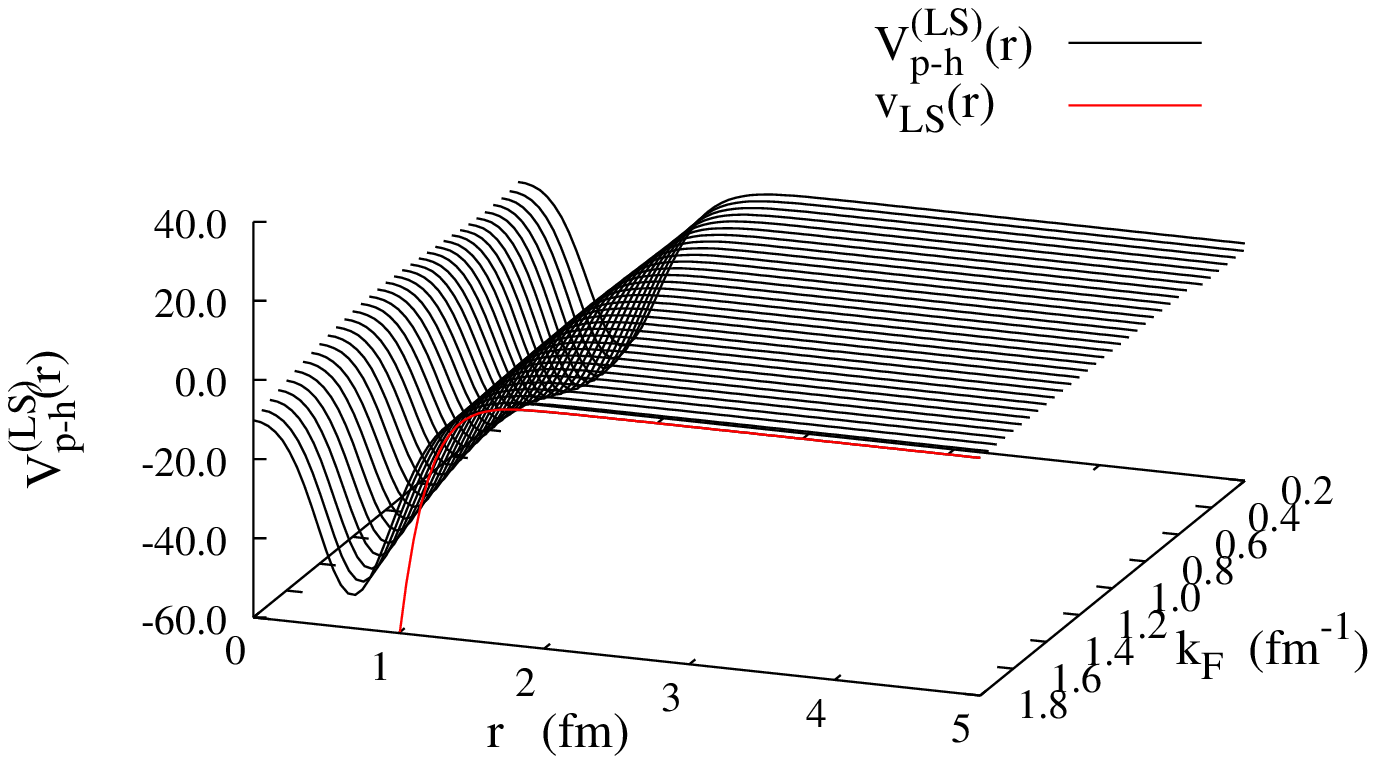}}
  \caption{(color online) The particle-hole interaction for the
    Argonne $v_8$ interaction in the four projection operator
    channels. The red line shows the bare interaction in the
    corresponding channels. Note the different energy
    scales.\label{fig:Vphargo}}
  \end{figure}

\begin{figure}[H]
  \subfigure[\label{fig:Vphceft_s}]%
            {\includegraphics[width=0.25\columnwidth,angle=-90]%
              {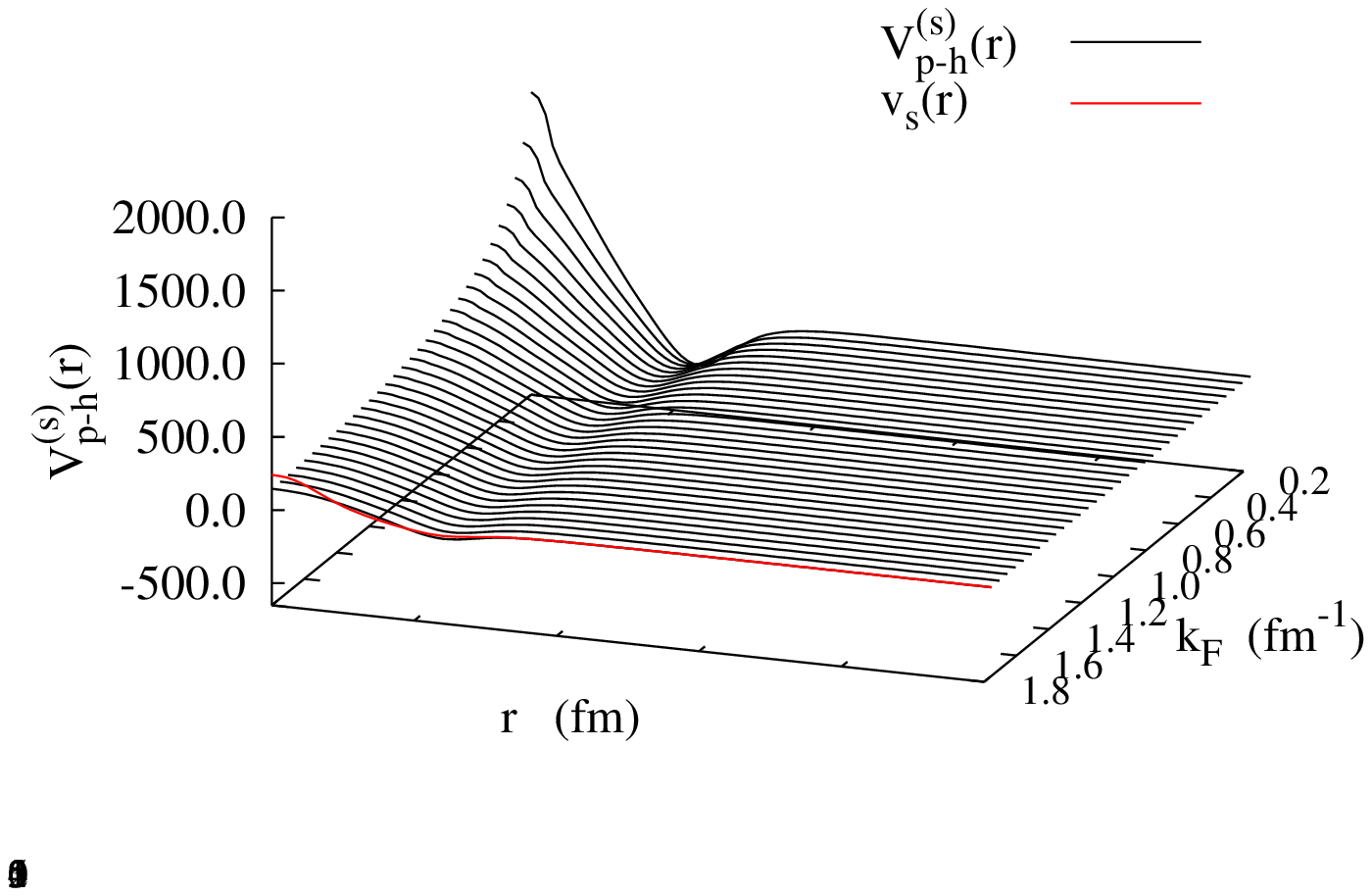}}\hfill
  \subfigure[\label{fig:Vphceft_tp}]%
            {\includegraphics[width=0.25\columnwidth,angle=-90]%
              {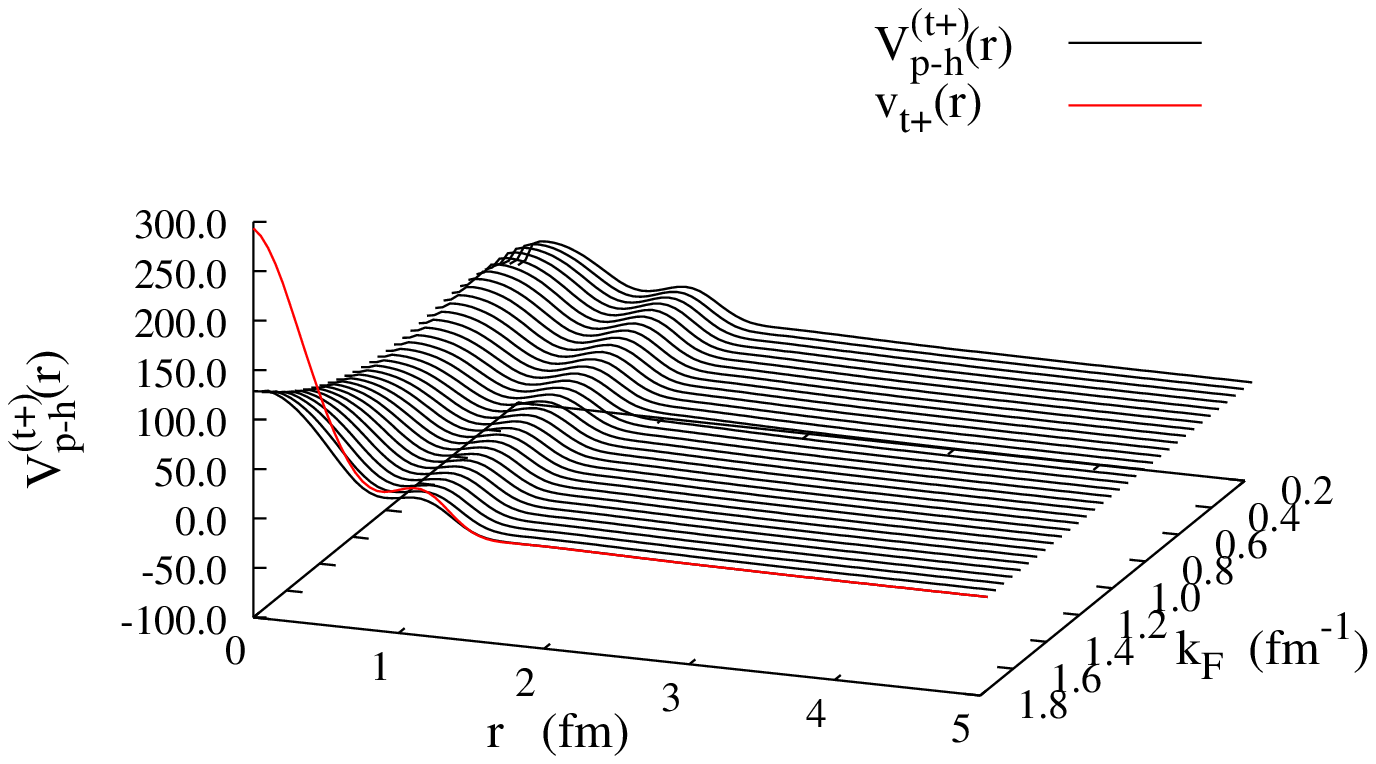}}\\
  \subfigure[\label{fig:Vphceft_tm}]%
            {\includegraphics[width=0.25\columnwidth,angle=-90]%
              {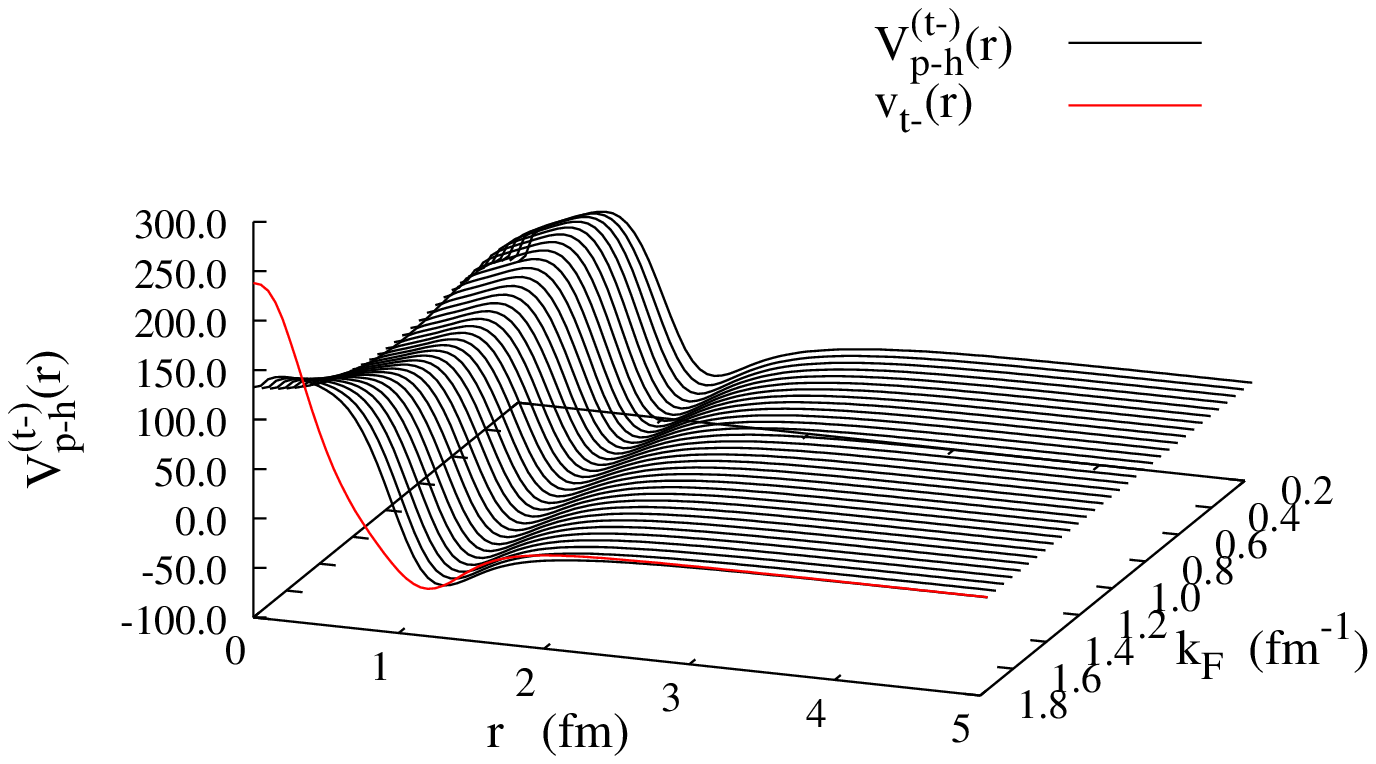}}\hfill
  \subfigure[\label{fig:Vphceft_ls}]%
            {\includegraphics[width=0.25\columnwidth,angle=-90]%
              {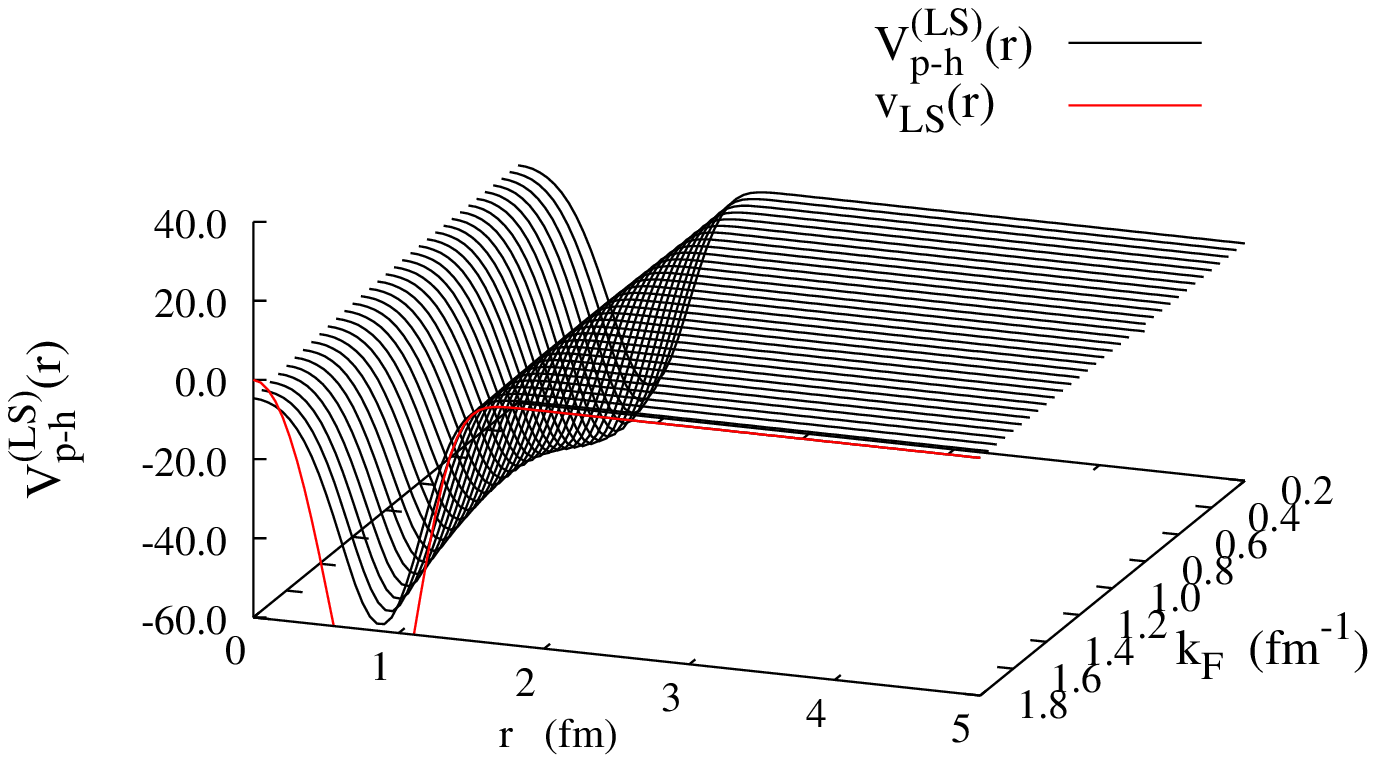}}
  \caption{Same as Fig. \ref{fig:Vphargo} for the CEFT $v_8$ interaction.
              \label{fig:Vphceft}}
  \end{figure}

To summarize our findings from this section, we show in
Figs. \ref{fig:Vphargo} and \ref{fig:Vphceft} the particle-hole
interactions in the four projector channels for the Argonne $v_8$ and
the CEFT interactions. The results from two versions of the Reid
interaction are close to those of Argonne $v_8$ and not shown.

Let us first focus on the Argonne interaction as a representative for
conventional interaction models. A number of phenomenological
considerations how local effective interactions should compare to the
underlying microscopic potentials have been spelled out very early in
the pseudopotentials theory for the helium liquids
\cite{Aldrich,ALP78}. These are short-ranged screening, at
intermediate distances a ``swelling'' of the repulsive core caused by
the kinetic energy cost from bending the pair wave function due to the
repulsive core, and an enhancement of the attractive well due to the
presence of other particles. Developing pseudopotentials is of course
much more complicated in neutron and nuclear matter due to the absence
of extensive experimental data; phenomenological theories
\citep{Wam93} therefore suffer from more ambiguities.

These effects are, within our microscopic analysis, represented
by the individual terms in Eq. \eqref{eq:FermiVphhat}
\begin{itemize}
\item{} Short-ranged screening is described by the pair wave function
  $\psi^{(s)}(r)$ going to zero for short distances, see the left panel
  of Fig, \ref{fig:Gddkf100}.
\item{} The enhanced repulsion is caused by the kinetic
  energy  $\frac{\hbar^2}{m}\left|\nabla\psi^{(s)}(r)\right|^2$,
\item{} The additional attraction is caused by the fact that the
  singlet pair wave function $\nabla\psi^{(s)}(r)$ has a strong
  peak around the location of the potential minimum which
  causes that $\psi^{(s)*}(r) v_s(r)\psi^{(s)}(r)$ is more attractive
  than $v_s(r)$.
  \end{itemize}
We also note, as already pointed out above, that the system
becomes more strongly correlated at lower densities, see Fig.
\ref{fig:psi_s_3d}.

In the two triplet channels, we see the same short-ranged damping.
The ``swelling'' of the repulsive core is much less pronounced since
the pair wave function is longer ranged, and there is no enhancement of
the attraction because $\psi^{(t\pm)}(r)$ is always less than 1.

None of these observations carry over to the spin-orbit contribution,
shown in Fig. \ref{fig:Vphargo_ls}.  Following our analysis of section
\ref{ssec:spinorbit}, the strongly repulsive core of the spin-triplet
interactions tends to suppress the bare spin-orbit interactions which
has, hence, little resemblance to the bare interaction.

The aforementioned physical effects are much less visible for the CEFT
interaction shown in Figs. \ref{fig:Vphceft}.  In particular there is
not much short-ranged screening in the spin-singlet channel, we have
observed and discussed this already in section \ref{ssec:gof0}. The
triplet channels are more repulsive which leads to some screening but
also again to no enhanced attraction at intermediate distances. The
effective spin-orbit interaction is actually quite similar to that of
the Argonne interaction albeit the underlying bare potential is quite
different.

We have above described the prescription of localized parquet diagram
summation to approximate the energy dependent effective interaction
$\hat W(q,\omega)$, Eq. \eqref{eq:Wnonlocal} by an energy-averaged
interaction $\hat W(q)$, \eqref{eq:Wlocal}. This procedure guarantees that
the static effective interaction reproduces the static structure
function which is, in principle, experimentally accessible. Of course,
this does not mean that the static  $\hat W(q)$ is a good approximation
in all cases. In particular, the effective interaction is a direct
input to the pairing calculations discussed in Section \ref{sec:BCS}
which is a low-energy phenomenon. In that case, the zero-energy
version $\hat W(q,\omega=0)$ is the more appropriate quantity.
We will return to this point in section \ref{sec:BCS} where we will
also discuss further corrections to the effective interactions.

\section{Linear Response and the Single Particle Spectrum\label{sec:dynamics}}

\subsection{Dynamic response and the dynamic structure function}
\label{ssec:response}

The response functions describe the response of a system to external
perturbations \cite{FetterWalecka}. An external field may be a scalar
field or a longitudinal or a transverse spin-dependent field
\begin{equation}
  h^{\rm (c)}_{\rm ext}(\qvec), \qquad h^{\rm (L)}_{\rm ext}(\qvec)\hat\qvec\cdot\bsigma,
  \quad\text{and}\quad h^{\rm (T)}_{\rm ext}(\qvec)(\hat\qvec\times\bsigma)
  \label{eq:extfields}\,,
\end{equation}
which induces density or spin-density fluctuations. The connection between
the external field and the induced (spin--)density fluctuations are
characterized by the response function $\chi(q,\omega)$.

A complete characterization of the dynamic response function requires
information on all excited states of the system under investigation.
Such a knowledge is, in practice, neither required nor useful because
one is concerned mostly with low-lying excitations which can be
classified by the quantum numbers of 1, 2, 3, $\dots$ particles.  The
simplest type of excitations is then the one that can be characterized
by the quantum numbers of a single particle.

The cleanest description of what these individual excitations describe
is provided for the case of interacting bosons which we describe here
briefly.  Beginning with Feynman \cite{Feynman} and Feynman-Cohen
\cite{FeynmanBackflow} as well as Jackson and Feenberg
\cite{JaFe,JaFe2,Jackson71,JacksonSkw} the approach has been
formulated in different ways; we choose here a derivation that
corresponds to the one to be used below.

The time-dependent wave function of the excited states is
written in the form
\begin{eqnarray}
\ket{\Psi(t)} &=&
         \exp\bigl[-\I E_0 t/\hbar\bigr]\,
        \ket{\Psi_0(t)}\;,\nonumber\\
\ket{\Psi_0(t)}
&=& \frac{1}{ I^{1/2}(t)}\exp\Bigl[\textstyle\frac{1}{2}\displaystyle\delta U(t)\Bigr]\,
\ket{\Phi_0}
\label{eq:BoseExcWaveFunction}\\
I(t)&=&\bra{\Phi_0}
        \exp\Bigl[\textstyle\frac{1}{2}\displaystyle \delta U^\dagger(t)\Bigr]
        \exp\Bigl[\textstyle\frac{1}{2}\displaystyle \delta U(t)\Bigr]
        \ket{\Phi_0} \;,\nonumber
\end{eqnarray}
where $\ket{\Phi_0}$ is the exact ground state and
the excitation operator can be written in full generality as
\begin{equation}
  \delta U(t) \equiv  \sum_i\delta u^{(1)}(\rvec_i;t)
  + \sum_{i<j}\delta u^{(2)}(\rvec_i,\rvec_j;t) + \dots
\equiv \delta U_1(t) + \delta U_2(t) + \dots\,.\label{eq:UopBose}
\end{equation}
The excitation amplitudes $\delta u^{(n)}(t)$ are determined
by the action principle
\begin{equation}
{\cal S} \left[\delta u^{(n)},\delta u^{(n)*}\right] = \int\!\! dt \; {\cal L}(t) \;,
\label{eq:action}
\end{equation}
with the Lagrangian \cite{KermanKoonin,KramerSaraceno,LDavid,rings}
\begin{eqnarray}
  {\cal L}(t) &=& \Big\langle\Psi(t)  \Big| \; H +
  H_{\rm ext}(t)- \I\, \hbar \frac{\partial}{
  \partial t}\; \Big| \Psi(t) \Big\rangle\nonumber\\
        &=& \Big\langle\Psi_0(t)  \Big| \; H ' +
  H_{\rm ext}(t)- \I\, \hbar\frac {\partial}{
  \partial t}\; \Big| \Psi_0(t) \Big\rangle \,.
  \label{eq:Lagrange1}
\end{eqnarray}
To derive linear equations of motion, the Lagrangian
(\ref{eq:Lagrange1}) is expanded to second order in the excitation
operator $U(t)$. When manipulating the Lagrangian, we can utilize that
it always appears in the action integral \eqref{eq:action}. For the
procedure to be meaningful, one must as usual ensure that the state
$\ket{\Phi_0}$ is stationary. This means that the first order terms must
vanish. These are
\begin{equation}
  \delta {\cal L}^{(1)}(t) = \Re \bra{\Psi_0} (H-E_0) 
    \delta U(t)\ket{\Phi_0}
  \label{eq:Bagrangian1}\,.
\end{equation}
The term proportional to $\dot U(t)$ does not contribute since it is
a total time derivative.  Evidently, the expression
\eqref{eq:Bagrangian1} is zero if $\ket{\Phi_0}$ is the exact ground
state. If the expansion \eqref{eq:UopBose} of the excitation
operator is truncated at the component $\delta u^{(n)}(t)$, then it is
sufficient that $\bra{\Phi_0}$ is a Feenberg wave function with
optimized correlations up to the level $n$.

The second-order term can be written as
\begin{eqnarray}\delta {\cal L}^{(2)}(t)\ &=&
  \frac{1}{4}
  \bra{\Phi_0}
  \delta U^*(t)\Re \delta U(t) H'+H'\delta U(t)\Re\delta U(t)
  \ket{\Phi_0}\nonumber\\
  &+&\frac{1}{8}
  \bra{\Phi_0}
        \left[ \delta U^*(t),\left[H',\delta U(t)\right]\right]
        \ket{\Phi_0}
\nonumber\\
        &-&\frac{\I\hbar}{8}\bra{\Psi_0}
          \delta \dot U(t) \delta U^*(t)
            -\delta \dot U^*(t) \delta U(t)\ket{\Psi_0}
\nonumber\\
        &+&\frac{1}{2}
        \bra{\Phi_0}\delta U^*(t)\delta H(t)
            +  H_{\rm ext}(t)\delta U(t)
                \ket{\Psi_0}
\label{eq:Bagrangian2}
\end{eqnarray}
Again, the term in the first line vanishes for the exact ground state
or if $\ket{\Phi_0}$ is a Feenberg wave function with optimized
correlations up to the level $2n$. Note also that, for a local
interaction, $\left[\delta U^*(t),\left[H',
    \delta U(t)\right]\right]=\left[\delta U^*(t),\left[T,\delta U(t)\right]\right]$.
The last three lines of Eq.
\eqref{eq:Bagrangian2} are the starting point for calculations of
the dynamic response of Bose fluids.

The purpose of the above discussion is to highlight both the physical
meaning of the individual fluctuations $\delta u^{(n)}(t)$ {\em and\/}
the importance of building a theory of excitations upon a sufficiently
accurate ground state.

The one-body component $\delta u^{(1)}(t)$ is in an extended system a
plane wave. If the excitation operator is truncated at that level,
the dispersion relation becomes \cite{Feynman}
\begin{equation}
  \hbar\omega(k) = \frac{\hbar^2 k^2}{2m S(k)}\,.
  \label{eq:Feynman}
  \end{equation}
It is influenced only by the mean field of the underlying many-particle system.

Comparing the excitation function \eqref{eq:UopBose} with the
correlated wave function \eqref{eq:Jastrow} we see that $\delta
u^{(2)}(\rvec_i,\rvec_j;t)$ adds a time-dependent component to the
pair correlation function $u_2(\rvec_i,\rvec_j)$. Since the latter
describes mostly short-ranged correlations, $\delta
u^{(2)}(\rvec_i,\rvec_j;t)$ describes the time-dependence of the wave
functions {\em at wave lengths that are comparable to the average
  particle distance.\/} An intuitive example for pair fluctuations is
the Feynman-Cohen backflow \cite{FeynmanBackflow}; a systematic
evaluation \cite{JaFe,JaFe2,Jackson71,JacksonSkw} has been carried out
by Jackson and Feenberg; both led to a significant improvement of the
theoretical predictions for the zero-sound dispersion relation in
\he4.  Going beyond pair fluctuations has to date led to an
unprecedented agreement
\cite{eomIII,skw4lett,He4Dispersion,Encyclopedia} between theory and
experiment in \he4.

From both Eqs. \eqref{eq:Bagrangian1} and \eqref{eq:Bagrangian2} it
should be obvious that a good approximation for the ground state wave
function is mandatory for both the stability of the excitations --
Eq. \eqref{eq:Bagrangian1} -- and the correct excitation Hamiltonian
-- Eq. \eqref{eq:Bagrangian2}. This is done by default in a strongly
interacting Bose system, but rarely possible for fermions.

To generalize the {\em ansatz\/} \eqref{eq:BoseExcWaveFunction} for
the excitation wave functions to Fermions, we define
\begin{eqnarray}
\Big|\Psi_0(t)\Big\rangle
&=& \frac{1}{ I^{1/2}(t)}\hat F\exp\Bigl[\textstyle\frac{1}{2}\displaystyle\delta U(t)\Bigr]\,
\ket{\Phi_0}
\label{eq:FermiExcWaveFunction}\\
I(t)&=&\bra{\Phi_0}
        \exp\Bigl[\textstyle\frac{1}{2}\displaystyle \delta U^\dagger(t)\Bigr]
        \hat F^\dagger\,\hat F\exp\Bigl[\textstyle\frac{1}{2}\displaystyle\delta U(t)\Bigr]
        \ket{\Phi_0};,\nonumber
\end{eqnarray}
where $\ket{\Phi_0}$ is the model state as defined in
Eq. \eqref{eq:PsiFPhi}, $E_0$ is the energy expectation value with
respect to the correlated wave function, and the excitation operator
is
\begin{eqnarray}
  \delta U(t) &\equiv&  \sum_{ph}\; \delta u^{(1)}_{ph} (t)\;
  \creat p \annil h 
    + \frac{1}{ 2}\;\sum_{pp'hh'}\;
   \delta u^{(2)}_{pp'hh'} (t)\; \creat p \creat {p'}
        \annil{h'} \annil{h}+\ldots\nonumber\\
&\equiv& \delta U_1(t) + \delta U_2(t)+\ldots\,.\label{eq:UopFermi}
\end{eqnarray}
The form \eqref{eq:FermiExcWaveFunction} deviates from the boson form
\eqref{eq:BoseExcWaveFunction} in the sense that the excitation
operator $U(t)$ acts on the model state and not on the exact ground
state.  The reason for this choice is that one wants to keep the
particle-hole structure of the Fermion system, but there is no
practical way to determine how a second-quantized operator
\eqref{eq:UopFermi} acts on either the exact ground state or a
correlation operator $\hat F$ of the form \eqref{eq:Jastrow} or
\eqref{eq:f_prodwave}.

The first order terms in the fluctuations are
\begin{eqnarray}
  \delta {\cal L}^{(1)}(t) &=& \frac{\Re \bra{\Phi_0}\hat F^\dagger(H-E_0)\hat F
    \delta U(t)\ket{\Phi_0}}{\bra{\Phi_0}\hat F^\dagger\hat F\ket{\Phi_0}}
  \nonumber\\
  &=&\Re \sum_n\sum_{\genfrac{}{}{0pt}{1}{p_1\ldots p_n}{h_1\ldots h_n}}
  \delta u^{(n)}_{p_1\ldots p_n,h_1\ldots h_n}(t)
  \frac{\bra{\Phi_0}\hat F^\dagger(H-E_0)\hat F\creat{p_1}\dots\creat{p_n}
    \annil{h_n}\dots\annil{h_1}\ket{\Phi_0}}{\bra{\Phi_0}\hat F^\dagger\hat F\ket{\Phi_0}}\nonumber\\
  \label{eq:Lagrangian1}\,.
\end{eqnarray}
Since the state to be perturbed should be stationary, the first order
term $\delta {\cal L}^{(1)}(t)$ should be zero. This condition
translates into
\begin{equation}
\frac{\bra{\Phi_0}\hat F^\dagger(H-E_0)\hat F\creat{p_1}\dots\creat{p_n}
    \annil{h_n}\dots\annil{h_1}\ket{\Phi_0}}{\bra{\Phi_0}\hat F^\dagger\hat F\ket{\Phi_0}} \genfrac{}{}{0pt}{1}{!}{=} 0\,.\label{eq:Brillouin}
\end{equation}
We shall refer to the conditions \eqref{eq:Brillouin} as the
Brillouin conditions.

Truncating the excitation operator at the level $U_1(t)$, the
equations of motion lead to the so-called ``random-phase
approximation'' (RPA) including exchange corrections. The Brillouin
condition for $n=1$ is automatically satisfied because the state
$U_1\ket{\Phi_0}$ has a finite momentum and is, hence, orthogonal to
the ground state.

Omitting exchange corrections in the RPA, the response to the three
external field \eqref{eq:extfields} ({\em plus\/} possible isospin
components) has the form
\begin{eqnarray}
   &&\chi_\alpha(q,\omega) = \frac{\chi_0(q,\omega)}
       {1-\tilde V_{\rm p-h}^{(\alpha)}(q,\omega)\chi_0(q,\omega)}\nonumber\\
       &=& \chi_0(q,\omega) +
       \chi_0(q,\omega)\tilde W^{(\alpha)}(q,\omega)\chi_0(q,\omega)
 \label{eq:chialpha}
 \end{eqnarray}
with the effective interactions \eqref{eq:W135}. We can ignore the
terms \eqref{eq:W79} because these have been found to be numerically
negligible.

The most important input to the calculations is the particle-hole
interaction.  If the correlation operator $\hat F$ is omitted, the
particle-hole interaction becomes the bare interaction.  Of course,
the non-interacting Fermi sea $\ket{\Phi_0}$ is nowhere close to the
exact ground state, hence the interaction must, in practical
applications, be replaced by effective interactions that are somehow
fitted to reproduce experiments. Applications of this procedure are
abundant, for a review see Ref. \citenum{CoRPA}. A phenomenological
approach that comes closest to what the method of correlated wave
functions predicts is the formulation in terms of pseudopotentials
\cite{ALP78,Wam93,Pines:1988gik}.

To go beyond the RPA model one includes multi-pair amplitudes in the
excitation operator. The procedure is known under different names in
various fields. The simplest version is, in nuclear physics, known as
``second RPA (SRPA)''.
\cite{PhysRev.126.2231,SRPA83,SRPA87,Wambach88,PhysRevC.81.024317}. It
has shown promise for describing nuclear giant resonance when it is
applied with realistic potentials \cite{PPSRPA09}. In an early work on
SRPA \cite{NDW1988}, the correlated nuclear ground state was described
by second-order perturbation theory, which subsequently was shown to
overestimate the effect of correlations \cite{VMS1992}. Many-body
effects have been included at several levels; a method that is
particularly popular in atomic and molecular systems is the coupled
cluster method (CCM) \cite{CK,KLZ,Day81,BartlettBook}. The
diagrammatically most complete evaluation of many-body quantities is
provided by parquet-diagram summations.  The method has been termed
``Dynamic Many-Body Theory'' (DMBT) and has led to excellent agreement
between theory and experiments in \he3
\cite{Encyclopedia,eomIII,2p2h,Nature_2p2h}.

Unlike for Bosons, satisfying the Brillouin condition for Fermion pair
excitations is more difficult. Using a non-interacting Fermi sea as
model ground state violates, in a realistic system, the Brillouin
condition massively. In fact, the operator \eqref{eq:UopFermi} with
{\em stationary\/} particle-hole amplitudes $u_{pp',hh'}$ acting on a
Slater determinant is the starting point of coupled cluster theory
\cite{CK,KLZ} that was originally designed to deal with the ground
state of nuclear systems. The extension of the coupled cluster method
to excited states \cite{Emrich1,Emrich2,RN236} builds the excitations on a
ground state that contains the same particle-hole amplitudes; it
satisfies the Brillouin conditions automatically. Jastrow-Feenberg or
local parquet theory provides, at the expense of the local
approximations discussed in section \ref{sec:central}, the most
complete summation of Feynman diagrams. The optimization condition
\eqref{eq:euler} for the pair correlations can be written in momentum
space as \cite{2p2h}
\begin{equation}
0 =\left[{\delta E \over \delta \tilde u_2({\bf r})}\right]^{\mathcal F}(\qvec)=
 \sum_{hh'}
    \frac{\bra{\Phi_0} F^{\dagger}H' F
          \ket{\creat{\hvec'+\qvec}\creat{\hvec-\qvec}\annil{\hvec}\annil{\hvec'}\Phi_0}}
   {\bra{\Phi_0}  F^{\dagger}F \ket{\Phi_0}}
\label{eq:zweicoll}
\end{equation}
\ie for optimized correlation functions, the 2-body Brillouin
condition is satisfied in the Fermi-sea average. A consequence of the
complete summation of ring and ladder diagrams is that the method can
easily deal with the very strongly correlated helium liquids without
the need for cutoff procedures to tame the hard core of the
interactions.

The actual evaluation of the equations of motion is rather complicated
and technical, we skip these developments which may be found in Ref.
\citenum{2p2h}.

Beginning with Eqs. (5.7) of Ref. \citenum{2p2h}, we write
\begin{equation}
  \chi(q; \omega) = N(q; \omega) / D(q; \omega)\label{eq:chidmbt}
  \end{equation}
with
\begin{eqnarray}
  N(q,\omega)&=&\kappa^{\phantom{*}}_0(q; \omega)
  \left[1-2\kappa_0^*(q; -\omega)\sigma_-(q)\widetilde W^*_{\!_{\rm A}}(q; -\omega)
   \right]\nonumber\\
  &+& \kappa_0^*(q; -\omega)
  \left[1-2\kappa^{\phantom{*}}_0(q; \omega)\sigma_+(q)\widetilde W _{\!_{\rm A}}(q;  \omega)\right]\label{eq:chinum}
\end{eqnarray}
and
\begin{eqnarray}
D(q; \omega)
        &=& 1
        - \chi(q; \omega)\left[\tilde V_{\rm p-h}(q)+\sigma_+^2(q)\widetilde W  _{\!_{\rm A}}(q; \omega)+\sigma_-^2(q)\widetilde W  _{\!_{\rm A}}^*(q;-\omega)\right]
 \nonumber\\
 &+&  \kappa^{\phantom{*}}_0(q; \omega) \kappa_0^*(q; -\omega)
 \frac{S(q)}{\SF(q)}\left[\sigma_+(q)\widetilde W  _{\!_{\rm A}}(q; \omega)-\sigma_-(q)\widetilde W  _{\!_{\rm A}}^*(q;-\omega)\right]\times\nonumber\\
 &\times&\left[2V_{\rm p-h}(q)+\sigma_+(q)\widetilde W  _{\!_{\rm A}}(q; \omega)+\sigma_-(q)\widetilde W^*_{\!_{\rm A}}(q;-\omega)\right]\,.\label{eq:chiden}
\end{eqnarray}
with the positive-energy Lindhard function
\begin{equation}
\kappa^{\phantom{*}}_0(q;\omega) \equiv \frac{1}{N}\sum_h
\frac{\bar n_{\bf p} n_{\bf h}}{\hbar\omega - t(p)-t(h) + \I\eta}\,,\qquad\pvec=\hvec+\qvec\,,
\label{eq:chi0pmdef}
\end{equation}
and $\sigma^{\pm}(q) \equiv [\SF(q)\pm S(q)]/2S(q)$.

\begin{figure}[H]
  \centerline{\includegraphics[width=0.7\textwidth]{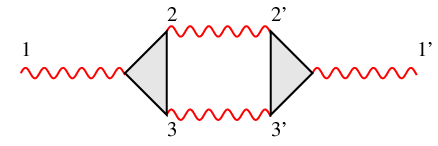}}
  \caption{The basic type of processes described by the dynamic
    interaction $\widetilde W_{\!\rm A}(q;\omega)$.  The red wavy
    lines are modified response functions $\kappa(q;\omega)$
    \eqref{eq:kappa_qomega} and the shaded triangle a three-body
    vertex $\tilde K_{q,q'q''}$.\label{fig:pair_propagator} }
\end{figure}

The key new element is the dynamic interaction correction
\begin{equation}
 \widetilde W_{\!\rm A}(q;\omega) =
 \frac{1}{2N}\sum_{{\bf q}'{\bf q}''}
 |\tilde K_{q,q'q''}|^2 \, \tilde E^{-1}(q',q''; \omega)\label{eq:resultWA2} \\
\end{equation}
which describes the splitting and recombination of two ``elementary''
excitations.  The pair energy denominator has the
form \begin{equation} \tilde E^{-1}(q_1,q_2; \omega) \;=\;
  -\!\int\limits_{-\infty}^{\infty}\!\frac{d\hbar\omega'}{2\pi\I}\>
  \kappa(q_1;\omega')\> \kappa(q_2;\omega\!-\!\omega')
 \label{eq:Einvqfromkappa}
\end{equation}
with
\begin{equation}
 \kappa(q;\omega) \>\equiv\> \frac{1}{N}\sum_{hh'} \kappa_{ph,p'h'}(\omega)
 \>=\> \frac {\kappa_0(q;\omega)}{1+
              \hbar\omega\tilde\Gamma_{\rm dd}(q)\kappa_0(q;\omega)}\,.
\label{eq:kappa_qomega}
\end{equation}
The processes are schematically described in Fig. \ref{fig:pair_propagator}.

Conventionally, the splitting is described in ``convolution approximation''
by the 3--body coupling matrix element
\begin{equation}
\tilde K_{q,q'q''}
        = \displaystyle \frac{\hbar^2}{2m} \,
                \frac{1}{\SF(q)}
                \left[
                  \qvec\!\cdot\qvec' \,
                  \tilde \Gamma_{\rm dd}(q')\frac{S(q'')}{\SF(q'')} +
                  \qvec\!\cdot\qvec''\,\tilde \Gamma_{\rm dd}(q'')
                  \frac{S(q')}{\SF(q')}
                \right]\,.
\label{eq:K3CA}
\end{equation}
Three-body correlations that are necessary in \he3 can be ignored in
neutron matter. In the limit of a Bose liquid, the form
\eqref{eq:chidmbt} reduces to the CBF form derived by Jackson and
Feenberg \cite{JaFe,JaFe2}.

The specific form \eqref{eq:K3CA} permits further simplifications.
The only new aspect for state-dependent correlations is that links in
the three-body vertices are operators that must be symmetrized.
Typically, we need to look at terms of the

\begin{equation}
\hat X_3(1,2,3)=\int d^3r_4\Tr_{\sigma_4}{\cal S} \hat F_1(1,4)\hat
F_2(2,4)\hat F_3(3,4)\label{eq:X3}
\end{equation}
where the ${\cal S}$ stands for the symmetrization of the spin
operators, and the $\hat F_\alpha(i,j)$,
can be any pair operator such as $\hat X_{\rm dd}$ or
$\hat S(q)/\SF(q)$.

\begin{figure}[H]
  \centerline{\includegraphics[width=0.95\textwidth]{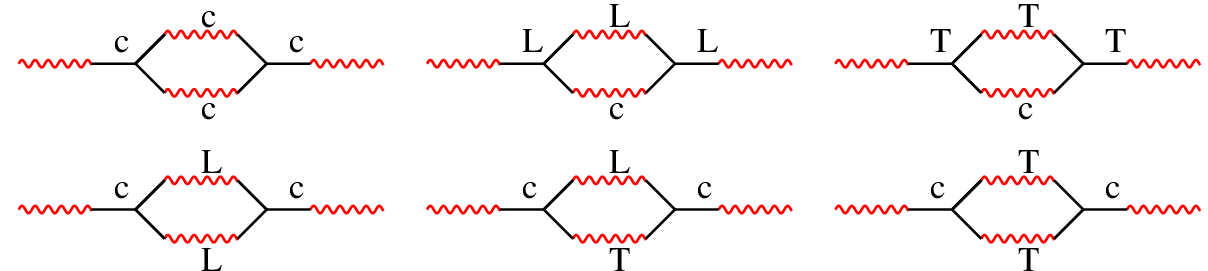}}
  \caption{This is the basic type of processes we must include in convolution
    approximation. The red wavy line is a modified response
    function that contains all the operators. The solid lines are
    $\hat W(i,j)$.\label{fig:pair_propagator_CA}
  }
\end{figure}

The paths $(1,2,2',1')$ and $(1,3,3',1')$ are convolutions, \ie that
separate in the $\{\1, \hat L, \hat T\}$ channels.  So at the end we
just need the 6 different cases shown in Fig.
\ref{fig:pair_propagator_CA}

  The left vertex involves
\begin{subequations}
  \begin{eqnarray}
    {\cal S}\left[\hat L(12)\hat L(13)\right] &=& {\cal S}\left[(\sigma_1\cdot\hat\rvec_{12})
      (\sigma_2\cdot\hat\rvec_{12})(\sigma_1\cdot\hat\rvec_{13})
      (\sigma_3\cdot\hat\rvec_{13})\right]\nonumber\\
    &=& \hat\rvec_{12}\cdot\hat\rvec_{13}(\sigma_2\cdot\hat\rvec_{12})
    (\sigma_3\cdot\hat\rvec_{13})\\
    {\cal S}\left[\hat L(12)\hat T(13)\right] &=&
    {\cal S}\left[(\sigma_1\cdot\hat\rvec_{12})
      (\sigma_2\cdot\hat\rvec_{12})\left[(\sigma_1\cdot\sigma_3)-(\sigma_1\cdot\hat\rvec_{13})
      (\sigma_3\cdot\hat\rvec_{13})\right]\right]\nonumber\\
      &=& (\sigma_2\cdot\hat\rvec_{12})(\sigma_3\cdot\hat\rvec_{12})
      -\hat\rvec_{12}\cdot\hat\rvec_{13}(\sigma_2\cdot\hat\rvec_{12})
      (\sigma_3\cdot\hat\rvec_{13})\\
      {\cal S}\left[\hat T(12)\hat T(13)\right] &=&\sigma_2\cdot\sigma_3
      -(\sigma_2\cdot\hat\rvec_{12})(\sigma_3\cdot\hat\rvec_{12})\nonumber\\
      &-&(\sigma_2\cdot\hat\rvec_{13})(\sigma_3\cdot\hat\rvec_{13})
      + \hat\rvec_{12}\cdot\hat\rvec_{13}(\sigma_2\cdot\hat\rvec_{12})(\sigma_3\cdot\hat\rvec_{13})
  \end{eqnarray}
\end{subequations}

Next we need to complete the chains and close them at the point
$\rvec_1'$.  Since the $\hat L$ and $\hat T$ operators are orthogonal,
it just means to set in the above 3 equations
$\rvec_2=\rvec_3=\rvec_1'$.  That means we get the 3 terms
\begin{subequations}
\begin{eqnarray}
    {\cal S}\left[\hat L(11')\hat L(11')\right] &=&
     \hat\rvec_{11'}\cdot\hat\rvec_{11'}(\sigma_1'\cdot\hat\rvec_{11'})
    (\sigma_1'\cdot\hat\rvec_{11'})=1\\
    {\cal S}\left[\hat L(11')\hat T(11')\right]
    &=& (\sigma_1'\cdot\hat\rvec_{11'})(\sigma_1'\cdot\hat\rvec_{11'})
    -\hat\rvec_{11'}\cdot\hat\rvec_{11'}(\sigma_1'\cdot\hat\rvec_{11'})
    (\sigma_1'\cdot\hat\rvec_{11'})\nonumber\\
&=& 0\\
      {\cal S}\left[\hat T(11')\hat T(11')\right] &=&\sigma_1'\cdot\sigma_1'
      -(\sigma_1'\cdot\hat\rvec_{11'})(\sigma_1'\cdot\hat\rvec_{11`})\nonumber\\
      &-&(\sigma_1'\cdot\hat\rvec_{11'})(\sigma_1'\cdot\hat\rvec_{11'})
      + \hat\rvec_{11'}\cdot\hat\rvec_{11'}(\sigma_1'\cdot\hat\rvec_{11'})
      (\sigma_{1}'\cdot\hat\rvec_{11'})\nonumber\\
      &=& 2\,.
  \end{eqnarray}
\end{subequations}
If there was no tensor force, the last 3 terms would reduce to only
one which can be easily interpreted: A phonon comes in, splits into
two spin-fluctuations, these recombine and a phonon comes out.

\subsection{Dynamic structure function}
\label{ssec:skw}

The dynamic structure function of nuclear and neutron matter has been
the subject of intense studies literally for decades, see among others
Ref. \citenum{ALBERICO1982429} for very early work.  Calculations were
typically at the RPA level, interactions were taken either
semi-phenomenologically using simplified nucleon-nucleon interactions
\cite{ALBERICO1982429,Hensel1983,PhysRevC.40.960}, effective Skyrme
interactions
\cite{PhysRevC.80.024314,PhysRevC.84.059904,PhysRevC.100.064301,%
  PhysRevC.89.044302,JPG41_2014,PhysRep536,AnnPhys214} or
pseudopotentials \cite{Pines:1988gik}. There are only a few low-order
calculations that attempt to include correlation effects in the
dynamic response \cite{NaiThesis,Benhar2009,Lovato2013}.

In our microscopic calculations, the response of the system to the
external fields \eqref{eq:extfields} is naturally formulated in terms
of the operators $\1$, $\hat L$ and $\hat T$, see
Eq. \eqref{eq:chialpha}. The tensor force breaks the degeneracy in
$\hat L$ and $\hat T$.

We have carried out calculations for the four interaction models
discussed in section \ref{sec:nucleons}. As noted already in
Ref. \citep{v4}, the results for all of these cases are very similar
and nothing can be learned from a comparison. We therefore show here
only results for the CEFT interaction, see Ref. \citep{v4} for
those of the Argonne potential.

\begin{figure}[H]
  \centerline{
    \includegraphics[width=0.36\columnwidth,angle=-90]{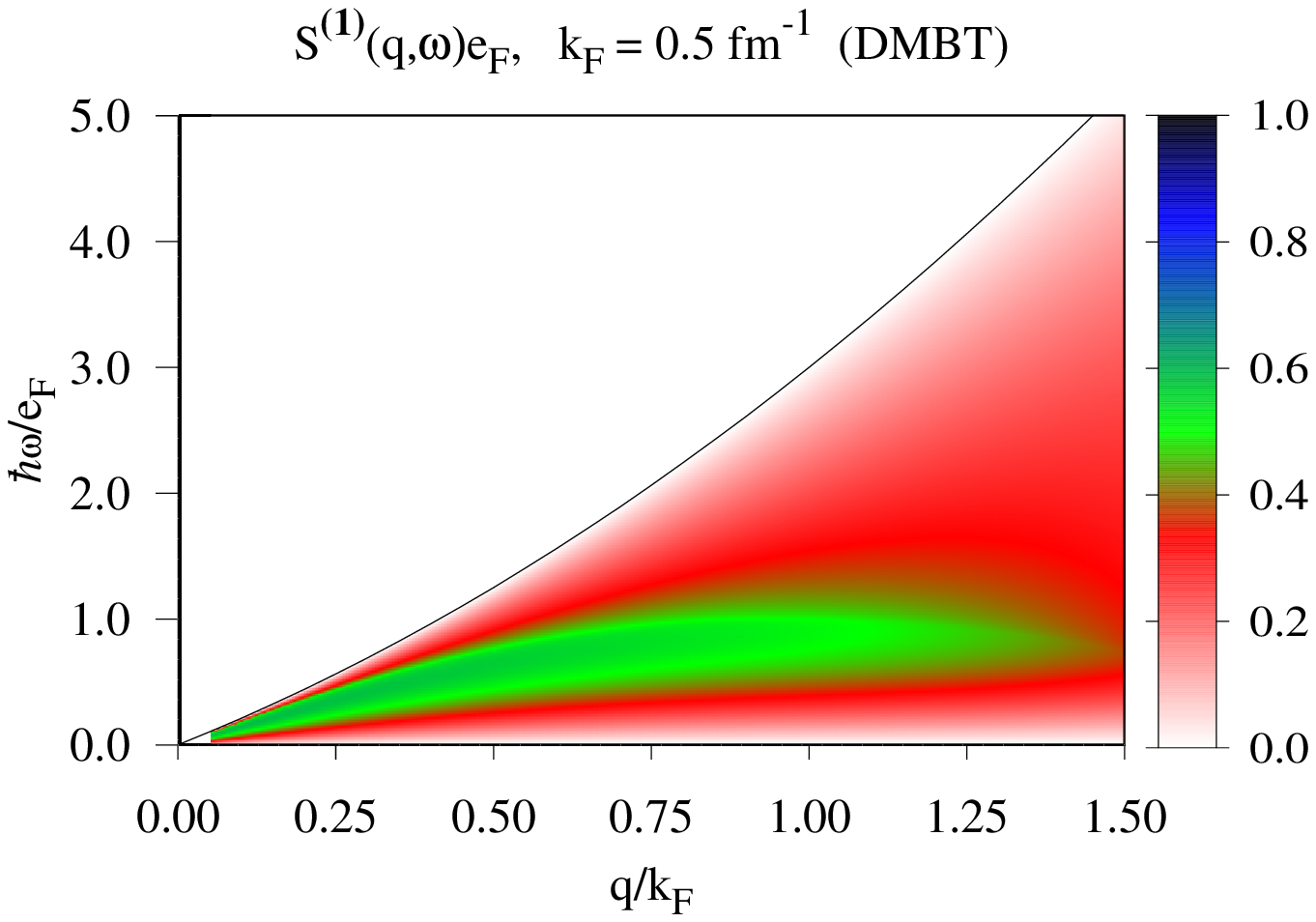}
\includegraphics[width=0.36\columnwidth,angle=-90]{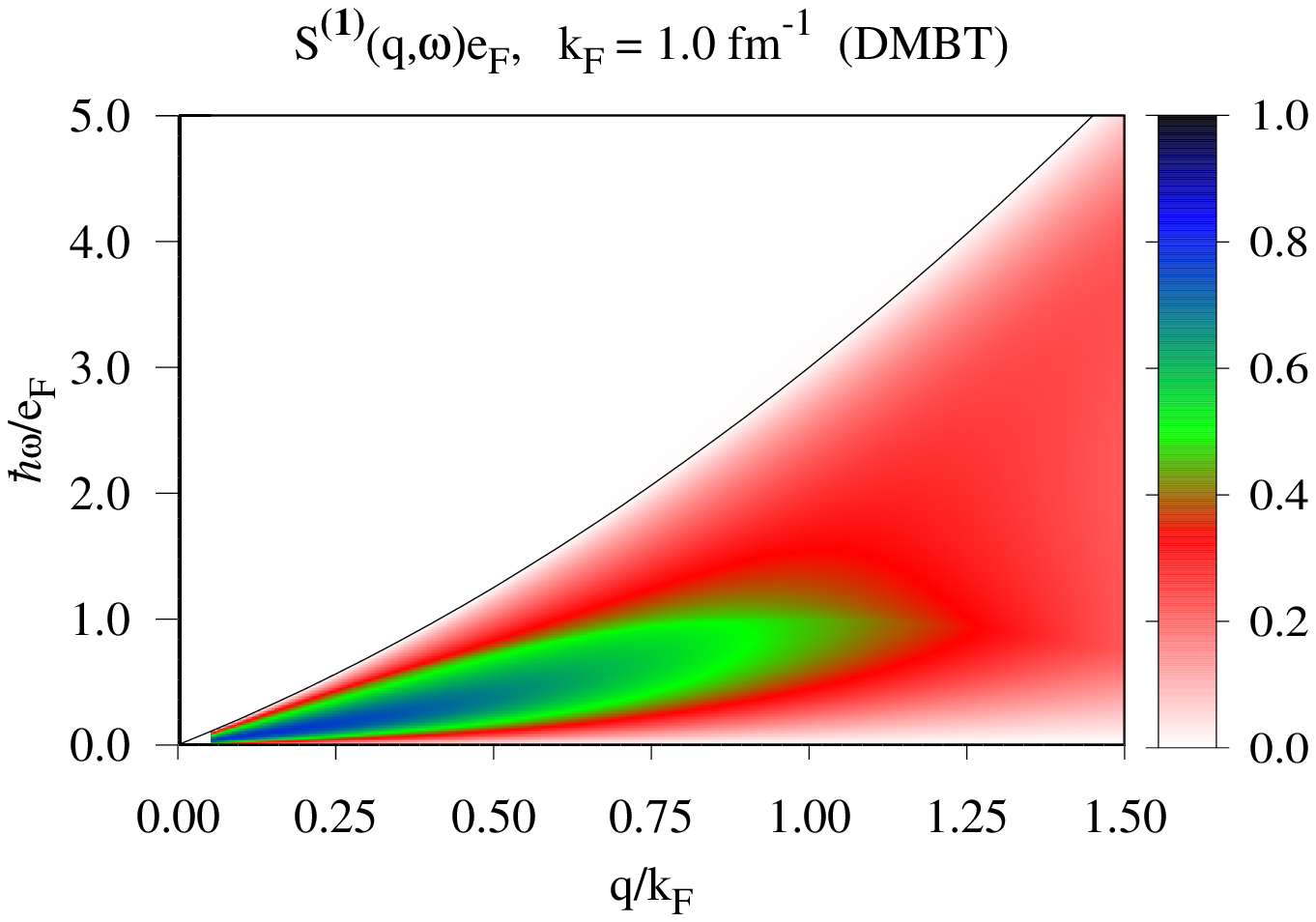}}
  \caption{(color online) The figure shows a color map of the density
    channel $S^{\rm (\1)}(q,\omega)$ of the dynamic structure function at
    $\KF = 0.5 \,\text{fm}^{-1}$ and $\KF = 1.0 \,\text{fm}^{-1}$.
    The solid line is the upper boundary,
    of the particle-hole continuum. \label{fig:SKW3DC}}
\end{figure}

\begin{figure}[H]
  \centerline{
    \includegraphics[width=0.36\columnwidth,angle=-90]{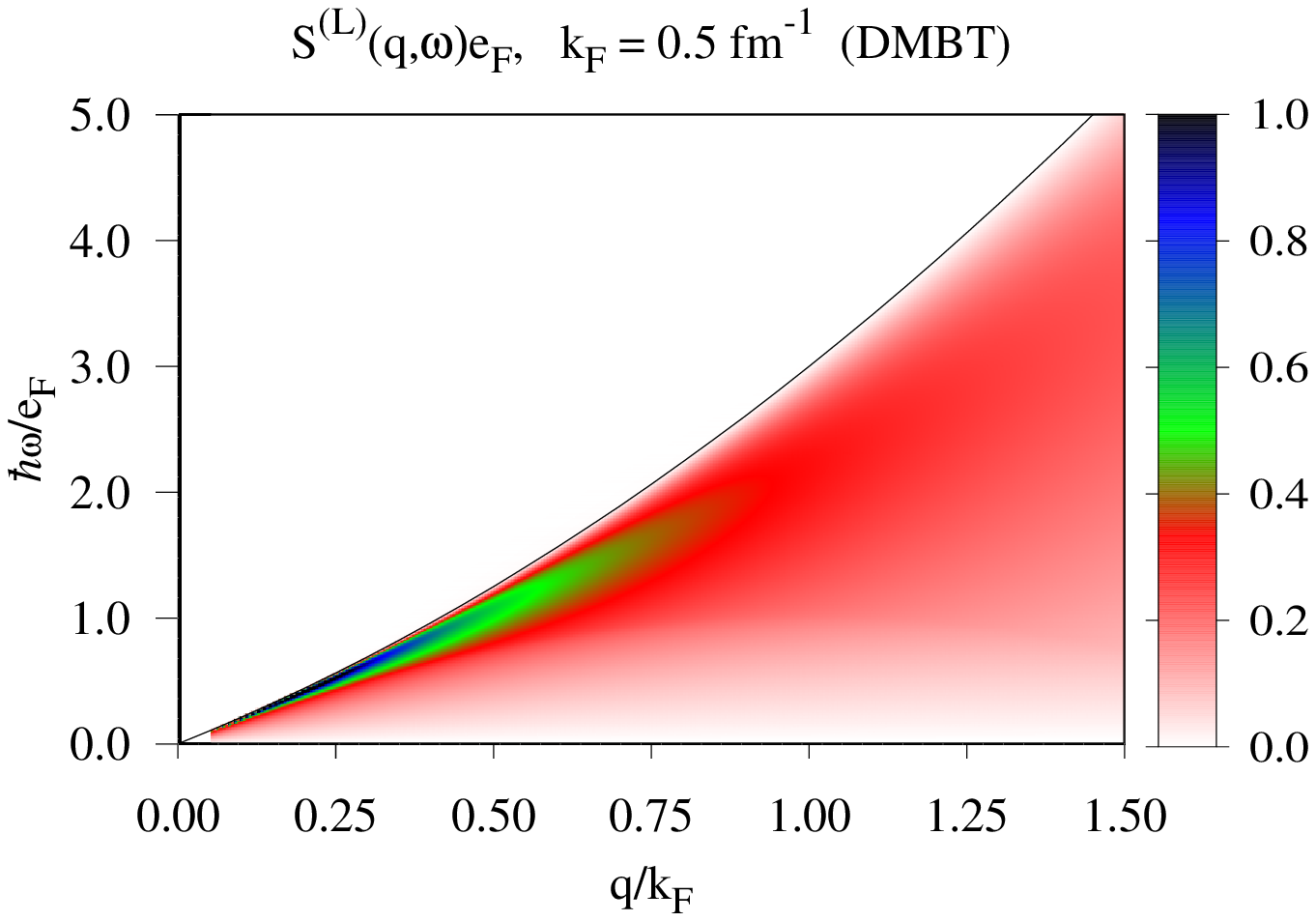}
\includegraphics[width=0.36\columnwidth,angle=-90]{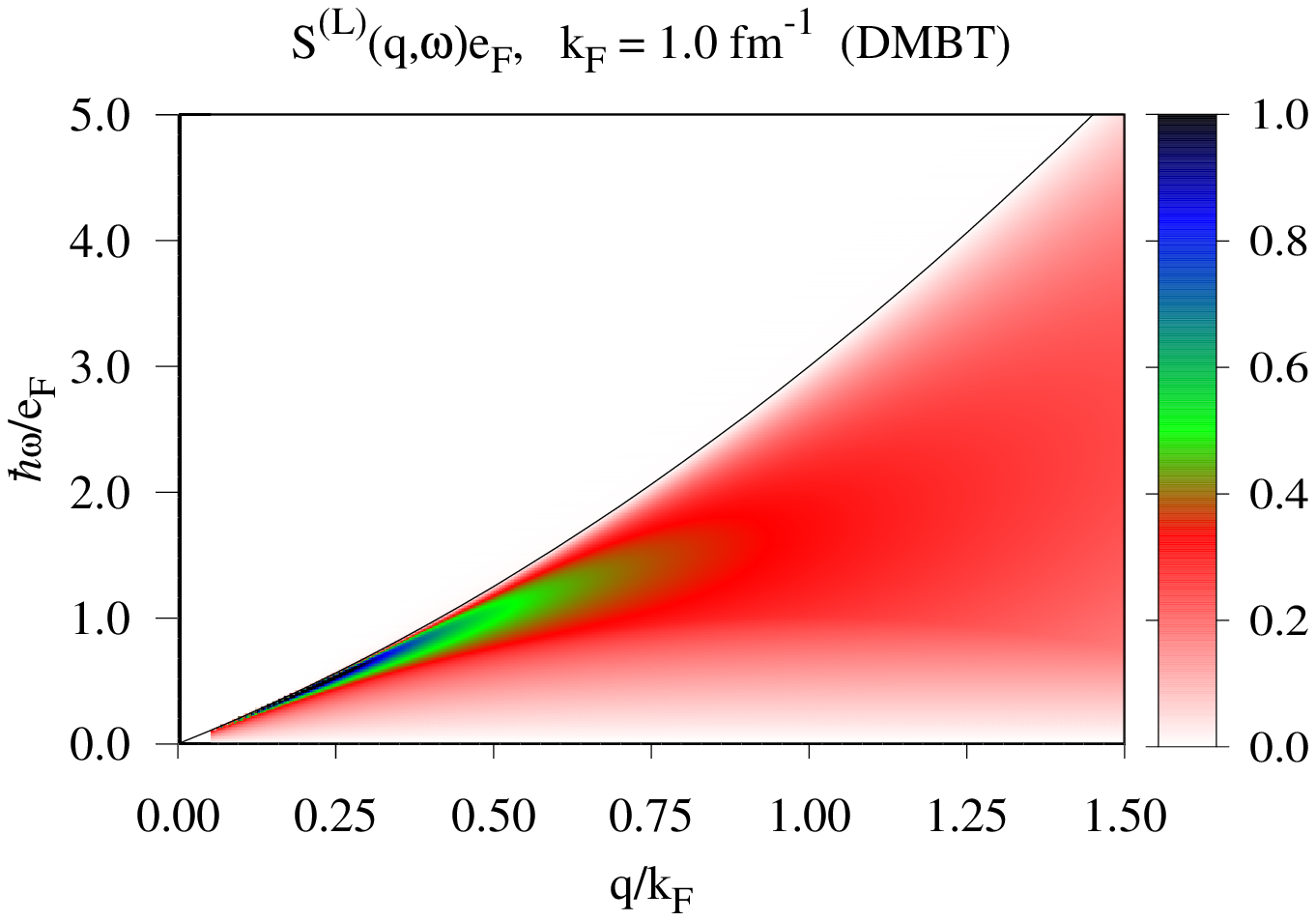}}
  \caption{(color online) Same as fig. \ref{fig:SKW3DC} for the
    $L$-channel.\label{fig:SKW3DL}}
\end{figure}

\begin{figure}[H]
  \centerline{
    \includegraphics[width=0.36\columnwidth,angle=-90]{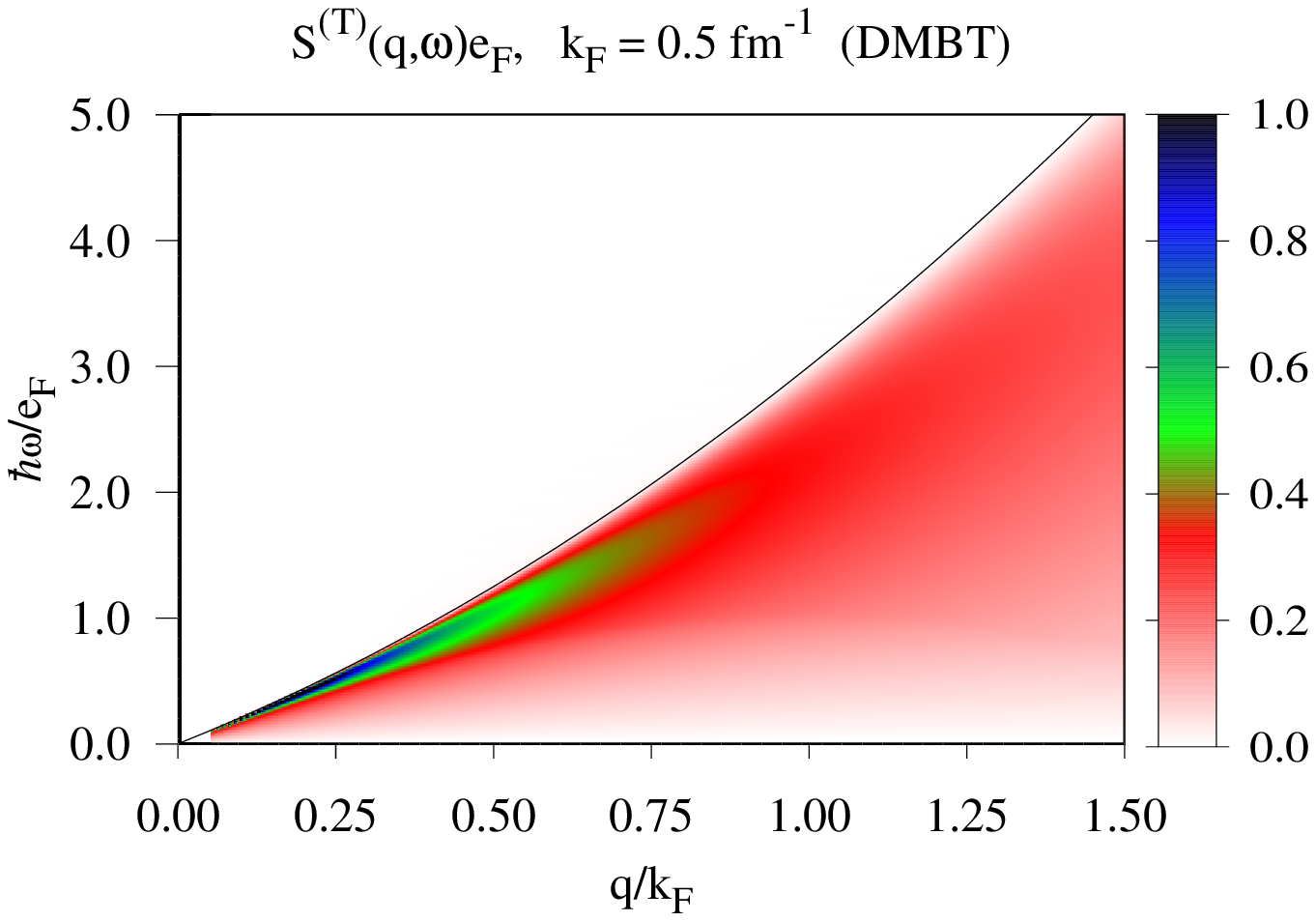}
\includegraphics[width=0.36\columnwidth,angle=-90]{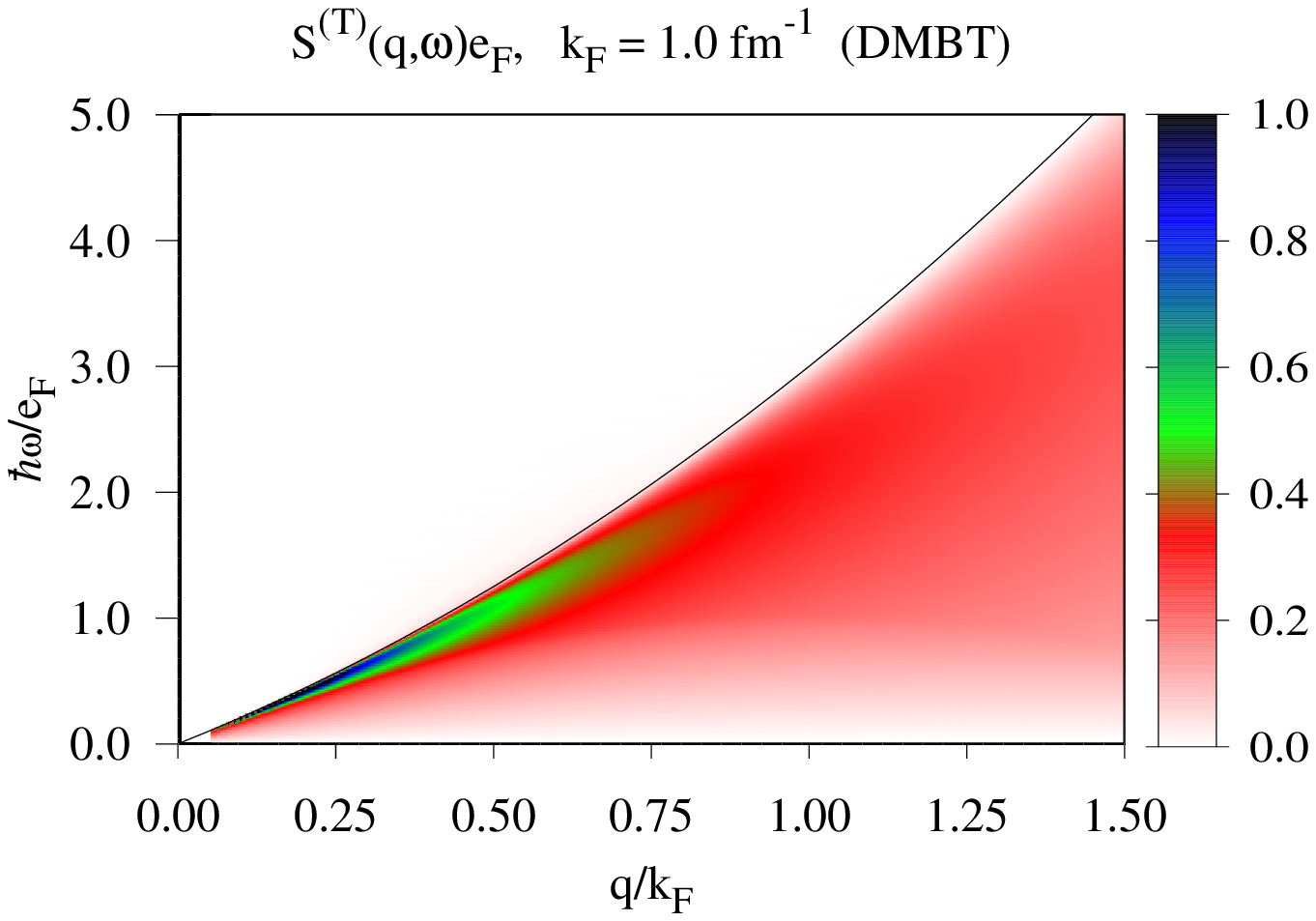}}
  \caption{(color online) Same as fig. \ref{fig:SKW3DC} for the
    $T$-channel. \label{fig:SKW3DT}}
\end{figure}

An overview of our results at low densities is given in
Figs. \ref{fig:SKW3DC}-\ref{fig:SKW3DT}. The most evident difference
between the distinct channels is that the strength of $S^{\rm
  (\1)}(q,\omega)$ is mostly in the middle of the particle-hole
continuum whereas $S^{\rm (L)}(q,\omega)$ and $S^{\rm (T)}(q,\omega)$
show, at long wavelengths, significant strength just below the
boundary of the particle-hole continuum. The DMBT version and the RPA
version are quite similar; hence we show only the DMBT results.
The $L$ and the $T$ channel are rather close, indicating that the
effect of the tensor force is small.


\begin{figure}[H]
  \centerline{
    \includegraphics[width=0.36\columnwidth,angle=-90]{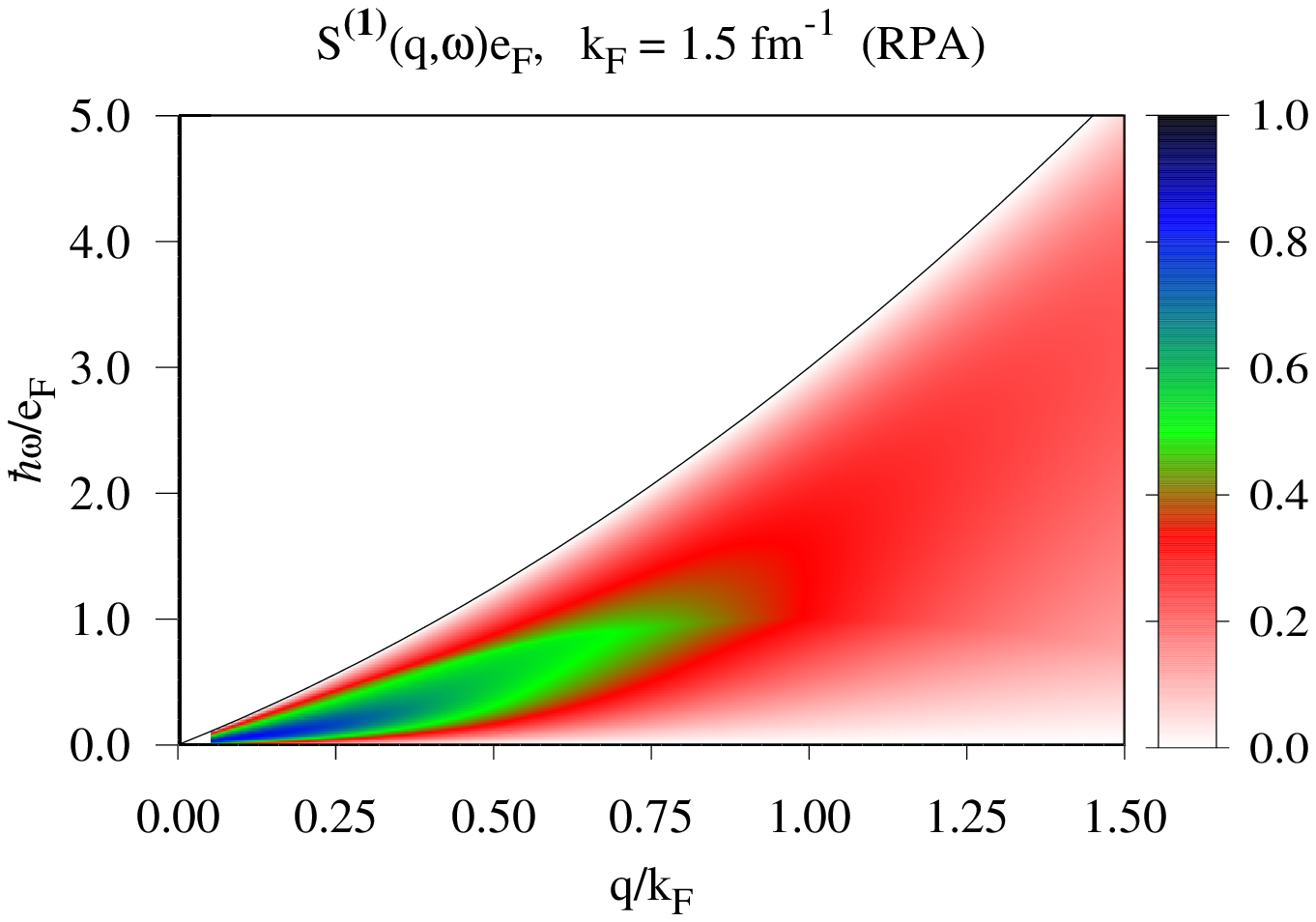}
\includegraphics[width=0.36\columnwidth,angle=-90]{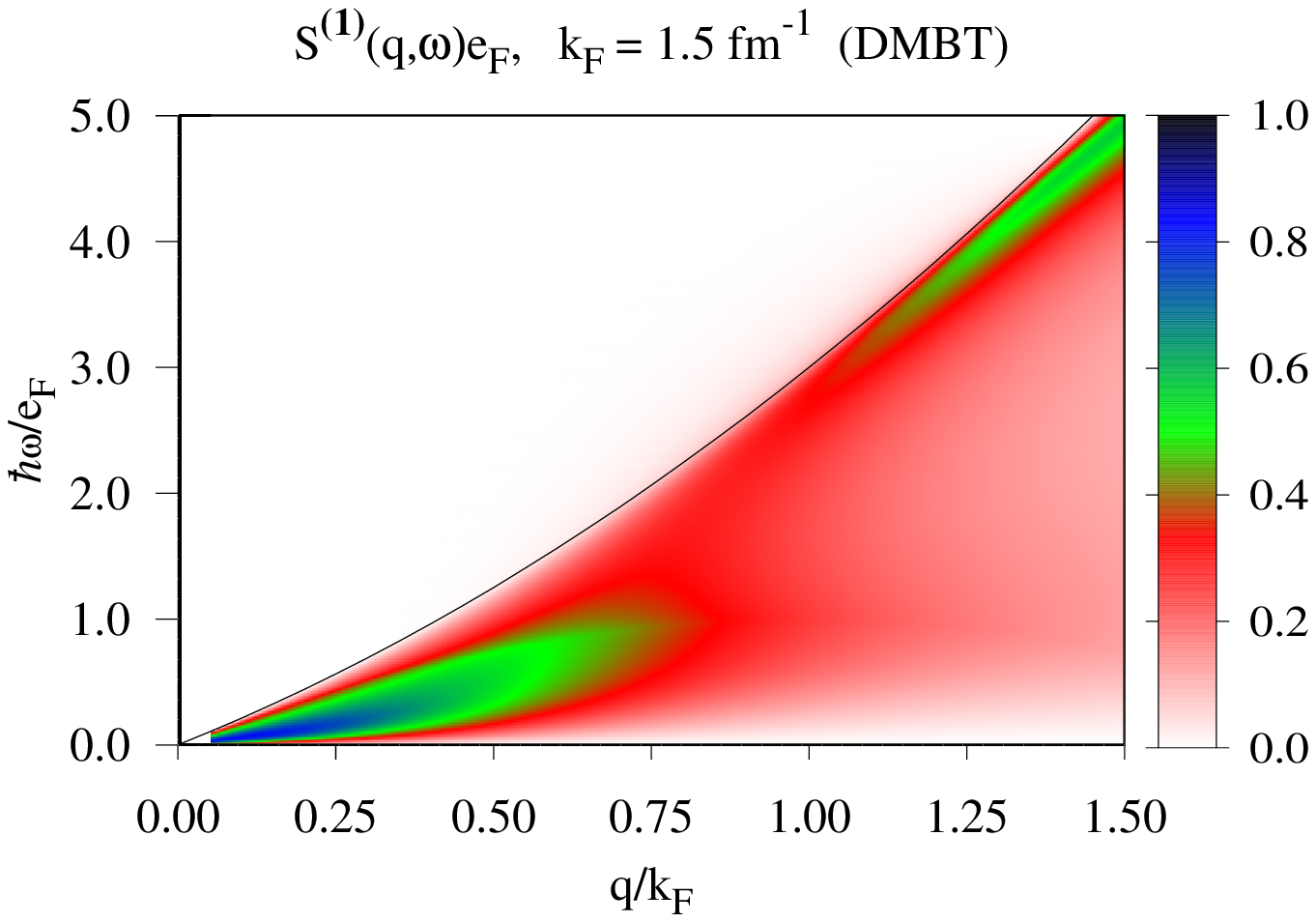}}
  \caption{(color online) The figure shows color maps 
    of the density
    channel $S^{\rm (\1)}(q,\omega)$ of the dynamic structure function at
    $\KF = 1.5 \,\text{fm}^{-1}$ in RPA (left figure) and DMBT (right figure).
    The solid line is the upper boundary
    of the particle-hole continuum. \label{fig:SKW3D150C}}
\end{figure}

\begin{figure}[H]
  \centerline{
    \includegraphics[width=0.36\columnwidth,angle=-90]{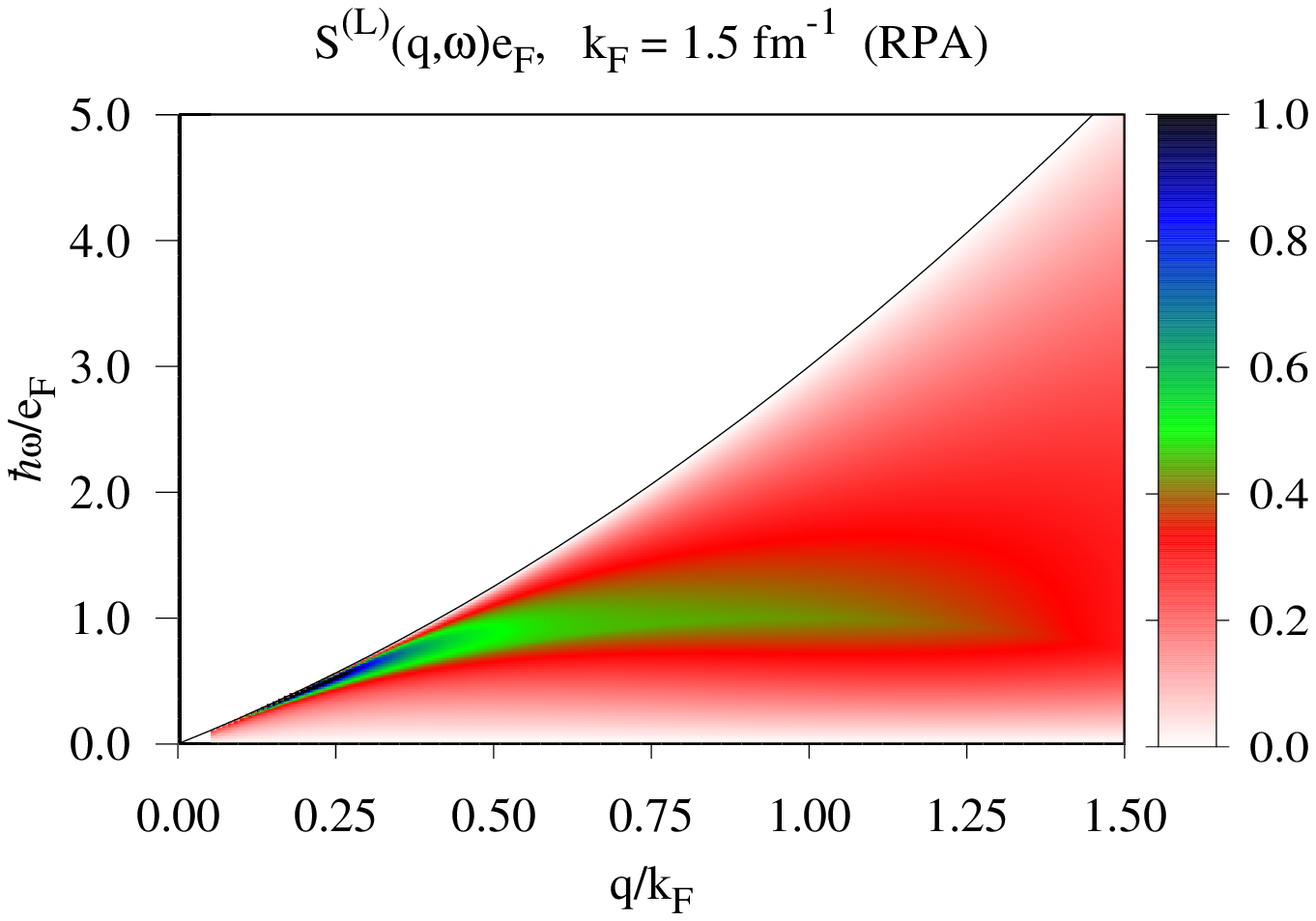}
\includegraphics[width=0.36\columnwidth,angle=-90]{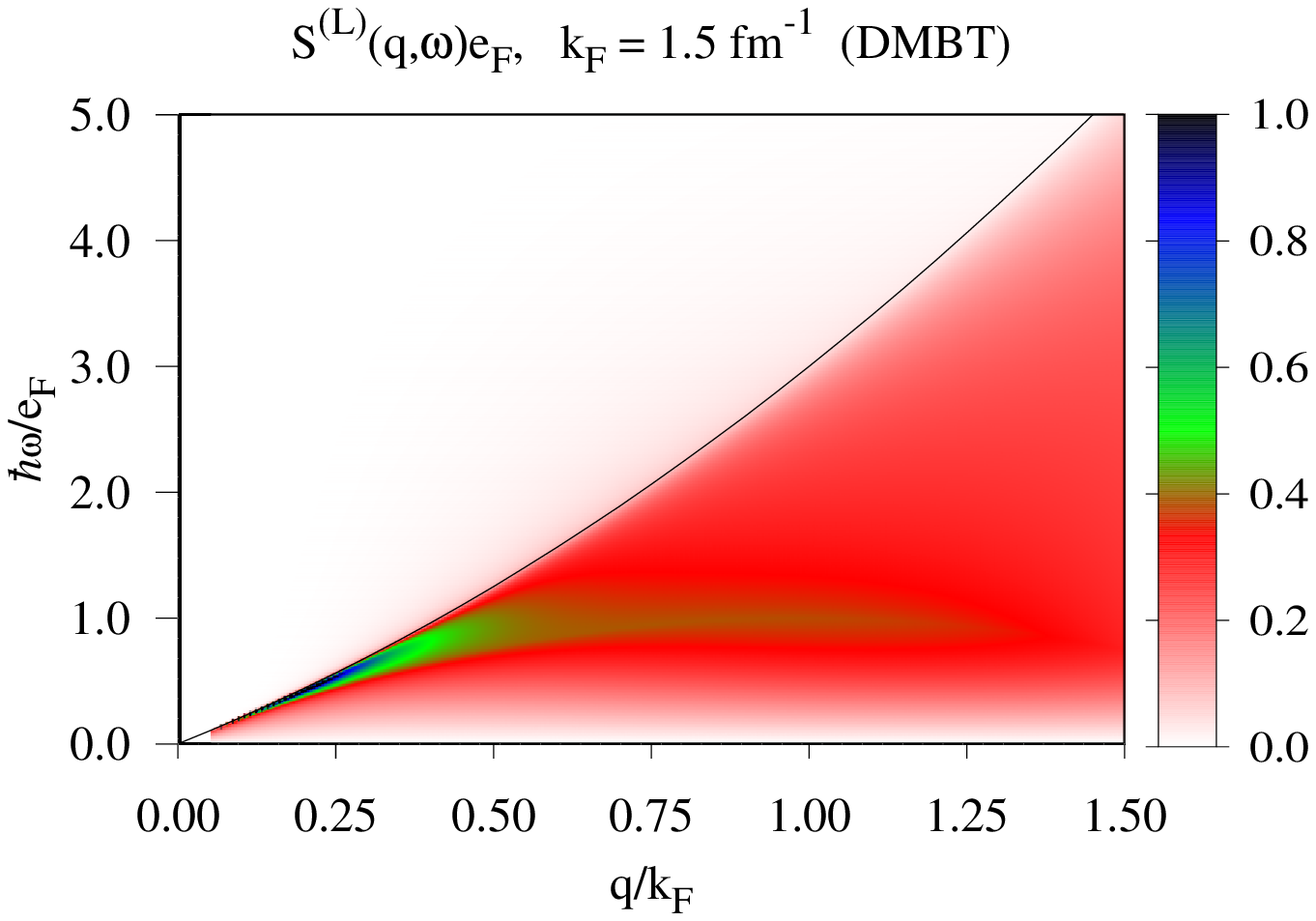}}
  \caption{(color online) Same as fig. \ref{fig:SKW3D150C} for the
    $L$-channel.\label{fig:SKW3D150L}}
\end{figure}

\begin{figure}[H]
  \centerline{
    \includegraphics[width=0.36\columnwidth,angle=-90]{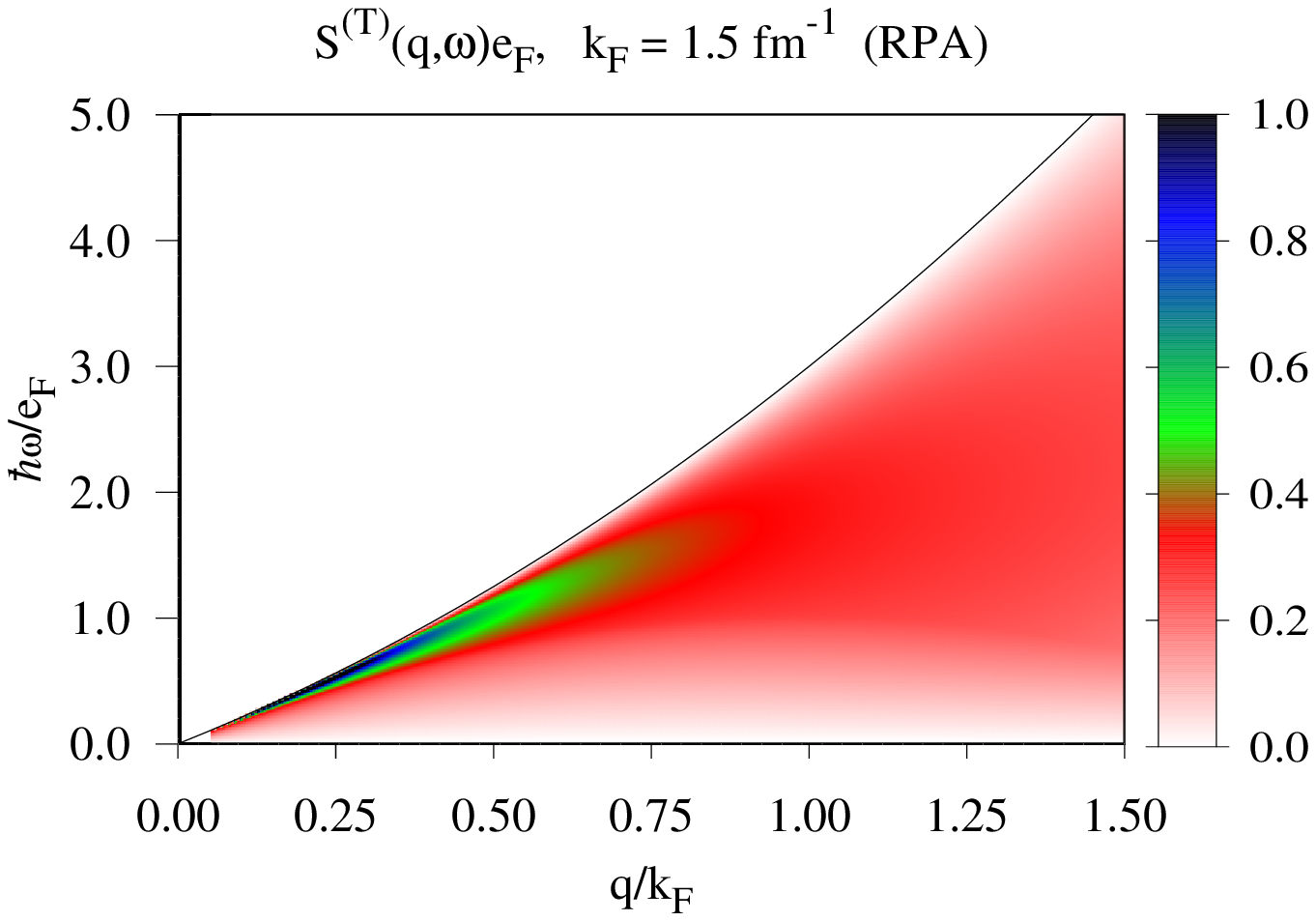}
\includegraphics[width=0.36\columnwidth,angle=-90]{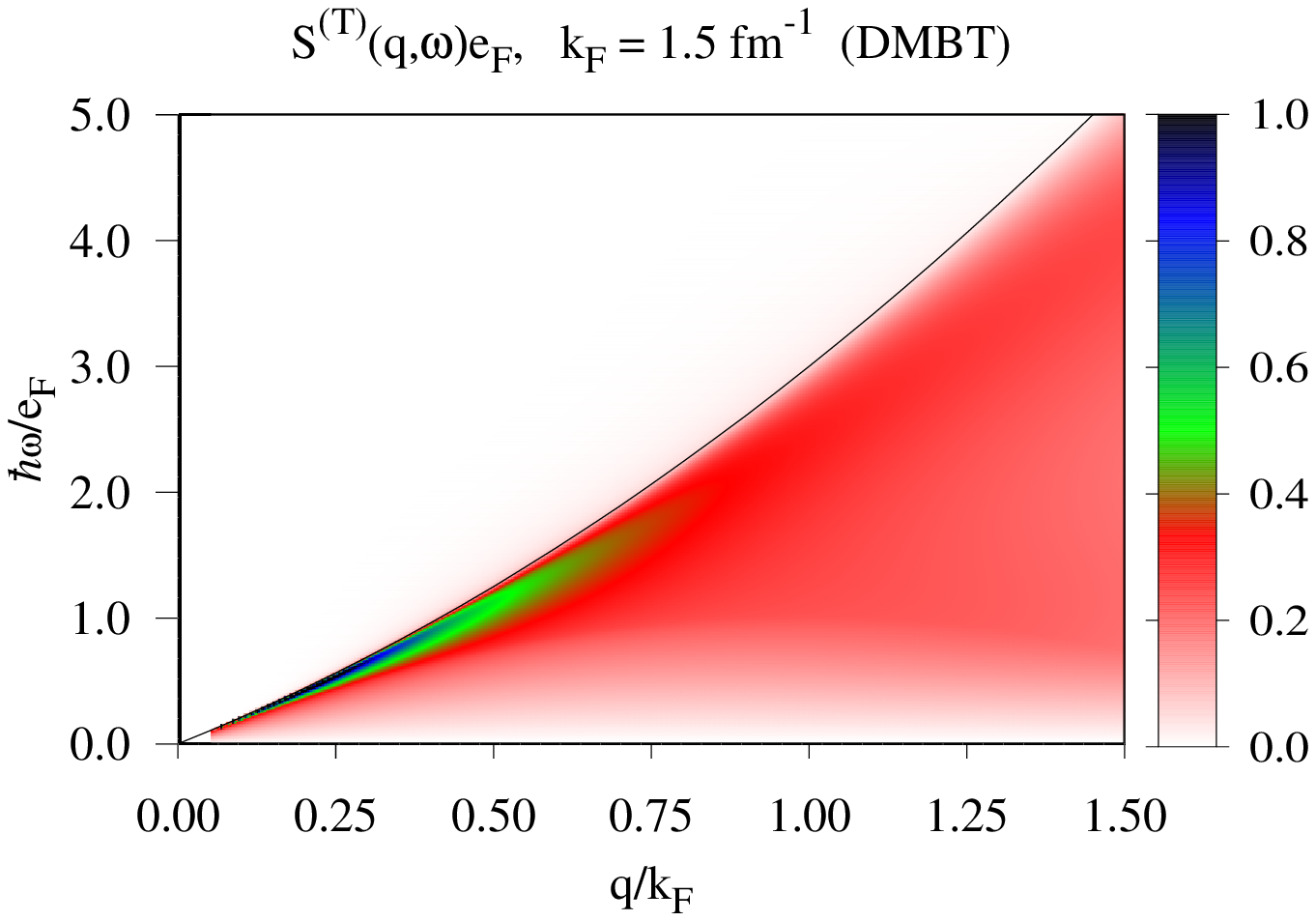}}
  \caption{(color online) Same as fig. \ref{fig:SKW3D150C} for the
    $T$-channel. \label{fig:SKW3D150T}}
\end{figure}

The picture changes when we go to higher densities as can be seen in
Figs. \ref{fig:SKW3D150C}-\ref{fig:SKW3D150T}. As discussed above; the
pair fluctuation corrections should become visible as the wave length
of the excitation becomes comparable to the correlation length.  In
the density channel we observe indeed significant strength around
wave lengths of the order $q/\KF \approx 1.5$. This is reminiscent of
the fact that pair excitations lower the spectrum of the helium
liquids in the roton regime by about a factor of 2. No such effect is
seen in the two spin channels. This is also expected as
spin-fluctuations are less affected by short-ranged correlations. A
noteworthy effect is that the tensor force breaks the degeneracy of
the two spin-channels visibly.


The peak in $S^{\rm (L)}(q,\omega)$ and $S^{\rm (T)}(q,\omega)$ is caused by a
node of the real part of the denominators in Eqs. \eqref{eq:chialpha},
\ie by the solution of

\begin{equation}
  \Re D^{(\alpha)}(q,\omega_{\rm zs}(q))
  = 0\,,\label{eq:ezsdef}
\end{equation}
with $\omega_{\rm zs}(q)$ being the zero-sound mode.

\begin{figure}[H]
  \centerline{\includegraphics[width=0.6\columnwidth,angle=-90]{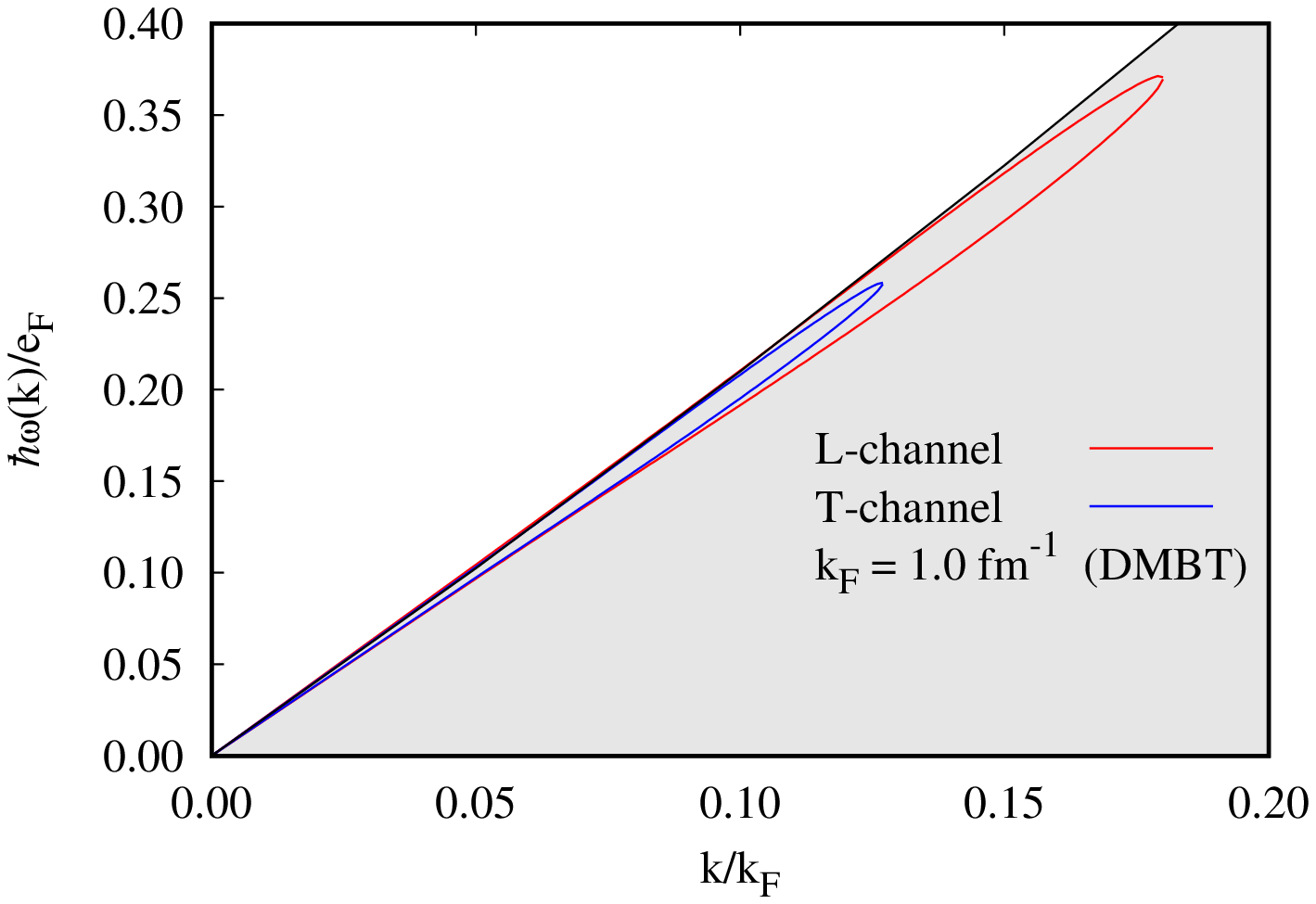}}
  \caption{(color online) The figure shows the location of the
    solution of Eq. \eqref{eq:ezsdef} for $\alpha={\rm L}$ and
    $\alpha={\rm T}$ at $\KF = 1.0 \,\text{fm}^{-1}$.  The gray shaded
    area is the particle-hole continuum. \label{fig:ezsTkf100}}
\end{figure}

In our case, the solutions of Eq. \eqref{eq:ezsdef} are inside the
continuum, that means the imaginary part is non-zero but evidently
very small. Fig. \ref{fig:ezsTkf100} shows, for $\KF =
1.0\,\text{fm}^{-1}$, the location of the ``zero sound pole'' for
longitudinal and transverse excitations.

\subsection{Self-energy}
\label{ssec:selfen}

The single-particle spectrum characterize both the thermodynamics and
potential superfluid phase transitions of the system.  Single-particle
properties are in perturbation theory normally discussed in terms of
the Dyson-Schwinger equation \cite{Dyson49,Schwinger51a,Schwinger51b}.
The relevant quantity is the single-particle Green's function
$G(k,\omega)$.  $G(k,\omega)$ is expressed in terms of the proper
self-energy $\Sigma^*(k,\omega)$ through the Dyson equation
\begin{equation}
G_{\sigma\sigma'}
(k,\omega) = \frac{\delta_{\sigma\sigma'}}
 { \hbar\omega - t(k) - \Sigma^*(k,\omega)}\,;
\label{eq:Dyson}
\end{equation}
the physical excitation spectrum is obtained by finding the pole of
the Green's function in the $(k,\omega)$-plane. Thus, the task of
many-body theory is the calculation of the proper self-energy
$\Sigma^*(k,\omega)$. Approximations are necessary to make the theory
useful. A popular approximation is the so-called G0W approximation
\cite{Hedin65,Rice65,FetterWalecka,FrimanBlaizot} for the self-energy
\begin{equation}
\Sigma(k,E) = u(k) + \I\sum_\alpha w_\alpha \int \frac{d^3 q \,
  d(\hbar\omega)}{ (2\pi)^4} G_0({\bf k}-{\bf q},E-\hbar\omega)
\left[\tilde V_{\rm p-h}^{(\alpha)}(q)\right]^2 \chi_\alpha(q,\omega)
\label{eq:G0W}
\end{equation}
with the weight factors $w_{c,t+,t-} = 1, 1, 2$ and
\begin{equation}
G_0(k,\omega) = \frac{\bar n(k)}{\hbar\omega - t(k) + \I\eta}
+ \frac{n(k)}{\hbar\omega - t(k) - \I\eta}
\label{eq:green}
\end{equation}
is the single-particle Green's function of non-interacting
fermions. $u(k)$ is a static field, the Fock term in Hartree-Fock
approximation, or the Brueckner-Hartree-Fock term in Brueckner
theory \cite{MahauxReview}.  Above, $\alpha \in \{c,\hat L, \hat T\}$
labels the spin channel, $\tilde V_{\rm p-h}^{\rm (\alpha)}(q)$ is the
particle-hole interaction in the appropriate spin channel, and
$\chi_\alpha(q,\omega)$ is the (spin-) density response function.  In
the term $\left[\tilde V_{\rm
    p-h}^{(\alpha)}(q)\right]^2\chi_\alpha(q,\omega)$ we recover the
RPA for the energy dependent induced interactions
\begin{equation}\hat w_{\rm  I}(q,\omega)
  = \hat W(q,\omega)-\hat V_{\rm p-h}(q)\,,
  \end{equation}
\ie we can write
\begin{equation}
\Sigma(k,E) = u(k) + \I\sum_\alpha w_\alpha \int \frac{d^3 q \,
  d(\hbar\omega)}{ (2\pi)^4} G_0({\bf k}-{\bf q},E-\hbar\omega)
\tilde w_I^{(\alpha)}(q,\omega).
\label{eq:G0Wa}
\end{equation}
Technically the self-energy is calculated by Wick-rotation in the
complex $\hbar\omega$ plane: That way, the energy integral
is decomposed into two terms, a smooth ``background-'' or
``line-term'', which consists of the frequency integral along the
imaginary $\hbar\omega$ axis, and a second, ``pole term'' from the residue
of the single-particle Green's function, \ie

\begin{equation}
\Sigma (k, E ) =  \Sigma_{\rm line}( k, E) + \Sigma_{\rm pole} (k,E)
\end{equation}
with
\begin{equation}
\Sigma_{\rm line} ( k, E) = -
\int_{ -\infty }^{\infty } \frac{ d (\hbar\omega)}{2\pi }
\int \frac{ d^3 q }{ ( 2\pi )^3\rho } \tilde w_I^{(\alpha)}(q,\I\omega)
{ E - e( | {\bf k} - {\bf q} | )\over
[ E - e( | {\bf k} - {\bf q} | ) ]^2 + \hbar^2\omega^2 }
\label{eq:SigmaLine}
\end{equation}
and
\begin{eqnarray}
\Sigma_{\rm pole} (k, E ) &=&
\int { d^3 q \over ( 2\pi )^3\rho }
\tilde w_I^{(\alpha)} (q, E - e( | {\bf k} - {\bf q} | ))
 [\Theta ( E - e(| {\bf p} - {\bf q} | ))
- \Theta ( \EF - e(| {\bf p} - {\bf q} | )) ]\, .\nonumber\\
\label{eq:SigmaPole}
\end{eqnarray}
 Since $\Sigma(k,\omega)$ sums
all chain (or particle-hole reducible) diagrams, the static field
$U(k)$ should be the Fock term generated by the particle-hole {\em
  irreducible\/} interaction, \ie by $\hat V_{\rm p-h}(q)$.

The excitation energies in the variational theory are simply given by
the diagonal matrix elements (\ref{eq:epscbf}).  In the level of FHNC
theory we are using here these energies are simply the Hartree-Fock
energies of the static effective interaction \eqref{eq:W135}
\cite{CBF2}.  Replacing the dynamic induced interaction $\hat
w_I(q,\omega)$ by the static approximation \eqref{eq:Scond}, we can
carry out the frequency integration and obtain a {\em static\/}
self-energy simply in the form of a Fock term in terms of the effective
interaction $\hat W(q)$.

The study of the self-energy highlights another limitation of locally
correlated wave functions: The self-energy $\Sigma(k,\omega)$ depends
on energy and momentum. Approximating this function by an ``average''
energy-independent function misses the important non-analytic
structure of the self-energy around the Fermi surface. This has the
well-known consequence of an enhancement of the effective mass in
nuclei around the Fermi surface \cite{BrownGunnGould}. The effect is
quite dramatic in $^3$He \cite{PethickMass,ZaringhalamMass} where a
Jastrow-Feenberg wave function predicts an effective mass ratio $m^*/m
< 1$ in massive contrast to the experimental value around $m^*/m
\approx 3$ \cite{GRE83,GRE86}. The experimental values are well
reproduced when the full self-energy is calculated
\cite{Bengt,he3mass}.

The single particle spectrum is ideally calculated by solving
\begin{equation}
  \varepsilon(k) = t(k) + \Sigma(k,\varepsilon(k))\,.\label{eq:spectrum}
\end{equation}
Most important for the thermodynamics of the system and for the
pairing phenomena discussed in the next section are low-lying single
particle states. These can be characterized by the effective mass
\begin{equation}
  \frac{\hbar^2\KF}{m^*} \equiv \left.\frac{d\varepsilon(k)}{dk}\right|_{\KF}\,.
\end{equation}
If Eq. \eqref{eq:spectrum} has been solved self-consistently,
the effective mass can be calculated from Eq. \eqref{eq:spectrum} as
\begin{equation}
  \frac{m^*}{m} =
  \frac{m_k^*}{m}\frac{m^*_E}{m},\quad \frac{m^*_k}{m}
  =\left[1+\frac{m}{k}\frac{\partial\Sigma}{\partial k}\right]^{-1}
      ,\quad\frac{m^*_E}{m}=1-\frac{\partial\Sigma}{\partial \hbar\omega}
      \end{equation}\label{eq:mEmk}
This is, however, in practice rarely done. It also should be noted that
the self-consistent solution of Eq. \eqref{eq:spectrum} renormalizes the
Green's function $G_0(k,\omega)$ appearing explicitly in Eq. \eqref{eq:G0W},
but not the Green's function that appears implicitly in the response
functions $\chi_\alpha(q,\omega)$. It is therefore a legitimate alternative
to use the same spectrum in $G_0(k,\omega)$ and  $\chi_\alpha(q,\omega)$
which is, in our case, the free particle spectrum $t(k)$.
This defines the ``on-shell mass''
\begin{equation}
  \frac{\hbar^2\KF}{m^*_{OS}} = \frac{d}{dk}\left[t(k)+\Sigma(k,t(k))\right]_{\KF}\,.\label{eq:mos}
\end{equation}
The difference between the two calculations $m^*/m$ can then serve as
an estimate for how accurately the free spectrum approximates the
solution of Eq. \eqref{eq:spectrum}.

There are typically four effects that contribute to the effective
mass: These are the static interaction, fermion exchange, hydrodynamic
backflow and possibly the coupling to low-lying excitations. The first
two are already represented in the static contribution
\eqref{eq:epscbf} which is, in the weakly interacting limit, the Fock
term of the bare interaction and in our case either the Fock term of
the effective interaction $\hat W(q)$ or, when the dynamic self-energy
is included, of the particle-hole interaction
$\hat V_{\rm p-h}(q)$. Coupling to low-lying excitations are described by
the second term.

\begin{figure}[H]
  \centerline{\includegraphics[width=0.65\textwidth,angle=270]{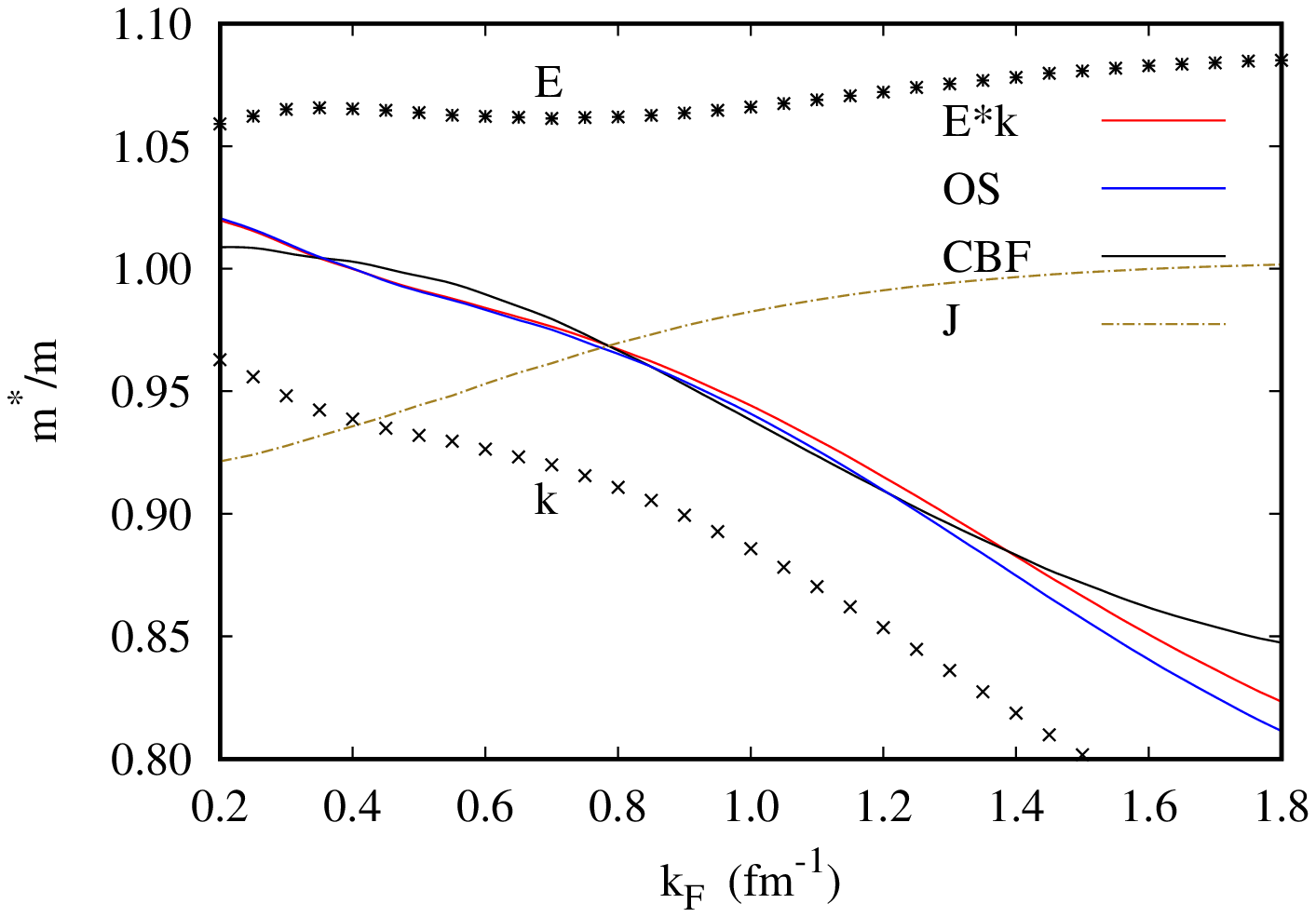}}
  \caption{(color online) The figure shows the effective mass as
    calculated for the Argonne $v_8$ interaction. The black line
    (marked with ``CBF'') is the effective mass obtained from the CBF
    spectrum \eqref{eq:epscbf}, the result from Eq. \eqref{eq:mEmk}
   (red line) is marked with ``E*k'' and the ``on-shell''
    effective mass defined in Eq. \eqref{eq:mos} (blue line) is marked
    by ``OS''.  We also show the individual contributions $m^*_k/m$
    (crosses) and $m^*_E/m$ (stars) for comparison as well as the
    result from state-independent Jastrow correlations (Dash-dotted
    line marked with ``J''). \label{fig:effmass}}
\end{figure}

Fig. \ref{fig:effmass} shows our results obtained from the CBF
spectrum \eqref{eq:epscbf} (black line), from Eq. \eqref{eq:mEmk} (red
line) and the ``on-shell'' effective mass defined in
Eq. \eqref{eq:mos} (blue line) for the Argonne $v_8$
interaction. Evidently these are all very close; we note that the
results for the other three interactions are practically
indistinguishable from those shown. We also show the individual
contributions $m^*_k/m$ and $m^*_E/m$ for comparison.

In previous work \cite{JLTP189_470} we have examined a very simple
model of neutron matter with a state-independent Jastrow wave function
\eqref{eq:Jastrow}. In that work, we found for $v_4$ type interactions
that the effective mass ratio $m^*/m \approx 1$ in the whole density
regime under consideration. We have repeated these calculations for
the present $v_8$ versions of the interactions and basically confirmed
the result of that work, albeit we have obtained a somewhat different
density dependence.

The good agreement between the CBF approximation \eqref{eq:epscbf}
for the effective mass and the dynamic calculations  \eqref{eq:mEmk}
or \eqref{eq:mos} is in sharp contrast to \he3.

Returning to the self-energy we note that both the static field term
$u(k)$ and the line term \eqref{eq:SigmaLine} are real whereas the
pole term \eqref{eq:SigmaPole} picks up the coupling to single-particle
excitations and collective modes and can, therefore, be complex.
Hence, $\Sigma(k,\omega)$ is a non-analytic function of the energy
in the vicinity of the Fermi energy \cite{BrownGunnGould,ZaringhalamMass}.
This non-analyticity leads, in nuclear matter and even more in \he3,
to a pronounced ``knee'' in the single particle spectrum.

\begin{figure}[H]
  \centerline{
    \includegraphics[width=0.35\textwidth,angle=270]{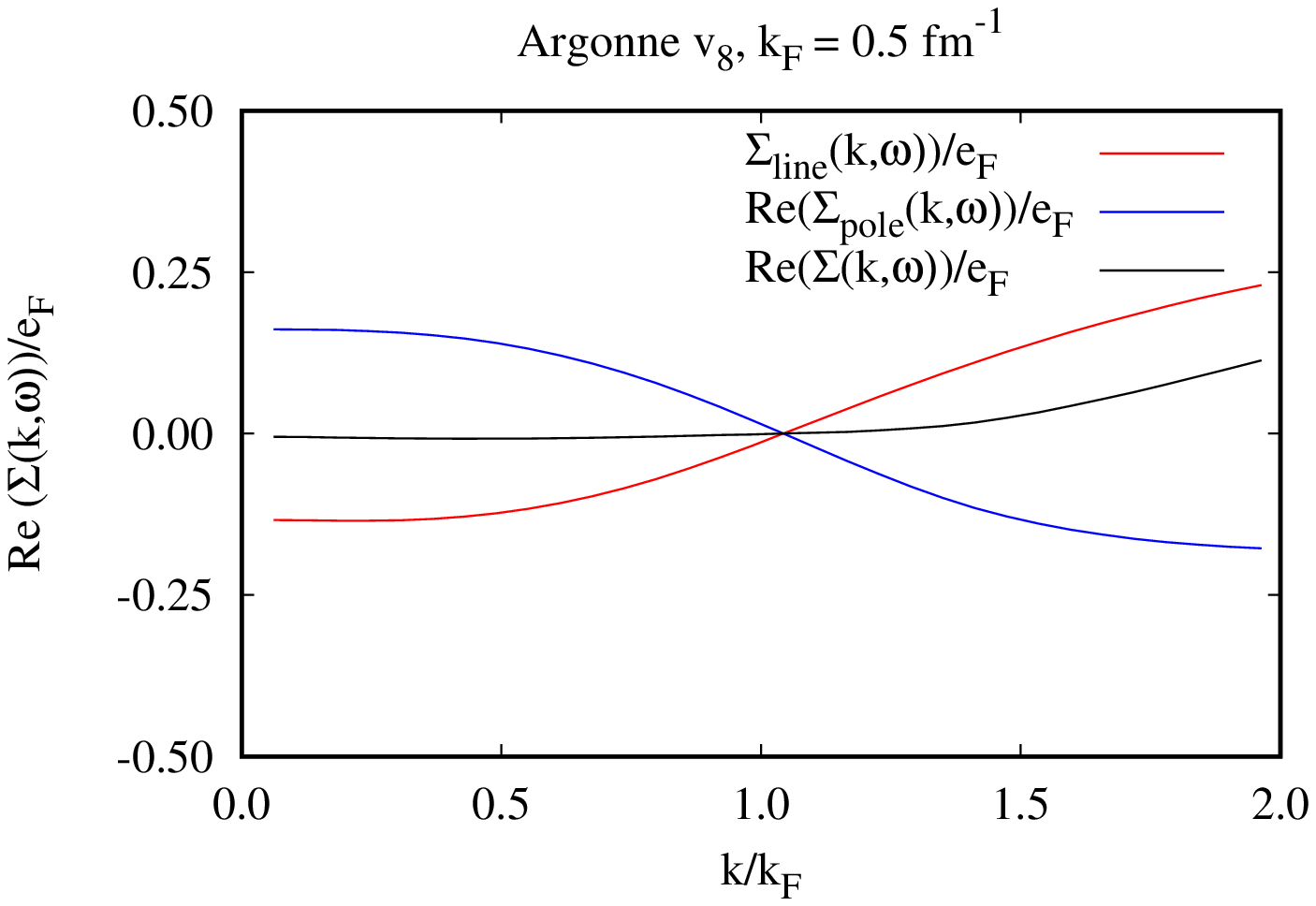}
    \includegraphics[width=0.35\textwidth,angle=270]{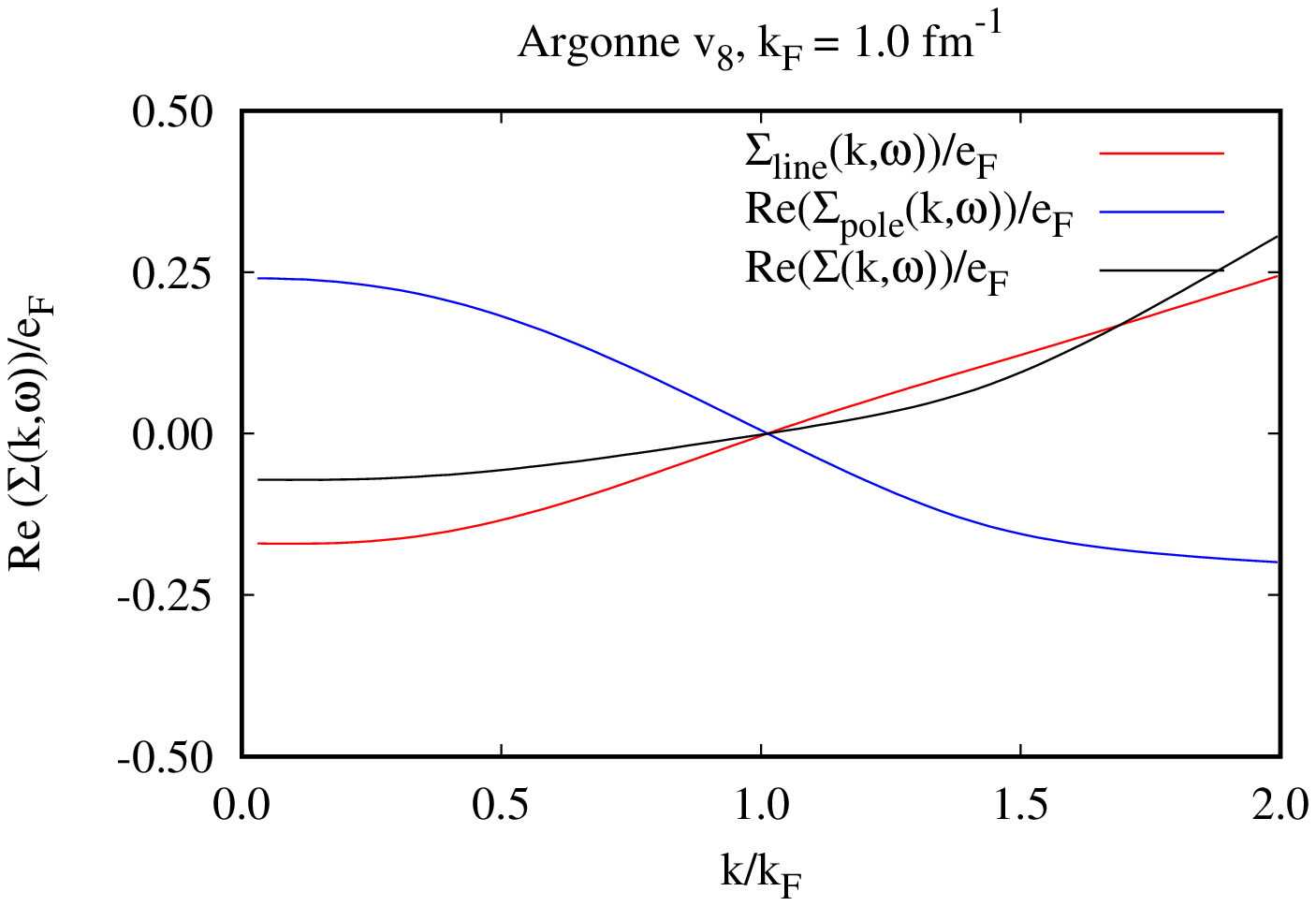}
    }
  \caption{(color online) The figures show the individual real parts
    $\Sigma_{\rm line}(k,t(k))$ and  $\Sigma_{\rm pole}(k,t(k))$
    of the on-shell self-energy as well as their sum
    $\Sigma(k,t(k))$ for the Argonne $v_8$ interaction for
    the densities $\KF = 0.5\,{\rm fm}^{-1}$ and $\KF = 1.0\,{\rm fm}^{-1}$.
 \label{fig:selfReplot}}
\end{figure}

Figs. \ref{fig:selfReplot} show selected results for the Argonne $v_8$
interaction. We also see the aforementioned ``knee'' around $\KF$, in
the pole term.  The effect is, however, minimal and does not lead to a
visible enhancement of the effective mass as seen in nuclear matter
and \he3.

\begin{figure}[H]
  \centerline{
    \includegraphics[width=0.35\textwidth,angle=270]{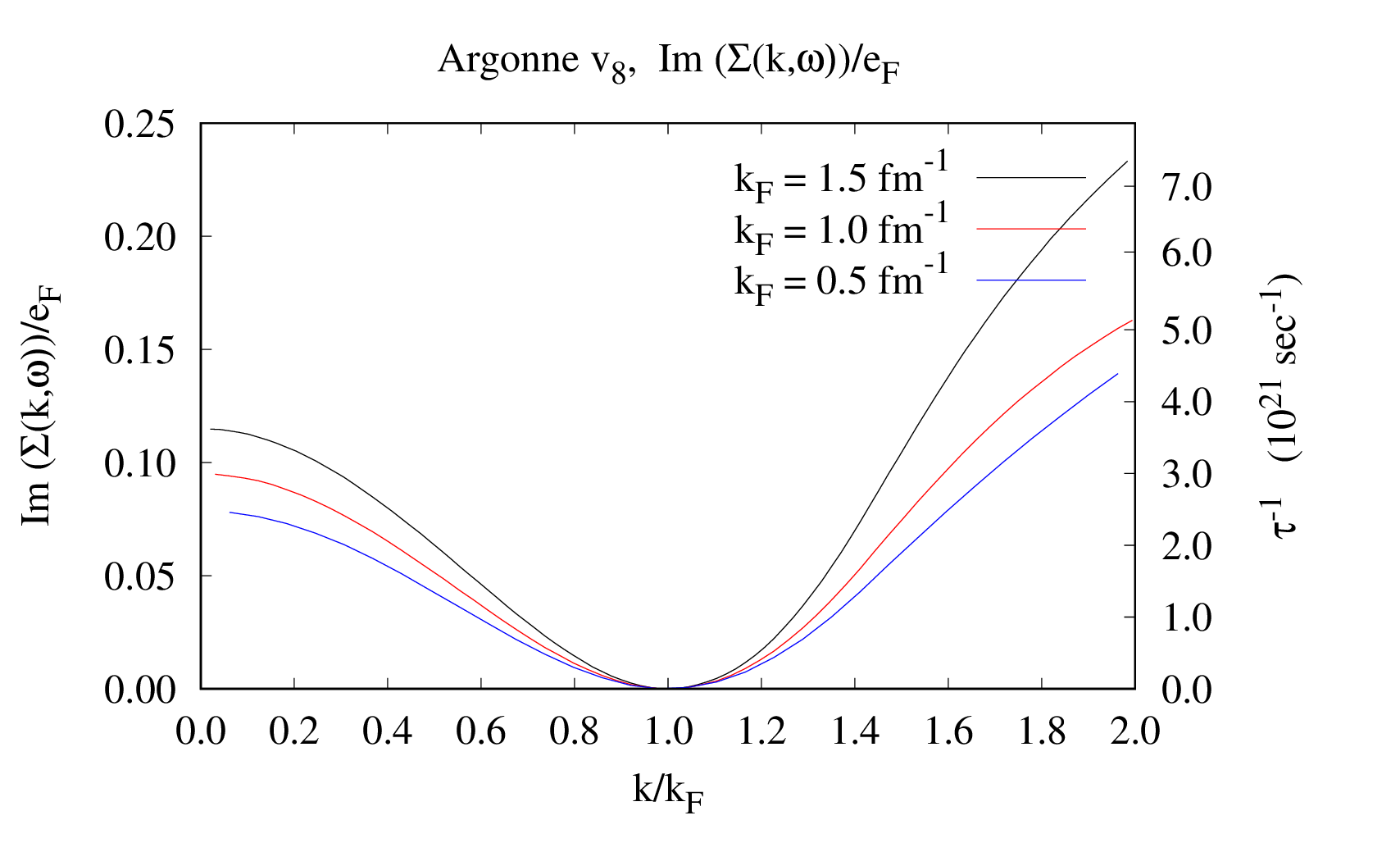}
    }
  \caption{(color online) The figures show the imaginary part
   of the on-shell self-energy
    $\Sigma(k,t(k))$ for the Argonne $v_8$ interaction for
   the densities $\KF = 0.5\,{\rm fm}^{-1}$, $\KF = 1.0\,{\rm fm}^{-1}$
   and $\KF = 1.5\,{\rm fm}^{-1}$. The right scale shows the inverse
   lifetime $\tau^{-1}$ for $\KF  = 1.0\,{\rm fm}^{-1}$.
 \label{fig:selfImplot}}
\end{figure}

Finally, we show in Fig. \ref{fig:selfImplot} the imaginary part of
the self-energy for the same interaction and densities as in
Fig. \ref{fig:selfReplot} which determines the lifetime of the single
particle excitations through
\begin{equation}
  \tau(k) = \hbar\, \Im [\Sigma(k,\hbar\varepsilon(k))]^{-1}
\label{eq:tau}
\end{equation}

To summarize, we have again obtained the result of
Ref. \citep{JLTP189_470} that the effective mass ratio is close to 1
throughout the whole density regime albeit we have obtained a somewhat
stronger and different density dependence. This finding justifies {\em
  a-posteriory\/} some of the approximations made here: A more
sophisticated calculation would, for example, use in Eq. \ref{eq:G0W}
the DMBT version of the response functions derived in the first part
of this section; our result shows that this is not necessary.  The
same is true for the pairing calculations to be presented in the next
section.

\section{Pairing in S and P states} 
\label{sec:BCS}
 
The issue of superfluid pairing phenomena in neutron and neutron star
matter has been discussed for decades
\cite{StellarSuperfluids,Gezerlis2014}.  From properties of the bare
interaction it has been concluded that, at low densities, neutrons
pair in $^1S_0$ states and, at higher densities, in $^3P_2$ or
$^3P_2$-$^3F_2$ states \cite{Tamagaki70}.

Pairing is investigated within the framework of BCS theory.  The
theory as originally formulated \cite{BCS,BCS50book} is intrinsically
a mean field theory; that leaves the question open to what extent such
a theory could capture the physics of a strongly interacting
system. This issue was first addressed by the introduction of
Jastrow-Feenberg correlation factors
\cite{YangNC,YangClarkBCS,YangThesis}.  When implemented in
a BCS extension, these advances have made possible the development of
a rigorous correlated BCS (CBCS) theory (\citenum{CBFPairing}, see
also Ref.~\citenum{HNCBCS}) that respects the U(1) symmetry-breaking
aspect of the superfluid state -- \ie the non-conservation of particle
number.

Recent in-depth studies of correlations in the low-density Fermi gas
\cite{cbcs,fullbcs}, with emphasis on the presence of Cooper pairing
and dimerization, document the power of the Euler-Lagrange-FHNC
approach adopted in the present work.  These calculations employed
simple state-independent correlation functions which makes the method
suitable for simple interactions. Improvements must be sought for
realistic nuclear Hamiltonians. We have therefore extended in Refs.
\citep{v3twist} and \citep{v3bcs} the correlated wave functions method
to realistic nuclear interactions, including spin-orbit forces
\cite{v4}.  We have reviewed this work in section \ref{sec:nucleons}.

The conclusion of our and all the previous work is that $S$-wave pairing
is basically understood: There is pairing in $^1S_0$ states with a
gap energy of a few MeV in the density regime $\KF < 1.5\,$fm$^{-1}$.
Quantitative corrections beyond the basic BCS correlations are
polarization effects \cite{CKY76,SchwenkFriman2004} and corrections
from the twisted-chain diagrams \cite{v3bcs}.  All of these
corrections change the magnitude of the $^1S_0$ gap by perhaps a
factor of 2 -- see below -- but have no dramatic consequences.

$P$-wave superfluidity is expected \cite{Tamagaki70} at higher densities
where, many-body effects become more important. Moreover, as already
pointed out in Ref. \citenum{Gezerlis2014}, the spin-orbit interaction
plays a crucial role.  We have addressed the task of including the
spin-orbit interaction in parquet theory in section
\ref{ssec:parquet}.  The second task is the generalization of the
correlated wave function for superfluid systems \cite{HNCBCS,cbcs} to
$P$-wave pairing. Ref. \citenum{v43p2} has already anticipated the
structure of a correlated BCS theory for pairing in arbitrary angular
momentum states. We include here the complete derivation.

\subsection{BCS theory with correlated wave functions}
\label{ssec:cbcs}

The mean-field theory of pairing in general angular momentum states
has been worked out in detail and applied to nuclear problems in the
pioneering work by Tamagaki, Takatsuka, and collaborators
\cite{Tamagaki70,10.1143/PTP.48.1517}.  Conventionally
\cite{PhysRev.131.1553}, the superfluid state is generated by a
generalized Bogoliubov transformation
\cite{Bogoliubov,BeliaevLesHouches}
\begin{subequations}
  \label{eq:PBCS}
\begin{eqnarray}
  \ket{\rm BCS} &=& e^{\I S}\ket{0},\\
  \I S &=& \frac{1}{2}\sum_{\kvec,\sigma_1,\sigma_2}\biggl[
      \theta_{\sigma_1\sigma_2}(\kvec)
      \creat{\kvec,\sigma_1}\creat{-\kvec,\sigma_2} 
      - \theta_{\sigma_1\sigma_2}^*(\kvec)\annil{-\kvec,\sigma_2}\annil{\kvec,\sigma_1}
      \biggr]\,,\\
  \theta_{\sigma_1\sigma_2}(\kvec) &=& \sqrt{2}\sum_{M_J}\Bigl(\frac{1}{2}\,\frac{1}{2}\,\sigma_1\,\sigma_2\Bigr|S\,M_S\Bigr)\bigl(L\,S\,M_L\,M_S\bigr|J\,M_J\bigr)Y_{L,M_L}(\Omega_\kvec)\chi_{JLSM_J}(k)\,.\nonumber\\
\end{eqnarray}
\end{subequations}
The above unitary transformation defines quasiparticle operators
\cite{Tamagaki70}
\begin{equation}
  \balpha_{\kvec} = e^{\I S} {\bf a}_{\kvec} e^{-\I S} = {\bf U}(\kvec) {\bf a}_{\kvec}-{\bf V}(\kvec) {\bf a}^{\dagger}_{\kvec}
\label{eq:balphaops}
\end{equation}
where  ${\bf U}(\kvec)$ and ${\bf V}(\kvec)$ are 2$\times$2 matrices
and
\begin{equation}
  {\bf a}_{\kvec}
  = \begin{pmatrix}\annil{\kvec\uparrow}\\ \annil{\kvec\downarrow}\end{pmatrix}\,.
\label{eq:particleops}
\end{equation}
The superconducting/superfluid ground state is then the state that is
annihilated by the quasiparticle destruction operators
$\balpha_{\kvec}$, and the amplitudes
$\theta_{\sigma_1\sigma_2}(\kvec)$ are determined by diagonalizing the
Hamiltonian in the quasiparticle basis.

An equivalent method that is compatible with the variational method is
to minimize the expectation value of the (zero temperature) Landau
potential
\begin{equation}
  \left\langle \hat H-\mu \hat N\right\rangle_s=
  \bra{\rm BCS} \hat H-\mu\hat N  \ket{\rm BCS}\,.
  \label{eq:EBCS}
\end{equation}
with respect to the amplitudes $\theta_{\sigma_1\sigma_2}(\kvec)$.
For that purpose, the paired state is written as
\begin{equation}
\ket{\rm BCS} = \prod_{\genfrac{}{}{0pt}{1}{\kvec}{k_z>0}}\left[
  u_\kvec^2 + u_\kvec v_\kvec \bcreat{\kvec}+v_\kvec^2
    \creat{\kvec\uparrow}\creat{-\kvec\downarrow}\creat{\kvec\downarrow}\creat{-\kvec\uparrow}\right]\ket{0}\label{eq:bcsdef}
\end{equation}
where
\begin{equation}
  \bcreat{\kvec}=\sum_{\sigma_1\sigma_2}\Lambda_{\sigma_1\sigma_2}(\kvec)
  \creat{\kvec\sigma_1}\creat{\kvec\sigma_2}
  \label{eq:bcreatdef}
  \end{equation}
is the pair creation operator, and $\Lambda_{\sigma_1\sigma_2}(\kvec)$
is the unitary  matrix
  \begin{equation}
    \Lambda_{\sigma_1\sigma_2}(\kvec)=\frac{\Theta_{\sigma_1\sigma_2}}{\Theta_D(\kvec)}\,.
 \label{eq:Lambdames}
\end{equation}
Above, $\Theta_D(\kvec)$ is a normalization factor.
Time-reversal invariance and isotropy of the system implies, for
triplet pairing, that $\Lambda_{\downarrow\uparrow}(\kvec) =
  \Lambda^*_{\downarrow\uparrow}(\kvec)
  =\Lambda_{\uparrow\downarrow}(\kvec)$ and
  $\Lambda_{\uparrow\uparrow}(\kvec)=-\Lambda^*_{\downarrow\downarrow}(\kvec)$,
  hence the matrix $\Lambda_{\sigma_1\sigma_2}(\kvec)$ has the form
\begin{equation}
  {\bm \Lambda} = \begin{pmatrix}c(\kvec) & d(\kvec)\\d(\kvec)& -c^*(\kvec)\\\end{pmatrix}\qquad\left|c(\kvec)\right|^2 + \left|d(\kvec)\right|^2=1\,,
\label{eq:Lambdamat}
\end{equation}
the last identity being due to unitarity. For singlet pairing we simply have
\begin{equation}
  {\bm \Lambda} = \begin{pmatrix}0 & 1\\-1& 0\\
  \end{pmatrix}\,.
\end{equation}
The energy in the paired state is then
\begin{eqnarray}
  &&\left\langle\hat H-\mu\hat N\right\rangle_s =\sum_{\kvec\sigma}v_\kvec^2(t(k)-\mu)+
  \frac{1}{2}\sum_{\kvec_i,\sigma_i}v^2_{\kvec_1} v^2_{\kvec_2}
  \bra{\kvec_1\sigma_1,\kvec_2\sigma_2}V\ket{\kvec_1\sigma_1,\kvec_2\sigma_2}_a
  \nonumber\\
  &+&\frac{1}{2}\sum_{\kvec_i,\sigma_i}u_{\kvec_1}v_{\kvec_1}\Lambda^*_{\sigma_1\sigma_1'}(\kvec_1)
    \bra{\kvec_1\sigma_1,-\kvec_1\sigma_1'}V\ket{\kvec_2\sigma_2,-\kvec_2\sigma_2'}\Lambda_{\sigma_2\sigma_2'}(\kvec_2)u_{\kvec_2}v_{\kvec_2}\nonumber\\
   \,. \label{eq:ebcsmf}  \end{eqnarray}

In the above, we have followed the standard derivations of a
mean-field theory of pairing in arbitrary angular momentum states
\cite{Tamagaki70,10.1143/PTP.48.1517}. One more step will be utilized
later which is not generally necessary but useful: The energy
functional \eqref{eq:ebcsmf} leads to a single-particle spectrum of
the form
\begin{equation}
  \varepsilon_k^s = t_k + \sum_{\kvec',\sigma'}v_{\kvec'}^2
  \bra{\kvec\sigma,\kvec'\sigma'}
  \hat V\ket{ \kvec\sigma,\kvec'\sigma'}_a\label{eq:epssup}\,.
\end{equation}
In the so-called ``decoupling approximation'' \cite{Kennedy1968}, the
spectrum $\varepsilon_k^s$ is approximated by the Hartree-Fock
spectrum of a normal system
\begin{equation}
  \varepsilon_k^{\rm HF} = t_k + \sum_{\kvec',\sigma'}n(k')
  \bra{\kvec\sigma,\kvec'\sigma'}
  \hat V\ket{ \kvec\sigma,\kvec'\sigma'}_a\,.\label{eq:epsnor}
\end{equation}
This approximation is a very good one because $N_0 =
\sum_{\kvec\sigma}v_\kvec^2= \sum_{\kvec\sigma}n(k)$ where $N_0$ is
the average particle number which is the same as the particle number
of the normal state,
\[N_0 \equiv \bra{\rm BCS}\hat N\ket{\rm BCS} = \bra{\Phi_0}\hat N \ket{\Phi_0}\,,\]
and $v_{\kvec'}^2$ is integrated with a smoothly varying function.

An energy functional leading to the decoupling approximation is
obtained by expanding \eqref{eq:ebcsmf} to first order in the quantity
$v_\kvec^2-n(k)$; it has the form
\begin{eqnarray}
  &&\bra{\rm BCS} \hat H-\mu\hat N \ket{\rm BCS} ={\rm const.} +
  \sum_{\kvec\sigma}v_\kvec^2(\varepsilon_k^{\rm HF}-\mu)
  \nonumber\\
  &+&\frac{1}{2}\sum_{\kvec_i,\sigma_i}u_{\kvec_1}v_{\kvec_1}\Lambda^*_{\sigma_1\sigma_1'}(\kvec_1)
    \bra{\kvec_1\sigma_1,-\kvec_1\sigma_1'}\hat V\ket{\kvec_2\sigma_2,-\kvec_2\sigma_2'}\Lambda_{\sigma_2\sigma_2'}(\kvec_2)u_{\kvec_2}v_{\kvec_2}\,.\nonumber
    \label{eq:ebcsmf1}  \end{eqnarray}

Thus far we have merely reviewed the derivation of a general pairing
theory \cite{Bogoliubov,BeliaevLesHouches,Tamagaki70} and have brought
it in a form that is suitable for the inclusion of correlations.  The
idea of a correlated wave functions theory of superfluidity
\cite{HNCBCS,cbcs} is to begin with the {\em normal, correlated\/}
ground state and generate one Cooper pair at a time out of the normal
system \cite{LeggettQFS2018}.  A {\em correlated\/} state is
constructed by applying a correlation operator \eqref{eq:f_prodwave}
to the superfluid model state. Since the superfluid state does not
have a fixed particle number, we must write the correlated state in
the form
\begin{equation}
\ket{\rm CBCS} =  \sum_{{\bf m},N} \ket {\Psi_{\bf m}^{(N)}}
\ovlp{{\bf m}^{(N)}}{\rm BCS} , 
\label{eq:CBCS}
\end{equation}
\ie , as an expansion in the $N-$body correlated basis wave
functions defined in Eq. (\ref{eq:basis}), where the expansion
coefficients are given by the overlap between the corresponding
uncorrelated normal state and the BCS state. The Landau potential is
calculated as
\begin{equation}
  \left\langle H'\right\rangle_c = \frac{\bra{\mathrm{CBCS}} \hat H'
    \ket{\mathrm{CBCS}}}
  {\bigl\langle\mathrm{CBCS}\big|\mathrm{CBCS}\bigr\rangle}\,.
  \label{eq:CEBCS}
\end{equation}

The choice \eqref{eq:CBCS} of the correlated wave function is not
completely obvious. In our earlier work \cite{HNCBCS} and that of
other authors \cite{Fantonipairing,IrvineBCS} a slightly different
normalization of the intermediate states was chosen which has the
consequence of awkward normalization terms. This is a part of the
reason why we need to redo the derivation of the gap-equation for
$P$-wave pairing from Ref. \cite{IrvineBCS}.

\subsection{Weakly coupled superfluids}

If the superfluid gap is small compared to the Fermi energy, it is
legitimate to simplify the problem by expanding $\left\langle
H'\right\rangle_c$, Eq. \eqref{eq:EBCS}, in the {\em deviation\/} of
the Bogoliubov amplitudes $u_{\kvec}$, $v_{\kvec}$ from their normal
state values $u^{(0)}({\kvec})= \bar n(k)$,
$v^{(0)}_{\sigma_1\sigma_2} ({\kvec})=n(k)$.  Adopting such an
expansion in the number of Cooper pairs, all correlation functions and
effective interactions can be taken from the normal system.

Carrying out the above expansion in terms of the deviation
of the Bogoliubov amplitudes $u_{\kvec}$, $v_{\kvec}$ from their
normal state values,
\begin{equation}
  \ket{\rm BCS}=\ket{\Phi_0}+\ket{{\rm BCS}^{(1)}}+\ket{{\rm BCS}^{(2)}}
  +\ldots
\end{equation}
lets us write, to the desired order,
\begin{equation}
  \ket{{\rm BCS}^{(1)}}=\sum_{\kvec}\begin{cases}\left[v_\kvec^2 \creat{\kvec\uparrow}\creat{-\kvec\downarrow}\creat{\kvec\downarrow}\creat{-\kvec\uparrow} +u_\kvec v_\kvec\bcreat{\kvec}
    - v^2_\kvec\right]\ket{\Phi_0}&k>\KF\\
    \left[u_\kvec^2\annil{-\kvec\uparrow}\annil{\kvec\downarrow}
    \annil{-\kvec\downarrow}\annil{\kvec\uparrow}-u_\kvec v_\kvec\bannil{\kvec}-u_\kvec^2\right]\ket{\Phi_0}&k\le\KF
        \end{cases}
  \label{eq:dbcs1}
\end{equation}
and
\begin{equation}
  \ket{{\rm BCS}^{(2)}}=\frac{1}{2}
  \sum_{\kvec,\kvec'}u_\kvec v_\kvec u_{\kvec'}v_{\kvec'}\times
    \begin{cases}\phantom{-}\bcreat{\kvec}\bcreat{\kvec'}\ket{\Phi_0}&k>\KF,k'>\KF\\
      -    \bcreat{\kvec}\bannil{\kvec'}\ket{\Phi_0}&k>\KF,k'\le\KF\\
      \phantom{-}\bannil{\kvec}\bannil{\kvec'}\ket{\Phi_0}&k\le\KF,k'\le\KF
      \,.\end{cases}\label{eq:dbcd2}
\end{equation}
The correlated states are then
\begin{equation}
  \sum_{{\bf m},N}\ket{\Psi_{\bf m}^{(N)}}\ovlp{{\bf m}^{(N)}}{{\rm BCS}^{(1)}}
  = \sum_{\kvec}\begin{cases} v_\kvec^2
    \displaystyle \frac{F_{N_0+4}\creat{\kvec\uparrow}\creat{-\kvec\downarrow}\creat{\kvec\downarrow}\creat{-\kvec\uparrow}
      \ket{\Phi_0}}{\displaystyle I^{(N_0+4)}_\kvec}\\
    \quad + u_\kvec v_\kvec\displaystyle\frac{F_{N_0+2}\bcreat{\kvec}\ket{\Phi_0}}{I^{(N_0+2)}_\kvec}
    - v^2_\kvec\ket{\Psi_0}&k>\KF\\
    u_\kvec^2\displaystyle\frac{F_{N_0-4}\annil{-\kvec\uparrow}\annil{\kvec\downarrow}
        \annil{-\kvec\downarrow}\annil{\kvec\uparrow}\ket{\Psi_0}}{I^{(N_0-4)}_\kvec}\\ \quad
      -u_\kvec v_\kvec\displaystyle\frac{F_{N_0-2}\bannil{\kvec}\ket{\Phi_0}}{I^{(N_0-2)}_\kvec}
      -u_\kvec^2\ket{\Phi_0}&k\le\KF
        \end{cases}
  \label{eq:Dbcs1}
\end{equation}
and
\begin{equation}
  \sum_{{\bf m},N}\ket{\Psi_{\bf m}^{(N)}}\ovlp{{\bf n}^{(N)}}{{\rm BCS}^{(2)}}
  = \frac{1}{2}
  \sum_{\kvec,\kvec'}u_\kvec v_\kvec u_{\kvec'}v_{\kvec'}\times
    \begin{cases}\phantom{-}\displaystyle\frac{F_{N_0+4}\bcreat{\kvec}\bcreat{\kvec'}\ket{\Phi_0}}{I^{(N_0+4)}_\kvec}&k>\KF,k'>\KF\\
      -    \displaystyle\frac{F_{N_0}\bcreat{\kvec}\bannil{\kvec'}\ket{\Phi_0}}{I^{(N_0)}}&k>\KF,k'\le\KF\\
      \phantom{-}\displaystyle\frac{F_{N_0-4}\bannil{\kvec}\bannil{\kvec'}\ket{\Phi_0}}{I^{(N_0-4)}_\kvec}&k\le\KF,k'\le\KF
      \,,\end{cases}\label{eq:Dbcd2}
\end{equation}
where the $I^{(N)}$ are the appropriate $N$-particle normalization
constants, see Eq. \ref{eq:basis}.

We are now ready to expand the Landau-potential \eqref{eq:CEBCS} to
the desired order. Abbreviate $N'=N-N_0$.
\begin{eqnarray}
  &&\frac{\bra{\rm CBCS}\hat H-\mu\hat N\ket{\rm CBCS}}
  {\ovlp{\rm CBCS}{\rm CBCS}}=
  E_0-\mu N_0\nonumber\\&+&\sum_{{\bf m},{\bf n},N}\ovlp{{\rm BCS}^{(1)}}{{\bf m}^{(N)}}
  \left[H^{'(N)}_{{\bf m},{\bf n}} - \mu N'I^{(N)}_{{\bf m},{\bf n}}
    \right]\ovlp{{\bf n}^{(N)}}{{\rm BCS}^{(1)}}\nonumber\\
  &+&\sum_{{\bf m},N}\left[H^{'(N)}_{{\bf 0},{\bf n}} -
    \mu N' I^{(N)}_{{\bf 0},{\bf n}}\right]\ovlp{{\bf m}^{(N)}}{{\rm BCS}^{(2)}}+ {\rm c.c.}+\ldots\label{eq:ebcs1}
\end{eqnarray}
Retaining only terms of first order on $v_\kvec^2-n_0(k)$, only one term
survives in each of the two terms in Eq. \eqref{eq:ebcs1}:
For example, for $k>\KF$, $k'\le\KF$ we have
\begin{eqnarray}
&&\sum_{{\bf m},N}\left[H^{'(N)}_{{\bf 0},{\bf n}} -
    \mu N'I^{(N)}_{{\bf 0},{\bf n}}\right]\ovlp{{\bf m}^{(N)}}{{\rm BCS}^{(2)}}\nonumber\\
  &=&\left[W_{{\bf 0},{\bf n}} -\frac{1}{2}\left[H_{{\bf n},{\bf n}}-E_0\right]
    N_{{\bf 0},{\bf n}}
 \right]\ovlp{{\bf m}^{(N)}}{{\rm BCS}^{(2)}}\nonumber\\
  &=&\sum_{\genfrac{}{}{0pt}{1}{\kvec,\kvec'}{k>\KF,k'\le \KF}}\sum_{\sigma_i}
  u_\kvec v_\kvec u_{\kvec'}v_{\kvec'}\Lambda^*_{\sigma_1\sigma_2}(\kvec)
  {P}_{\kvec,\kvec'}(\S)\Lambda^*_{\sigma_1\sigma_2}(\kvec')\label{eq:term1} , 
\end{eqnarray}
where the integrand ${\cal P}_{\kvec\kvec'}(\S)$ takes the form  
\begin{eqnarray} 
  {\cal P}_{\kvec\kvec'}(\S)
    &=&
    \bra{\kvec'\sigma_1,-\kvec'\sigma_1'}
  {\cal W}(1,2)\ket{\kvec\sigma_2,-\kvec\sigma_2'}\nonumber\\
  &+& (|\varepsilon_k-\mu|+ |\varepsilon_{k'}-\mu|)
  \bra{\kvec'\sigma_1,-\kvec'\sigma_1'}
  {\cal N}(1,2)\ket{\kvec\sigma_2,-\kvec\sigma_2'} \nonumber \\ 
   & \equiv &  \bra{\kvec'\sigma_1,-\kvec'\sigma_1'} {\cal P}(1,2)  \ket{\kvec\sigma_2,-\kvec\sigma_2'} \nonumber \\ 
   & \equiv &  {\cal W}_{\kvec\kvec'}(\S)+(|e_{\kvec}- \mu | 
+ |e_{\kvec'}- \mu |){\cal N}_{\kvec\kvec'}(\S) 
            \label{eq:Pdef}\,, 
\end{eqnarray}
where $\S$ is the total spin.

The diagonal terms $\kvec=\kvec'$ in the first term of
Eq. \eqref{eq:ebcs1} must be treated separately, for example for
$k>\KF$:
  
        \begin{eqnarray}
          &&\sum_{\genfrac{}{}{0pt}{1}{\kvec}{k_z>0}}u_\kvec^2 v_\kvec^2
        \sum_{{\bf m}}
    \ovlp{\Psi_0 \bannil{\kvec}}{{\bf m}^{(N+2)}}\left[H^{'(N+2)}_{{\bf m},{\bf m}}
       -\mu N'\right]
      \ovlp{{\bf m}^{(N+2)}}{\bcreat{\kvec}\Psi_0}\nonumber\\
      &=&\sum_{\genfrac{}{}{0pt}{1}{\kvec}{k_z>0}}u_\kvec^2 v_\kvec^2\sum_{\sigma\sigma'}\left|\Lambda_{\sigma\sigma'}(\kvec)\right|^2\left(2\varepsilon_k-2\mu\right)\nonumber\\
      &=& \sum_{\kvec\sigma}v_\kvec^2\left(\varepsilon_k-\mu\right)
        \end{eqnarray}

The remaining off-diagonal term is calculated analogously to
Eq. \eqref{eq:term1}, leading to the final expression for the energy
of the superfluid state
\begin{eqnarray}
  \langle \hat H-\mu\hat N\rangle_c &=& E_0 - \mu N_0 + \sum_{\kvec,\sigma}
  v_{\kvec}^2 (e_{\kvec} - \mu )\nonumber\\
  &+&\sum_{\kvec,\kvec'}\sum_{\sigma_i}
  u_\kvec v_\kvec u_{\kvec'}v_{\kvec'}\Lambda^*_{\sigma_1\sigma_1'}(\kvec)
  {P}_{\kvec,\kvec'}(\S)\Lambda^*_{\sigma_2\sigma_2'}(\kvec')\label{eq:Ebcs}
\end{eqnarray}
where the pair interaction ${\cal P}_{\kvec\kvec'}(\S)$ is defined in
Eq. \eqref{eq:Pdef}. Evidently, the expression \eqref{eq:Ebcs} has the
same structure as the mean field result Eq. \eqref{eq:ebcsmf1}.  Only
the single-particle spectrum and the effective interaction are
redefined. That way, we have mapped the theory of a strongly
interacting system onto that of a weakly interacting one.

The remaining manipulations are therefore textbook material:
The Bogoliubov amplitudes $u_\kvec $, $v_\kvec $ are obtained in the
standard way by variation of the energy expectation (\ref{eq:Ebcs}).
This leads to the familiar gap equation
\begin{equation}
\Delta(\kvec) = -\frac{1}{2}\sum_{\kvec'} {\cal P}_{\kvec\kvec'}
\frac{\Delta({\kvec'})}{\sqrt{(e_{\kvec'}-\mu)^2 + \Delta({\kvec'})^2}}\,.
\label{eq:gap}
\end{equation}%
Replacing $ {\cal P}_{\kvec\kvec'}$ with the bare interaction, one
recovers the conventional, mean-field BCS gap equation.

For pairing in states other than $S$-waves, the angle dependence of
the gap function must be taken into account.  Here we follow the
standard ``angle-average'' approximation, as formulated, for example,
in Ref.  \citenum{Baldo3P23F2}.  In general, the gap equation can
couple different angular momenta and becomes a matrix equation of the
form
\begin{equation}
  \Delta^{(\ell)}(k) =\label{eq:multigap}
  -\frac{1}{2}\sum_{\ell'}\int \frac{d^3k'}{(2\pi)^3}
  P_{\ell\,\ell'}(k,k')\frac{\Delta^{(\ell')}(k')}{
\sqrt{(\varepsilon_{k'}-\mu)^2 + D^2(k')}}\,, 
\end{equation}
where $D^2(k) = \sum_\ell \left|\Delta^{(\ell)}(k)\right|^2$,
and we have suppressed spin degrees of freedom because we are always
working in either an $S=0$ or an $S=1$ state.

\subsection{Analysis of the pairing interaction}
\label{ssec:Lvpair}

The appearance of the ``energy numerator'' term in the pairing
interaction matrix element \eqref{eq:Pdef} might be unfamiliar to the
reader who is used to mean-field theories, but it appears quite
naturally when the gap equation is expressed in terms of the
$T$-matrix \cite{PethickSmith}.  This section is devoted to a
discussion of the importance of the energy numerator term which arises
in the expansion of the correlated BCS state \eqref{eq:CBCS} in the
number of Cooper pairs.

The effective interaction components ${\cal W}(1,2)$ and ${\cal
  N}(1,2)$ can be expressed rigorously in terms of cluster expansions
of the FHNC method \cite{CBF2}. Their leading terms are local
functions ${\cal W}(r)$ and ${\cal N}(r)$. The correlation correction
is readily identified with the direct correlation function
\begin{equation}
  {\cal N}(r) = \Gamma_{\rm dd}(r)
   \label{eq:NGdd}
\end{equation}
and the local part of ${\cal W}(r)$ can be identified with the static
effective interaction \eqref{eq:Fermiwind}.

If the gap at the Fermi surface is small, we can replace the pairing
interaction $\tilde{\cal W}(q)$ by its $S$-wave matrix element at the
Fermi surface,
\begin{equation}
\tilde {\cal W}_F \equiv \frac{1}{2 \KF^2}\int_0^{2\KF} q dq \widetildeto{\Gamma}{W}(q)
= N{\cal W}_{\KF,\KF}\,.
\label{eq:V1S0}
\end{equation}
Then we can write the gap equation as
\begin{equation}
1 = - \tilde {\cal W}_F\int\frac{ d^3k'}{(2\pi)^3\rho}
\Bigg[\frac{1}{\sqrt{(e_{k'}-\mu)^2 + \Delta^2_{\KF}}}  \label{eq:gaplowdens}
  -\frac{|e_{k'}-\mu|}{\sqrt{(e_{k'}-\mu)^2 + \Delta^2_{\KF}}}
\frac{\SF(k')}{t(k')} \Bigg]\,,
\end{equation}
where $\Delta_{\KF}$ is the gap amplitude at the Fermi surface.  The
above expression is almost identical to Eq.~(16.91) in
Ref.~\citenum{PethickSmith}.  In particular, the second term, which
originates from the energy numerator generated in Eq.~(\ref{eq:gap}),
has the function of regularizing the integral for large $k'$.  Thus,
the above reformulation of the gap equation allows a clean treatment
of zero-range effective interactions without the need of cutoff
parameters.

This observation leads us to two conclusions: First, the effective
interaction $\widetildeto{\Gamma}{\cal W}(k)$ should be viewed as a
local approximation to the $T$-matrix. This is also evident because
its diagrammatic structure contains both particle-particle and
particle-hole reducible diagrams. Second, a correct balance between
the energy numerator and the interaction term is essential to
guarantee the convergence of the integral.

When the gap is large, one can no longer argue that the energy
numerator, which vanishes at the Fermi surface, is negligible.  The
convergence of the integrals is, in this case, guaranteed by the fact
that the interactions fall off for large $k'$. To demonstrate that, we
can again study the behavior of the integrand (\ref{eq:Pdef}) for large $k'$:
\begin{equation}
{\cal P}_{\kvec\kvec'} 
\rightarrow {\cal W}_{0,\kvec'}+ t(k'){\cal N}_{0,\kvec'}
\end{equation}
From Eqs. \eqref{eq:Wfermi} and \eqref{eq:NGdd} we can now conclude
that these two terms cancel for large arguments.

The cancellation of these two terms is a consequence of either the
functional optimization of the correlations, or the parquet diagram
summations. It is therefore expected that the actual value of the gap
depends sensitively on how the energy numerator is treated.  This also
applies to the question of how one should deal with a non-trivial
single-particle spectrum; comments on this are found in
Ref. \citenum{ectpaper}. Similar concerns apply to calculations that
use state-independent correlation functions of the form
\eqref{eq:Jastrow}, including our own work \cite{ectpaper}: The
correlations are optimized for the central channel of the interaction,
but the paring interaction is calculated in the
singlet-$S$-channel. Hence, the cancellations between energy numerator
and interaction term are violated. The alternative, namely calculating
the correlations for a model where the singlet-$S$ channel is taking
as state-independent interaction is not a viable one because such a
system would become unstable against infinitesimal density
fluctuations at densities much smaller than those of interest here.

Identifying the local pairing interaction $\widetildeto{\Gamma}{\cal
  W}(q)$ with a static approximation for the energy-dependent
effective interaction \eqref{eq:Wnonlocal} now also allows us to go
beyond the variational model. In particular, when the gap is small
compared to the Fermi energy, it is more appropriate to use
$\widetildeto{\Gamma}{W}(q,\omega=0)$ in the pairing interaction.
 
\subsection{Strongly coupled superfluids}
\label{ssec:fullbcs}

The above treatment of a superfluid state has the appealing feature
the theory can be mapped onto an ordinary BCS theory, where the
effective interactions and, if applicable, the single particle
spectrum, are calculated for the normal system.

The basic assumption of the ``weak coupling'' approximation is that
the superfluid gap at the Fermi surface is small compared to the Fermi
energy. This assumption is not met in low-density neutron matter where
the gap energy can indeed be of the order of half of the Fermi energy.
This is a common feature of practically all neutron matter gap
calculations since the 1970s \cite{YangClarkBCS} until recently
\cite{Gandolfi2009b,ectpaper}. To examine this problem we have derived
in Ref. \citenum{fullbcs} the full FHNC theory for a superfluid state
of the form \eqref{eq:CBCS}.  Without going into the complex details
of this derivation, we outline here only the essential steps for the
benefit of those who are familiar with the techniques of cluster
expansions and summation methods for wave functions of the type
\eqref{eq:Jastrow} \cite{Johnreview,polish,KroTrieste}.

According to the analysis of Ref. \citenum{fullbcs}, the cluster
expansion of the superfluid energy \eqref{eq:CEBCS} can be obtained
from that of the normal system as follows:

The cluster expansion of a normal system can be represented as
expansion in terms of correlation lines $h(r_{ij})$
(Eq. \eqref{eq:hdef}) and exchange lines $\ell(r_{ij}\KF)$
(Eq. \eqref{eq:eldef}). The expansion of the corresponding superfluid
system is then generated by replacing each exchange loop by the sum of
two loops formed by the functions
\begin{equation}
  \ell_v(r) \equiv \frac{\nu}{N_0}
  \sum_{\kvec} v_{\kvec}^2e^{\I\kvec\cdot\rvec}\qquad\mathrm{and}
  \quad\ell_u(r) \equiv \frac{\nu}{N_0}
	   \sum_{\kvec} u_{\kvec}v_{\kvec}e^{\I\kvec\cdot\rvec}
	  \label{eq:luvdef}\,,
\end{equation}
respectively.

The resulting gap equation is the same as Eq. \eqref{eq:gap}, the only
difference being that the ingredients ${\cal W}(1,2)$ and ${\cal
  N}(1,2)$ should be determined self-consistently for a superfluid
system and depend implicitly on the $\ell_v(r)$ and $\ell_u(r)$.
Therefore, one does not expect a large effect from this modification.

An immediate consequence of this derivation deserves further
discussion: To see the problem, return to the simplest version of the
Euler equation, Eq. \eqref{eq:FermiPPA}, which is the same in the
superfluid system; a small additive term coming from exchange diagrams
\cite{fullbcs} does not change our analysis. The static structure
function of the non-interacting system has the form
\begin{equation}
  \SF(q) = 1 - \frac{\rho}{\nu}
	  \int d^3 r e^{\I \qvec\cdot\rvec}\left[
	    \ell_v^2(r) - \ell_u^2(r)\right]
	\,.\label{eq:SFdef}
\end{equation}
It follows immediately from the definitions \eqref{eq:luvdef} that the
long-wavelength limit of $\SF(q)$ is
\begin{equation}
  \SF (0+) = 2\frac{\sum_{\kvec} u_{\kvec}^2 v_{\kvec}^2}{\sum_{\kvec}v_{\kvec}^2} > 0\,.
\label{eq:SF0}
\end{equation}
Hence, $\SF(0+) > 0$ for the superfluid system. {\em As a consequence,
  Eq. \eqref{eq:FermiPPA} and, hence, the optimization problem
  \eqref{eq:euler}, has no sensible solution if the corresponding
  Landau parameter $F_0^s = \tilde V_{\rm p-h}/m {c_{\rm F}^*}*2 < 0$
  even for an infinitesimally small but finite gap.}

The problem is readily solved by abandoning the collective
approximation \eqref{eq:ChiColl}, or, in other words moving from the
pure Jastrow-Feenberg wave function to the parquet summations. It is
then also appropriate to use superfluid particle-hole propagators.
There have been several suggestions for a Lindhard function for a
superfluid system
\cite{Schrieffer1999,PhysRevB.61.9095,Steiner2009,Vitali2017}. The
most frequently used form for $T=0$ is given below. In the superfluid
case, $\chi_0(q,\omega)$ also depends on the spins. In terms of the
usual relationships of BCS theory,
\begin{eqnarray}
  u_k^2 &=& \frac{1}{2}\left(1+\frac{\xi_\kvec}{E_\kvec}\right)\,,
  \nonumber\\
  v_k^2 &=& \frac{1}{2}\left(1-\frac{\xi_\kvec}{E_\kvec}\right)\,,
\end{eqnarray}
with $\xi_{\kvec} = t(k)-\mu$ and $E_{\kvec} =
\sqrt{\xi_{\kvec}^2+\Delta_{\kvec}^2}$
we have \cite{Schrieffer1999,Kee1998,Kee1999,PhysRevB.61.9095}

\begin{equation}
  \chi_0^{(\rho,\sigma)}(\qvec,\omega) = \frac{\nu}{N}\sum_{\kvec}
  b_{\kvec,\qvec}^{(\rho,\sigma)}\Biggl[\frac{1}
    {\hbar\omega-E_{\kvec+\qvec}-E_{\kvec}+\I\eta} -
    \frac{1}{\hbar\omega+E_{\kvec+\qvec}+E_{\qvec}+\I\eta}\Biggr]
  \label{eq:BCSLindha}
\end{equation}
with
\begin{eqnarray}
  b_{\kvec,\qvec}^{(\rho,\sigma)} &=&\frac{1}{4}\left[\left(1-
    \frac{\xi_{\kvec}}{E_{\kvec}}\right)
    \left(1+\frac{\xi_{\kvec+\qvec}}{E_{\kvec+\qvec}} \right)
    \pm\frac{\Delta_{\kvec}}{E_{\kvec}}
    \frac{\Delta_{\kvec+\qvec}}{E_{\kvec+\qvec}}\right]
  \nonumber\\ &=&v_{\kvec}^2u_{\kvec+\qvec}^2 \pm u_{\kvec} v_{\kvec}
  u_{\kvec+\qvec} v_{\kvec+\qvec}\,,
\end{eqnarray}
where the upper sign applies to the density channel, and the lower
to the spin channel, respectively.

There is, of course, no reason for not using the full Lindhard
functions for defining the effective interaction
$\widetildeto{\Gamma}{W}(q)$ in \eqref{eq:Wlocal}. This can be done
using the Lindhard function for the normal system, or
\eqref{eq:BCSLindha}. It turns out that the use of
\eqref{eq:BCSLindha} makes little difference for $\tilde V_{\rm
  p-h}(q)$. We have therefore used in our ground state calculations
the Lindhard function for the normal system.

This is different for the effective interaction ${\cal W}(1,2)$ used
in the gap equation for which we have used the local forms given in
Eqs. \eqref{eq:W135}. We have there used either the normal
\eqref{eq:chi0} or the superfluid system \eqref{eq:BCSLindha} Lindhard
function at $\omega=0$ for the effective interactions in the pairing
calculation. This is more appropriate for these low-energy phenomena

\subsection{$S$-wave pairing\label{ssec:SwavePairing}}

We have used in all of our calculations a free single-particle
spectrum; this is justified by our findings on the self-energy
discussed in section \ref{ssec:selfen}, see also
Ref. \citep{JLTP189_470}.

\begin{figure}
  \centerline{
    \includegraphics[width=0.4\textwidth,angle=270]{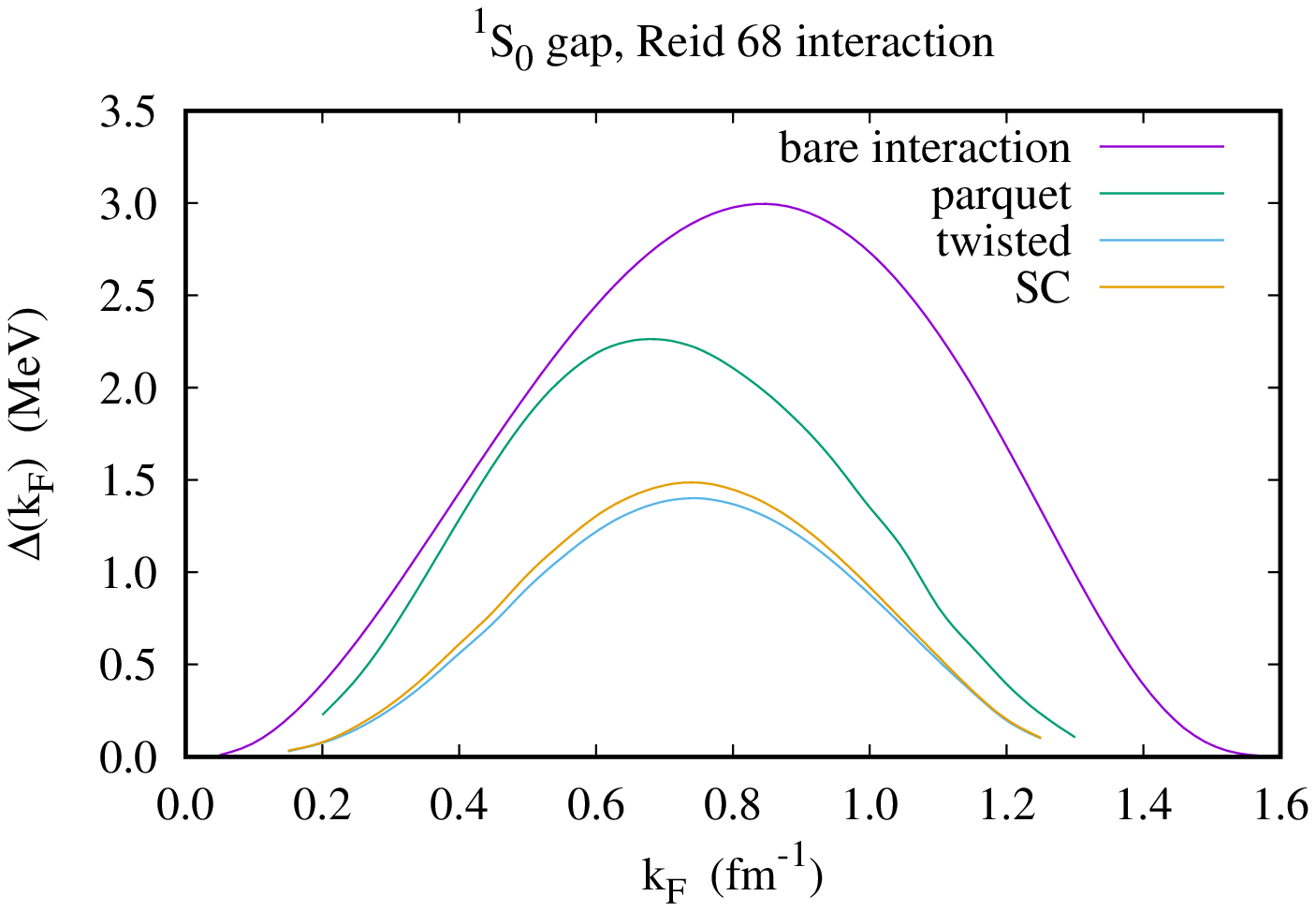}
    \includegraphics[width=0.4\textwidth,angle=270]{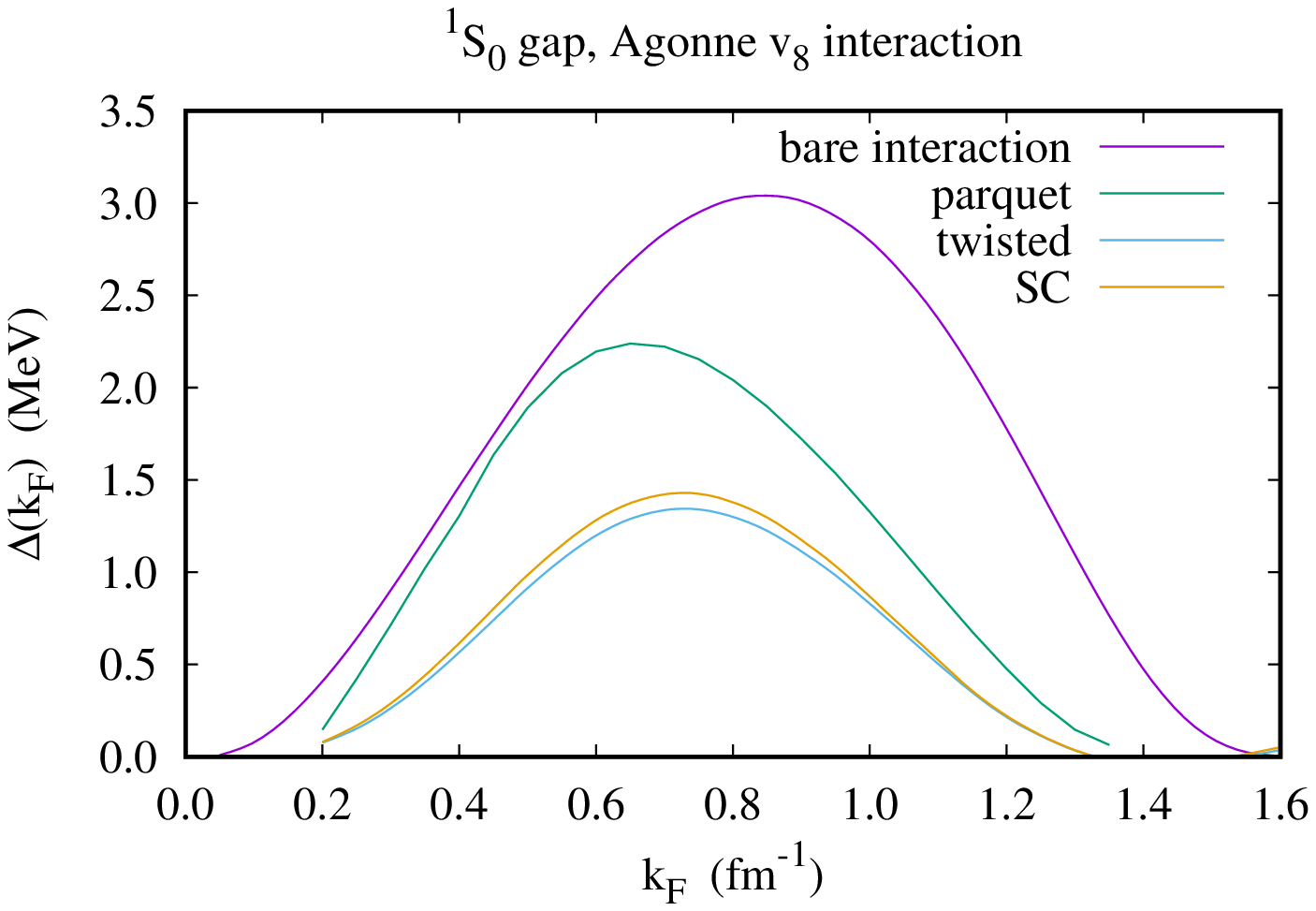}}
  \centerline{
    \includegraphics[width=0.4\textwidth,angle=270]{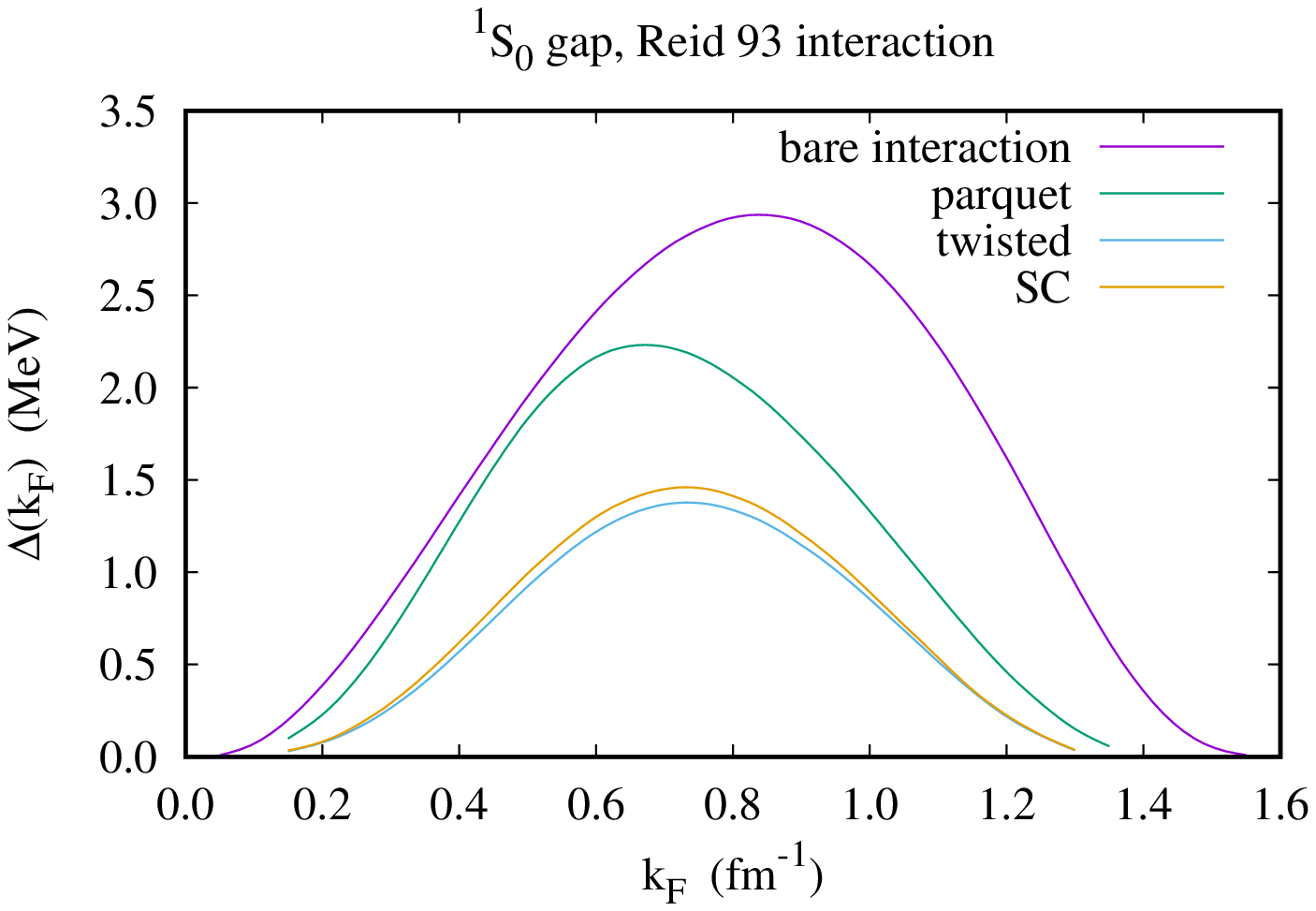}
    \includegraphics[width=0.4\textwidth,angle=270]{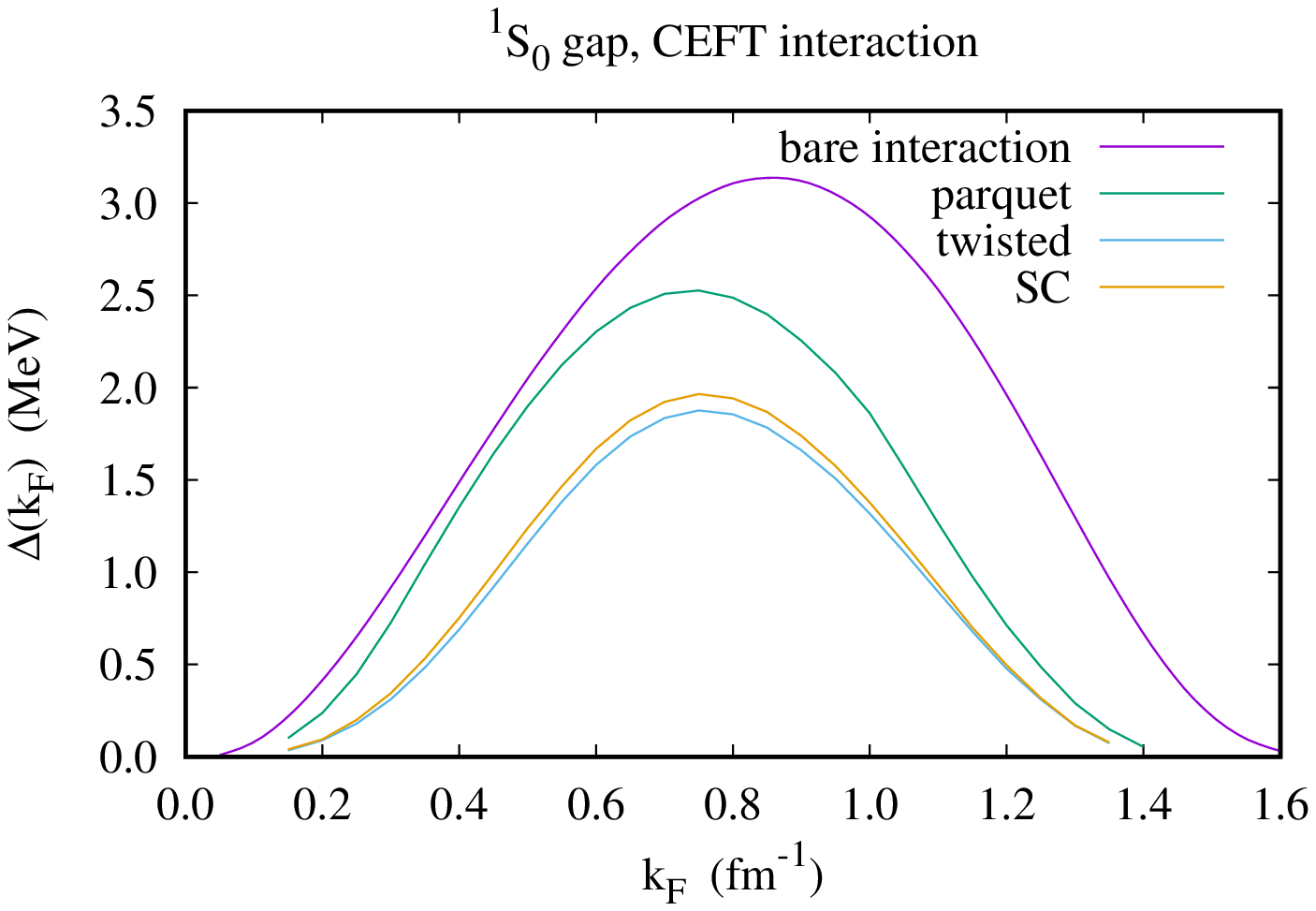}}
  \caption{Results for the $^1S_0$ gaps using our four interaction 
  models in various approximations as discussed in the text.
  \label{fig:1S0gaps}}
\end{figure}

We show in figs.~\ref{fig:1S0gaps} results for the $^1S_0$
gaps using our four interaction models.

The figures show our results for the bare interaction, the parquet
calculation, the calculation including twisted-chain diagrams, and
the calculation using the superfluid particle-hole propagators
\eqref{eq:BCSLindha}. The findings for all interactions are basically
the same:
    \begin{itemize}
    \item{} Correlations ({\em i.e. among others polarization corrections\/})
      reduce the gap by about a factor of 2 compared to the
      ``mean field'' approximation using the bare interaction,
 \item{} The twisted-chain diagrams reduce the $^1S_0$ gap by
   another factor of 2,
 \item{} Using a superfluid particle-hole propagator makes little
   difference for $S$-wave pairing.
  \end{itemize}

\subsection{P wave pairing\label{ssec:PwavePairing}}

\begin{figure}[H]
    \centerline{
      \includegraphics[width=0.3\textwidth,angle=270]{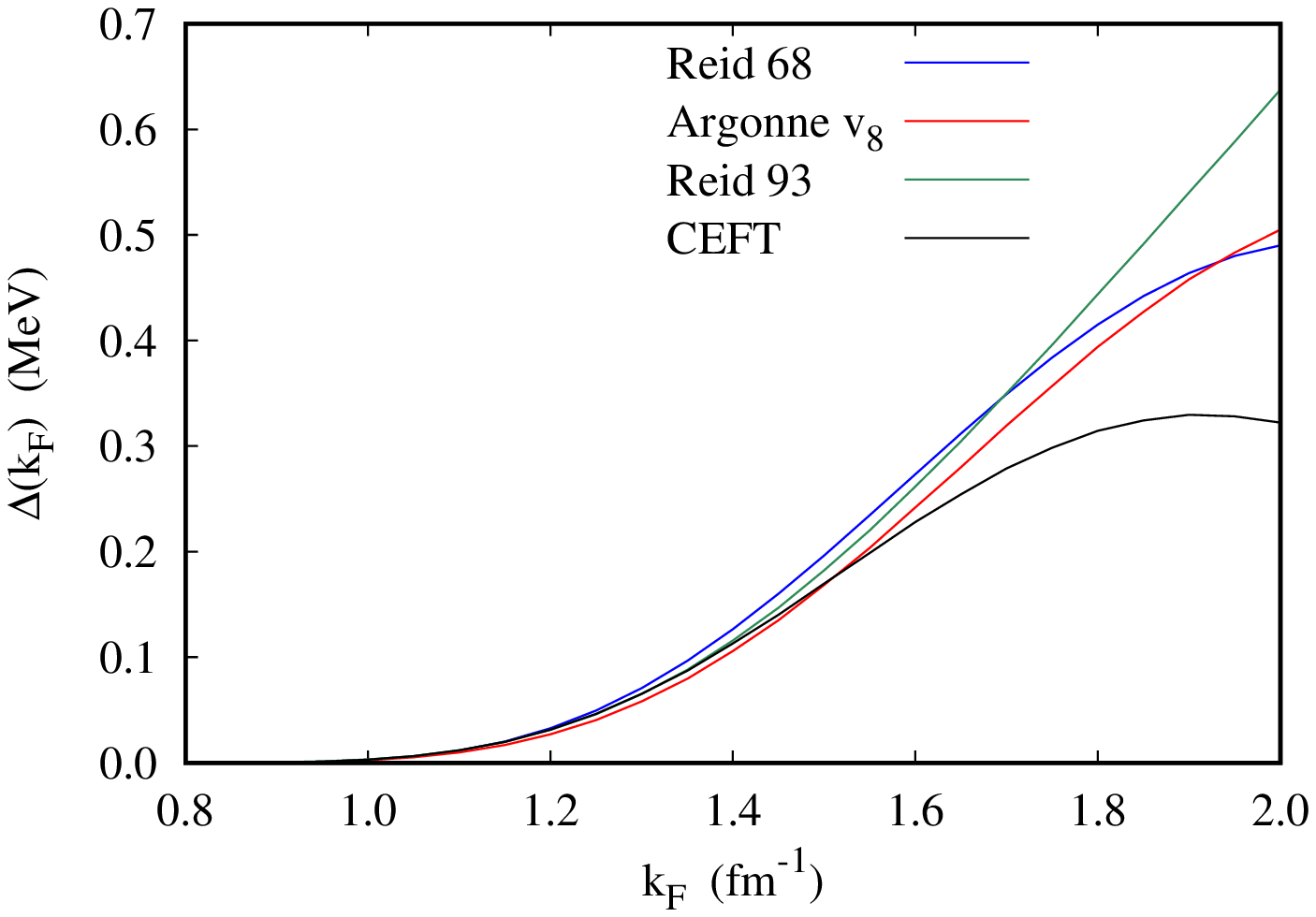}
      \includegraphics[width=0.3\textwidth,angle=270]{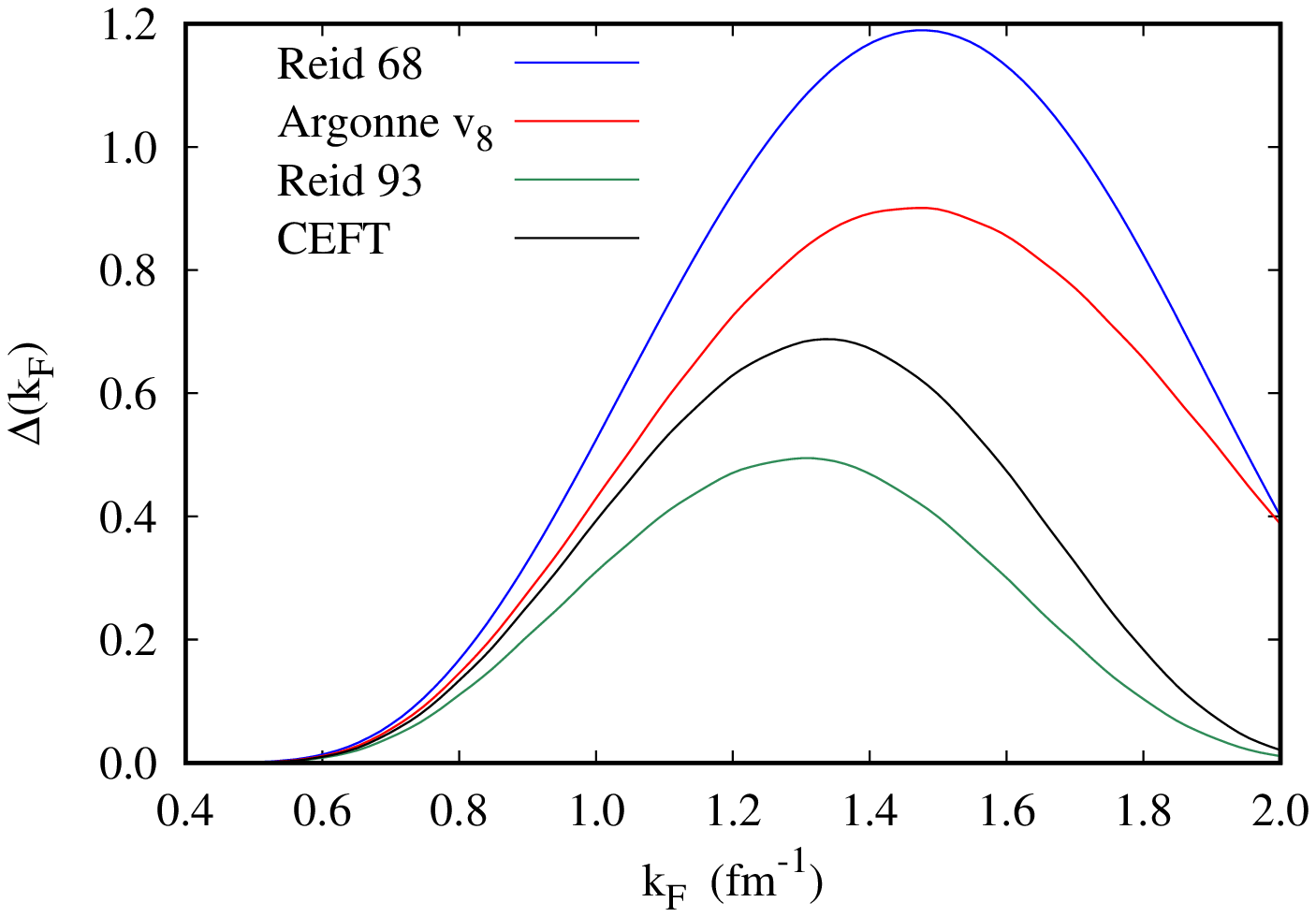}}
    \caption{The left figure shows, for all four interactions, the
     $^3P_2$-$^3F_2$ gap $D(\KF)$ as functions of $\KF$.
      The right figure shows the $^3P_0$ gap $\Delta(\KF)$ for
      these interaction models when the spin-orbit force has
      been turned off.\label{fig:baregaps}}
    \end{figure}

The situation is somewhat different for $P$-wave pairing. While the
results for $S$-wave pairing were almost identical for all four
potential models studied here; there are visible differences for
$P$-wave pairing.  Since the pioneering work of Tamagaki \etal
\citep{Tamagaki70,10.1143/PTP.48.1517} there was a general
understanding that, when many-body effects are neglected,$^3P_2$
and$^3P_2$-$^3F_2$ prevail in neutron
matter. Figs. \ref{fig:baregaps} shows this for our four interaction
models. In that calculation we did not go beyond $\KF =
2.0\,$fm$^{-1}$ because the many-body calculations we are about to
compare with become less trustworthy at high densities.  Evidently the
predictions of these interaction models are already quite
different. We also note that there is no or only minimal pairing in
$^3P_2$ and $^3P_0$ states.

It has been pointed out \cite{Gezerlis2014} that the existence of
$^3P_2$-$^3F_2$ pairing is due to the spin-orbit interaction and
that  $^3P_0$ would be preferred if there were no spin-orbit force.
We have verified this assertion for our four interaction models. Indeed,
when the spin-orbit force is turned off, we have pairing in $^3P_0$
states whereas there is no such effect in $^3P_2$-$^3F_2$ or
$^3P_2$ states, see Fig. \ref{fig:baregaps}, right panel.

The results shown in Figs. \ref{fig:baregaps} have to be kept in mind
when we turn to the discussion of the effect of many-body
correlations: Evidently, the uncertainty in the mean-field
approximation for $P$-wave pairing propagates to the many-body
treatment. The major contribution of many-body physics is the
suppression of the spin-orbit interaction through short-ranged
correlations as discussed in section \ref{ssec:spinorbit} which is, in
the above mean-field model, entirely {\em ad-hoc.}

\begin{figure}[H]
  \centerline{
    \includegraphics[width=0.3\textwidth,angle=270]{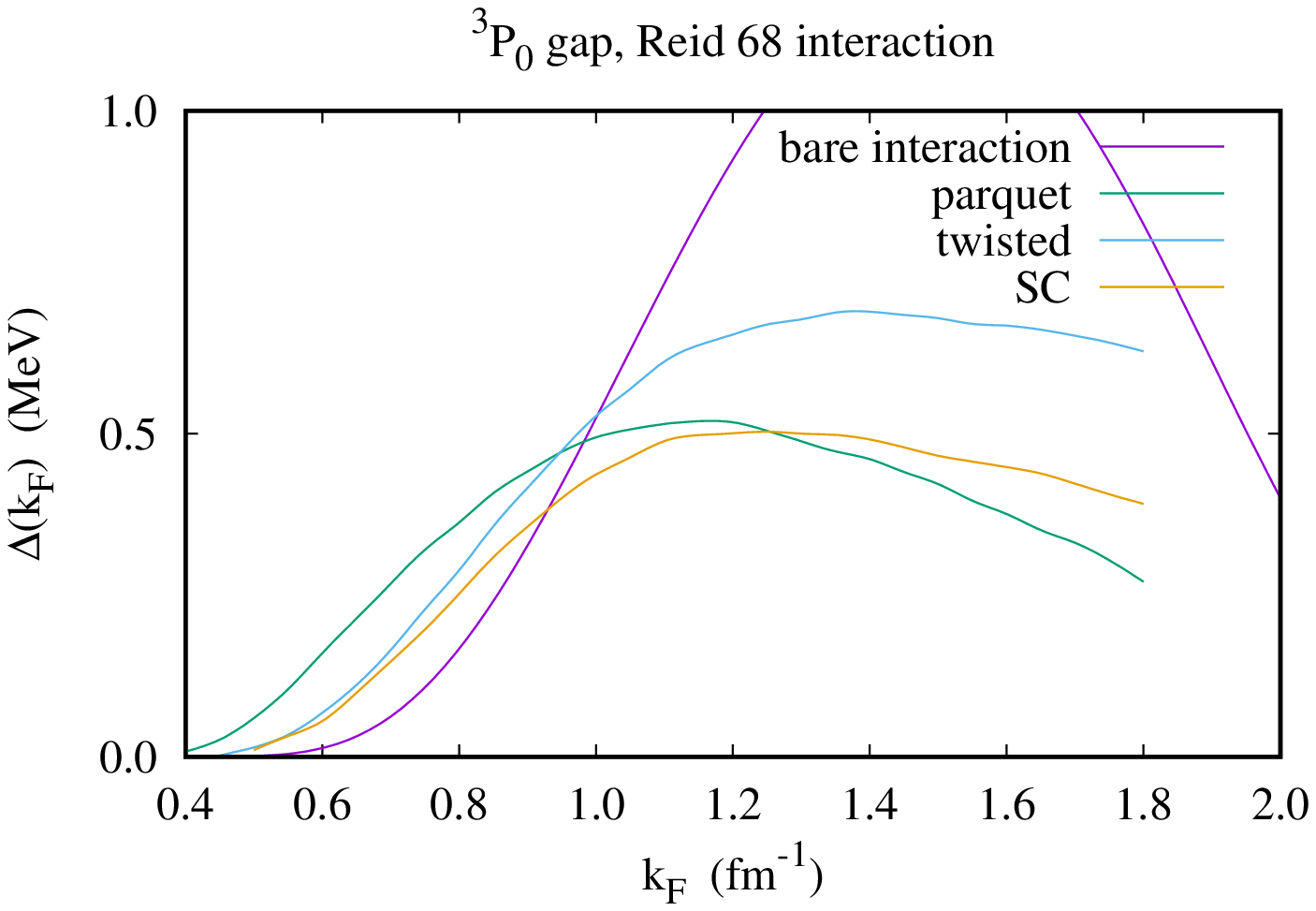}
   \includegraphics[width=0.3\textwidth,angle=270]{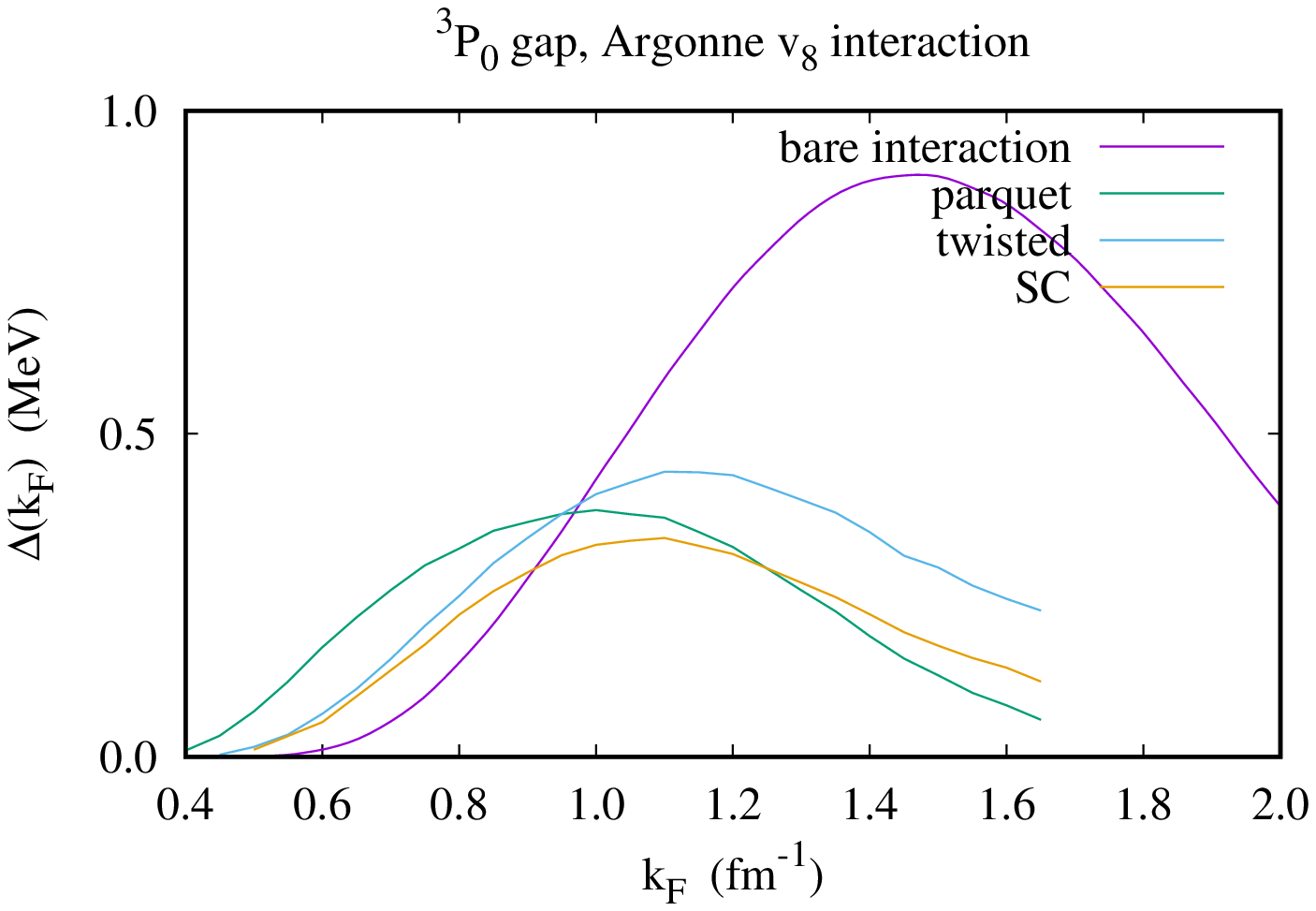}}
  \centerline{
    \includegraphics[width=0.3\textwidth,angle=270]{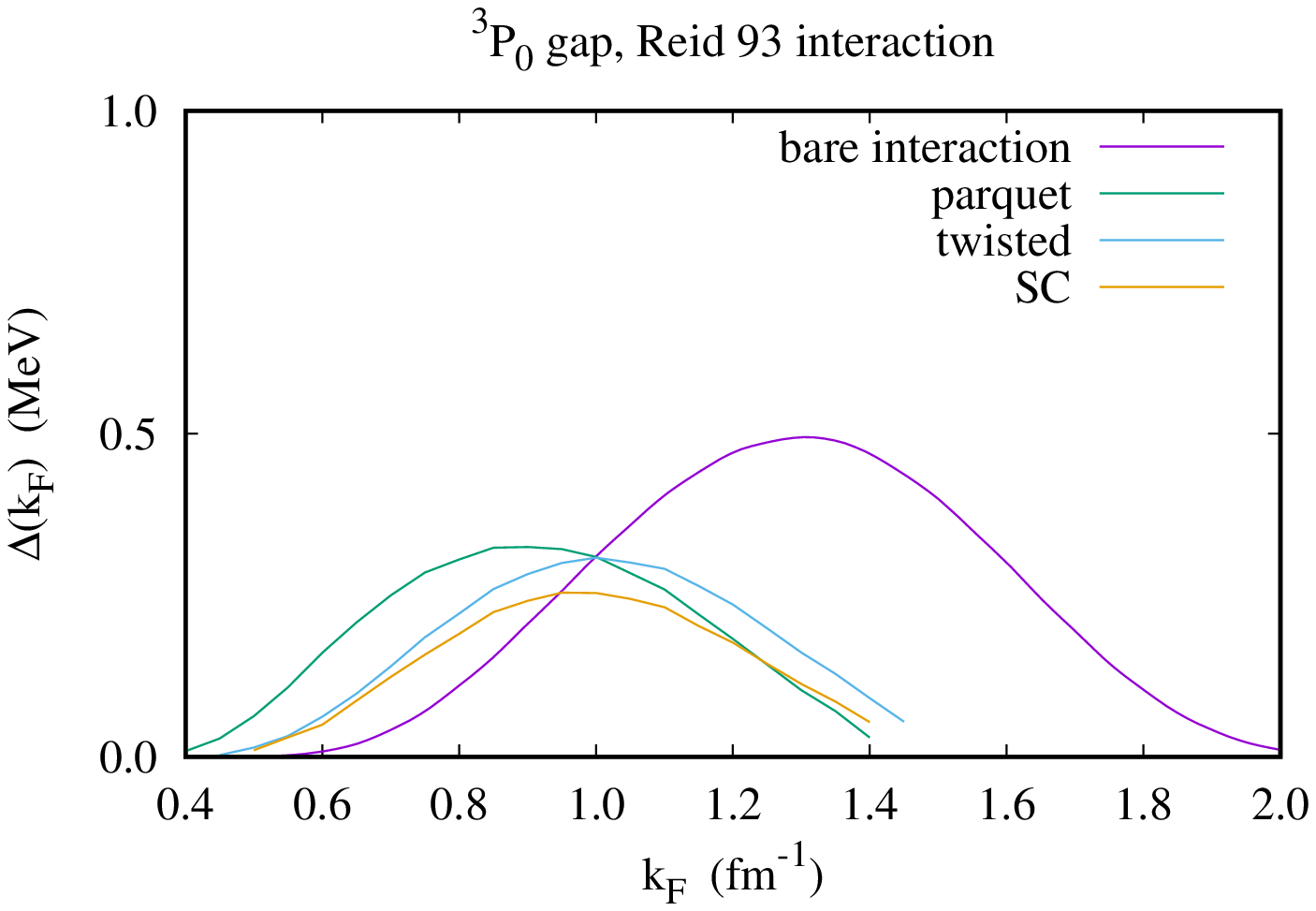}
    \includegraphics[width=0.3\textwidth,angle=270]{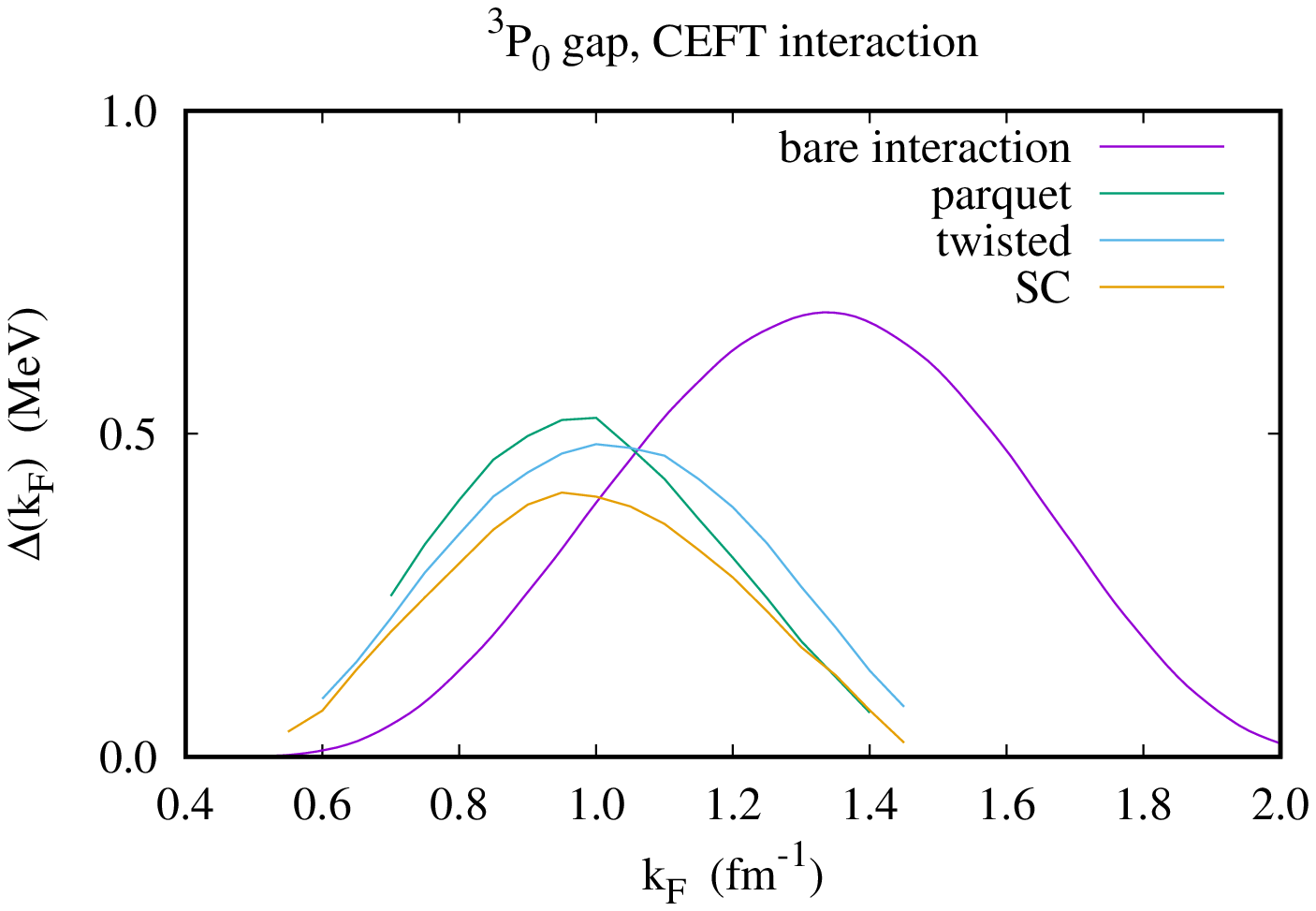}}
  \caption{Results for the $^3P_0$ gaps using our four interaction models
    for the mean-field approximation omitting the spin-orbit
    interaction, and for three versions of the parquet calculation.
    \label{fig:3P0gaps}}
\end{figure}

Our results for the $^3P_0$ gap for our four potential models are
shown in Figs. \ref{fig:3P0gaps}. Keeping the above statement about
the generic uncertainty stemming from different potential models in
mind, we find that many-body corrections play a more important role in
this case compared with $S$-wave pairing. Twisted-chain corrections
are somewhat less important for spin-triplet pairing. This is
understandable because the non-parquet diagrams mix the repulsive
spin-triplet interaction into the spin-singlet interaction, hence the
attraction in spin-singlet channels is reduced. On the other hand, in
triplet states, the admixture of the singlet interaction is, similar
to the spin-orbit force, largely damped by the short-ranged
correlation hole.

A bit unexpected is the more visible correction stemming from using
the superfluid particle-hole propagator. 

\begin{figure}[H]
  \centerline{\includegraphics[width=0.5\textwidth,angle=270]{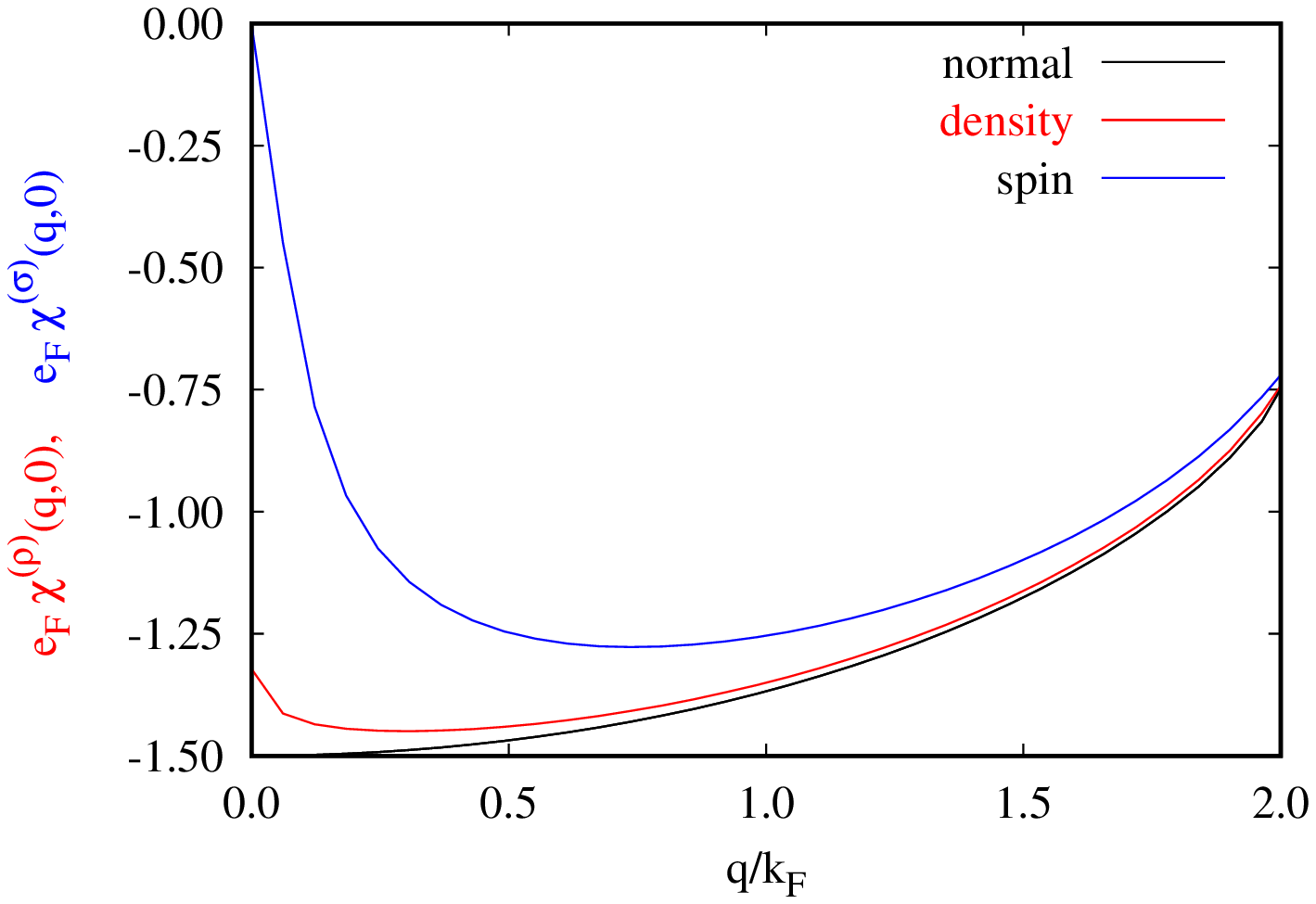}}
  \caption{(color online) The figure shows the particle-hole propagator
    of a normal system (black line) as well as those for a superfluid system
    (red and blue lines). The Bogoliubov amplitudes
    $u_\kvec$, $v_\kvec$ are taken from a  $^1S_0$ gap calculation
    for the Argonne potential at $\KF = 1.0\,$fm$^{-1}$.
    \label{fig:chi0plot}}
\end{figure}

An explanation of the findings of Figs. \ref{fig:1S0gaps} and
\ref{fig:3P0gaps} requires the examination of the effective
interactions entering the gap equations. Besides containing
conventional corrections like polarization effects to the pairing
interaction, our calculations contain two components that have so far
been neglected.  These are the twisted-chain corrections and the
use of the superfluid particle-hole propagator
\eqref{eq:BCSLindha}. We show in Fig.  \ref{fig:chi0plot} a comparison
of the zero-energy Lindhard function $\chi_0(q,0)$
(Eq. \eqref{eq:chi0}) of the normal phase with the density- and
spin-Lindhard functions for the superfluid system. The specific case
is the $^1S_0$ gap for the Argonne potential at $\KF =
1.0\,$fm$^{-1}$, see Fig. \ref{fig:1S0gaps}. As already observed in
Ref. \citep{fullbcs} we find a suppression of the Lindhard function in
the spin-channel whereas the density channel changes very little.

\begin{figure}[H]
  \centerline{
    \includegraphics[width=0.35\textwidth,angle=270]{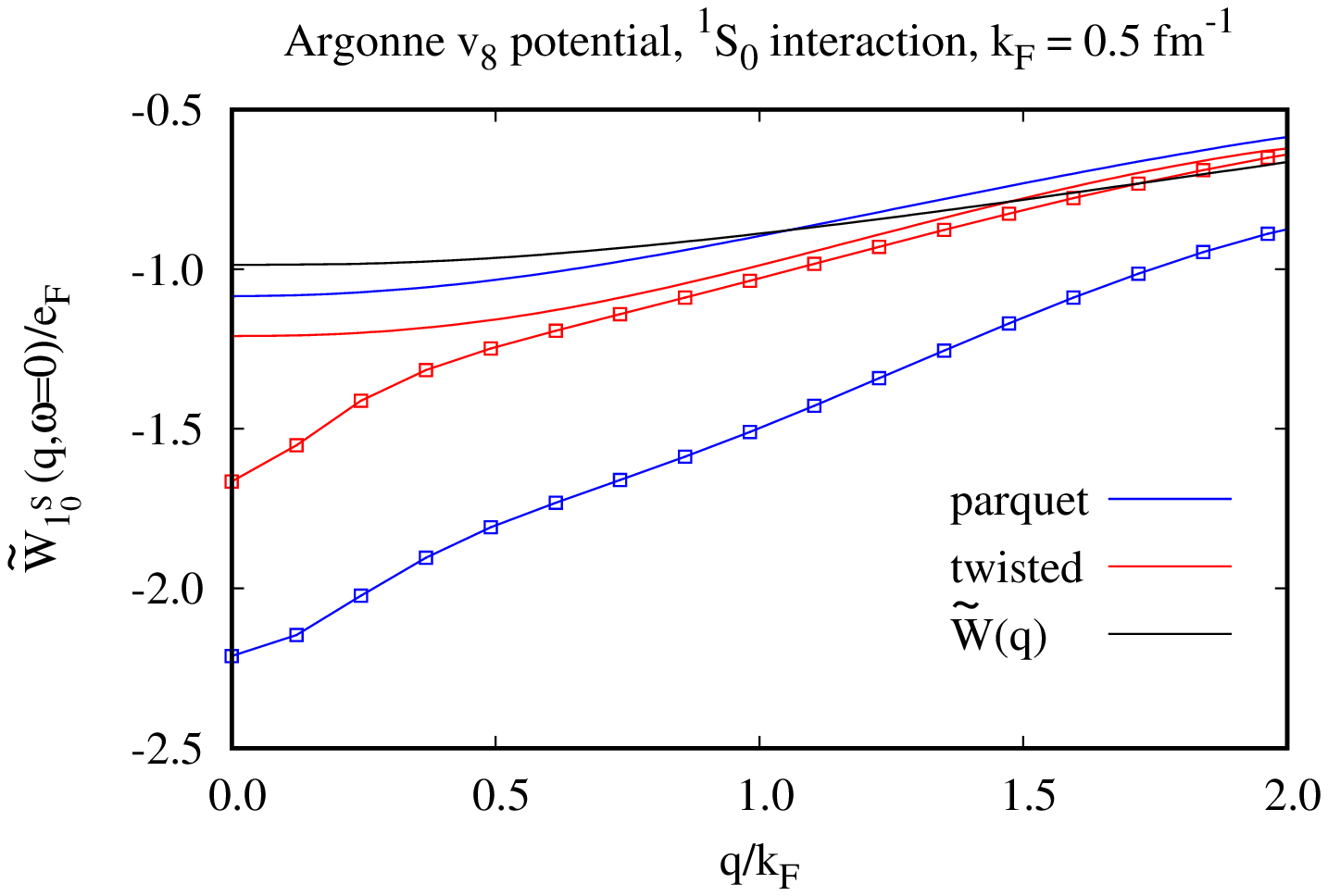}
\includegraphics[width=0.35\textwidth,angle=270]{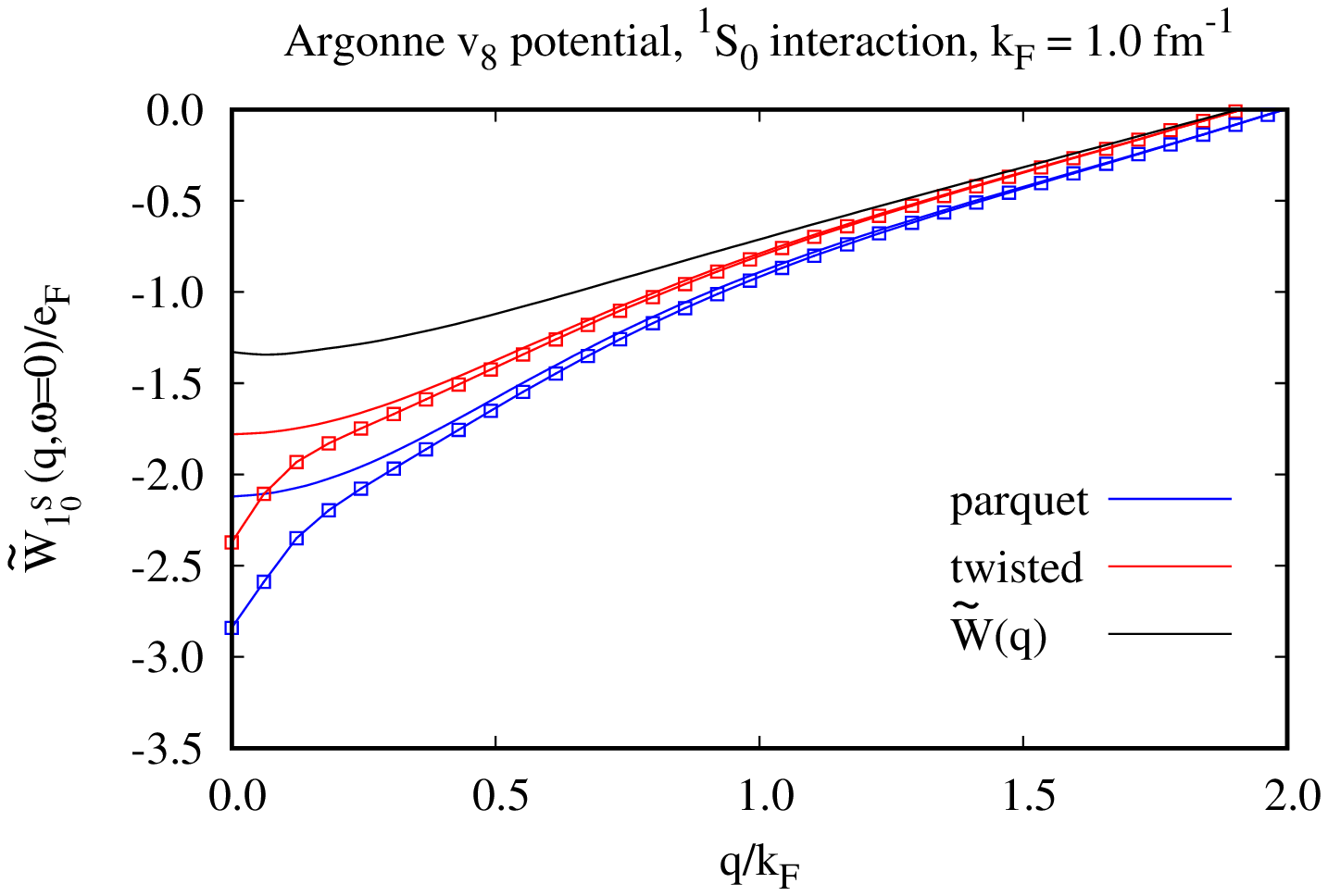}
  }
  \caption{(color online) The figure shows the momentum-space form of
    the $^1S_0$ pairing interaction in various approximations: the
    solid lines show the parquet results without (red) and with (blue)
    twisted-chain corrections using the normal system particle-hole
    propagator. The lines with markers contain the same quantities
    calculated with superfluid particle-hole propagators for $\KF =
    0.5\,$fm$^{-1}$ and $\KF = 1.0\,$fm$^{-1}$ for the Argonne $v_8$
    interaction. We also show, for comparison, the energy independent
    interaction $\tilde W(q)$, Eq. \ref{eq:Wlocal}.
    \label{fig:1S0parts}}
\end{figure}

\begin{figure}[H]
  \centerline{
    \includegraphics[width=0.35\textwidth,angle=270]{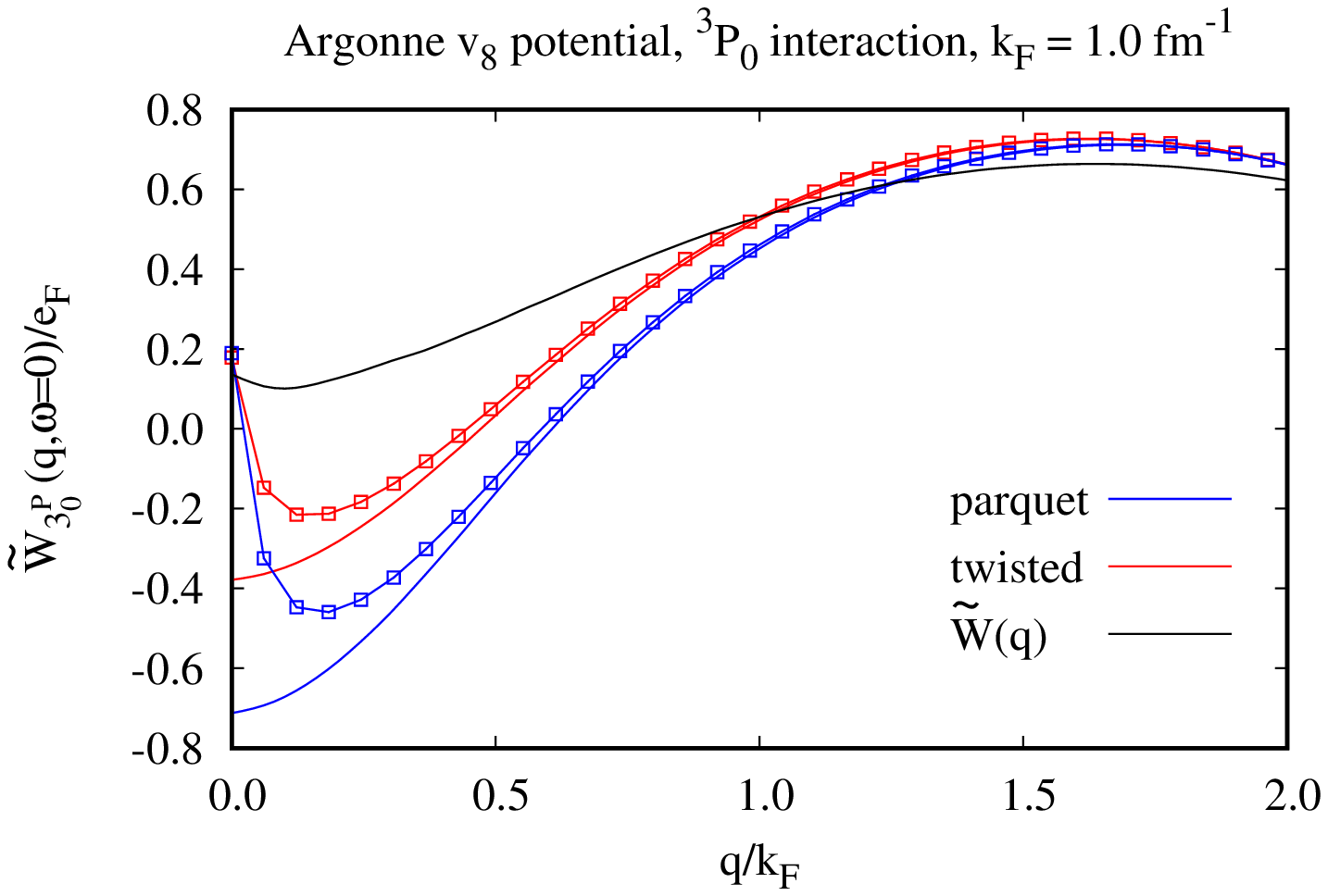}
\includegraphics[width=0.35\textwidth,angle=270]{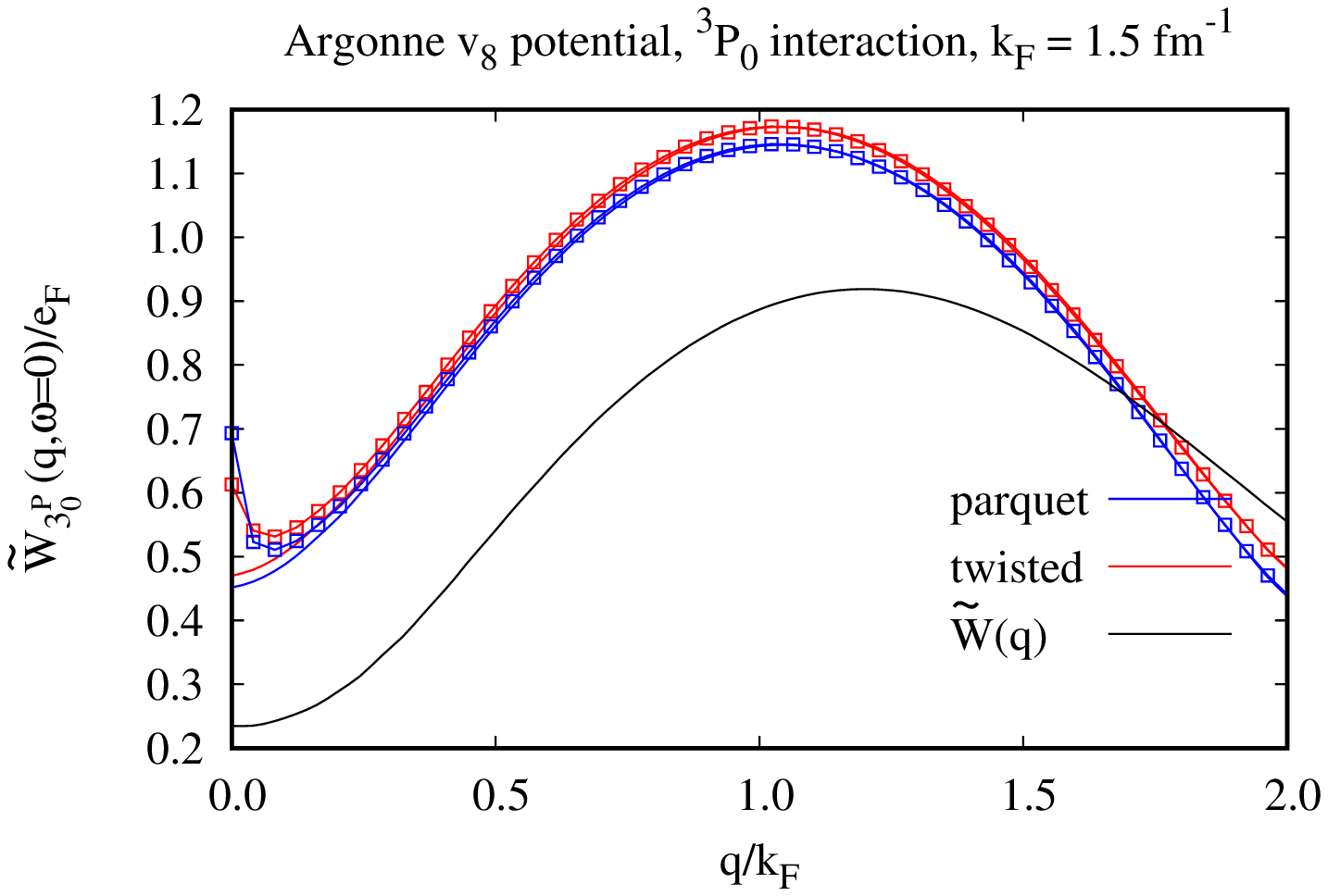}
  }
  \caption{(color online) Same as Figs. \ref{fig:1S0parts} for
    the $^3P_0$ pairing interaction for  $\KF =
    1.0\,$fm$^{-1}$ and $\KF = 1.5\,$fm$^{-1}$ 
    \label{fig:3P0parts}}
\end{figure}

We have highlighted the importance of twisted-chain corrections in
section \ref{ssec:vtwist} and shown that these are of similar or
higher magnitude than the induced interactions, see
Figs. \ref{fig:Vtwist}. Figs. \ref{fig:1S0parts} and
\ref{fig:3P0parts} show, at two representative densities, different
versions of the $^1S_0$ and the $^3P_0$ pairing interactions in
the most important momentum regime $0 \le q \le 2\KF$:
\begin{subequations}
\begin{eqnarray}
  \widetilde {\cal W}_{^1\!S_0}(q) &=& \tilde W_c(q) - \tilde W_L(q) - 2\tilde W_T(q)\\
  \widetilde {\cal W}_{^3\!P_0}(q) &=& \tilde W_c(q) - \tilde W_L(q) + 2\tilde W_T(q)\,.
\end{eqnarray}
\end{subequations}
Evidently the influence of various contributions is rather different:
Whereas twisted-chain corrections contribute a rather smooth
correction, the modification due to the superfluid particle-hole
propagator are -- consistent with Fig. \ref{fig:chi0plot} mostly in
the long-wavelength regime. It is also noteworthy that these
corrections go in the opposite direction in the singlet and triplet
case. Another relevant observation is that the correction arising from
twisted-chain diagrams is just as large as that from going from the
localized effective interaction to the zero-energy version. Only at
the highest density $\KF = 1.5\,$fm$^{-1}$ the twisted-chain
correction is negligible; we have already discussed above why that
correction should be less important for triplet interactions.

\section{Summary\label{sec:summary}}

We have in this paper reviewed and summarized the developments of
parquet-diagram/variational theory for state-dependent interactions
that are typical for nuclear systems. Our work is mainly of technical
nature. Neutron matter has served as a system for demonstration
because it is stable at low density where relatively simple
implementations of the theory, which best display its physical content
but would be inadequate in higher density systems like \he3 or nuclear
matter, can give reasonably trustworthy results.

We hope that we have provided ample evidence for the fruitful
combination of ideas of Green's functions based many-body theories and
variational methods. After the work of Refs. \citep{parquet1,parquet2}
it appeared as if Green's functions based method had become obsolete:
After all, it took a lengthy derivation to arrive a set of generic
many-body equations that had been known for two decades and could be
derived in a few lines from a Jastrow-Feenberg {\em ansatz\/} for the
wave function \cite{FeenbergBook,LanttoSiemens}. In its essence, the
fact that the wave function \eqref{eq:Jastrow} provides a sequence of
rigorous upper bounds for the energy, defines specific approximations
to Feynman diagrams that provided the best accuracy for the
calculational effort one is willing to pay.

The problem is more complicated for Fermions since the wave function
\eqref{eq:Jastrow} is not even in principle exact. Of course, the
upper bound property still holds and, hence, approximations to Feynman
diagrams suggested by the variational {\em ansatz\/} are expected to
be good albeit not exact. We have outlined some specific
approximations above in section \ref{ssec:rings} for the ring
diagrams, in section \ref{ssec:ladders} for the Bethe-Goldstone
equation, and in section \ref{ssec:selfen} for corrections to the
single-particle propagator, see also section
\ref{ssec:exchanges}. Realizing the connection one is, of course, at
liberty to {\em choose\/} to use these approximations or not depending
on the physical quantity one wants to describe. For example the
``collective approximation'' for $S(q)$, Eq. \eqref{eq:FermiPPA},
which follows from \eqref{eq:Chi0Coll}, is accurate at the percent
level compared to the result from the energy integration
\eqref{eq:SRPA} but of course totally invalid for the calculation of
$S(q,\omega)$.  Two very instructive examples are our calculations of
the single-particle spectrum and the pairing gap: In the
single-particle sprectum the static approximation misses the imaginary
part and is, in \he3, totally inadequate.  For the effective
interaction entering the gap equation we have considered it more
reasonable to use the zero-energy limit $\hat W(q,\omega=0)$ instead
of the energy averaged effective interaction $\hat W(q)$
\eqref{eq:Wlocal}. The situation becomes more drastic if we treat BCS
correlations self-consistently: In that case, the combined Euler and
BCS equations for a Jastrow-Feenberg wave function have no sensible
solution, see the argument around Eq. \eqref{eq:SF0}. Abandoning the
collective approximation and using the correct Lindhard functions
\eqref{eq:BCSLindha} eliminates this problem.

A particularly instructive example for the fruitful combination of
ideas of perturbation theory and variational wave functions is our
analysis of ``beyond parquet'' or ``commutator'' corrections. From the
point of perturbative many-body theory it is obvious that the exchange
of spin-fluctuations can mix spin-singlet and spin-triplet states in
the Bethe-Goldstone equation.  From the point of view of variational
wave functions, ``commutator corrections'' were largely considered a
technical nuisance, but it is similarly obvious that these corrections
mix the short-ranged structure of the spin-singlet and spin-triplet
interactions. Including corrections such as the third diagram in
Fig. \ref{fig:ladders} in the Bethe-Goldstone equation is not trivial
due to the complicated structure of the energy denominators in the
exact expression and we are not aware that these effects have been
discussed in the past. The relationship to variational wave functions
suggest practical ways of calculating these processes. In that
connection we have shown that, for short distances, the twisted-chain
corrections are quantitatively more important than all other many-body
effects, see Figs. \ref{fig:Vtwist}.

We have extended in section \ref{sec:dynamics} our previous work
\cite{eomIII,2p2h} to state-dependent correlations. Many-body dynamics
is often discussed in terms of the random-phase approximations which
describes excited states by the superposition of 1-particle-1-hole
excitations of the non-interacting ground state. Momentum conservation
guarantees that these states are orthogonal to the ground state. Of
course, a non-interacting ground state is normally a poor description
of the reality and, hence, typically effective interactions are
introduced. Going beyond the RPA, one can introduce 2-particle-2-hole
excitations; the approach is termed ``second RPA'' in nuclear physics.
Of course, the same remark holds on the quality of a non-interacting
ground state as reference, an additional problem is that
2-particle-2-hole excitations are not orthogonal to a non-interacting
ground state.  Like in the RPA, the problem has been solved by using
correlated states. Section \ref{sec:dynamics} has also described
calculation of the nucleon self-energy and the effective mass which is
potentially important for pairing phenomena.

In the final section we have turned our attention to $S$-wave and
$P$-wave pairings in neutron matter. We have shown that the
singlet-triplet mixing by the twisted-chain corrections can
significantly reduce the predicted $^1S_0$ pairing gap. A further
notable result is a suppression of the $^3P_2$-$^3F_2$ pairing gap
and an enhancement of $^3P_0$ pairing, which differs from the
mean-field prediction that $^3P_2$-$^3F_2$ is dominant. Such a
remarkable difference comes from the fact that the spin-orbit
interaction can couple with the repulsive spin-triplet interaction in
the many-body correlations. These results are important for the
analysis of neutron star cooling
\citep{PageReddy2006,StellarSuperfluids,YaH2003}. Additionally, they
further highlight the importance of beyond-mean-field calculations for
strongly interacting many-body systems.

\appendix

\section{Construction of a $v_8$ interaction  from partial waves\label{app:v8}}
For the Reid 93 interaction we determine the operator channels
$\alpha \in \{({\rm c}), (\sigma), ({\rm S}), ({\rm LS})\}$ in neutron
matter as follows:

\begin{itemize} 
\item
In the spin-singlet case we only
have a central interaction, hence
\begin{equation}
v(S=0,T=1)(r) \equiv V_{^1S_0}(r) = v_{\rm c,0}(r)=v_{(c)}(r)-3v_{(\sigma)}(r)
\label{eq:nmsinglet}
\end{equation}
\item
In the spin-triplet case, the operator structure
is
\begin{equation}
\hat v(S=1,T=1)(r) = v_{\rm c,1}(r) \1 + v_{(S)}(r)\hat S_{12}(\rvec)
+ v_{LS}(r) {\bf L}\cdot{\bf S}
\label{eq:nmtriplet}
\end{equation}
where $v_{\rm c,0}(r)=v_{(c)}(r)-3v_{(\sigma)}(r)$, $v_{\rm c,1}(r)=v_{(c)}(r)+v_{(\sigma)}(r)$ are partial-wave central channels with spin-0 and spin-1, represented by the operator channels. Also, for each partial wave
\begin{equation}
{\bf L}\cdot{\bf S}=\begin{pmatrix}
j-1 & 0 \\
0 & -j-2 \end{pmatrix}\,.\label{eq:LS}
\end{equation}
and
\begin{equation}
S_{ij} = \begin{pmatrix}
\frac{-2(j-1)}{2j+1} & \frac{6\sqrt{j(j+1)}}{2j+1}\\
\frac{6\sqrt{j(j+1)}}{2j+1} & \frac{-2(j+2)}{2j+1}\end{pmatrix}\,.\label{eq:Sij}
\end{equation}
The tensor interaction can be determined from the $^3P_2$-$^3F_2$
interaction,
\begin{equation}
  v_{(S)}(r) = \frac{5}{6\sqrt{6}} v_{^3P_2- ^3F_2}(r)\,.
  \label{eq:Vtensor}
  \end{equation}
If we want to reproduce the $^3P_0$-$^3P_0$ and the
$^3P_2$-$^3P_2$ phase shifts we get
\begin{subequations}\label{eq:R93a}
  \begin{eqnarray}
  v_{\rm c,1}(r) &=& \frac{1}{3}\left(2v_{^3{\rm P}_2- ^3{\rm P}_2}(r)+
  v_{^3{\rm P}_0- ^3{\rm P}_0}(r)\right)+\frac{8}{5}v_{\rm S}(r)\,,\\
  v_{\rm LS}(r) &=& \frac{1}{3}\left(v_{^3{\rm P}_2- ^3{\rm P}_2}(r)-
  v_{^3{\rm P}_0- ^3{\rm P}_0}(r)\right)-\frac{6}{5}v_{\rm S}(r)\,.
  \end{eqnarray}
\end{subequations}
\end{itemize}

\section{Elements of Correlated Basis Functions Theory
\label{app:cbf}}

We list here the most important definitions of correlated basis
functions (CBF) theory. For details, the reader is referred to review
articles \cite{Johnreview} and pedagogical material
\cite{KroTriesteBook}.  The CBF method uses the correlation operator
$\hat {F}_N$ to generate a non-orthogonal basis of the Hilbert space
\begin{equation}
  \ket{\Psi_{\bf m}} = \frac{1}{I_{\bf m}^{1/2}}
  \hat F_N\ket{\bf m},\qquad I_{\bf m}= \bra{\bf m}\hat F_N^\dagger
  \hat F_N^{\phantom{\dagger}}\ket{\bf m} ,
  \label{eq:basis}
\end{equation}
where $\ket{\bf m}$ is the state of a non-interacting system
characterized by the set of quantum numbers ${\bf m}$, and we identify
$\ket{\bf o}\equiv\ket{\Phi_0}$.
Alternatively, and in
view of the application of the correlated wave functions for superfluid
systems discussed in section \ref{sec:BCS}, we shall also use the notation
\begin{equation}
  \ket{\Psi_{p}^{(N+1)}} = \frac{1}{I_{p}^{(N+1)}}
  \hat F_{N+1}\ket{\creat{p}\Phi_0},\quad I_{p}^{(N+1)} =
  \bra{\Phi_{ o}\annil{{p}}}\hat F_{N+1}^\dagger \hat F_{N+1}^{\phantom{\dagger}}
    \ket{\creat{p}\Phi_0}^{1/2}
  \label{eq:phbasis}
\end{equation}
and analogously for hole states, particle-hole, \dots\, states. 
Matrix elements of the Hamiltonian are
\begin{equation}
  H_{{\bf m},{\bf n}} = \bra{\Psi_{\bf m}}H\ket{\Psi_{\bf n}},
  \qquad  H'_{{\bf m},{\bf n}} = \bra{\Psi_{\bf m}}H-E_0\ket{\Psi_{\bf n}}\,.
  \label{eq:Hmn}
\end{equation}
and the unit matrix
\begin{equation}
  I_{{\bf m},{\bf n}} = \ovlp{\Psi_{\bf m}}{\Psi_{\bf n}}
  \equiv \delta_{{\bf m},{\bf n}}+(1-\delta_{{\bf m},{\bf n}})J_{{\bf m},{\bf n}}\,.
  \label{eq:Imn}
\end{equation}
A natural decomposition of the off-diagonal matrix elements
$H'_{{\bf m},{\bf n}}$ is  \cite{CBF2}
\begin{equation}
  H'_{{\bf m},{\bf n}} = W_{{\bf m},{\bf n}}+ \frac{1}{2}\left(H_{{\bf m},{\bf m}}' +
  H_{{\bf n},{\bf n}}'\right)I_{{\bf m},{\bf n}}\,.\label{eq:Wmn}
\end{equation}
The states $\ket{\bf m}$ and $\ket{\bf n}$ differ normally only in a
few quantum numbers, {\em i.e.\/} we can write
\begin{equation}
\label{eq:defwave}
\ket{{\bf m}} = \creat{m_1}\creat{m_2}
\cdots
\creat{m_d} \; \annil{n_d} \cdots \annil{n_2}\annil{n_1}
\ket{{\bf n}} \;.
\end{equation}
To leading order in the particle number, the above off-diagonal matrix
elements can be written as antisymmetrized matrix elements of $d$-body
operators
\begin{subequations}
\begin{eqnarray}
  W_{{\bf m},{\bf n}} &\equiv& \bra{ m_1\, m_2 \, \ldots m_d \,}
{\cal W}(1,2,\ldots d) \,\ket{n_1\,
  n_2 \, \ldots n_d}_a \,,\\
  J_{{\bf m},{\bf n}} &\equiv& \bra{ m_1\, m_2 \, \ldots m_d \,}
{\cal N}(1,2,\ldots d) \,\ket{n_1\,
  n_2 \, \ldots n_d}_a \,.
\end{eqnarray}
\end{subequations}
The subscript $a$ above stands for antisymmetrization.

For the diagonal matrix elements $H'_{{\bf m},{\bf m}}$, assume that
$\ket{{\bf m}} = \creat{m_1}\creat{m_2}
\cdots
\creat{m_d} \; \annil{n_d} \cdots \annil{n_2}\annil{n_1}
\ket{\Phi_0}$. Then one obtains, in leading order in the particle
number
\begin{equation}
  H'_{{\bf m},{\bf m}}=\sum_{i=1}^d(\varepsilon_{m_i}-\varepsilon_{n_i})
  \label{eq:epscbf}
\end{equation}
where
\begin{equation}
  \varepsilon_{m_i}-\varepsilon_{n_i}=
  \frac{\bra{\Phi_0\annil{m_i}\creat{n_i}}F^\dagger \hat H F
    \ket{\annil{n_i}\creat{m_i}\Phi_0}}
  {\bra{\Phi_0\annil{m_i}\creat{n_i}}F^\dagger F
    \ket{\annil{n_i}\creat{m_i}\Phi_0}}
\label{eq:epscbfexp}
\end{equation}
are the single particle energies of CBF theory which are the direct
generalizations of the Hartee-Fock single particle energies in the
case that correlations are omitted.

\subsubsection*{Acknlowledgments} 
We thank Panagiota Papakonstantinou for comments and suggestions on
earlier versions of this manuscript. We also thank Alexandros Gezerlis
for providing the code for the CEFT interaction. JW thanks the support
from College of Arts and Sciences, University at Buffalo -- SUNY.
\newpage
\bibliography{papers,Emrich}
\bibliographystyle{sn-mathphys}

\end{document}